\documentclass{article}
\usepackage{amssymb}
\usepackage{amsmath}
\usepackage{cprotect}
\usepackage{cite}
\usepackage{hyperref}
\usepackage{amsfonts}
\usepackage{comment}
\usepackage{soul}
\usepackage{tikz}
\usepackage{tikz-cd}
\usepackage{subcaption}
\usetikzlibrary{arrows,decorations.markings,decorations.pathreplacing}
\usepackage{bbm}

\input xy
\xyoption{all}
\usepackage{enumitem}

\usepackage{graphicx} % Required for inserting images

\newcommand{\be}{\begin{equation}}
	\newcommand{\ee}{\end{equation}}

\newcommand{\al}{\alpha}
\renewcommand{\d}{\delta}
\newcommand{\e}{\epsilon}

\newcommand{\g}{\gamma}

\newcommand{\m}{\mu}

\newcommand{\om}{\omega}

\newcommand{\id}{\mathbbm{1}}
\newcommand{\mcO}{\mathcal{O}}
\newcommand{\C}{\mathbb{C}}
\newcommand{\Z}{\mathbb{Z}}
\newcommand{\mcC}{\mathcal{C}}
\newcommand{\Hom}{\operatorname{Hom}}
\newcommand{\Rep}{\operatorname{Rep}}

\renewcommand{\H}{\mathcal{H}}
\newcommand{\hlf}{\frac{1}{2}}
\newcommand{\F}{\mathcal{F}}

\newcommand{\non}{\nonumber}

\newcommand{\rr}{\rightarrow}

\newcommand{\GL}{\operatorname{GL}}

\newcommand{\U}{\operatorname{U}}

\newcommand{\lp}{\left(}
\newcommand{\rp}{\right)}
\newcommand{\ls}{\left[}
\newcommand{\rs}{\right]}
\newcommand{\mcA}{\mathcal{A}}

\def\sqr#1#2{{\vcenter{\vbox{\hrule height.#2pt
            \hbox{\vrule width.#2pt height#1pt \kern#1pt
                  \vrule width.#2pt}\hrule height.#2pt}}}}

\def\square
 {\mathop{\mathchoice{\sqr{12}{15}}{\sqr{9}{12}}
{\sqr{6.3}{9}}{\sqr{4.5}{9}}}}

\def\sqra#1#2#3{{\vcenter{\vbox{\hrule height.#2pt
            \hbox{\vrule width.#2pt height#1pt \kern5pt %\kern#1pt
#3
                  \vrule width.#2pt}\hrule height.#2pt}}}}

\tikzset{
	partial ellipse/.style args={#1:#2:#3}{
		insert path={+ (#1:#3) arc (#1:#2:#3)}
	}
}

\numberwithin{equation}{section}

\numberwithin{table}{section}

\begin{document}

\begin{center}

{\large\bf Notes on gauging noninvertible symmetries, part 1: Multiplicity-free cases}

\vspace*{0.2in}

A.~Perez-Lona$^1$, D.~Robbins$^2$, E.~Sharpe$^1$, T.~Vandermeulen$^3$, X.~Yu$^1$

        \vspace*{0.1in}
        
        \begin{tabular}{cc}
                {\begin{tabular}{l}
                $^1$ Department of Physics MC 0435\\
                                850 West Campus Drive\\
                                Virginia Tech\\
                                Blacksburg, VA  24061 \end{tabular}}
                                 &
                {\begin{tabular}{l}
                $^2$ Department of Physics\\
                                University at Albany\\
                                Albany, NY 12222 \end{tabular}}
        \end{tabular}
        
                {\begin{tabular}{l}
                                $^3$ George P.~and Cynthia W.~Mitchell Institute\\
                                for Fundamental Physics and Astronomy\\
                                Texas A\&M University\\
                                College Station, TX 77843 \end{tabular}}
                        
        \vspace*{0.2in}

        {\tt aperezl@vt.edu},
        {\tt dgrobbins@albany.edu},
        {\tt ersharpe@vt.edu},
        {\tt tvand@tamu.edu},
        {\tt xingyangy@vt.edu}
        
\end{center}

In this paper we discuss gauging noninvertible zero-form symmetries in two dimensions.  We specialize to certain gaugeable cases, specifically, fusion categories of the form Rep$({\cal H})$ for
${\cal H}$ a suitable Hopf algebra (which includes the special case Rep$(G)$ for $G$ a finite group).  We also specialize to the case that the fusion category is multiplicity-free.  We discuss how to construct a modular-invariant partition function from a choice of Frobenius algebra structure on ${\cal H}^*$.  We discuss how ordinary $G$ orbifolds for finite groups $G$ are a special case of the construction, corresponding to the fusion category Vec$(G) = {\rm Rep}({\mathbb C}[G]^*)$.  For the cases Rep$(S_3)$, Rep$(D_4)$, and Rep$(Q_8)$, we construct the crossing kernels for general intertwiner maps.  We explicitly compute partition functions in the examples of Rep$(S_3)$, Rep$(D_4)$,
Rep$(Q_8)$, and Rep$({\cal H}_8)$, and discuss applications in $c=1$ CFTs.  We also discuss decomposition in the special case that the entire noninvertible symmetry group acts trivially.

\begin{flushleft}
    November 2023
\end{flushleft}

\newpage

\tableofcontents

\newpage

\section{Introduction}

Recently, there has been a great deal of interest in 
noninvertible symmetries, see e.g.~\cite{Cordova:2022ruw,Schafer-Nameki:2023jdn,Shao:2023gho} for recent reviews, and see also e.g.~\cite{Carqueville:2012dk,Carqueville:2017aoe,Carqueville:2018sld,Carqueville:2023aak,Carqueville:2023qrk} for related work in for example three-dimensional TFTs, which give conceptual insights into gaugeability of noninvertible symmetries in physical theories via the `sandwich' construction,
and also see e.g.~\cite{Fuchs:2004dz,Frohlich:2004ef,Frohlich:2009gb,Fuchs:2007tx}. 
The purpose of this paper is to explicitly gauge noninvertible global symmetries in some examples, to construct concrete expressions for partition functions and discuss applications such as decomposition, following up pioneering papers such as
\cite{Fuchs:2002cm,Bhardwaj:2017xup} which discuss basics of such gaugings.  

As has been discussed elsewhere in the literature, (finite) noninvertible symmetries are described by a fusion category.  We will review later how not every fusion category can be gauged (see e.g.~\cite{Kaidi:2023maf,Zhang:2023wlu,Cordova:2023bja,Antinucci:2023ezl} for other recent discussions).  Gaugeability imposes constraints on the fusion category, and to satisfy those constraints, we will restrict to fusion categories of the form
Rep$(G)$ (the category of representations of a group $G$), for finite $G$, and to generalizations 
Rep$({\cal H})$ where ${\cal H}$ will be a semisimple finite-dimensional Hopf algebra (which we will define later).
(See also \cite{Mulevicius:2023jhg} for a recent discussion of Rep(${\cal H}$) fusion categories.)
This will turn out to include ordinary orbifolds as special cases, for which the fusion category
is Vec$(G) = {\rm Rep}({\mathbb C}[G]^*)$, the category of $G$-graded vector spaces), for $G$ the
(finite) orbifold group, as we shall discuss.  (We emphasize that these are not the most general possible fusion categories arising in two-dimensional theories as describing global symmetries, as has also been remarked by e.g.~\cite{Bhardwaj:2017xup}.)

Now, to gauge the noninvertible symmetry, to construct a modular-invariant partition function, we will see that we need to specify additional structure, namely a (special, symmetric) Frobenius algebra.  Many choices are possible; we will describe pertinent choices in terms of the regular representation of $G$ or ${\cal H}$.  In practice, for categories of representations of finite groups $G$,
the Frobenius algebra is defined on the dual ${\mathbb C}[G]^*$ or ${\cal H}^*$
(where we have used the fact that Rep$(G)$ = Rep$({\mathbb C}[G])$, for ${\mathbb C}[G]$ the group algebra of $G$, to describe Rep$(G)$ as a special
case of Rep$({\cal H})$).

One can gauge along any number of Frobenius algebras; we emphasize that when we speak of gauging the (entire) noninvertible symmetry, we mean that we are gauging the Frobenius algebra associated with the regular representation.  Further generalizations may be possible; we do not address the question of
constructing Frobenius algebras from more general fusion categories, or their gauging.

Intuitively, the reader might compare to the case of ordinary group orbifolds, and ask why the additional step of specifying a Frobenius algebra is required.  After all,
standard orbifold constructions only require specifying a group and its action.
Our intuition is that in ordinary orbifolds there exists a canonical choice of Frobenius algebra
(as we shall discuss later in section~\ref{sect:genl-algebra}), which one works with
implicitly.  Similarly, when gauging a (gaugeable) noninvertible symmetry, there exists a canonical choice of Frobenius algebra structure (constructed on the regular
representation of the group $G$ or Hopf algebra ${\cal H}$).

The construction of partition functions we describe here is closely related to that
described in \cite{Fuchs:2002cm}, which formulated rational conformal field theory in terms of a symmetric special Frobenius algebra.

In this paper, we will focus on gauging noninvertible symmetries in multiplicity-free cases, meaning that spaces of junction operators are one-dimensional, as we shall explain later.  We will describe gauging in more general cases in our
subsequent work \cite{toappear}.

We begin in section~\ref{sec:general} by describing the basic principles underlying the gauging of noninvertible symmetries.  We define partial traces for noninvertible symmetry analogues of orbifolds, the building blocks of partition functions, and observe in an example that not every noninvertible symmetry can be gauged, as it is not always possible to construct a modular-invariant partition function.  We discuss necessary conditions for gauging, and restrict to a class of gaugeable noninvertible symmetries, defined by fusion categories of the form
Rep(${\cal H}$) for ${\cal H}$ a finite-dimensional semisimple Hopf algebra.  (This includes as special cases both Rep$(G)$ for $G$ a finite group,
as well as ordinary orbifolds, which are described by the fusion category Vec$(G) = {\rm Rep}( {\mathbb C}[G]^*)$.)
For simplicity, in this paper we restrict to `multiplicity-free' cases, which we explain.
In order to perform a gauging, in order to construct a modular invariant partition function, one must specify a special symmetric Frobenius algebra (constructed from the dual of the Hopf algebra ${\cal H}^*$, as we discuss).
We give formal expressions for partition functions (at genus one and higher), and check modular invariance.
We also discuss in detail how ordinary orbifolds arise as special cases.

In section~\ref{sect:examples} we compute partition functions for Frobenius algebras arising in several examples,
namely Rep$(S_3)$, Rep$(D_4)$, Rep$(Q_8)$, and Rep$({\cal H}_8)$.  For the first three, we give first-principles computations, deriving the general form of associators and crossing kernels / F-symbols for general intertwiner maps, expressions for modular transformations of partial traces, and compute partition functions for every relevant choice of Frobenius algebra, checking modular invariance in each case.  For the last example,
Rep$({\cal H}_8)$, we use standard results for crossing kernels (for fixed intertwiner values), and only compute the partition function for one Frobenius algebra (corresponding to the regular representation), and not more general cases.  We also briefly outline analogues of discrete torsion that can arise in these theories.

In section~\ref{sect:apps} we discuss applications. We focus on $c=1$ theories which enjoy a rich structure of noninvertible symmetries. We discuss how to gauge these noninvertible symmetries on certain points of the orbifold branch. Specifically, we construct noninvertible duality defects arising from gauging noninvertible symmetries in $c=1$ theories.

Finally, in section~\ref{sect:decomp} we discuss decomposition.  Recall decomposition is the statement that a $d$-dimensional local quantum field theory with a global $(d-1)$-form symmetry is equivalent to a disjoint union of theories.  In two dimensions, decomposition often arises in gauge theories in which a subgroup of the gauge group acts trivially.  In this section we discuss what it means for a noninvertible symmetry to act trivially, then make a proposal for the form of decomposition in gauged noninvertible symmetries, in the special case that the entire noninvertible symmetry acts trivially.  We check this conjecture in details for Rep$(S_3)$, Rep$(D_4)$, and Rep$(Q_8)$.

As both Hopf and Frobenius algebras play an important role in this paper, and they may be obscure to the reader, in appendix~\ref{app:algebras} we summarize the definitions of both.  In appendix~\ref{app:ty} we review modular transformations in ${\mathbb Z}_2 \times {\mathbb Z}_2$ Tambara-Yamagami examples, which include
Rep$(D_4)$, Rep$(Q_8)$, and Rep$({\cal H}_8)$, albeit in each case for specific choices of intertwiners.
In appendix~\ref{app:disjoint-union} we briefly discuss disjoint unions of spaces as simple playgrounds for both $G$ orbifolds and Rep$(G)$ quantum symmetries.

Note added: While finalizing this work, \cite{CLS23} appeared, which also discusses  gauging noninvertible symmetries in general as well as the specific example of gauging Rep$(H_8)$ for $c=1$ CFTs. Our work is complementary to theirs: we provide a first-principle derivation starting from intertwiners, discuss further examples such as Rep$(S_3)$ and Rep$(Q_8)$. We also discuss examples where noninvertible self-duality defects can be built via half-gauging a Frobenius subalgebra of a categorical symmetry, compared to the case in \cite{CLS23} where a half-gauging is performed for the full categorical symmetry, and we also discuss decomposition arising in gauging trivially-acting noninvertible symmetries.

As this paper was nearing publication, we were informed that related results will appear in \cite{Wang:2023}.

To assist the reader, we summarize below the notation we use in this paper:
\begin{itemize}
    \item ${\cal A}$ denotes algebra objects and Frobenius algebras.
    \item ${\cal H}$ denotes a Hopf algebra.
    \item $\mu, \Delta$ denote multiplication, comultiplication, respectively.
    \item $u, u^o$ denote unit, counit, respectively.
    \item $\epsilon, \gamma$ denote evaluation, coevaluation, respectively.
    \item $\tilde{K}$ and $F$ denote associators/crossing kernels
    \item $\alpha$ denotes an associator.
    \item $L, M, N$ denote generic lines / simple objects in any fusion category,
    \item $i, j$ denote junction operators.
    \item $R$ denotes a generic irreducible representation of a group.
     \item For all groups, $1$ denotes the trivial irreducible representation.  (Some references use $0$ instead.)
    \item $e_{Ri}$ (for various integers $i$) denotes elements of a basis of a vector space on which the representation $R$ acts.
    \item $a, b, c, m$ denotes irreducible representations of $D_4$, $Q_8$, ${\cal H}_8$. 
    \item $\beta$ denotes coefficients in intertwiners.  
    \item $\phi$ denotes an intertwiner map.
    \item $G$ denotes a group.
    \item $H$ denotes a subgroup (not necessarily normal).
    (Note $H \neq {\cal H}$, the latter of which denotes a Hopf algebra.)
    \item $K$ denotes a normal subgroup of a group.
    \item $I$ denotes an ideal or coideal of a (Hopf) algebra.
    \item $\tau$ denotes either (depending upon context) the modular parameter, or a parameter specifying a Tambara-Yamagami category.
\end{itemize}

\section{Gauging: general principles}\label{sec:general}

In this section we will discuss general aspects of gauging noninvertible zero-form symmetries.
In ordinary orbifolds, much insight can be gleaned from the study of the genus-one partition function, which for
an orbifold by a group $G$, has the form
\begin{equation} \label{eq:Z:ord-orb}
    Z \: = \: \frac{1}{|G|} \sum_{g,h} Z_{g,h},
\end{equation}
where we will refer to the $Z_{g,h}$ 
as partial traces.  We begin with a discussion of how partial traces are defined for noninvertible zero-form symmetries
in subsection~\ref{sect:defn:partialtrace},
and then discuss modular transformations of those partial traces in subsection~\ref{sect:genl:modulartrans}.  Now, in general, for a random fusion category, there may not exist a modular-invariant combination of partial traces; no partition function may exist, which we discuss in subsection~\ref{sect:gaugeability}.
This has been previously discussed in the literature, see e.g.~\cite{Fuchs:2002cm}. 
A solution is to restrict to fusion categories from which one can derive
a Frobenius algebra, and then that Frobenius algebra is used to construct partition functions, as we shall describe
in subsections~\ref{sect:genl-algebra}, \ref{sect:genl-genus1}.
Briefly, given a Frobenius algebra ${\cal A}$, the genus-one partition function~(\ref{eq:Z:ord-orb}) generalizes to
an expression 
\begin{equation}  \label{eq:genus1Z}
    Z \: = \: Z_{ {\cal A}, {\cal A}} \: = \: \sum_{L_1, L_2, L_3} \mu_{L_1, L_2}^{L_3} \Delta_{L_3}^{L_2,L_1} Z_{L_1,L_2}^{L_3}.
\end{equation}
(The $\mu_{L_1,L_2}^{L_3}$, $\Delta_{L_3}^{L_2,L_1}$ are derived from the Frobenius algebra ${\cal A}$; the
partial traces $Z_{L_1,L_2}^{L_3}$ can be constructed solely from the fusion category.)
We demonstrate that the resulting genus one partition functions~(\ref{eq:genus1Z}) are modular-invariant in
subsection~\ref{sect:modinv}, and go on to construct higher-genus partition functions in
subsection~\ref{ssec:genus2}.  In subsection~\ref{sect:formal-state-spaces} we make some formal observations regarding state spaces, and argue formally in subsection~\ref{sect:formal-proj} that the partition function encodes a projector, just as happens in ordinary orbifolds.  In subsection~\ref{sect:ord-orbifold} we verify that all of these formal considerations correctly specialize to ordinary orbifolds by finite groups.

\subsection{Fusion categories, associators, and crossing kernels}

Before defining partial traces and orbifold partition functions, we begin with a brief
review of some terminology and notation. Our main source for definitions relevant to fusion categories is \cite{eno}. 
Noninvertible symmetries generalizing finite groups 
are usually\footnote{
For a slight generalization, see e.g.~\cite{JF22} for a discussion on the role of \textit{multi}-fusion categories as $2$d symmetry categories.
} expected to be described by mathematical structures known as \textit{fusion categories}. Technically, fusion categories are $\Bbbk$-linear semisimple rigid tensor categories with a finite number of isomorphism classes of simple objects. Unpacking this definition, this means these are categories such that 
\begin{enumerate}
\item (linear) the morphisms between any two objects form a $\Bbbk$-vector space, 
\item (semisimple) all objects are isomorphic to a finite direct sum of distinguished objects called simple objects, 
\item (rigid) every object has a corresponding dual object, 
\item (tensor) there is a notion of product between any two objects that distributes over direct sums, in analogy to the tensor product of vector spaces, and this includes an identity simple object $1=L_0$, and 
\item that the simple objects define a finite number $n\in\mathbb{N}$ of isomorphism classes. 
\end{enumerate}
By definition, simple objects $\{L_i\}_{i\in\{0,1,\cdots,n-1\}}$ (representing distinct isomorphism classes) satisfy the property that
\begin{equation}
    \text{dim}_{\Bbbk}(\text{Hom}(L_i,L_j))=\delta_{i,j}.
\end{equation}
Throughout this article, we will exclusively consider the case $\Bbbk=\C$.

Physically, these axioms are justified by the interpretation that the objects in the category correspond to topological operators that can be inserted along codimension $1$ submanifolds, thus creating \textit{defects} in the theory. In particular, the tensor product is interpreted as the fusion of defects.  The morphisms will live at junctions between lines.

An important piece of the structure that comes with a consistent definition of a tensor product and that will play an important role in this discussion is the \textit{associator} \cite{Kel82}. The associator $\alpha$ of the tensor product $\otimes$ of a fusion category $\mathcal{C}$ is a natural collection of isomorphisms
\begin{equation}
    \alpha_{x,y,z}: (x\otimes y)\otimes z \xrightarrow{\cong} x\otimes (y \otimes z)
\end{equation}
for $x,y,z\in\text{ob}(\mathcal{C})$ objects of the fusion category. In simple terms, the existence of an associator is the statement that the tensor product $\otimes$ is associative \textit{up to} isomorphism. Diagrammatically, the associator is simply a natural isomorphism:
\begin{equation}
\begin{tikzcd}
\mathcal{C}\times \mathcal{C}\times \mathcal{C} \arrow[dd, "\otimes\times 1_{\mathcal{C}}"'] \arrow[rr, "1_{\mathcal{C}}\times\otimes"] &  & \mathcal{C}\times \mathcal{C} \arrow[dd, "\otimes"] \\
                                                                                                                                        &  &                                                     \\
\mathcal{C}\times \mathcal{C} \arrow[rruu, "\alpha", Rightarrow] \arrow[rruu, "\cong"', Rightarrow] \arrow[rr, "\otimes"']              &  & \mathcal{C}                                        
\end{tikzcd}
\end{equation}
Associators are required to satisfy an identity known as the \textit{pentagon identity}. In components, for arbitrary objects $a,b,c,d\in\text{ob}(\mathcal{C})$, this is the commutative diagram:
\begin{equation}  \label{eq:associator-pentagon}
    \begin{tikzcd}
     & (a\otimes b)\otimes (c\otimes d) \arrow[rd, "{\alpha_{a,b,c\otimes d}}"] &  
     \\
((a\otimes b)\otimes c)\otimes d \arrow[ru, "{\alpha_{a\otimes b,c,d}}"] \arrow[d, "{\alpha_{a,b,c}\otimes \text{Id}_d}"'] 
&  & a\otimes (b\otimes (c\otimes d))  
\\
(a\otimes (b\otimes c))\otimes d \arrow[rr, "{\alpha_{a,b\otimes c,d}}"']   
& & a\otimes ((b\otimes c)\otimes d) \arrow[u, "{\text{Id}_a\otimes\alpha_{b,c,d}}"']
\end{tikzcd}
\end{equation}

Associators are defined for general monoidal categories. However, for fusion categories, this information can be specified by a generating set called $F$-symbols. The generating set is obtained by looking at the \textit{hom-spaces} of simple objects. Let $\{L_i\}_{i\in I}$ be the simple objects. Then the associator
\begin{equation}
    \alpha_{i,j,k}:(L_i\otimes L_j)\otimes L_k\to L_i\otimes (L_j\otimes L_k)
\end{equation}
becomes a $\C$-linear map of hom-spaces after applying the $\text{Hom}(-,L_l)$ functor:
\begin{equation}
    F_{ijk}^l:=\text{Hom}(-,L_l)(\alpha_{i,j,k}):\text{Hom}((L_i\otimes L_j)\otimes L_k,L_l)\to \text{Hom}(L_i\otimes (L_j\otimes L_k),L_l) 
\end{equation}
which we call the $F$-matrices. By fixing a basis $\{\lambda_{ij}^{k\,\alpha}\}$ for $\text{Hom}(L_i\otimes L_j,L_k)$, we can obtain the elements of the $F$-matrix. To do this, first note 
\begin{eqnarray}
    \text{Hom}((L_i\otimes L_j)\otimes L_k,L_l) = \bigoplus_{p}\text{Hom}(L_i\otimes L_j,L_p)\otimes\text{Hom}(L_p\otimes L_k,L_l)\\
    \text{Hom}(L_i\otimes (L_j\otimes L_k),L_l) = \bigoplus_q\text{Hom}(L_j\otimes L_k,L_q)\otimes\text{Hom}(L_i\otimes L_q,L_l).
\end{eqnarray}
This refines the $F$-matrices to linear maps of tensors of hom-spaces of simple objects as
\begin{equation}
    (F_{ijk}^l)_p^q:\text{Hom}(L_i\otimes L_j,L_p)\otimes\text{Hom}(L_p\otimes L_k,L_l)\to \text{Hom}(L_j\otimes L_k,L_q)\otimes\text{Hom}(L_i\otimes L_q,L_l).
\end{equation}
We now use the fixed basis to get matrix elements as
\begin{equation}
    (F_{ijk}^l)_{p}^{q}(\lambda_{ij}^{p\,\beta}\otimes\lambda_{pk}^{l\,\alpha})=(F_{ijk}^l)_{\alpha p\beta}^{\gamma q\delta} \ \lambda_{jk}^{q\,\delta}\otimes\lambda_{iq}^{l\,\gamma}
\end{equation}
for $(F_{ijk}^l)_{\alpha p\beta}^{\gamma q\delta}\in\C$. Note that if in particular the fusion category is \textit{multiplicity-free}, meaning that
\begin{equation}
    \text{dim}_{\C}(\text{Hom}(L_i\otimes L_j,L_k))\leq 1,
\end{equation}
then $(F_{ijk}^l)_p^q$ in the chosen basis is simply a scalar. Then these coefficients (for which the Greek letter indices are redundant) are often referred to as $6j$-symbols.

A consequence of the definitions above and the pentagon identity~(\ref{eq:associator-pentagon}) of associators is that the $F$ symbols obey \cite[Lemma 3.4]{Usher19}
    \begin{equation}
      \sum_{\epsilon} (F_{mkl}^p)_{\beta n \chi}^{\delta q \epsilon} (F_{ijq}^p)_{\alpha m \epsilon}^{\delta s \gamma} = \sum_{t,\eta,\phi\kappa} (F_{ijk}^n)_{\alpha m \beta}^{\eta t \phi} (F_{itl}^p)_{\phi n\chi}^{\kappa s\gamma} (F_{jkl}^s)_{\eta t \kappa}^{\delta q\phi},
    \end{equation}
    which in the multiplicity-free case reduces to the more familiar identity for $6j$-symbols:
    \begin{equation} \label{eq:6jreln}
       (F_{mkl}^p)_{n}^{q} (F_{ijq}^p)_{m}^{s} = \sum_{t} (F_{ijk}^n)_{m}^{t} (F_{itl}^p)_{n}^{s} (F_{jkl}^s)_{t}^{q}.
    \end{equation}

In the remainder of this paper, we will focus on multiplicity-free fusion categories.
We will return to the general case in our subsequent work \cite{toappear}.

A fusion category, by virtue of having duals, comes with evaluation maps $\epsilon_i\in\text{Hom}(\overline{L}_i\otimes L_i,L_0)$ and $\overline{\epsilon}_i\in\text{Hom}(L_i\otimes\overline{L}_i,L_0)$, and co-evaluation maps $\gamma_i\in\text{Hom}(L_0,L_i\otimes\overline{L}_i)$ and $\overline{\gamma}_i\in\text{Hom}(L_0,\overline{L}_i\otimes L_i)$, for each simple object $L_i$, and $L_0$ the monoidal unit.  Here $\overline{L}$ mathematically denotes the \textit{dual} of $L$ (which exist in any fusion category by definition), and physically represents $L$ with the opposite orientation, where
\begin{equation}
    \overline{L_1 \oplus L_2} \: = \: \overline{L}_1 \oplus \overline{L}_2,
    \: \: \:
    \overline{ L_1 \otimes L_2} \: = \: \overline{L_2} \otimes \overline{L}_1.
\end{equation}
The evaluation and coevaluation homomorphisms should satisfy relations which ensure that a line can be ``unfolded'' (see~\cite{Bhardwaj:2017xup}), specifically
\begin{equation}
    (\overline{\epsilon}_i\otimes 1_{L_i})\circ\alpha_{i,\overline{i},i}^{-1}\circ(1_{L_i}\otimes\overline{\gamma}_i(1))=1_{L_i},\qquad (1_{L_i}\otimes\epsilon_i)\circ\alpha_{i,\overline{i},i}\circ(\gamma_i(1)\otimes 1_{L_i})=1_{L_i}.
\end{equation}
In many cases (such as when the fusion category comes from representations), as we will see, a choice of the basis vectors $\lambda_{ij}^k$ induces evaluation maps $\epsilon_i$ and $\overline{\epsilon}_i$, and then the relations above determine the forms of $\gamma_i$ and $\overline{\gamma}_i$.

Now we need to discuss some of our conventions for diagrams, which stand in for correlation functions in our physical 2D theories.  Topological line operators are labeled by objects of the fusion category, and include an orientation indicated by an arrow on the line.  Lines meet at junctions which should have an insertion of a point operator.  For tri-valent junctions of simple lines we'll include a counter-clockwise blue arrow which will give an ordering to the three lines, as shown below.
\begin{equation}
    \begin{tikzpicture}
        \draw[very thick, ->] (1.25,2)--(1.25,3);
        \draw[very thick] (1.25,3)--(1.25,4);
        \draw[very thick] (0.6,0.96)--(1.25,2);
        \draw[very thick,->] (0,0)--(0.6,0.96);
        \draw[very thick] (1.25,2)--(1.9,0.96);
        \draw[very thick,->] (2.5,0)--(1.9,0.96); \draw[thick,color=blue,->,opacity=0.5] (1.25,2) [partial ellipse=-120:90:0.2cm and 0.2cm];
        \node at (2.7,0.3) {$L_j$};
        \node at (0.45,0.2) {$L_i$};
        \node at (1.6,3) {$L_k$};
    \end{tikzpicture}
\end{equation}
To such a junction, with the lines in the orientations shown, we will associate the vector space 
$$\text{Hom}(L_i\otimes L_j,L_k),$$ 
and if we have chosen basis vectors $\lambda_{ij}^k$ (since we are restricting to the multiplicity-free case, we can drop the index $\alpha$), then unless otherwise noted, that is the vector placed at such a junction in a correlation function.  If any of the lines have their orientation reversed, then we will replace that line by its dual via insertion of an evaluation or coevaluation map, and then associate the fusion $\text{Hom}$ space, so for instance if the $L_j$ line above was outgoing rather than incoming, we would insert a coevaluation map as shown below, and then associate the space $\text{Hom}(L_i\otimes\overline{L}_j,L_k)$ to the resulting tri-valent junction.

\begin{equation}
    \begin{tikzpicture}
        % Left picture:
        \draw[very thick,->] (1.25,0)--(1.25,1);
        \draw[very thick] (1.25,1)--(1.25,2);
        \draw[very thick,->] (1.25,2)--(0.6,3.04);
        \draw[very thick] (0.6,3.04)--(0,4);
        \draw[very thick,->] (1.25,2)--(1.9,3.04);
        \draw[very thick] (1.9,3.04)--(2.5,4);
        %\node at (1.25,2.2) {$\times$};
        \draw[thick,color=blue,->,opacity=0.5] (1.25,2) [partial ellipse=-90:110:0.2cm and 0.2cm];
        \node at (2.7,3.7) {$L_j$};
        \node at (0.45,3.7) {$L_k$};
        \node at (1.6,1) {$L_i$};
        \node at (4.0,2) {$ = $};
        %
        % Right picture:   x + 4.5
        \draw[very thick,->] (6.25,0)--(6.25,1);
        \draw[very thick] (6.25,1)--(6.25,2);
        \draw[very thick,->] (6.25,2)--(5.6,3.04);
        \draw[very thick] (5.6,3.04)--(5.0,4);
        \draw[very thick,->] (6.9,0.96)--(6.575,1.48);
        \draw[very thick] (6.575,1.48)--(6.25,2);
        \draw[very thick,->] (6.9,0.96)--(7.55,2);
        \draw[very thick] (7.55,2)--(8.8,4);
        %\node at (6.05,1.68) {$\times$};
        \draw[thick,color=blue,->,opacity=0.5] (6.25,2) [partial ellipse=-90:110:0.2cm and 0.2cm];
        \node at (9,3.7) {$L_j$};
        \node at (6.8,1.7) {$\overline{L}_j$};
        \node at (5.45,3.7) {$L_k$};
        \node at (5.9,1) {$L_i$};
        \node at (7,0.75) {$\overline{\gamma}_j$};
    \end{tikzpicture}
\end{equation}

Finally, using the evaluation or coevaluation maps we can construct isomorphisms between the fusion homomorphisms and co-fusion homomorphisms.  Explicitly, define an isomorphism $\phi_{ijk}:\text{Hom}(L_i\otimes L_j,L_k)\rightarrow\text{Hom}(L_i,L_k\otimes\overline{L}_j)$ by its action on $v\in\text{Hom}(L_i\otimes L_j,L_k)$,
\begin{equation}
    \phi_{ijk}(v)=(v\otimes 1_{\overline{L}_j})\circ\alpha_{i,j,\overline{j}}^{-1}\circ(1_{L_i}\otimes\gamma_j(1)),
\end{equation}
with inverse
\begin{equation}
    \phi_{ijk}^{-1}(u)=(1_{L_k}\otimes\epsilon_j)\circ\alpha_{k,\overline{j},j}\circ(u\otimes 1_{L_j}).
\end{equation}
In particular, to the junction above we can alternatively map the vector space to $\text{Hom}(L_i,L_k\otimes\overline{L}_j)$.  If we have chosen basis vectors $\lambda_{ij}^k$ for $\text{Hom}(L_i\otimes L_j,L_k)$, then we get an associated choice of basis vectors $\delta_i^{jk}$ for $\text{Hom}(L_i,L_j\otimes L_k)$ by
\begin{equation}
    \delta_i^{jk}=\phi_{i\overline{k}j}(\lambda_{i\overline{k}}^j)=(\lambda_{i\overline{k}}^j\otimes 1_{L_k})\circ\alpha_{i,\overline{k},k}^{-1}\circ(1_{L_i}\otimes\gamma_{\overline{k}}(1)).
\end{equation}

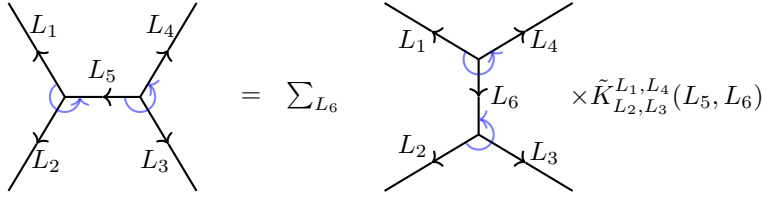
\begin{figure}
\centering
\begin{tikzpicture} % (2.5,2.5) box for each diagram
        \draw[thick,->] (1.75,1.25)--(1.25,1.25);
        \draw[thick] (1.25,1.25)--(0.75,1.25);
        \draw[thick,->] (0.75,1.25)--(0.375,1.875);
        \draw[thick] (0.375,1.875)--(0,2.5);
        \draw[thick,->] (0.75,1.25)--(0.375,0.625);
        \draw[thick] (0.375,0.625)--(0,0);
        \draw[thick,->] (1.75,1.25)--(2.125,1.875);
        \draw[thick] (2.125,1.875)--(2.5,2.5);
        \draw[thick,->] (1.75,1.25)--(2.125,0.625);
        \draw[thick] (2.125,0.625)--(2.5,0);
        \node at (1.25,1.6) {$L_5$};
        \node at (0.475,2.2) {$L_1$};
        \node at (2.025,2.2) {$L_4$};
        \node at (0.5,0.4) {$L_2$};
        \node at (1.95,0.4) {$L_3$};
        \draw[thick,color=blue,->,opacity=0.5] (0.75,1.25) [partial ellipse=-235:0:0.2cm and 0.2cm];
		\draw[thick,color=blue,->,opacity=0.5] (1.75,1.25) [partial ellipse=-180:60:0.2cm and 0.2cm];
        \node at (3.75,1.25) {$ = \: \: \: \sum_{L_6}$};
        %
        % Right-hand side:  (shift x by 5)
        \draw[thick,->] (6.25,1.75)--(6.25,1.25);
        \draw[thick] (6.25,1.25)--(6.25,0.75);
        \draw[thick,->] (6.25,0.75)--(6.875,0.375);
        \draw[thick] (6.875,0.375)--(7.5,0);
        \draw[thick,->] (6.25,0.75)--(5.625,0.375);
        \draw[thick] (5.625,0.375)--(5,0);
        \draw[thick,->] (6.25,1.75)--(6.875,2.125);
        \draw[thick] (6.875,2.125)--(7.5,2.5);
        \draw[thick,->] (6.25,1.75)--(5.625,2.125);
        \draw[thick] (5.625,2.125)--(5,2.5);
        \node at (6.6,1.25) {$L_6$};
        \node at (5.375,2.0) {$L_1$};
        \node at (5.375,0.6) {$L_2$};
        \node at (7.125,2.0) {$L_4$};
        \node at (7.125,0.5) {$L_3$};
        \draw[thick,color=blue,->,opacity=0.5] (6.25,1.75) [partial ellipse=-215:35:0.2cm and 0.2cm];
		\draw[thick,color=blue,->,opacity=0.5] (6.25,0.75) [partial ellipse=-145:90:0.2cm and 0.2cm];
        \node at (8.75,1.25) { $\times \tilde{K}_{L_2, L_3}^{L_1, L_4}(L_5, L_6)$};
\end{tikzpicture}
\caption{\label{fig:crossing-kernel} Defining property of the crossing kernel $\tilde{K}$,
taken from \cite[figure 13]{Chang:2018iay}. The sum is over simple objects.}
\end{figure}

Our notation for fusion categories primarily
follows \cite{Chang:2018iay}.  One particular object which will play a crucial role for
us is the crossing kernel $\tilde{K}$, which is another notation for $F$.  In the multiplicity-free case,
they are related by
\begin{equation}
    \tilde{K}^{i,\overline{m}}_{j,k}(p,q) \: = \: 
    (F^{m}_{ijk})_p^q 
\end{equation}
Graphically, the crossing kernel is defined\footnote{
The careful reader will observe a slight issue of notation.  Ordinarily, a linear map $f: a \rightarrow b$ might be said to act by multiplication by a number $c$ as $f: a \mapsto b = ac$.  
However, the conventions of \cite{Chang:2018iay} instead define $c$ by $a = bc$, as their analysis is primarily graphical.
In any event, we will follow their conventions in this paper.
} by the relation shown in figure~\ref{fig:crossing-kernel}. 

As previously derived in the language of $F$ symbols,
the crossing kernels obey the pentagon identity~(\ref{eq:6jreln}),
which for later use we rewrite here in the notation of crossing kernels $\tilde{K}$,
to explicitly\footnote{
To make the relation completely clear, we include the following table to convert to the indices of \cite{Chang:2018iay}:
\begin{center}
    \begin{tabular}{cc|cc|cc}
    Here & \cite{Chang:2018iay} & Here & \cite{Chang:2018iay} & Here & \cite{Chang:2018iay} \\ \hline
    $i$ & $j1$ &  $m$ & $j$ &  $k$ & $j4$ \\
    $j$ & $j2$ & $n$ & $\overline{j3}$ & $\ell$ & $k4$ \\
    $q$ & $k2$ & $p$ & $\overline{k1}$ & $s$ & $\overline{k3}$ \\
    & & & & $t$ & $j'$
    \end{tabular}
\end{center}
}
match \cite[equ'n (2.9)]{Chang:2018iay}:
\begin{equation}
    \tilde{K}^{i,\overline{p}}_{j,q}(m,s) \,
    \tilde{K}^{m,\overline{p}}_{k,l}(n,q)
    \: = \:
    \sum_{t} \tilde{K}^{j,\overline{s}}_{k,l}(t,q) \,
    \tilde{K}^{i,\overline{p}}_{t,l}(n,s) \,
    \tilde{K}^{i,\overline{n}}_{j,k}(m,t),
\end{equation}
The admissible crossing kernels are precisely those that obey the pentagon identity above.

\subsection{Definition of partial traces for noninvertible symmetries}
\label{sect:defn:partialtrace}

As reviewed earlier, in an ordinary orbifold by a finite group $G$,
the
genus one partition function has the form
\begin{equation}
    Z([X/G]) \: = \: \frac{1}{|G|} \sum_{gh=hg} Z_{g,h},
\end{equation}
where $Z_{g,h}$ represents the `partial trace,' 
schematically,
\begin{equation}
    {\scriptstyle g} \square_h \, ,
\end{equation}
the contribution to the $T^2$ partition function
from a worldsheet with a pair of branch cuts defined by $g, h \in G$.

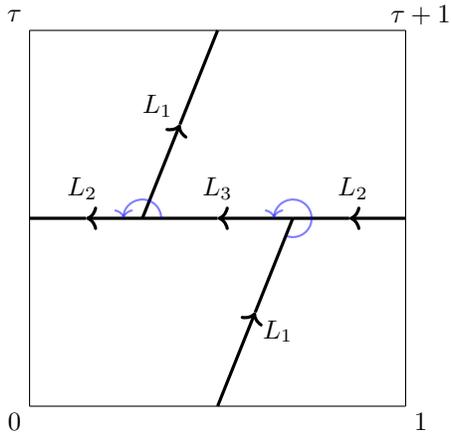
\begin{figure}
		\centering
		\begin{tikzpicture}
			\draw[thin] (0,0)--(5,0);
			\draw[thin] (5,0)--(5,5);
			\draw[thin] (0,0)--(0,5);
			\draw[thin] (0,5)--(5,5);
			\draw[very thick,->] (2.5,0)--(3,1.25);
			\draw[very thick] (3,1.25)--(3.5,2.5);
			\draw[very thick,->] (1.5,2.5)--(2,3.75);
			\draw[very thick] (2,3.75)--(2.5,5);
			\draw[very thick,->] (5,2.5)--(4.25,2.5);
			\draw[very thick] (4.25,2.5)--(3.5,2.5);
			\draw[very thick,->] (3.5,2.5)--(2.5,2.5);
			\draw[very thick] (2.5,2.5)--(1.5,2.5);
			\draw[very thick,->] (1.5,2.5)--(0.75,2.5);
			\draw[very thick] (0.75,2.5)--(0,2.5);
			\node at (3.3,1) {$L_1$};
			\node at (1.7,4) {$L_1$};
			\node at (0.7,2.9) {$L_2$};
			\node at (4.3,2.9) {$L_2$};
			\node at (2.5,2.9) {$L_3$};
			\node at (-0.2,5.2) {$\tau$};
            \node at (5.2,-0.2) {$1$};
            \node at (5.2,5.2) {$\tau+1$};
            \node at (-0.2,-0.2) {$0$};

            \draw[thick,color=blue,->,opacity=0.5] (3.5,2.5) [partial ellipse=-110:180:0.25cm and 0.25cm];
		      \draw[thick,color=blue,->,opacity=0.5] (1.5,2.5) [partial ellipse=0:180:0.25cm and 0.25cm];
		\end{tikzpicture}
  \caption{\label{fig:noninv-partial-trace}
  Diagrammatic definition of the partial trace $Z^{L_3}_{L_1,L_2}$, the noninvertible analogue of the partial trace $Z_{g,h}$ appearing in genus one orbifold partition functions $[X/G]$ for $G$ an ordinary group.  Our conventions for partial traces essentially follow \cite[appendix A]{Robbins:2019zdb}.
  }
  \end{figure}

Now, in the noninvertible case, to make sense of the partial traces $Z_{g,h}$, one must work harder.  If we proceed naively and consider analogues of $Z_{g,h}$ defined by pairs of lines $L_1$, $L_2$, one problem we quickly
encounter is that the 4-point junction at which a pair of lines intersections is not uniquely
defined in the noninvertible theory, because $L_1 \otimes L_2$ can receive contributions from a linear combination of several lines.  To make the contribution well-defined, we must resolve the
4-valent junction into a pair of 3-valent junctions, and specify the line joining the
two three-point junctions.  Instead of $Z_{g,h}$, we have $Z_{L_1,L_2}^{L_3}$, where $L_1$ and $L_2$ are simple lines, and $L_3$ is a simple line that appears in the fusion $L_1\otimes L_2$, as is illustrated
in Figure~\ref{fig:noninv-partial-trace}.  In this paper, we will for the most part restrict to the so-called multiplicity-free case, where each simple line in the fusion product of two simple lines appears at most once.

In principle, the diagram illustrated in Figure~\ref{fig:noninv-partial-trace} does not yet specify a correlation function, since one must also specify a choice of junction operators at each of the triple intersections above.
For example, the junction operators at the vertex on the right in Figure~\ref{fig:noninv-partial-trace} are elements of Hom$(L_1 \otimes L_2, L_3)$.  (Later in this paper, we will specialize to fusion categories of the form Rep$(G)$,
and for such, these Hom spaces are (projections of) the space of intertwiners.)  In any event, in this paper we specialize to multiplicity-free cases, for which those Hom spaces have dimension either zero or one.  If the dimension is zero, the vertex does not exist at all; if the dimension is one, then up to rescaling there is a single operator that can be placed at the vertex.  Thus, in this paper, the matter of choosing junction operators is moot.  We will consider more general cases in our followup paper \cite{toappear}.

In the special case that the noninvertible symmetry group is an actual group $G$,
in the noninvertible partial trace $Z_{L_g,L_h}^{L_k}$, for $g, h, k \in G$,
the value of $L_k$ (the intermediate line) is uniquely determined by (a commuting pair) $g$ and $h$, and is given by $k = gh = hg$.  In this case, 
\begin{equation}
    Z_{g,h} \: = \: Z_{L_g,L_h}^{L_{gh}},
\end{equation}
where the $Z_{g,h}$ on the left is the partial trace of the ordinary orbifold,
and the $Z_{L_g,L_h}^{L_{gh}}$ on the right is its analogue in the noninvertible orbifold.

Returning to noninvertible cases,
it remains to find modular-invariant combinations of the partial traces $Z_{L_1,L_2}^{L_3}$.
In subsection~\ref{sect:genl:modulartrans}, we shall discuss modular transformations of these partial traces.
Later in section~\ref{sect:genl-algebra}
we will discuss systematic computational tools for constructing modular-invariant combinations, to form
physical partition functions, which will turn out to correspond to (special symmetric) Frobenius algebras.

\subsection{Modular transformations}
\label{sect:genl:modulartrans}

In this section we will argue that
\begin{equation} \label{eq:modtrans:T}
Z^{L_3}_{L_1, L_2}(\tau + 1, \overline{\tau} + 1) \: = \:
\sum_{L_4} \tilde{K}^{\overline{L}_1, L_2}_{\overline{L}_2, L_1}(\overline{L}_3,\overline{L}_4) \, Z^{L_2}_{L_1, L_4}(\tau,\overline{\tau}),
\end{equation}
\begin{equation} \label{eq:modtrans:S}
Z^{L_3}_{L_1, L_2}(-1/\tau, -1/\overline{\tau}) \: = \:
\sum_{L_4} \tilde{K}^{\overline{L}_1,1}_{\overline{L}_4, L_2}(\overline{L}_2, L_1)  \, \tilde{K}^{\overline{L}_1, L_2}_{\overline{L}_2, L_1}(\overline{L}_3,\overline{L}_4) \,
Z^{L_4}_{L_2,\overline{L}_1}(\tau,\overline{\tau}).
\end{equation}
Note in the above that we are following a variation of the conventions of
\cite[appendix A]{Robbins:2019zdb}, where the modular transformations
of boundary conditions are distinct from those of partial traces.
(In the language of the former, the $T$ transformation above might
instead be interpreted as a relation for $\tau-1$ instead of $\tau+1$,
for example.)

As a consistency check, let us compare to the group-like case.
There,
$Z_{L_1, L_2}(\tau) = Z^{L_3}_{L_1, L_2}(\tau)$ for $L_3 = L_1 L_2$, and the associator
$\tilde{K}^{L_1, L_4}_{L_2, L_3}(L_5, L_6)$ is nonzero (and equal to $1$) only for
\begin{equation}
L_4 \: = \: (L_1 L_2 L_3)^{-1}, \: \: \:
L_5 \: = \: L_1 L_2, \: \: \: {\rm and } \: \: \:
L_6 \: = \: L_2 L_3.
\end{equation}
Then, applying the equations above, one finds
\begin{eqnarray}
Z^{L_1 L_2}_{L_1, L_2}(\tau+1) & = & Z^{L_2}_{L_1,L_1^{-1} L_2}(\tau),
\\
Z^{L_1 L_2}_{L_1, L_2}(-1/\tau) & = & Z^{L_1^{-1} L_2}_{L_2,L_1^{-1}}(\tau).
\end{eqnarray}
The reader should note that in our conventions for the group-like case, $\overline{L} = L^{-1}$
(where $\overline{L}$ denotes the dual in the fusion category).

The key role in these transformations is played by transforming one figure into another figure,
using the crossing property, shown in figure~\ref{fig:crossing-kernel}.

\begin{figure}
	\begin{subfigure}{\textwidth}
		\centering
		\begin{tikzpicture}
			\draw[thin] (0,0)--(5,0);
			\draw[thin] (5,0)--(5,5);
			\draw[thin] (0,0)--(0,5);
			\draw[thin] (0,5)--(5,5);
			\draw[very thick,->] (2.5,0)--(3,1.25);
			\draw[very thick] (3,1.25)--(3.5,2.5);
			\draw[very thick,->] (1.5,2.5)--(2,3.75);
			\draw[very thick] (2,3.75)--(2.5,5);
			\draw[very thick,->] (5,2.5)--(4.25,2.5);
			\draw[very thick] (4.25,2.5)--(3.5,2.5);
			\draw[very thick,->] (3.5,2.5)--(2.5,2.5);
			\draw[very thick] (2.5,2.5)--(1.5,2.5);
			\draw[very thick,->] (1.5,2.5)--(0.75,2.5);
			\draw[very thick] (0.75,2.5)--(0,2.5);
			\node at (3.3,1) {$L_1$};
			\node at (1.7,4) {$L_1$};
			\node at (0.7,2.9) {$L_2$};
			\node at (4.3,2.9) {$L_2$};
			\node at (2.5,2.9) {$L_3$};
			\node at (-0.2,5.2) {$\tau$};
             \node at (5.2,-0.2) {$1$};
            \node at (5.2,5.2) {$\tau+1$};
            \node at (-0.2,-0.2) {$0$};

            \draw[thick,color=blue,->,opacity=0.5] (3.5,2.5) [partial ellipse=-110:180:0.25cm and 0.25cm];
		      \draw[thick,color=blue,->,opacity=0.5] (1.5,2.5) [partial ellipse=0:180:0.25cm and 0.25cm];
		\end{tikzpicture}\\
  	\caption{}
		\label{pt1}
	\end{subfigure}
	\begin{subfigure}{\textwidth}
		\centering
		\begin{tikzpicture}
			\draw[thin] (0,0)--(5,0);
			\draw[thin] (5,0)--(5,5);
			\draw[thin] (0,0)--(0,5);
			\draw[thin] (0,5)--(5,5);
			\draw[very thick,->] (2.5,0)--(3,1.25);
			\draw[very thick] (3,1.25)--(3.5,2.5);
			\draw[very thick,->] (1.5,2.5)--(2,3.75);
			\draw[very thick] (2,3.75)--(2.5,5);
			\draw[very thick,->] (5,2.5)--(4.25,2.5);
			\draw[very thick] (4.25,2.5)--(3.5,2.5);
			\draw[very thick,->] (3.5,2.5)--(2.5,2.5);
			\draw[very thick] (2.5,2.5)--(1.5,2.5);
			\draw[very thick,->] (1.5,2.5)--(0.75,2.5);
			\draw[very thick] (0.75,2.5)--(0,2.5);
			\draw[thin,dashed] (0,0)--(-5,5);
			\draw[thin,dashed] (5,0)--(0,5);
			\draw[thin,dashed] (0,5)--(-5,5);
			\draw[thin,dotted] (0,2.5)--(-2.5,2.5);
			%\draw[thin,dotted] (-1.5,2.5)--(-2.5,5);
            \draw[thin,dotted] (-1.786,1.786)--(-1.5,2.5);
            \draw[thin,dotted] (-3.214,3.214)--(-2.5,5);
			\node at (3.3,1) {$L_1$};
			\node at (1.7,4) {$L_1$};
			\node at (0.7,2.9) {$L_2$};
			\node at (4.3,2.9) {$L_2$};
			\node at (2.5,2.9) {$L_3$};
			\node at (-0.2,5.2) {$\tau+1$};
			\node at (-5.2,5.2) {$\tau$};
            \node at (5.2,5.2) {$\tau+2$};
            \node at (5.2,-0.2) {$1$};

            \draw[thick,color=blue,->,opacity=0.5] (3.5,2.5) [partial ellipse=-110:180:0.25cm and 0.25cm];
		      \draw[thick,color=blue,->,opacity=0.5] (1.5,2.5) [partial ellipse=0:180:0.25cm and 0.25cm];
		\end{tikzpicture}
		\caption{}
		\label{fig:modtransT:pt2}
	\end{subfigure}
 \begin{subfigure}{\textwidth}
\centering
		\begin{tikzpicture}
  %          Left-hand side picture
			\draw[thin] (0,0)--(5,0);
			\draw[thin] (5,0)--(5,5);
			\draw[thin] (0,0)--(0,5);
			\draw[thin] (0,5)--(5,5);
			\draw[very thick,->] (2.5,0)--(4.5,1.25);
            \draw[very thick] (4.5,1.25)--(5,1.5625);
			\draw[very thick] (0,1.5625)--(1.5,2.5);
			\draw[very thick] (3.5,2.5)--(5,3.4375);
			\draw[very thick,->] (0,3.4375)--(0.5,3.75);
            \draw[very thick] (0.5,3.75)--(2.5,5);
			\draw[very thick,->] (5,2.5)--(4.25,2.5);
			\draw[very thick] (4.25,2.5)--(3.5,2.5);
			\draw[very thick,->] (3.5,2.5)--(2.5,2.5);
			\draw[very thick] (2.5,2.5)--(1.5,2.5);
			\draw[very thick,->] (1.5,2.5)--(0.75,2.5);
			\draw[very thick] (0.75,2.5)--(0,2.5);
			\node at (1.7,4) {$L_1$};
			\node at (3.3,1) {$L_1$};
			\node at (0.7,2.9) {$L_3$};
			\node at (4.3,2.1) {$L_3$};
			\node at (2.5,2.9) {$L_2$};
			\node at (-0.2,5.2) {$\tau$};
            \node at (5.2,-0.2) {$1$};

            \draw[thick,color=blue,->,opacity=0.5] (3.5,2.5) [partial ellipse=0:180:0.25cm and 0.25cm];
		      \draw[thick,color=blue,->,opacity=0.5] (1.5,2.5) [partial ellipse=-145:180:0.25cm and 0.25cm];
   %
   %        Middle text
            \node at (7,2.5) {$ = \: \: \: \sum_{L_4}$};
    %
   %        Right-hand side picture:  x + 8     
   		\draw[thin] (8,0)--(13,0);
			\draw[thin] (13,0)--(13,5);
			\draw[thin] (8,0)--(8,5);
			\draw[thin] (8,5)--(13,5);
			\draw[very thick,->] (10.5,0)--(11,1.25);
			\draw[very thick] (11,1.25)--(11.5,2.5);
			\draw[very thick,->] (9.5,2.5)--(10,3.75);
			\draw[very thick] (10,3.75)--(10.5,5);
			\draw[very thick,->] (13,2.5)--(12.25,2.5);
			\draw[very thick] (12.25,2.5)--(11.5,2.5);
			\draw[very thick,->] (11.5,2.5)--(10.5,2.5);
			\draw[very thick] (10.5,2.5)--(9.5,2.5);
			\draw[very thick,->] (9.5,2.5)--(8.75,2.5);
			\draw[very thick] (8.75,2.5)--(8,2.5);
			\node at (11.3,1) {$L_1$};
			\node at (9.7,4) {$L_1$};
			\node at (8.7,2.9) {$L_4$};
			\node at (12.3,2.9) {$L_4$};
			\node at (10.5,2.9) {$L_2$};
			\node at (7.8,5.2) {$\tau$};
            \node at (13.2,-0.2) {$1$};

            \draw[thick,color=blue,->,opacity=0.5] (11.5,2.5) [partial ellipse=-110:180:0.25cm and 0.25cm];
		      \draw[thick,color=blue,->,opacity=0.5] (9.5,2.5) [partial ellipse=0:180:0.25cm and 0.25cm];
   %
   %         Right-most text:
            \node at (14.4,2.5) {$ \times \tilde{K}^{\overline{L}_1, L_2}_{\overline{L}_2, L_1}(\overline{L}_3,\overline{L}_4)$};
		\end{tikzpicture}
  \caption{}
  \label{fig:modtransT:pt3}
 \end{subfigure}
\caption{The effect of the modular $T$ transformation.  Diagram (a) displays the original $Z_{L_1, L_2}^{L_3}(\tau)$; (b) displays $Z_{L_1, L_2}^{L_3}(\tau+1)$, which is equivalent to diagram (c).  The left- and right-hand sides of diagram (c) are related using crossing as in %equation~(\ref{eq:crossing})
figure~\ref{fig:crossing-kernel}, with $L_3$ as the internal line before crossing.}
\label{fig:pt}
\end{figure}
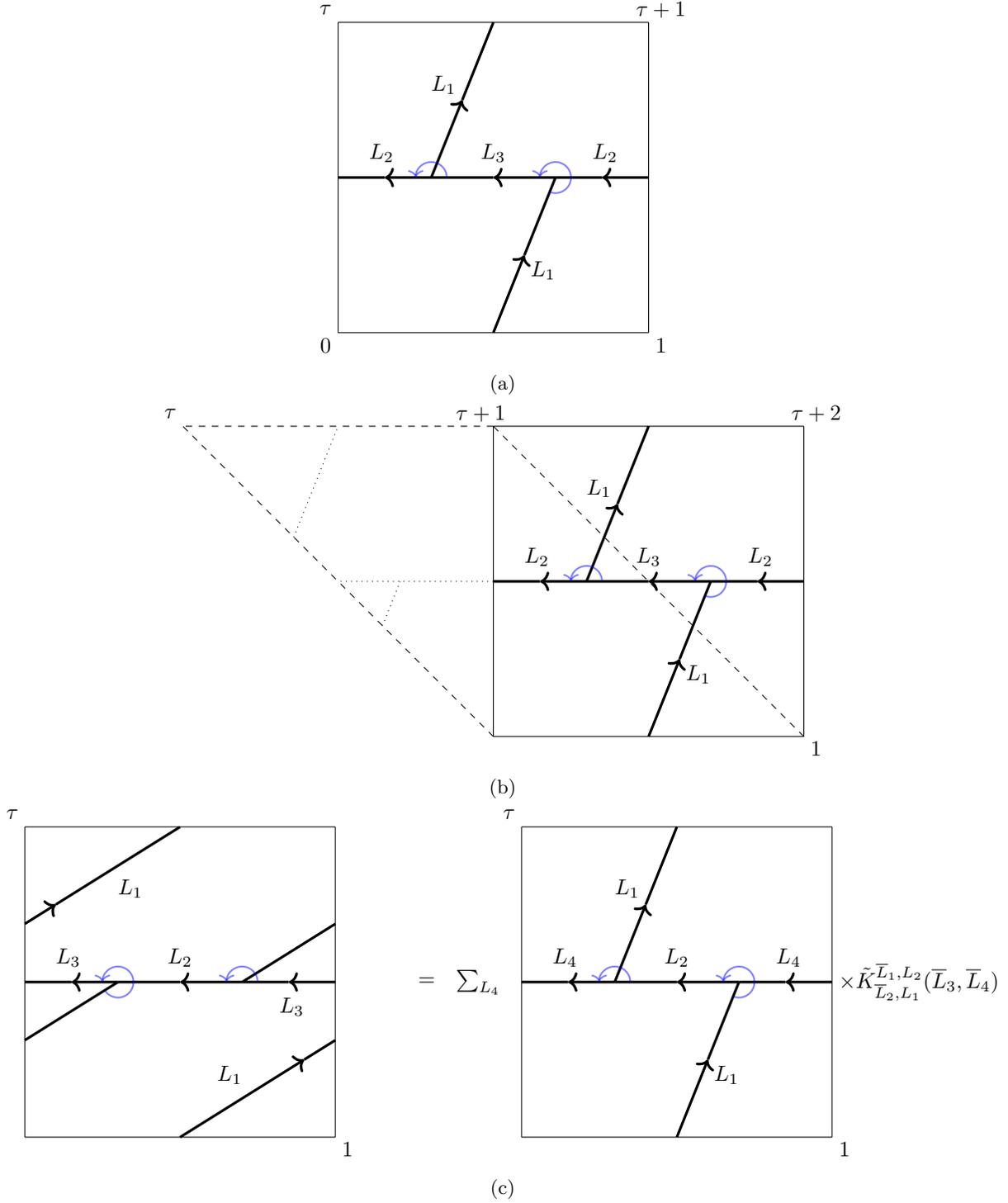

The T transformation is displayed schematically in figure~\ref{fig:pt}.
The original $Z_{L_1, L_2}^{L_3}(\tau)$ is shown in part (a), and part (b) shows the T transformation.
Part (c) of that figure is equivalent to $Z_{L_1, L_2}^{L_3}(\tau+1)$, which is related to another partial trace
using $\tilde{K}$, as in equation~(\ref{eq:modtrans:T}) above.

Before working out the modular S transformation, it will be useful to work out a cyclic transformation identity.  Inserting the identity line $1$ and using the crossing identity (figure~\ref{fig:crossing-kernel}),%~(\ref{eq:crossing}), 
we find
\begin{equation} 
    \begin{tikzpicture}
        % Left picture:
        \draw[very thick, ->] (1.25,2)--(1.25,3);
        \draw[very thick] (1.25,3)--(1.25,4);
        \draw[very thick,->] (1.25,2)--(0.6,0.96);
        \draw[very thick] (0.6,0.96)--(0,0);
        \draw[very thick,->] (1.25,2)--(1.9,0.96);
        \draw[very thick] (1.9,0.96)--(2.5,0);
        %\node at (1.25,2.2) {$\times$};
        \draw[thick,color=blue,->,opacity=0.5] (1.25,2) [partial ellipse=-120:90:0.2cm and 0.2cm];
        \node at (2.7,0.3) {$L_2$};
        \node at (0.45,0.2) {$L_1$};
        \node at (1.6,3) {$L_3$};
        \node at (3,2) {$=$};
        %
        % Middle picture:    x + 3.5
        %
        \draw[very thick,->] (4.75,2)--(4.75,2.5);
        \draw[very thick,->] (4.75,2.5)--(4.75,3.5);
        \draw[very thick] (4.75,3.5)--(4.75,4);
        \draw[dashed] (4.75,2.9)--(3.9,4);
        \draw[very thick,->] (4.75,2)--(4.1,0.96);
        \draw[very thick] (4.1,0.96)--(3.5,0);
        \draw[very thick,->] (4.75,2)--(5.4,0.96);
        \draw[very thick] (5.4,0.96)--(6,0);
        \draw[thick,color=blue,->,opacity=0.5] (4.75,3) [partial ellipse=-90:145:0.2cm and 0.2cm];
		\draw[thick,color=blue,->,opacity=0.5] (4.75,2) [partial ellipse=-130:90:0.2cm and 0.2cm];
        \node at (3.9,3.7) {$1$};
        \node at (6.2,0.3) {$L_2$};
        \node at (3.95,0.2) {$L_1$};
        \node at (5.1,2.5) {$L_3$};
        \node at (5.1,3.5) {$L_3$};
        \node at (7.5,2) {$ = \: \tilde{K}^{L_1 1}_{L_2 L_3}(\overline{L}_3, \overline{L}_1) \times$};
        %
        % Right picture:  x + 9   (for x of left)
        \draw[very thick, ->] (10.25,2)--(10.25,3);
        \draw[very thick] (10.25,3)--(10.25,4);
        \draw[very thick,->] (10.25,2)--(9.85,1.36);
        \draw[very thick,->] (9.85,1.36)--(9.45,0.72);
        \draw[very thick] (9.45,0.72)--(9,0);
        \draw[dashed] (9.65,1.04)--(9.0,4);
        \draw[very thick,->] (10.25,2)--(10.9,0.96);
        \draw[very thick] (10.9,0.96)--(11.5,0);
        \draw[thick,color=blue,->,opacity=0.5] (9.65,1.04) [partial ellipse=-110:100:0.2cm and 0.2cm];
		\draw[thick,color=blue,->,opacity=0.5] (10.25,2) [partial ellipse=-60:235:0.2cm and 0.2cm];
        \node at (9.2,3.8) {$1$};
        \node at (11.6,0.3) {$L_2$};
        \node at (9.35,0.2) {$L_1$};
        \node at (10.15,1.3) {$L_1$};
        \node at (10.5,3) {$L_3$};
    \end{tikzpicture}
\end{equation}
which yields the cyclic transformation identity
\begin{equation}  \label{eq:fig:cyclic}
    \begin{tikzpicture}
        % Left picture:
        \draw[very thick, ->] (1.25,2)--(1.25,3);
        \draw[very thick] (1.25,3)--(1.25,4);
        \draw[very thick,->] (1.25,2)--(0.6,0.96);
        \draw[very thick] (0.6,0.96)--(0,0);
        \draw[very thick,->] (1.25,2)--(1.9,0.96);
        \draw[very thick] (1.9,0.96)--(2.5,0);
        \draw[thick,color=blue,->,opacity=0.5] (1.25,2) [partial ellipse=-110:90:0.2cm and 0.2cm];
        \node at (2.7,0.3) {$L_2$};
        \node at (0.45,0.2) {$L_1$};
        \node at (1.6,3) {$L_3$};
        \node at (4.0,2) {$ = \: \tilde{K}^{L_1, 1}_{L_2, L_3}(\overline{L}_3, \overline{L}_1) \times$};
        %
        % Right picture:   x + 4.5
        \draw[very thick, ->] (6.25,2)--(6.25,3);
        \draw[very thick] (6.25,3)--(6.25,4);
        \draw[very thick,->] (6.25,2)--(5.6,0.96);
        \draw[very thick] (5.6,0.96)--(5.0,0);
        \draw[very thick,->] (6.25,2)--(6.9,0.96);
        \draw[very thick] (6.9,0.96)--(7.5,0);
        \draw[thick,color=blue,->,opacity=0.5] (6.25,2) [partial ellipse=-70:240:0.2cm and 0.2cm];
        \node at (7.7,0.3) {$L_2$};
        \node at (5.45,0.2) {$L_1$};
        \node at (6.6,3) {$L_3$};
        \end{tikzpicture}
\end{equation}

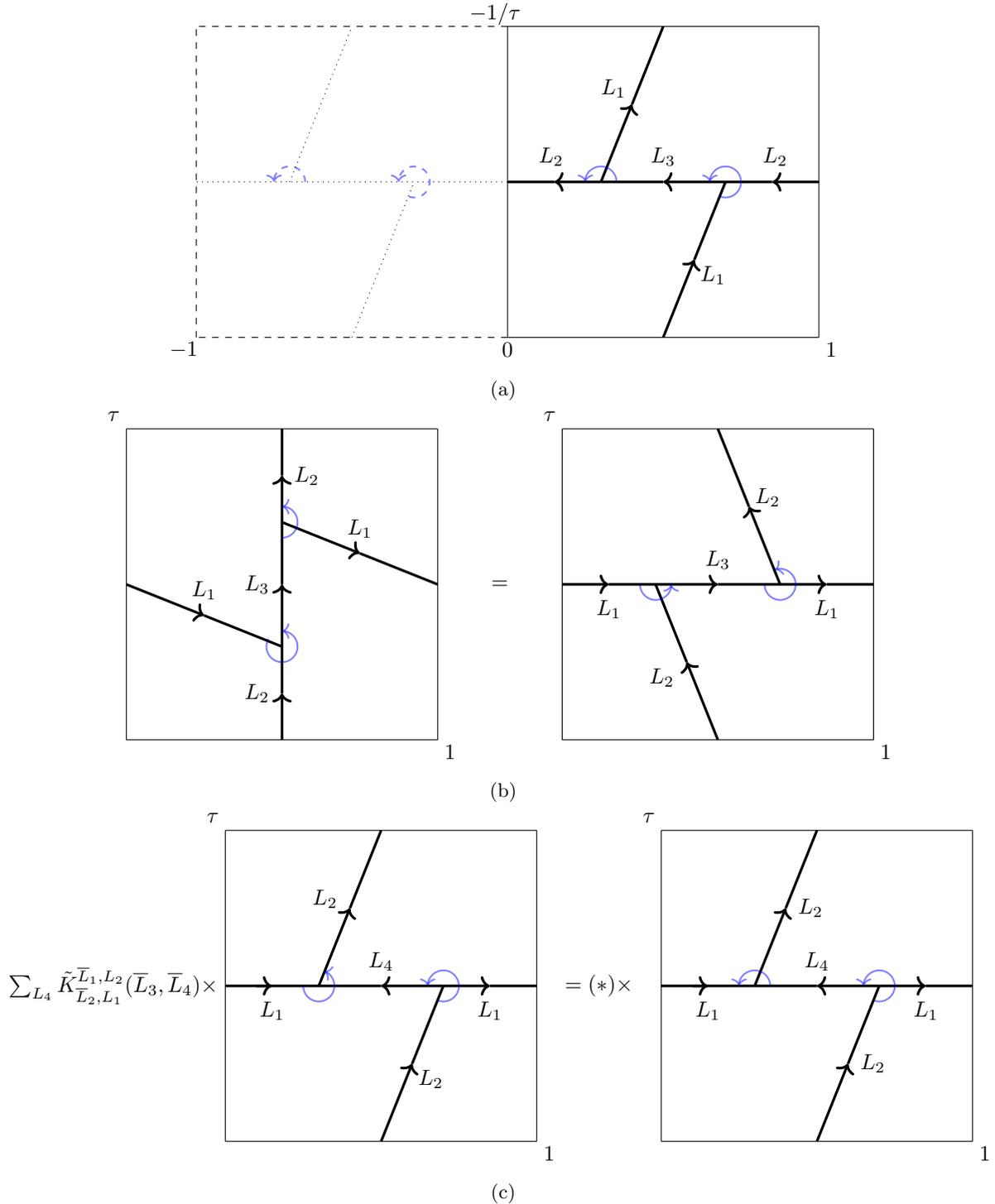
\begin{figure}
\begin{subfigure}{\textwidth}
		\centering
		\begin{tikzpicture}
			\draw[thin] (0,0)--(5,0);
			\draw[thin] (5,0)--(5,5);
			\draw[thin] (0,0)--(0,5);
			\draw[thin] (0,5)--(5,5);
			\draw[very thick,->] (2.5,0)--(3,1.25);
			\draw[very thick] (3,1.25)--(3.5,2.5);
			\draw[very thick,->] (1.5,2.5)--(2,3.75);
			\draw[very thick] (2,3.75)--(2.5,5);
			\draw[very thick,->] (5,2.5)--(4.25,2.5);
			\draw[very thick] (4.25,2.5)--(3.5,2.5);
			\draw[very thick,->] (3.5,2.5)--(2.5,2.5);
			\draw[very thick] (2.5,2.5)--(1.5,2.5);
			\draw[very thick,->] (1.5,2.5)--(0.75,2.5);
			\draw[very thick] (0.75,2.5)--(0,2.5);
			\draw[thin,dashed] (0,5)--(-5,5);
            \draw[thin,dashed] (0,0)--(-5,0);
            \draw[thin,dashed] (-5,0)--(-5,5);
			\draw[thin,dotted] (0,2.5)--(-5,2.5);
			\draw[thin,dotted] (-3.5,2.5)--(-2.5,5);
            \draw[thin,dotted] (-2.5,0)--(-1.5,2.5);
			\node at (3.3,1) {$L_1$};
			\node at (1.7,4) {$L_1$};
			\node at (0.7,2.9) {$L_2$};
			\node at (4.3,2.9) {$L_2$};
			\node at (2.5,2.9) {$L_3$};
            \draw[thick,color=blue,->,opacity=0.5] (3.5,2.5) [partial ellipse=-110:180:0.25cm and 0.25cm];
		      \draw[thick,color=blue,->,opacity=0.5] (1.5,2.5) [partial ellipse=0:180:0.25cm and 0.25cm];
            \draw[thick,color=blue,->,opacity=0.5,dashed] (-1.5,2.5) [partial ellipse=-110:180:0.25cm and 0.25cm];
		      \draw[thick,color=blue,->,opacity=0.5,dashed] (-3.5,2.5) [partial ellipse=0:180:0.25cm and 0.25cm];
			\node at (-0.2,5.2) {$-1/\tau$};
            \node at (-5.2,-0.2) {$-1$};
            \node at (0,-0.2) {$0$};
            \node at (5.2,-0.2) {$1$};
 %           %
		\end{tikzpicture}
		\caption{}
		\label{fig:modtransS:pt1}
	\end{subfigure}
\begin{subfigure}{\textwidth}
    \centering
    \begin{tikzpicture}
        \draw[thin] (0,0)--(5,0);
        \draw[thin] (0,0)--(0,5);
        \draw[thin] (0,5)--(5,5);
        \draw[thin] (5,5)--(5,0);
        \draw[very thick,->] (2.5,0)--(2.5,0.75); 
        \draw[very thick,->] (2.5,0.75)--(2.5,2.5);
        \draw[very thick,->] (2.5,2.5)--(2.5,4.25);
        \draw[very thick](2.5,4.25)--(2.5,5);
        \draw[very thick,->] (0,2.5)--(1.25,2);
        \draw[very thick] (1.25,2)--(2.5,1.5);
        \draw[very thick,->] (2.5,3.5)--(3.75,3);
        \draw[very thick] (3.75,3)--(5,2.5);
        \node at (2.1,2.5) {$L_3$};
        \node at (2.1,0.75) {$L_2$};
        \node at (2.9,4.25) {$L_2$};
        \node at (1.25,2.4) {$L_1$};
        \node at (3.75,3.4) {$L_1$};
        \draw[thick,color=blue,->,opacity=0.5] (2.5,1.5) [partial ellipse=-210:90:0.25cm and 0.25cm];
		\draw[thick,color=blue,->,opacity=0.5] (2.5,3.5) [partial ellipse=-90:90:0.25cm and 0.25cm];
        \node at (-0.2,5.2) {$\tau$};
        \node at (5.2,-0.2) {$1$};
        %
        % Middle:
        \node at (6,2.5) {$=$};
        %
        % Right-hand-side picture:   x+7
        \draw[thin] (7,0)--(12,0);
        \draw[thin] (7,0)--(7,5);
        \draw[thin] (7,5)--(12,5);
        \draw[thin] (12,5)--(12,0);
        \draw[very thick,->] (7,2.5)--(7.75,2.5);
        \draw[very thick,->] (7.75,2.5)--(9.5,2.5);
        \draw[very thick,->] (9.5,2.5)--(11.25,2.5);
        \draw[very thick] (11.25,2.5)--(12,2.5);
        \draw[very thick,->] (9.5,0)--(9,1.25);
        \draw[very thick] (9,1.25)--(8.5,2.5);
        \draw[very thick,->] (10.5,2.5)--(10,3.75);
        \draw[very thick] (10,3.75)--(9.5,5);
        \node at (9.5,2.9) {$L_3$};
        \node at (7.75,2.1) {$L_1$};
        \node at (11.25,2.1) {$L_1$};
        \node at (10.3,3.9) {$L_2$};
        \node at (8.6,1.0) {$L_2$};
        \draw[thick,color=blue,->,opacity=0.5] (10.5,2.5) [partial ellipse=-180:110:0.25cm and 0.25cm];
		\draw[thick,color=blue,->,opacity=0.5] (8.5,2.5) [partial ellipse=-180:0:0.25cm and 0.25cm];
        \node at (6.8,5.2) {$\tau$};
        \node at (12.2,-0.2) {$1$};
    \end{tikzpicture}
    \label{fig:modtransS:pt2}
    \caption{}
\end{subfigure}
\begin{subfigure}{\textwidth}
    \centering
    \begin{tikzpicture}
        \node at (-0.3,2.5) {$ \sum_{L_4}\tilde{K}^{\overline{L}_1, L_2}_{\overline{L}_2, L_1}(\overline{L}_3, \overline{L}_4) \times$};
        \draw[thin] (1.5,0)--(6.5,0);
        \draw[thin] (1.5,0)--(1.5,5);
        \draw[thin] (1.5,5)--(6.5,5);
        \draw[thin] (6.5,5)--(6.5,0);
        \draw[very thick,->] (1.5,2.5)--(2.25,2.5);
        \draw[very thick,->] (5,2.5)--(4,2.5);
        \draw[very thick] (4,2.5)--(2,2.5);
        \draw[very thick,->] (5,2.5)--(5.75,2.5);
        \draw[very thick] (5.75,2.5)--(6.5,2.5);
        \draw[very thick,->] (4,0)--(4.5,1.25);
        \draw[very thick] (4.5,1.25)--(5,2.5);
        \draw[very thick,->] (3,2.5)--(3.5,3.75);
        \draw[very thick] (3.5,3.75)--(4,5);
        \node at (4,2.9) {$L_4$};
        \node at (2.25,2.1) {$L_1$};
        \node at (5.75,2.1) {$L_1$};
        \node at (3.1,3.9) {$L_2$};
        \node at (4.8,1.0) {$L_2$};
        \draw[thick,color=blue,->,opacity=0.5] (5,2.5) [partial ellipse=-110:180:0.25cm and 0.25cm];
		\draw[thick,color=blue,->,opacity=0.5] (3,2.5) [partial ellipse=-180:70:0.25cm and 0.25cm];
        \node at (1.3,5.2) {$\tau$};
        \node at (6.7,-0.2) {$1$};
        \node at (7.5,2.5) {$ = (*) \times$};
        %
        % Right-hand-side:    x + 7
        \draw[thin] (8.5,0)--(13.5,0);
        \draw[thin] (8.5,0)--(8.5,5);
        \draw[thin] (8.5,5)--(13.5,5);
        \draw[thin] (13.5,5)--(13.5,0);
        \draw[very thick,->] (8.5,2.5)--(9.25,2.5);
        \draw[very thick,->] (12,2.5)--(11,2.5);
        \draw[very thick] (11,2.5)--(9,2.5);
        \draw[very thick,->] (12,2.5)--(12.75,2.5);
        \draw[very thick] (12.75,2.5)--(13.5,2.5);
        \draw[very thick,->] (11,0)--(11.5,1.25);
        \draw[very thick] (11.5,1.25)--(12,2.5);
        \draw[very thick,->] (10,2.5)--(10.5,3.75);
        \draw[very thick] (10.5,3.75)--(11,5);

        \node at (11,2.9) {$L_4$};
        \node at (9.25,2.1) {$L_1$};
        \node at (12.75,2.1) {$L_1$};
        \node at (11.9,1.25) {$L_2$};
        \node at (10.9,3.75) {$L_2$};
        \draw[thick,color=blue,->,opacity=0.5] (12,2.5) [partial ellipse=-110:180:0.25cm and 0.25cm];
		\draw[thick,color=blue,->,opacity=0.5] (10,2.5) [partial ellipse=0:180:0.25cm and 0.25cm];
        \node at (8.3,5.2) {$\tau$};
        \node at (13.7,-0.2) {$1$};
    \end{tikzpicture}
    \label{fig:modtransS;pt3}
    \caption{}
    \end{subfigure}
    \caption{Three successive views of $Z_{L_1,L_2}^{L_3}(-1/\tau)$.  Part (a) shows the original modular transformation.  Part (b) rewrites the result of the modular transformation.   The right-hand side of (c) is obtained from the left-hand side of (b) using the cyclic identity~(\ref{eq:fig:cyclic}).  Part (c) uses the crossing identity (figure~\ref{fig:crossing-kernel}) %~(\ref{eq:crossing}) 
    to write the result in terms of other $Z_{L_1,L_2}^{L_3}(\tau)$, where we use the abbreviation $* = \sum_{L_4} \tilde{K}^{\overline{L}_1, L_2}_{\overline{L}_2, L_1}(\overline{L}_3, \overline{L}_4) \tilde{K}^{\overline{L}_1, 1}_{\overline{L}_4, L_2} (\overline{L}_2,L_1)$.  Each part is an equivalent expression for $Z_{L_1,L_2}^{L_3}(-1/\tau)$.  \label{fig:modtransS}}
\end{figure}

The S transformation~(\ref{eq:modtrans:S}) is derived schematically in figure~\ref{fig:modtransS}.

\subsection{Gaugeability}
\label{sect:gaugeability}

So far, we have outlined noninvertible symmetries (defined by fusion categories, which generalize finite groups), noninvertible analogues of the partial traces used to construct orbifold partition functions, and also described modular transformations.
Clearly, the next step in constructing a noninvertible analogue of an orbifold is to construct modular-invariant partition functions, built as linear combinations of the partial traces constructed in section~\ref{sect:defn:partialtrace}.

However, in general, that is not always possible.  Only in some cases do modular-invariant partition functions exist,
and as we shall describe later in section~\ref{sect:genl-algebra}, their construction will involve specifying a (special symmetric)
Frobenius algebra (which we will define in section~\ref{sect:genl-algebra}).
At least morally, such obstructions correspond to analogues of gauge anomalies, though that description in the noninvertible
context may be more obscure.\footnote{Generally speaking, there can be two notions of anomalies: The obstruction of gauging a symmetry and the obstruction from a trivially gapped phase. In the group-like case, these two notions coincide. However, in the case of noninvertible symmetries, symmetries can be gaugeable even if it is incompatible with a trivially gapped phase. We refer the reader to \cite{Choi:2023xjw, Cordova:2023bja} for more details.}

The purpose of this section is to demonstrate an example in which a modular-invariant partition function cannot be constructed.

\paragraph{Non-gaugeable example: Ising CFT.}
The symmetry category for the Ising CFT includes three simple objects, identity line $1$, $\mathbb{Z}_2$ symmetry line $\eta$, and Kramers-Wannier noninvertible line $\mathcal{N}$. The fusion rules are given by
\begin{equation}
\begin{split}
     1\otimes \eta=\eta\otimes 1=\eta,\\
     1\otimes \mathcal{N}=\mathcal{N}\otimes 1=\mathcal{N},\\
     \eta\otimes \mathcal{N}=\mathcal{N}\otimes \eta=\mathcal{N},\\
     \eta\otimes \eta=1,\\
     \mathcal{N}\otimes \mathcal{N}=1\oplus \eta,
\end{split}
\end{equation}
which is also known as the $\mathbb{Z}_2$ Tambara-Yamagami fusion category \cite{ty}. Based on the above fusion rules, the set of all possible partial traces $Z_{L_1,L_2}^{L_3}$ from various insertions of topological lines $L_1$ and $L_2$ is
\begin{equation}
    Z_{1,1}^1,Z_{1,\eta}^\eta,Z_{\eta,1}^\eta,Z_{\eta,\eta}^1, Z_{1,\mathcal{N}}^\mathcal{N},Z_{\mathcal{N},1}^\mathcal{N},Z_{\eta,\mathcal{N}}^\mathcal{N},Z_{\mathcal{N},\eta}^\mathcal{N},Z_{\mathcal{N},\mathcal{N}}^1,Z_{\mathcal{N},\mathcal{N}}^\eta .
\end{equation}
The modular transformations of these twisted  partition functions follow the rule (\ref{eq:modtrans:T}) and (\ref{eq:modtrans:S}), where the $\tilde{K}$ matrix for the Ising model reads\footnote{
The $\sigma$, $\psi$ of \cite{Moore:1988qv} are denoted $\mathcal{N}$, $\eta$ in this paper.
} \cite[appendix D]{Moore:1988qv}
\begin{equation}
\begin{split}
  & \begin{pmatrix}
    \tilde{K}_{\mathcal{N},\mathcal{N}}^{\mathcal{N},\mathcal{N}}(1,1)& \tilde{K}_{\mathcal{N},\mathcal{N}}^{\mathcal{N},\mathcal{N}}(1,\eta)\\
    \tilde{K}_{\mathcal{N},\mathcal{N}}^{\mathcal{N},\mathcal{N}}(\eta,1)& \tilde{K}_{\mathcal{N},\mathcal{N}}^{\mathcal{N},\mathcal{N}}(\eta,\eta)
\end{pmatrix}=
\begin{pmatrix}
    \frac{1}{\sqrt{2}}& \frac{1}{\sqrt{2}}\\
    \frac{1}{\sqrt{2}}& -\frac{1}{\sqrt{2}},
\end{pmatrix},\\
&\tilde{K}_{\mathcal{N},\eta}^{\eta,\mathcal{N}}(\mathcal{N},\mathcal{N})=\tilde{K}_{\eta,\mathcal{N}}^{\mathcal{N},\eta}(\mathcal{N},\mathcal{N})=-1,
\end{split}
\end{equation}
with all other elements equal to 1. We compute the modular $S$ transformation for all twisted partition functions as 
\begin{equation}\label{eq: S transformation of ising twisted partition function}
\begin{split}
     &Z_{1,1}^{1}\rightarrow Z_{1,1}^1,  Z_{1,\eta}^{\eta}\leftrightarrow Z_{\eta,1}^\eta,Z_{\eta,\eta}^1\rightarrow Z_{\eta,\eta}^1,\\
     &Z_{1,\mathcal{N}}^\mathcal{N}\leftrightarrow Z_{\mathcal{N},1}^\mathcal{N}, Z_{\eta,\mathcal{N}}^\mathcal{N}\leftrightarrow -Z_{\mathcal{N},\eta}^\mathcal{N},\\
     &Z_{\mathcal{N},\mathcal{N}}^1\rightarrow \frac{1}{\sqrt{2}}Z_{\mathcal{N},\mathcal{N}}^1+\frac{1}{\sqrt{2}}Z_{\mathcal{N},\mathcal{N}}^\eta,\\
     &Z_{\mathcal{N},\mathcal{N}}^\eta\rightarrow \frac{1}{\sqrt{2}}Z_{\mathcal{N},\mathcal{N}}^1-\frac{1}{\sqrt{2}}Z_{\mathcal{N},\mathcal{N}}^\eta .
\end{split}
\end{equation}
If the full categorical symmetry is gaugeable, there should exist a set of non-trivial coefficients $\Lambda_{L_1,L_2}^{L_3}$, each of which for the twisted partition function $Z_{L_1,L_2}^{L_3}$, so that the 
\begin{equation}
Z_{\text{gauged}}=\sum_{L_1,L_2,L_3}\Lambda_{L_1,L_2}^{L_3}Z_{L_1,L_2}^{L_3}
\end{equation}
is  modular invariant partition function for the well-defined CFT after gauging. However, from (\ref{eq: S transformation of ising twisted partition function}) it is easy to check the coefficients for $Z_{\mathcal{N},\mathcal{N}}^1$ and $Z_{\mathcal{N},\mathcal{N}}^\eta$ have to be trivial in order to be modular S invariant. Therefore, there is an obstruction of summing over all twisted sectors with the presence of the Kramers-Wannier line $\mathcal{N}$, thus the categorical symmetry not gaugeable. 

We remark that though one cannot gauge the full categorical symmetry, but the $\mathbb{Z}_2$ subgroup generated by the identity $1$ and the $\mathbb{Z}_2$ line $\eta$ is indeed gaugeable. This can be seen by checking $Z_{1,1}^1+Z_{1,\eta}^1+Z_{\eta,1}^1+Z_{\eta,\eta}^1$ is indeed modular invariant, which leads to a $\mathbb{Z}_2$ orbifold partition function
\begin{equation}
    Z[\text{Ising}/\mathbb{Z}_2]\equiv \frac{1}{2}(Z_{1,1}^1+Z_{1,\eta}^1+Z_{\eta,1}^1+Z_{\eta,\eta}^1).
\end{equation}
One can further check that this gauged partition function is equal to that before the gauging, i.e. $Z[\text{Ising}/\mathbb{Z}_2]=Z_{1,1}^1$, which shows the self-duality of the Ising CFT under $\mathbb{Z}_2$ gauging.

\subsection{Formal specification of gauging:  algebra objects}
\label{sect:genl-algebra}

Given a fusion category acting as the symmetry category of a two-dimensional theory,
we need a systematic prescription for computing physical partition functions, when they exist -- and indeed, as the example in the previous section illustrates, modular-invariant partition functions will not always exist, not every fusion category can be gauged.

When a fusion category acting as the symmetry category of a two-dimensional theory can be gauged, we can gauge by a particular kind of algebra object derived from it, called a \textit{special symmetric Frobenius
algebra}, which we denote ${\cal A}$.  Intuitively, we think of ${\cal A}$ as an identity operator in the gauged theory, and the operation of gauging involves inserting a sufficiently fine mesh of lines of ${\cal A}$.
For example, the one-loop partition function can then, at least formally,
be written
\begin{equation}
    Z \: = \:  Z_{{\cal A},{\cal A}}^{\cal A}.
\end{equation}
We shall elaborate on the precise meaning of this expression later.
(See also e.g.~\cite{Fuchs:2002cm}.)

In the special case of gauging an ordinary (non-anomalous) group $G$, for example, the corresponding symmetry category before gauging is the fusion category $\text{Vec}(G)$, the category of $G$-graded vector spaces. In this category, there is a distinguished collection of objects $\{L_g\}_{g\in G}$, called simple objects, which correspond to one-dimensional vector spaces labeled by $g\in G$. These have the special property that any other object in $\text{Vec}(G)$ is isomorphic to a direct sum involving only such simple objects. In this category, a distinguished symmetric special Frobenius object ${\cal A}$ is the direct
sum of the simple objects $L_g$, which in the two-dimensional system act as line defects:
\begin{equation}
    {\cal A} \: = \: \bigoplus_{g \in G} L_g,
\end{equation}
The corresponding one-loop partition function becomes the familiar orbifold partition function
\begin{equation}
     Z_{{\cal A},{\cal A}}^{\cal A} \: = \: \frac{1}{|G|} \sum_{gh = hg} Z_{g,h}.
\end{equation}
For more general noninvertible symmetry groups, the $Z_{g,h}$ of ordinary orbifolds will be replaced by 
$Z_{L,M}^N$'s, in which the four-point junction at which the $g$, $h$ lines intersect is resolves into a pair of three-point junctions joined by a choice of intermediate state $c$, which is no longer uniquely determined by the
lines $L$, $M$.

Now, not all fusion categories are gaugeable; for example, as we saw in the previous section,  
one cannot always construct modular-invariant partition functions from the entire fusion category.  At least in general terms,
this issue has been addressed in the language of SymTFTs.  Specifically,
in \cite[Theorem 1]{Thorngren:2019iar}, \cite{Huang:2021zvu}, it was argued that the Turaev-Viro 3d TQFT defined by the symmetry fusion category $\mathcal{C}$ admits a gapped, non-degenerate $\mathcal{C}$-symmetric boundary condition, as needed to describe a Neumann boundary condition\footnote{
See \cite{FMT22} for a more general discussion of this fact.
} and hence gauge all of ${\cal C}$, if and only if $\mathcal{C}$ admits an exact, faithful tensor functor
\begin{equation}
    F:\mathcal{C}\to\text{Vec},
\end{equation}
whose target category in this case is $\text{Vec}$, the category of vector spaces. This functor is known as a fiber functor. (However, later in this paper we will sometimes only gauge subalgebras, and for those, a fiber functor is not required.  See also  \cite[section 5.2]{Bhardwaj:2017xup}, \cite{Huang:2021zvu,ostrik1} for other examples of this form.) 
Also, in general, a fusion category can have multiple fiber functors; however, in this paper, we will specialize
to fusion categories of the form Rep(${\cal H}$), for which there is a canonical fiber functor, namely the forgetful functor mapping to the vector spaces underlying any given representation.  Technically, a choice of semisimple Hopf algebra ${\cal H}$ is equivalent, by the Tannaka reconstruction theorem, to a fusion category Rep$({\cal H})$ together with a fiber functor.  (For example, ${\cal H}$ is given by the endomorphisms of the fiber functor.) In particular, there exist examples of groups $G_1 \neq G_2$ such that Rep($G_1$) = Rep($G_2$) as fusion categories, but which have different fiber functors \cite{tannaka}.

It is known (see e.g.~\cite[Corollary 2.22]{ENO02}) that any fusion category equipped with a fiber functor is equivalent to the category of representations of a connected semisimple weak Hopf algebra.  (A Hopf algebra ${\cal H}$ is a generalization of a group algebra, defined by both a multiplication $\mu: {\cal H} \otimes {\cal H} \rightarrow {\cal H}$ as well as a
comultiplication $\Delta: {\cal H} \rightarrow {\cal H} \otimes  {\cal H}$, plus additional structures, satisfying some conditions.  We give a technical definition in appendix~\ref{app:Hopf}.)
For this reason, in this paper we will specialize to fusion categories of the form
Rep$({\cal H}$), for ${\cal H}$ a connected semisimple finite-dimensional weak Hopf algebra.  (In fact, to satisfy an additional constraint, we will further specialize to
semisimple finite-dimensional Hopf algebras, which are special cases, as we describe momentarily.)

We can use this general principle to give a simple characterization of gaugeability of an entire fusion category.
A fiber functor sends every object $X$ to a vector space $F(X)$ whose dimension is the quantum dimension of $X$, meaning that, as observed in Theorem 2 in \cite{Thorngren:2019iar}, a necessary condition for a fusion category to admit a fiber functor (and hence be gaugeable) is that all objects have a non-negative integer quantum dimension. This already characterizes familiar categories, such as the \textbf{Ising} category, as non-gaugeable, and is a simple way to understand the obstruction we discovered in subsection~\ref{sect:gaugeability}. There, the Kramers-Wannier line in the Ising model has quantum dimension $\sqrt{2}$, which by the arguments here, excludes the existence of a suitable Frobenius algebra.  We repeat that if one is not gauging the entire fusion category (as we will sometimes do), a fiber functor on the entire category is not required.

We can also obtain a second necessary condition for gaugeability from these considerations.
Gauging a fusion category equiped with a fiber functor is then understood as taking the regular object $R$ defined as
\begin{equation}
    R=\bigoplus_{i=1}^n \text{dim}(N_i)N_i
\end{equation}
for $N_i$ and gauging by it. However, this is possible only as long as $R$ is a symmetric special Frobenius algebra (which we will define shortly). It is known that not all\footnote{See p. 6 in \cite{IK08} for an example.} weak Hopf algebras admit a Frobenius algebra structure \cite{IK08}. Although this situation is somewhat ameliorated for connected semisimple ones, (see e.g. Prop. 3.1.5 in \cite{Nik04}), for concreteness in the present paper we restrict to working with categories of representations of finite-dimensional semisimple Hopf algebras, where it was already shown that the regular object admits a symmetric special Frobenius algebra structure \cite{FSS11}. 
This class of representation categories is already quite general and includes many familiar examples, such as $\text{Vec}(G)\cong \text{Rep}((\C[G])^*)$, and $\text{Rep}(G)\cong \text{Rep}(\C[G])$. The process we describe allows not only to gauge the whole fusion category but also \textit{subcategories} of it determined by Hopf ideals.

Shortly we will explicitly construct modular-invariant partition functions using a special symmetric Frobenius algebra. Thus, this allows us to see the gaugeability directly at the level of partition function computations.

Next, we will construct a symmetric special Frobenius algebra in $\text{Rep}({\cal H})$. (To be clear, we begin by picking a finite-dimensional semisimple Hopf algebra ${\cal H}$, which is equivalent to\footnote{By the Tannaka reconstruction theorem.} a fusion category Rep(${\cal H}$) together with a fiber functor.) Before proceeding, we define the latter\footnote{A more classical definition of a Frobenius algebra is phrased in terms of a non-degenerate bilinear form on ${\cal A}$. These definitions are equivalent.}. We start with an algebra ${\cal A}$, in our case over $\Bbbk=\C$, with unit $u:{\cal A}\to\C$ and multiplication $\mu:{\cal A}\otimes {\cal A}\to \cal A$. We then endow the algebra ${\cal A}$ with a coalgebra structure, with counit $u^o_F:{\cal A}\to\C$ and comultiplication $\Delta_F:{\cal A}\to {\cal A}\otimes {\cal A}$. For ${\cal A}$ to be a Frobenius algebra, the algebra and coalgebra structures are required to satisfy the following commutative diagrams, called the Frobenius identities:
\begin{equation}  \label{eq:Frobenius-identities}
        \xymatrix{
        {\cal A} \otimes {\cal A} \ar[rr]^{\Delta_F \otimes \text{Id}_{\cal A}} \ar[d]_{\mu} 
        & & {\cal A} \otimes {\cal A} \otimes {\cal A} \ar[d]^{\text{Id}_{\cal A} \otimes \mu}
        \\
        {\cal A} \ar[rr]^{\Delta_F} & & {\cal A} \otimes {\cal A}
        },
        \: \: \: \: \:
        \xymatrix{
        {\cal A} \otimes {\cal A} \ar[rr]^{\text{Id}_{\cal A} \otimes \Delta_F} \ar[d]_{\mu} 
        & & {\cal A} \otimes {\cal A} \otimes {\cal A} \ar[d]^{\mu \otimes \text{Id}_{\cal A}}
        \\
        {\cal A} \ar[rr]^{\Delta_F} & & {\cal A} \otimes {\cal A}
        }
    \end{equation}

Next, we recall the definition of 
\textit{symmetric special} Frobenius algebras below \cite{Fuchs:2002cm}.

Given a Frobenius algebra $({\cal A},\mu,\Delta_F,u,u^o_F)$, we say it is a \textit{special} Frobenius algebra if
\begin{eqnarray} \label{eq:conds}
    u^o\circ u = \beta_1\text{Id}_1; & \mu \circ \Delta_F = \beta_{\cal A}\text{Id}_{\cal A}
\end{eqnarray}
for $\beta_1,\beta_{\cal A}\in \C^{\times}$.  We normalize $\Delta_F$ by taking $\beta_{\cal A} = 1$.

A Frobenius algebra is called \textit{symmetric} if the following equality of morphisms ${\cal A}\to {\cal A}^*$ holds
\begin{equation}  \label{eq:Fr:symmetric}
    ((u^o_F\circ\mu)\otimes \text{Id}_{{\cal A}^*})\circ(\text{Id}_{\cal A}\otimes \gamma_{\cal A})= (\text{Id}_{{\cal A}^*}\otimes (u^o_F\circ\mu ))\circ (\overline{\gamma}_{\cal A}\otimes\text{Id}_{\cal A})
\end{equation}
with coevaluation maps $\gamma_{\cal A}:{\mathbb C}\to {\cal A}\otimes {\cal A}^*$, $\overline{\gamma}_{\cal A}:{\mathbb C}\to {\cal A}^*\otimes {\cal A}$, which exist by definition of ${\cal A}^*$ being a dual of ${\cal A}$.

We are interested in realizing the regular representation of ${\cal H}$ as a symmetric special Frobenius algebra. (Later we will also consider quotient algebras.) Realizing the regular representation in this fashion is equivalent to endowing the dual Hopf algebra ${\cal H}^*$ with a symmetric special Frobenius algebra structure \cite{FSS11}. It is a theorem \cite{LS69} that any \textit{finite-dimensional} Hopf algebra can be endowed with a Frobenius algebra structure. Moreover, ${\cal H}^*$ as a Frobenius algebra is symmetric, special, and has trivial twist if and only if ${\cal H}$ is (finite-dimensional) semisimple (see Section 4 of \cite{FSS11}). It is for this reason that we are mainly interested in (finite-dimensional) semisimple Hopf algebras. We now walk through how this works, since this process is leveraged explicitly in computations.

Let us take a moment to review the construction \cite[chapter VI]{Sweedler69} of the dual Hopf algebra ${\cal H}^*$.
Since we will only be interested in finite-dimensional Hopf algebras, let $V, W$ be a finite-dimensional vector spaces, and $V^*=\text{Hom}(V,\C), W^*=\text{Hom}(W,\C)$ their corresponding duals. Then any linear map $f:V\to W$ induces a map $f^*:W^*\to V^*$ by precomposition. If $V$ furthermore has the structure of a bialgebra ($V, \mu_V, \Delta_V, u_V, u^o_V$), then one can show that the dual morphisms
\begin{eqnarray}
    \mu_*:=\Delta_V^*:&V^*\otimes V^*&\to V^*,    \label{eq:defn:mu-star}\\
    u_*:=(u^o_V)^*:& \C&\to V^*,\\
    \Delta^*:=\mu_V^*:& V^*&\to V^*\otimes V^*,\\
    u^o_*:=u_V^*:&V^*&\to \C,
\end{eqnarray}
endow $V^*$ with the structure of a bialgebra. For finite-dimensional vector spaces, there is an isomorphism $(V\otimes W)^*\cong V^*\otimes W^*$, so that there are no issues with respect to the domain and images of these maps. Furthermore, if $V$ is a Hopf algebra, so that it is also equipped with an antipode map $S:V\to V$, then the corresponding dual
\begin{equation}
    S^*: \: V^*\to V^*,
\end{equation}
then $(V^*,\mu_*,u_*,\Delta_*,u^o_*,S_*)$ is also a Hopf algebra.

Knowing ${\cal H}^*$ as a finite-dimensional Hopf algebra, one can then apply the integral construction of Larson and Sweedler \cite{LS69} to produce a Frobenius algebra. In this prescription, adapted in \cite{FSS11} for the present purposes, one retains the algebra structure $({\cal H}^*,\mu_*,u_*)$ and changes coalgebra structure ($\Delta, u^o$). The key object in this construction is that of (left-)\textit{integrals} and \textit{cointegrals}. Given any Hopf algebra ${\cal H}$ with counit $u^o:{\cal H}\to\C$, a left-integral is an element $\Lambda\in {\cal H}$ such that for all $h\in {\cal H}$ it holds that
$\mu(h,\Lambda)$ is given by scalar multiplication:
\begin{equation}
    \mu(h,\Lambda)=u^o(h) \, \Lambda.
\end{equation}
On the other hand, a right-cointegral is an element $\lambda\in {\cal H}^*$ such that
\begin{equation}
    (\lambda\otimes \text{Id}_{\cal H})(\Delta(h))=\lambda(h)\otimes\text{1}_{\cal H},
\end{equation}
where $1_{\cal H}=u(1)$ is the unit element of ${\cal H}$.

The main result of \cite{LS69} is that for finite-dimensional Hopf algebras, left-integrals always exist and span a one-dimensional vector space. The same applies to cointegrals, since in the finite-dimensional setting these are identified with the integrals of the dual Hopf algebra. We use this to endow ${\cal H}^*$ with the structure of a Frobenius algebra. Let $\lambda$ be a cointegral of ${\cal H}$, that is, an integral of ${\cal H}^*$, and $\Lambda$ an integral of ${\cal H}$, with the normalization condition
\begin{equation}
    \lambda\circ\Lambda=1
\end{equation}
as a morphism in $\text{Hom}(\C,\C)$. It is shown in \cite{FSS11} that the tuple $({\cal H}^*,\mu_*,u_*,\Delta_F,u^o_F)$ with $\mu_*$, $u_*$ as above,
\begin{eqnarray}
    \Delta_F&:=& k(\text{Id}_{\cal H}\otimes(\lambda\circ\mu))\circ (\text{Id}_{\cal H}\otimes S\otimes \text{Id}_{\cal H})\circ(\Delta\otimes\text{Id}_{\cal H}))^*,
\label{eq:DeltaF-defn}
    \\
    u^o_F&:=&\Lambda^*,
\end{eqnarray}
for $k\in\C^{\times}$ is a symmetric special Frobenius algebra in $\text{Rep}({\cal H})$\footnote{In fact, it is a Frobenius algebra in the category of $({\cal H},{\cal H})$-bimodules.} iff ${\cal H}$ is semisimple. Thus, ${\cal H}^*$ can be used to gauge. Throughout this paper we choose $k$ to get the normalization
\begin{equation}
    \mu_*\circ\Delta_F=\text{Id}_{{\cal H}^*}.
\end{equation}

An important consequence of this process is that it also allows us to identify other suitable objects, or subcategories, by which we can gauge. It suffices to find a Hopf ideal. It is a well-known fact that Hopf ideal $I\hookrightarrow {\cal H}$, that is, a subset $I$ that is both an ideal and coideal closed under the antipode map on ${\cal H}$. Since $I$ is a Hopf ideal, this defines a quotient Hopf algebra ${\cal H}/I$, so that we have a Hopf subalgebra $({\cal H}/I)^*\hookrightarrow {\cal H}^*$. This in fact induces a functor
\begin{equation}\label{eq:quotientinclusion}
    \imath: \text{Rep}({\cal H}/I)\hookrightarrow \text{Rep}({\cal H})
\end{equation}
such that $\imath(({\cal H}/I)^*)$ is a symmetric special Frobenius algebra, with ${\cal H}$-action that factors through ${\cal H}/I$. Of course, we do not obtain all possible Frobenius objects in this way, yet it is a useful criterion.

 Now, let us describe some concrete examples.
 The case of ordinary orbifolds, where the fusion category is ${\rm Vec}(G) = {\rm Rep}( {\mathbb C}[G]^*)$, will be discussed in detail in section~\ref{sect:ord-orbifold}, and we will demonstrate there that this technology does indeed correctly reproduce ordinary orbifolds.

 For the rest of this section, we specialize to the case that the fusion category is 
 \begin{equation}
     {\rm Rep}(G) \: = \: {\rm Rep}( {\mathbb C}[G]) \: = \: {\rm Rep}({\cal H}),
 \end{equation}
so that
${\cal H} = {\mathbb C}[G]$, for some finite group $G$. 
We are then going to build a Frobenius algebra structure on ${\cal A} = {\cal H}^* = 
{\mathbb C}[G]^*$.  This will correspond to the regular representation of $G$, and Frobenius structures on subalgebras of ${\cal A} = {\cal H}^*$ will correspond to certain other
representations of $G$, as we shall see.

The algebra structure $({\cal H},\mu,u)$ is a linear extension of the group product,
\begin{eqnarray}
    \mu: & g \otimes h  \mapsto  gh, \\
    u: & 1  \mapsto  1.
\end{eqnarray}
The coalgebra structure is
\begin{eqnarray}
    \Delta:& g\mapsto g\otimes g\\
    u^o:& g\mapsto 1
\end{eqnarray}
The antipode map corresponds on the basis elements to the inverse operation
\begin{eqnarray}
    S:&g\mapsto g^{-1}
\end{eqnarray}
In this case, ${\cal H}^*$ has underlying vector space ${\mathbb C}[G]^*$. We will use the dual basis characterized by the equation
\begin{equation}
    v_g(h)=\delta_{g,h}
\end{equation}
We can construct a Frobenius algebra on ${\cal H}^* = {\mathbb C}[G]^*$, as follows.  Some of the structure is inherited from the dual Hopf algebra, specifically,
the multiplication $\mu_*: {\cal H}^* \otimes {\cal H}^* \rightarrow {\cal H}^*$
\begin{equation}  \label{eq:Fr:mult}
    \mu_*(v_{g} \otimes v_{h}) \: = \: \delta_{g,h} \; v_{g},
\end{equation}
and unit $u_*: {\mathbb C} \rightarrow {\cal H}^*$
\begin{equation}
    u_*(1) \: = \: \sum_{g} v_{g},
\end{equation}
where in each case we have given an action on basis elements, which are extended linearly over
${\mathbb C}$.  We now choose integrals and cointegrals to construct the coalgebra structure. The integral element of $\C[G]$ in this case is
\begin{equation}
    \Lambda=\sum_{g\in G}g
\end{equation}
regarded as a linear function on ${\cal H}^*$. A cointegral of $\C[G]$ is
\begin{equation}
    \lambda = v_1 .
\end{equation}
On the basis elements, the counit $u^o_F:=\Lambda^*:{\cal H}^*\to\C$ (where we regard $\Lambda\in {\cal H}$ as a morphism $\Lambda:\C\to {\cal H}$) is
\begin{equation} \label{eq:Fr:counit}
    u^o_F(v_g)=v_g\left(\sum_{h\in G}h\right) = v_g(g) = 1.
\end{equation}
The comultiplication is the dual of the composition
\begin{eqnarray}
    g\otimes h\mapsto g\otimes g\otimes h\mapsto g\otimes g^{-1}\otimes h\mapsto g\otimes \delta_{g^{-1}h,1}
\end{eqnarray}
so that
\begin{eqnarray}
    \Delta_F&:=& v_g\mapsto v_g\otimes v_g .
\end{eqnarray}

It is straightforward to check that $\mu_* \circ \Delta_F  = {\rm Id}_{{\cal H}^*}$, hence it is correctly normalized (following the discussion after~(\ref{eq:conds})).

The algebra structures $(\mu_*, u_*, \Delta_F, u^0_F)$ define
the Frobenius algebra identified for $\text{Rep}(G)$ in \cite{Bhardwaj:2017xup}. It is straightforward to check that the 
$(\mu_*, u_*, \Delta_F, u^0_F)$ defined above satisfy all of the axioms of a symmetric special Frobenius algebra, as outlined in appendix~\ref{app:Frobenius}.
We illustrate below the key steps in the verification, as specialized to $(\mu_*, u_*, \Delta_F, u_F^o)$:
\begin{itemize}
    \item associativity:
    \begin{equation}
        \mu_*( \mu_*( v_{g} \otimes v_{h}) \otimes v_{k} ) ) \: = \: 
        \mu_*( v_{g} \otimes \mu_*( v_{h} \otimes v_{k})),
    \end{equation}
    \item unit axiom:
    \begin{equation}
        \mu_*( v_{g} \otimes u_*(1) ) \: = \: v_{g} \: = \: \mu_*( u_*(1) \otimes v_{g} ),
    \end{equation}
    \item coassociativity:
    \begin{equation}
        ((\text{Id}_{\mathcal{A}}\otimes \Delta_F)\circ \Delta_F)(v_g)=((\Delta_F\otimes \text{Id}_{\mathcal{A}})\circ\Delta_F)(v_g),
    \end{equation}
    \item counit axiom:
 \begin{equation}
 ((\text{Id}_{\mathcal{A}}\otimes u^o_F)\circ\Delta_F)(v_g) = v_g = ((u^o_F\otimes \text{Id}_{\mathcal{A}})\circ\Delta_F)(v_g),
    \end{equation}
    \item Frobenius identities~(\ref{eq:Frobenius-identities}):
    \begin{equation}
\begin{tikzcd}
v_g\otimes v_h \arrow[rr, "\Delta_F\otimes \text{Id}_{\mathcal{A}}", maps to] \arrow[dd, "\mu_*"', maps to] &  & v_g\otimes v_g\otimes v_h \arrow[dd, "\text{Id}_{\mathcal{A}}\otimes \mu_*", maps to] &  & v_g\otimes v_h \arrow[dd, "\mu_*"', maps to] \arrow[rr, "\text{Id}_{\mathcal{A}}\otimes\Delta_F", maps to] &  & v_g\otimes v_h\otimes v_h \arrow[dd, "\mu_*\otimes\text{Id}_{\mathcal{A}}", maps to] 
\\
   &  &           &  &            &  &   
   \\
{\delta_{g,h}v_g} \arrow[rr, "\Delta_F"', maps to]    
&  & {\delta_{g,h}\ v_g\otimes v_g}  
&  & {\delta_{g,h}v_h} \arrow[rr, "\Delta_F"', maps to]      &  & {\delta_{g,h}\ v_h\otimes v_h}   
\end{tikzcd}
    \end{equation}
    \item special~(\ref{eq:conds}):
    \begin{equation}
    (u^o_F \circ u_*) (1) \: = \:
        \Lambda\left(u_*(1)\right) \: = \: \left(\sum_{g\in G}v_g \right)\left(\sum_{h\in G}h\right) = \text{dim}({\cal H}^*)=|G|,
    \end{equation}
        \begin{equation}
        \mu_*\left( \Delta_F(v_{g}) \right) \: = \: \mu_*( v_{g} \otimes v_{g} ) \: = \: v_{g},
    \end{equation}
    \item symmetric~(\ref{eq:Fr:symmetric}):
    \begin{equation}
 \begin{tikzcd}
v_g \arrow[dd, maps to, "\text{Id}_{\cal A} \otimes \gamma_{\cal A}"] \arrow[rr, "\overline{\gamma}_{\cal A} \otimes \text{Id}_{\cal A}", maps to]                             
&  & \left(\sum_{h\in G}h\otimes v_h\right)\otimes v_g \arrow[dd, "\text{Id}_{{\cal A}^*}\otimes (u^o_F\circ\mu_* )", maps to] 
\\
  &  & 
  \\
v_g\otimes \left(\sum_{h\in G}v_h\otimes h\right) \arrow[rr,  "(u^o_F\circ\mu_*)\otimes \text{Id}_{{\cal A}^*}", maps to] 
&  &  g   
\end{tikzcd}
    \end{equation}
    using~(\ref{eq:Fr:mult}), (\ref{eq:Fr:counit}), and where
    \begin{equation} \label{eq:coeval:repg}
        \gamma_{\cal A}(1) \: = \: \sum_{h \in G} v_h \otimes h,
        \: \: \:
        \overline{\gamma}_{\cal A}(1) \: = \:  \sum_{h \in G} h \otimes v_h,
    \end{equation}
    as follows from the fact that $\text{ev}_{\mathcal{A}}\circ \gamma_{\mathcal{A}}=\text{dim}(\mathcal{A}) = |G|$, and that in this basis $\text{ev}_{\mathcal{A}}(v_g\otimes h)=\delta_{g,h}$.
\end{itemize}
See e.g.~appendix~\ref{app:Frobenius} for a general discussion of the key axioms for a special symmetric Frobenius
algebra.

Let us now use this explicit formulation to construct Frobenius algebras other than that corresponding to the regular representation. As explained previously (c.f.~Equation (\ref{eq:quotientinclusion})), Hopf ideals of the Hopf algebra $\mathcal{H}$ give other Frobenius algebras by which we can gauge. In the special case of $\mathcal{H}=\C[G]$, the Hopf ideals correspond precisely to the normal subgroups $K\trianglelefteq G$ of $G$. Concretely, this correspondence is realized by the fact that the Hopf ideals of $\C[G]$ are all of the form of augmentation ideals $\C[G](\C[K])^+$ of Hopf subalgebras $\C[K]$ of normal subgroups, where $(\C[K])^+:=\text{ker}(u^o\vert_{\C[K]}:\C[K]\to\C)$. Moreover, $\C[G]/(\C[G](\C[K])^+)\cong \C[G/K]$ as Hopf algebras.

In the usual orbifold case, for which the symmetry category is $\text{Vec}(G)$, one can actually gauge any subgroup $H\leq G$, not necessarily normal. Since $\text{Rep}(G)$ is the corresponding quantum symmetry fusion category, one would expect that any subgroup $H$ should give a gaugeable Frobenius object in $\text{Rep}(G)$ as well. We now show that this is indeed the case.

Let us start with a Hopf subalgebra of $\C[G]$. It is known that all Hopf subalgebras in this case are of the form $\C[H]$ for $H\leq G$ a subgroup, not necessarily normal. We again construct the augmentation ideal $\C[G](\C[H])^+$ of the Hopf subalgebra $\C[H]$. This is a left ideal and a coideal of $\C[H]$. Although this is not a Hopf ideal, such that the coset space does not admit a well-defined Hopf algebra structure, it is still a coideal, so that at least it gives rise to a coalgebra structure on the quotient. That is, we have a sequence of \textit{coalgebras}
\begin{equation}
    \C[G](\C[H])^+\hookrightarrow\C[G]\xrightarrow{\pi} \C[G/H]
\end{equation}
This means that the dual vector space $(\C[G/H])^*$ is endowed with an algebra structure as
\begin{equation}
    \mu_*: \: v_{gH}\otimes v_{g'H}\mapsto \delta_{gH,g'H}\ v_{gH},
    \: \: \: \: \:
    u_*: \: 1\mapsto \sum_{gH\in G/H} v_{gH},
\end{equation}
for $\{v_{gH}\}_{gH\in G/H}$ the dual basis of the vector space $\C[G/H]$. Moreover, this algebra also carries a $\C[G]$-action, given on the generators by
\begin{equation}
    \rho_H: \: g\cdot v_{g'H}\mapsto v_{\pi(g)g'H}
\end{equation}
This means that $(\C[G/H])^*$ is an algebra object in $\text{Rep}(\C[G])$. 

Let us pause to give a more elementary way to understand how one associates a representation.
Given a subgroup $H$, build a vector space by associating basis vectors to elements of the coset $G/H$, and then letting $G$ act on those basis elements by their action on cosets.
In this fashion, given a subgroup $H \subset G$, we can associate a representation of $G$.

In the table below, we illustrate this in the case
of $G = S_3$:  
\begin{center}
    \begin{tabular}{ccc}
    Coset & Subalgebra & Representation \\ \hline
    $G/\langle b \rangle$ & ${\rm Span}[v_{K}, v_{a K}]$, $K = \langle b \rangle$ & $1+X$
\\
$G/\langle a \rangle$ & ${\rm Span}[v_{H}, v_{b H}, v_{b^2 H}]$, $H = \langle a \rangle$ & $1+Y$
\\
$G/1$ & ${\mathbb C}[G]^*$ & $1+X+2Y$\\
& & (the regular representation)
\end{tabular}
\end{center}
We will return to this classification in the example of Rep$(S_3)$ in section~\ref{sect:S3-subalg}.

We can finally realize $( {\mathbb C}[G/H])^*$ as a symmetric special Frobenius object by endowing it with the obvious Frobenius coalgebra structure analogous to that of $(\C[G])^*$:
\begin{equation}
    \Delta_F: \: v_{gH}\mapsto v_{gH}\otimes v_{gH},
    \: \: \: \: \:
    u^o_F: \: v_{gH}\mapsto 1.
\end{equation}
All in all, this allows us to produce a gaugeable object in $\text{Rep}(G)$ for any subgroup $H\leq G$, just as for the $\text{Vec}(G)$ case. This seems to hint at a correspondence between the gaugeable objects in $\text{Rep}(\mathcal{H})$ and those in $\text{Rep}(\mathcal{H}^*)$, though it is not obvious how this would work in these more general cases.

To summarize, given a subgroup $H \leq G$, we have constructed a special symmetric Frobenius algebra on 
$( {\mathbb C}[G/H] )^*$.

\subsection{Genus one partition functions}  \label{sect:genl-genus1}

First, recall that in an ordinary orbifold by a finite group $G$, the
genus one partition function has the form
\begin{equation}
    Z([X/G]) \: = \: \frac{1}{|G|} \sum_{gh=hg} Z_{g,h},
\end{equation}
where $Z_{g,h}$ represents the `partial trace,' 
schematically,
\begin{equation}
    {\scriptstyle g} \square_h \, ,
\end{equation}
the contribution to the $T^2$ partition function
from a worldsheet with a pair of branch cuts defined by $g, h \in G$.

Given an algebra object ${\cal A}$, morally the genus one partition function for a gauged
noninvertible 0-form symmetry has the form
\begin{equation}
    Z \: = \: Z_{{\cal A},{\cal A}}^{\cal A},
\end{equation}
which (at least formally) closely mirrors the form above.

Given a Frobenius algebra structure, we can now define a modular-invariant combination of partial traces $Z_{L_1,L_2}^{L_3}$.
The genus-one partition function is
\begin{eqnarray}
    Z 
    & = & \sum_{L_1, L_2, L_3} \mu^{L_3}_{L_1,L_2} \, \Delta^{L_2,L_1}_{L_3} \, Z^{L_3}_{L_1,L_2},
    \label{eq:Z:genl1}
\end{eqnarray}
where $\mu_{L_1, L_2}^{L_3}$ indicates components
of $\mu_*$, and $\Delta_{L_3}^{L_2,L_1}$ above indicates components of $\Delta_F$, the comultiplication in the Frobenius algebra.
(Higher-genus partition functions are similar; we shall discuss
them later in subsection~\ref{ssec:genus2}).

The components are defined formally as follows.
Given $\mu_*: {\cal A} \otimes {\cal A} \rightarrow {\cal A}$, the components are defined by corresponding maps
\begin{equation}
    \mu_{L_1, L_2}^{L_3}: \:
    {\rm Hom}_{{\rm Rep}({\cal H})}( L_1, {\cal A}) \otimes 
    {\rm Hom}_{{\rm Rep}({\cal H})}( L_2, {\cal A}) \: \longrightarrow \:
    {\rm Hom}_{{\rm Rep}({\cal H})}( {\cal A}, L_3).
\end{equation}
From Schur's lemma, each Hom will only receive contributions from portions of ${\cal A}$ in isomorphic representations.  We can compute these components explicitly from $\mu_*$ by normalizing by the intertwiners in
Hom$(L_1 \otimes L_2, L_3)$, as we will discuss later in examples.
Similarly, given $\Delta_F: {\cal A} \rightarrow {\cal A} \otimes {\cal A}$, the components are defined by corresponding maps
\begin{equation}
    \Delta_{L_1}^{L_2, L_3}: \:
    {\rm Hom}_{{\rm Rep}({\cal H})}( L_1, {\cal A}) \: \longrightarrow \:
    {\rm Hom}_{{\rm Rep}({\cal H})}( {\cal A}, L_2) \otimes
    {\rm Hom}_{{\rm Rep}({\cal H})}( {\cal A}, L_3).
\end{equation}
These can be obtained from $\Delta_F$ by normalizing by a basis for Hom$(L_1, L_2 \otimes L_3)$,
which can be obtained from intertwiners using (co)evaluation maps in several possible ways, as we will discuss later in examples.

\begin{figure}[h]
\centering
	\begin{tikzpicture}
			\draw[thin] (0,0)--(5,0);
			\draw[thin] (5,0)--(5,5);
			\draw[thin] (0,0)--(0,5);
			\draw[thin] (0,5)--(5,5);
			\draw[very thick,->] (2.5,0)--(3,1.25);
			\draw[very thick] (3,1.25)--(3.5,2.5);
			\draw[very thick,->] (1.5,2.5)--(2,3.75);
			\draw[very thick] (2,3.75)--(2.5,5);
			\draw[very thick,->] (5,2.5)--(4.25,2.5);
			\draw[very thick] (4.25,2.5)--(3.5,2.5);
			\draw[very thick,->] (3.5,2.5)--(2.5,2.5);
			\draw[very thick] (2.5,2.5)--(1.5,2.5);
			\draw[very thick,->] (1.5,2.5)--(0.75,2.5);
			\draw[very thick] (0.75,2.5)--(0,2.5);
			\node at (3.3,1) {$L_1$};
			\node at (1.7,4) {$L_1$};
			\node at (0.7,2.9) {$L_2$};
			\node at (4.3,2.9) {$L_2$};
			\node at (2.5,2.9) {$L_3$};
	        \node at (3.5,2.8) {$\mu_*$};
            \node at (1.5,2.2) {$\Delta_F$};
            \node at (-0.2,5.2) {$\tau$};
            \node at (5.2,-0.2) {$1$};
            \node at (5.2,5.2) {$\tau+1$};
            \node at (-0.2,-0.2) {$0$};

            \draw[thick,color=blue,->,opacity=0.5] (3.5,2.5) [partial ellipse=-110:180:0.25cm and 0.25cm];
		      \draw[thick,color=blue,->,opacity=0.5] (1.5,2.5) [partial ellipse=0:180:0.25cm and 0.25cm];
		\end{tikzpicture}
  \caption{Resolution of four-point vertex in the Frobenius algebra ${\cal A}$ into a pair of three-point vertices.}
  \label{fig:genus1:Fr-resn}
\end{figure}
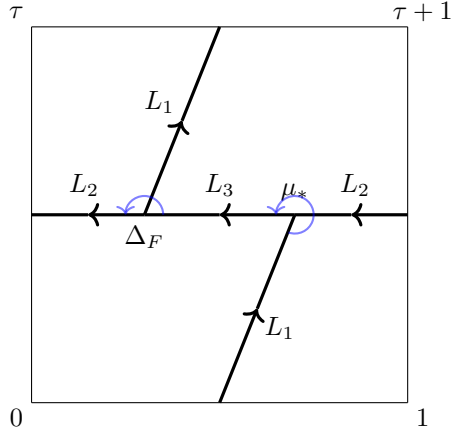

We can understand expression~(\ref{eq:Z:genl1}) diagramatically as arising naturally from the description of resolving four-point junctions of lines in the Frobenius algebra ${\cal A}$.  Schematically, if we write the
resolution in the form of figure~\ref{fig:genus1:Fr-resn},  
where we denote the components $\mu_{L_i,L_j}^{L_k}$ of $\mu_*$ by
\begin{equation}
    \begin{tikzpicture}
        \draw[very thick, ->] (1.25,2)--(1.25,3);
        \draw[very thick] (1.25,3)--(1.25,4);
        \draw[very thick] (0.6,0.96)--(1.25,2);
        \draw[very thick,->] (0,0)--(0.6,0.96);
        \draw[very thick] (1.25,2)--(1.9,0.96);
        \draw[very thick,->] (2.5,0)--(1.9,0.96); \draw[thick,color=blue,->,opacity=0.5] (1.25,2) [partial ellipse=-120:90:0.2cm and 0.2cm];
        \node at (2.7,0.3) {$L_j$};
        \node at (0.45,0.2) {$L_i$};
        \node at (1.6,3) {$L_k$};
        \node at (1.6,2) {$\mu_*$};
    \end{tikzpicture}
\end{equation}
and the components $\Delta_{L_i}^{L_j,L_k}$ of $\Delta_F$ by
\begin{equation}
    \begin{tikzpicture}
        \draw[very thick, ->] (1.25,0) -- (1.25,1); \draw[very thick] (1.25,1) -- (1.25,2);
        \draw[very thick, ->] (1.25,2) -- (0.6,3.04);  \draw[very thick] (0.6,3.04)--(0,4);
        \draw[very thick, ->] (1.25,2) -- (1.9,3.04); \draw[very thick] (1.9,3.04) -- (2.55,4);
        \draw[thick,color=blue,->,opacity=0.5] (1.25,2) [partial ellipse=-90:120:0.2cm and 0.2cm];
        \node at (1.6,1) {$L_i$};
        \node at (0.55,3.7) {$L_j$};
        \node at (1.9,3.7) {$L_k$};
        \node at (1.6,2) {$\Delta_F$};
    \end{tikzpicture}
\end{equation}
then we read off from Figure~\ref{fig:genus1:Fr-resn} the
\begin{equation}
    \mu_{L_1, L_2}^{L_3} \Delta_{L_3}^{L_2, L_1}
\end{equation}
factor, and then after removing the Frobenius algebra structures, we are left with the partial trace of figure~\ref{fig:noninv-partial-trace}, which defines $Z_{L_1, L_2}^{L_3}$.

In passing, we note that equation~(\ref{eq:Z:genl1}) omits the possibility of the junction vector spaces having dimension greater than one.  However, in this paper we are specializing to multiplicity-free fusion categories, and will return to more general cases in
our followup work \cite{toappear}.

In passing, implicit here is a choice of normalization of $\Delta_F$, which is defined by the
`special' part of the special symmetric Frobenius algebra axioms.  This also informs the implicit
normalization of the coevaluation map.

In section~\ref{sect:modinv}, we will check that this partition function is always modular invariant.

It is natural to ask whether there exists an analogue of discrete torsion 
\cite{Vafa:1986wx,Sharpe:2000ki} in noninvertible symmetry gaugings.
Recall that in ordinary orbifolds, discrete torsion is a set of modular-invariant phases that can be added
to weight the various partial traces.  For ordinary group orbifolds, realized as Vec$(G) = {\rm Rep}( {\mathbb C}[G]^*)$ gaugings for $G$ a finite group,
we will discuss how discrete torsion arises later in section~\ref{sect:ord-orbifold}.
For noninvertible symmetries, discrete torsion will arise when there are multiple nonequivalent Frobenius algebra structures we could choose for a given algebra object.  In this case we do not have a first-principles classification (as we do with second cohomology in the grouplike case), but we will see in examples
later in this paper that there certainly seem to exist modular-invariant phases that can be added to partition functions,
forming an analogue of discrete torsion.  (That said, we have not checked e.g.~multiloop factorization, so it is possible that some choices are not physically sensible.)  Also, the reference
\cite[theorem 2]{ostrik1} 
suggests that discrete torsion in a Rep$(G)$ quotient should match discrete torsion in a $G$ orbifold, which we will find to be consistent with our examples.  We hope to return to questions of discrete torsion in noninvertible symmetry gaugings in
future work.

\subsection{Modular invariance}
\label{sect:modinv}

\begin{figure}
	\begin{subfigure}{0.3\textwidth}
		\centering
		\begin{tikzpicture}
			\draw[->] (-1,-2) -- (-0.5,-1.5);
			\draw (-0.5,-1.5) -- (0,-1);
			\draw[->] (1,-2) -- (0.5,-1.5);
			\draw (0.5,-1.5) -- (0,-1);
			\draw[->] (0,-1) -- (0,0);
			\draw (0,0) -- (0,1);
			\draw[->] (0,1) -- (-0.5,1.5);
			\draw (-0.5,1.5) -- (-1,2);
			\draw[->] (0,1) -- (0.5,1.5);
			\draw (0.5,1.5) -- (1,2);
			\filldraw[black] (0,1) circle (2pt);
			\filldraw[black] (0,-1) circle (2pt);
			\node at (0,1.5) {$\Delta_F$};
			\node at (0,-1.5) {$\mu_*$};
			\draw[thick,color=blue,->,opacity=0.5] (0,-1) [partial ellipse=-135:90:0.2cm and 0.2cm];
			\draw[thick,color=blue,->,opacity=0.5] (0,1) [partial ellipse=-90:135:0.2cm and 0.2cm];
		\end{tikzpicture}
    \caption{}
	\label{frobassoc1}
	\end{subfigure}
	\begin{subfigure}{0.3\textwidth}
		\centering
		\begin{tikzpicture}
			\draw[->] (0,0) [partial ellipse=-90:-180:0.5cm and 0.5cm];
			\draw[->] (0,0) [partial ellipse=-90:0:0.5cm and 0.5cm];
			\draw (-0.5,0) -- (-0.5,1.5);
			\draw[->] (0,-1.5) -- (0,-1);
			\draw (0,-1) -- (0,-0.5);
			\draw (1,0) [partial ellipse=180:0:0.5cm and 0.5cm];
			\draw[->] (1.5,-1.5) -- (1.5,0);
			\draw[->] (1,0.5) -- (1,1);
			\draw (1,1) -- (1,1.5);
			\filldraw[black] (1,0.5) circle (2pt);
			\filldraw[black] (0,-0.5) circle (2pt);
			\node at (0,0) {$\Delta_F$};
			\node at (1,0) {$\mu_*$};
			\draw[thick,color=blue,->,opacity=0.5] (0,-0.5) [partial ellipse=-90:170:0.2cm and 0.2cm];
			\draw[thick,color=blue,->,opacity=0.5] (1,0.5) [partial ellipse=-170:90:0.2cm and 0.2cm];
		\end{tikzpicture}
    \caption{}
	\label{frobassoc2}
	\end{subfigure}
	\begin{subfigure}{0.3\textwidth}
		\centering
		\begin{tikzpicture}
			\draw[->] (0,0) [partial ellipse=-90:-180:0.5cm and 0.5cm];
			\draw[->] (0,0) [partial ellipse=-90:0:0.5cm and 0.5cm];
			\draw (-1,0) [partial ellipse=0:180:0.5cm and 0.5cm];
			\draw[->] (-1.5,-1.5) -- (-1.5,0);
			\draw[->] (0,-1.5) -- (0,-1);
			\draw (0,-1) -- (0,-0.5);
			\draw[->] (-1,0.5) -- (-1,1);
			\draw (-1,1) -- (-1,1.5);
            \draw (0.5,0) -- (0.5,1.5);
			\filldraw[black] (-1,0.5) circle (2pt);
			\filldraw[black] (0,-0.5) circle (2pt);
			\node at (-1,0) {$\mu_*$};
			\node at (0,0) {$\Delta_F$};
			\draw[thick,color=blue,->,opacity=0.5] (0,-0.5) [partial ellipse=-90:170:0.2cm and 0.2cm];
			\draw[thick,color=blue,->,opacity=0.5] (-1,0.5) [partial ellipse=-170:90:0.2cm and 0.2cm];
		\end{tikzpicture}
    \caption{}
	\label{frobassoc3}
	\end{subfigure}
 \caption{Three ways to resolve a four-way junction into a multiplication and a comultiplication.  All of the lines pictured are the algebra object $\mcA$.}
 \label{frobassoc}
\end{figure}
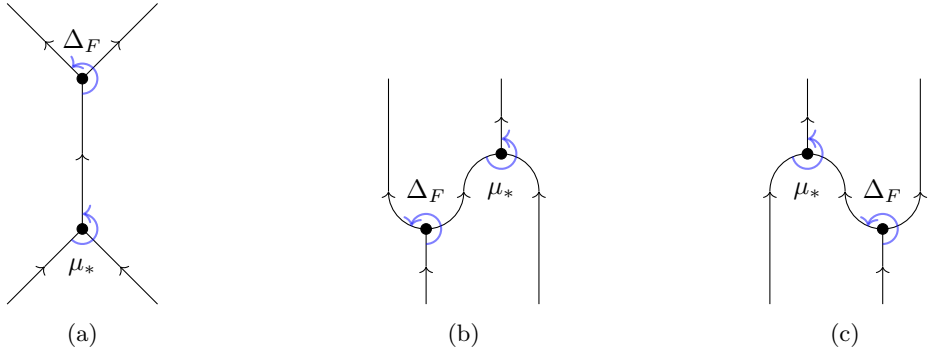

In this section we shall demonstrate that the general expression for partition functions given earlier in equation~(\ref{eq:Z:genl1}) is 
modular-invariant.  In order to do this we will need to write out the associativity condition satisfied by the Frobenius algebra in components.  Diagramatically, associativity of the Frobenius algebra can be expressed as the requirement that Figure~\ref{frobassoc1}, Figure~\ref{frobassoc2} and Figure~\ref{frobassoc3} are equivalent \cite[(3.29)]{Fuchs:2002cm} (more generally it is the statement that any such resolution of a four-way junction is equivalent).

In order to write these conditions in a computationally useful form, we break each line (labeled by $\mcA$) into components.  This gives a sum over diagrams labeled by simple objects with $\mu$ and $\Delta$ as coefficients.  In order to relate any of the two resulting expressions we need to apply a swap move, then equate the coefficients of matching diagrams.  Applying the swap shown in Figure~\ref{fig:crossing-kernel} to Figure~\ref{frobassoc1} and equating coefficients with those of Figure~\ref{frobassoc3} produces
\begin{equation}
\label{frobassoc_components}
	\sum_{L_3} \mu_{L_1, L_2}^{L_3} \Delta^{L_2, L_1}_{L_3} \, 
	\tilde{K}^{\overline{L}_1, L_2}_{\overline{L}_2, L_1}(\overline{L}_3,
	\overline{L}_4)
	\: = \:
	\mu_{L_1, L_4}^{L_2} \Delta^{L_4, L_1}_{L_2}.
\end{equation}

\begin{figure}
    \begin{subfigure}{0.3\textwidth}
    \centering
    \begin{tikzpicture}
        \draw[->](-1,-1) -- (-0.5,-0.5);
        \draw (-0.5,-0.5) -- (0,0);
        \node at (-1,-0.5) {$\mcA$};
        \draw[->] (1,-1) -- (0.5,-0.5);
        \draw (0.5,-0.5) -- (0,0);
        \node at (1,-0.5) {$\mcA$};
        \draw[->] (0,0) -- (0.5,0.5);
        \draw (0.5,0.5) -- (1,1);
        \node at (1,0.5) {$\mcA$};
        \draw[->] (0,0) -- (-0.5,0.5);
        \draw (-0.5,0.5) -- (-1,1);
        \node at (-1,0.5) {$\mcA$};
    \end{tikzpicture}
    \caption{}
    \label{algob_4junc1}
    \end{subfigure}
    \begin{subfigure}{0.3\textwidth}
    \centering
    \begin{tikzpicture}
		\draw[->] (-1,-2) -- (-0.5,-1.5);
		\draw (-0.5,-1.5) -- (0,-1);
        \node at (-1,-1.5) {$L_1$};
		\draw[->] (1,-2) -- (0.5,-1.5);
		\draw (0.5,-1.5) -- (0,-1);
        \node at (1,-1.5) {$L_2$};
		\draw[->] (0,-1) -- (0,0);
        \node at (0.5,0) {$L_5$};
		\draw (0,0) -- (0,1);
		\draw[->] (0,1) -- (-0.5,1.5);
		\draw (-0.5,1.5) -- (-1,2);
        \node at (-1,1.5) {$L_4$};
		\draw[->] (0,1) -- (0.5,1.5);
		\draw (0.5,1.5) -- (1,2);
        \node at (1,1.5) {$L_3$};
		\filldraw[black] (0,1) circle (2pt);
		\filldraw[black] (0,-1) circle (2pt);
		\node at (0,1.5) {$\Delta$};
		\node at (0,-1.5) {$\mu$};
		\draw[thick,color=blue,->,opacity=0.5] (0,-1) [partial ellipse=-135:90:0.2cm and 0.2cm];
		\draw[thick,color=blue,->,opacity=0.5] (0,1) [partial ellipse=-90:135:0.2cm and 0.2cm];
	\end{tikzpicture}
    \caption{}
    \label{algob_4junc2}
    \end{subfigure}
    \begin{subfigure}{0.3\textwidth}
    \centering
    \begin{tikzpicture}
		\draw[->] (-1,-2) -- (-0.75,-1.75);
        \draw[] (-0.75,-1.75) -- (-0.5,-1.5);
        \filldraw[black] (-0.5,-1.5) circle (2pt);
        \node at (-0.25,-1.75) {$\epsilon$};
		\draw[] (-0.5,-1.5) -- (-0.25,-1.25);
        \draw[->] (0,-1) -- (-0.25,-1.25);
        \node at (-1,-1.5) {$L_1$};
        \draw[dashed] (-0.5,-1.5) -- (-1,-1);
        \node at (-0.5,-1.0) {$\overline{L}_1$};
		\draw[->] (1,-2) -- (0.5,-1.5);
		\draw (0.5,-1.5) -- (0,-1);
        \node at (1,-1.5) {$L_2$};
		\draw[->] (0,-1) -- (0,0);
        \node at (0.5,0) {$L_6$};
		\draw (0,0) -- (0,1);
		\draw[->] (0,1) -- (-0.5,1.5);
		\draw (-0.5,1.5) -- (-1,2);
        \node at (-1,1.5) {$L_4$};
		\draw[] (0,1) -- (0.25,1.25);
        \filldraw[black] (0.5,1.5) circle (2pt);
        \node at (0.25,1.75) {$\Gamma$};
        \draw[->] (0.5,1.5) -- (0.25,1.25);
        \node at (0.5,1.0) {$\overline{L}_3$};
        \draw[dashed] (0.5,1.5) -- (1,1);
        \draw[->] (0.5,1.5) -- (0.75,1.75);
		\draw (0.75,1.75) -- (1,2);
        \node at (1,1.5) {$L_3$};
		\filldraw[black] (0,1) circle (2pt);
		\filldraw[black] (0,-1) circle (2pt);
		\node at (-0.4,0.9) {$\mu$};
		\node at (0.5,-1.0) {$\Delta$};
		\draw[thick,color=blue,->,opacity=0.5] (0,-1) [partial ellipse=-45:225:0.2cm and 0.2cm];
		\draw[thick,color=blue,->,opacity=0.5] (0,1) [partial ellipse=-90:135:0.2cm and 0.2cm];
	\end{tikzpicture}
    \caption{}
    \label{algob_4junc3}
    \end{subfigure}
\caption{Two ways of resolving a four-way junction of algebra object lines.}
\label{algob_4junc}
\end{figure}
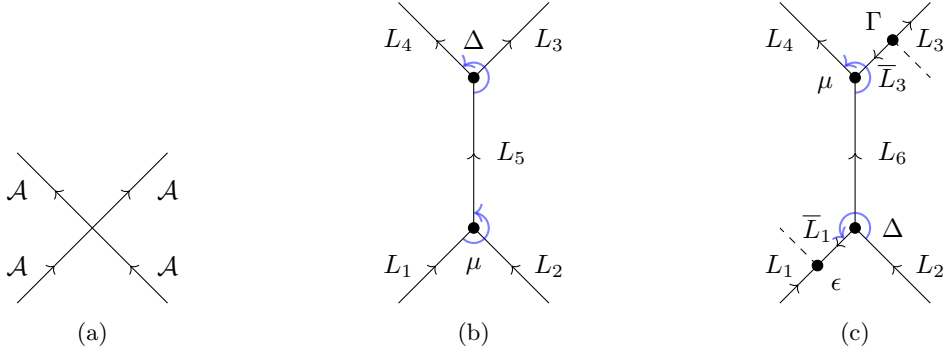

We will need one additional relation.  Figure~\ref{algob_4junc1} shows a four-way junction of algebra objects, which can be broken up into two three-way junctions of simple lines in multiple ways.  Two of these ways are shown in Figures \ref{algob_4junc2} and \ref{algob_4junc3}.  As before, breaking the $\mcA$ lines into components yields a sum over diagrams with one $\mu$ and one $\Delta$ as components.  In order to match up the diagrams in each sum, we need to apply the cyclic transformation (\ref{eq:fig:cyclic}) to the lower junction in Figure~\ref{algob_4junc2}.  Doing so and equating the coefficients of identical diagrams produces the identity
\be
\label{mcmrel}
\mu_{L_1,L_2}^{L_5}\Delta_{L_5}^{L_4,L_3}\tilde{K}^{\overline{L}_1,1}_{\overline{L}_2,L_5}(\overline{L}_5,L_1)=\mu_{L_5,\overline{L}_3}^{L_4}\Delta_{L_2}^{\overline{L}_1,L_5}\epsilon_{L_1}\gamma_{L_3}.
\ee
This expression includes coefficients arising from applying the evaluation map $\epsilon$ and coevaluation map $\gamma$.  Below when we use this equation we will take $L_1=L_3$ and these coefficients will cancel each other.

Now we can apply these relations to modular transformations.  The general form of the modular $T$ transformation of the partition function~(\ref{eq:Z:genl1}) is
\begin{eqnarray}
	Z_{T^2}(\tau + 1, \overline{\tau} + 1) & = &
	\sum_{L_1, L_2, L_3} \mu_{L_1, L_2}^{L_3} \Delta^{L_2, L_1}_{L_3} \,
	\sum_{L_4} \tilde{K}^{\overline{L}_1, L_2}_{\overline{L}_2, L_1}(\overline{L}_3,
	\overline{L}_4) \,
	Z_{L_1, L_4}^{L_2}(\tau, \overline{\tau}),
 \\
 & = &
 \sum_{L_1, L_2, L_4} \mu^{L_2}_{L_1, L_4} \Delta^{L_4,L_1}_{L_2} \, Z^{L_2}_{L_1, L_4}(\tau,\overline{\tau}),
 \\
 & = & Z_{T^2}(\tau,\overline{\tau}),
\end{eqnarray}
where we have used~(\ref{frobassoc_components}).  Thus, the partition function~(\ref{eq:Z:genl1}) is invariant under modular $T$ transformations.

We can check invariance under modular $S$ transformations similarly.  The partition function~(\ref{eq:Z:genl1}) transforms as
\begin{eqnarray}
    Z_{T^2}(-1/\tau, -1/\overline{\tau}) & = & \sum_{L_1, L_2, L_3} \mu^{L_3}_{L_1, L_2} \Delta^{L_2, L_1}_{L_3} \, Z^{L_3}_{L_1, L_2}(-1/\tau, -1/\overline{\tau}),
    \\
    & = &  \sum_{L_1, L_2, L_3} \mu^{L_3}_{L_1, L_2} \Delta^{L_2, L_1}_{L_3} \, \sum_{L_4} \tilde{K}_{\overline{L}_4, L_2}^{\overline{L}_1, 1}( \overline{L}_2, L_1) \tilde{K}_{\overline{L}_2, L_1}^{\overline{L}_1, L_2}(\overline{L}_3, \overline{L}_4) \, Z^{L_4}_{L_2, \overline{L}_1}(\tau, \overline{\tau}),
    \\
    & = & \sum_{L_1, L_2, L_4} \mu^{L_2}_{L_1, L_4} \Delta^{L_4, L_1}_{L_2} \, \tilde{K}_{\overline{L}_4, L_2}^{\overline{L}_1, 1}( \overline{L}_2, L_1)  \, Z^{L_4}_{L_2, \overline{L}_1}(\tau, \overline{\tau}),
    \\
    & = & \sum_{L_1,L_2,L_4} \mu_{L_2,\overline{L}_1}^{L_4}\Delta_{L_4}^{\overline{L}_1,L_2} Z^{L_4}_{L_2, \overline{L}_1}(\tau, \overline{\tau}),
     \\
 & = & Z_{T^2}(\tau,\overline{\tau}),
\end{eqnarray}
using~(\ref{frobassoc_components}) and (\ref{mcmrel}).

\subsection{Higher genus partition functions and change of triangulation} \label{ssec:genus2}

So far we have concentrated on computations pertaining to a genus $1$ worldsheet. The formalism used presently is not exclusive to this but also extends to more general worldsheets. In the present section, we briefly discuss these computations for a genus $2$ worldsheet, and observe how the familiar discrete torsion factors naturally arise in the special case of a group-like symmetry. 

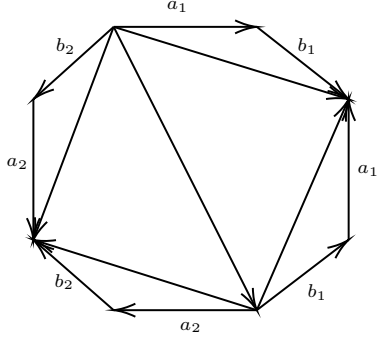
\begin{figure}
    \centering
\tikzset{every picture/.style={line width=0.75pt}} %set default line width to 0.75pt        

\begin{tikzpicture}[x=0.75pt,y=0.75pt,yscale=-1,xscale=1]

%Straight Lines [id:da8077614003171456] 
\draw    (95.5,74) -- (165.67,74) ;
\draw [shift={(167.67,74)}, rotate = 180] [color={rgb, 255:red, 0; green, 0; blue, 0 }  ][line width=0.75]    (10.93,-3.29) .. controls (6.95,-1.4) and (3.31,-0.3) .. (0,0) .. controls (3.31,0.3) and (6.95,1.4) .. (10.93,3.29)   ;
%Straight Lines [id:da36313262239688515] 
\draw    (167.67,74) -- (212.43,109.26) ;
\draw [shift={(214,110.5)}, rotate = 218.23] [color={rgb, 255:red, 0; green, 0; blue, 0 }  ][line width=0.75]    (10.93,-3.29) .. controls (6.95,-1.4) and (3.31,-0.3) .. (0,0) .. controls (3.31,0.3) and (6.95,1.4) .. (10.93,3.29)   ;
%Straight Lines [id:da9983809556046528] 
\draw    (214,181.5) -- (214,112.5) ;
\draw [shift={(214,110.5)}, rotate = 90] [color={rgb, 255:red, 0; green, 0; blue, 0 }  ][line width=0.75]    (10.93,-3.29) .. controls (6.95,-1.4) and (3.31,-0.3) .. (0,0) .. controls (3.31,0.3) and (6.95,1.4) .. (10.93,3.29)   ;
%Straight Lines [id:da02072819793485392] 
\draw    (167.67,217) -- (212.41,182.72) ;
\draw [shift={(214,181.5)}, rotate = 142.54] [color={rgb, 255:red, 0; green, 0; blue, 0 }  ][line width=0.75]    (10.93,-3.29) .. controls (6.95,-1.4) and (3.31,-0.3) .. (0,0) .. controls (3.31,0.3) and (6.95,1.4) .. (10.93,3.29)   ;
%Straight Lines [id:da511852036978925] 
\draw    (167.67,217) -- (97.5,217) ;
\draw [shift={(95.5,217)}, rotate = 360] [color={rgb, 255:red, 0; green, 0; blue, 0 }  ][line width=0.75]    (10.93,-3.29) .. controls (6.95,-1.4) and (3.31,-0.3) .. (0,0) .. controls (3.31,0.3) and (6.95,1.4) .. (10.93,3.29)   ;
%Straight Lines [id:da7788868303119147] 
\draw    (95.5,217) -- (56.5,182.82) ;
\draw [shift={(55,181.5)}, rotate = 41.24] [color={rgb, 255:red, 0; green, 0; blue, 0 }  ][line width=0.75]    (10.93,-3.29) .. controls (6.95,-1.4) and (3.31,-0.3) .. (0,0) .. controls (3.31,0.3) and (6.95,1.4) .. (10.93,3.29)   ;
%Straight Lines [id:da10674626913638674] 
\draw    (55,110.5) -- (55,179.5) ;
\draw [shift={(55,181.5)}, rotate = 270] [color={rgb, 255:red, 0; green, 0; blue, 0 }  ][line width=0.75]    (10.93,-3.29) .. controls (6.95,-1.4) and (3.31,-0.3) .. (0,0) .. controls (3.31,0.3) and (6.95,1.4) .. (10.93,3.29)   ;
%Straight Lines [id:da0434370329192979] 
\draw    (95.5,74) -- (56.49,109.16) ;
\draw [shift={(55,110.5)}, rotate = 317.97] [color={rgb, 255:red, 0; green, 0; blue, 0 }  ][line width=0.75]    (10.93,-3.29) .. controls (6.95,-1.4) and (3.31,-0.3) .. (0,0) .. controls (3.31,0.3) and (6.95,1.4) .. (10.93,3.29)   ;
%Straight Lines [id:da5883991488540037] 
\draw    (95.5,74) -- (212.09,109.91) ;
\draw [shift={(214,110.5)}, rotate = 197.12] [color={rgb, 255:red, 0; green, 0; blue, 0 }  ][line width=0.75]    (10.93,-3.29) .. controls (6.95,-1.4) and (3.31,-0.3) .. (0,0) .. controls (3.31,0.3) and (6.95,1.4) .. (10.93,3.29)   ;
%Straight Lines [id:da004361021413345378] 
\draw    (95.5,74) -- (55.71,179.63) ;
\draw [shift={(55,181.5)}, rotate = 290.64] [color={rgb, 255:red, 0; green, 0; blue, 0 }  ][line width=0.75]    (10.93,-3.29) .. controls (6.95,-1.4) and (3.31,-0.3) .. (0,0) .. controls (3.31,0.3) and (6.95,1.4) .. (10.93,3.29)   ;
%Straight Lines [id:da9213106859147207] 
\draw    (167.67,217) -- (56.91,182.1) ;
\draw [shift={(55,181.5)}, rotate = 17.49] [color={rgb, 255:red, 0; green, 0; blue, 0 }  ][line width=0.75]    (10.93,-3.29) .. controls (6.95,-1.4) and (3.31,-0.3) .. (0,0) .. controls (3.31,0.3) and (6.95,1.4) .. (10.93,3.29)   ;
%Straight Lines [id:da0034025853974857245] 
\draw    (167.67,217) -- (213.2,112.33) ;
\draw [shift={(214,110.5)}, rotate = 113.51] [color={rgb, 255:red, 0; green, 0; blue, 0 }  ][line width=0.75]    (10.93,-3.29) .. controls (6.95,-1.4) and (3.31,-0.3) .. (0,0) .. controls (3.31,0.3) and (6.95,1.4) .. (10.93,3.29)   ;
%Straight Lines [id:da7878114803915275] 
\draw    (95.5,74) -- (166.77,215.21) ;
\draw [shift={(167.67,217)}, rotate = 243.22] [color={rgb, 255:red, 0; green, 0; blue, 0 }  ][line width=0.75]    (10.93,-3.29) .. controls (6.95,-1.4) and (3.31,-0.3) .. (0,0) .. controls (3.31,0.3) and (6.95,1.4) .. (10.93,3.29)   ;

% Text Node
\draw (120.78,59.4) node [anchor=north west][inner sep=0.75pt]  [font=\scriptsize]  {$a_{1}$};
% Text Node
\draw (186.78,78.4) node [anchor=north west][inner sep=0.75pt]  [font=\scriptsize]  {$b_{1}$};
% Text Node
\draw (217,142.4) node [anchor=north west][inner sep=0.75pt]  [font=\scriptsize]  {$a_{1}$};
% Text Node
\draw (191.78,202.4) node [anchor=north west][inner sep=0.75pt]  [font=\scriptsize]  {$b_{1}$};
% Text Node
\draw (127.44,221.4) node [anchor=north west][inner sep=0.75pt]  [font=\scriptsize]  {$a_{2}$};
% Text Node
\draw (64.11,197.07) node [anchor=north west][inner sep=0.75pt]  [font=\scriptsize]  {$b_{2}$};
% Text Node
\draw (40,138.73) node [anchor=north west][inner sep=0.75pt]  [font=\scriptsize]  {$a_{2}$};
% Text Node
\draw (64.78,78.4) node [anchor=north west][inner sep=0.75pt]  [font=\scriptsize]  {$b_{2}$};
\end{tikzpicture}
    \caption{Triangulation of a genus $2$ surface \cite{Aspinwall:2000xv}.}
    \label{fig:g2triangulation}
\end{figure}

\begin{figure}
    \centering
\tikzset{every picture/.style={line width=0.75pt}} %set default line width to 0.75pt        

\begin{tikzpicture}[x=0.75pt,y=0.75pt,yscale=-1,xscale=1]

%Straight Lines [id:da09879049727609068] 
\draw    (119.44,39) -- (178.68,79.21) ;
\draw [shift={(180.33,80.33)}, rotate = 214.17] [color={rgb, 255:red, 0; green, 0; blue, 0 }  ][line width=0.75]    (10.93,-3.29) .. controls (6.95,-1.4) and (3.31,-0.3) .. (0,0) .. controls (3.31,0.3) and (6.95,1.4) .. (10.93,3.29)   ;
%Straight Lines [id:da022679162896319216] 
\draw    (240.44,39) -- (181.98,79.2) ;
\draw [shift={(180.33,80.33)}, rotate = 325.49] [color={rgb, 255:red, 0; green, 0; blue, 0 }  ][line width=0.75]    (10.93,-3.29) .. controls (6.95,-1.4) and (3.31,-0.3) .. (0,0) .. controls (3.31,0.3) and (6.95,1.4) .. (10.93,3.29)   ;
%Straight Lines [id:da6784132674816423] 
\draw    (180.33,80.33) -- (180.33,117.67) ;
\draw [shift={(180.33,119.67)}, rotate = 270] [color={rgb, 255:red, 0; green, 0; blue, 0 }  ][line width=0.75]    (10.93,-3.29) .. controls (6.95,-1.4) and (3.31,-0.3) .. (0,0) .. controls (3.31,0.3) and (6.95,1.4) .. (10.93,3.29)   ;
%Straight Lines [id:da641402190441402] 
\draw    (180.33,119.67) -- (238.44,119.67) ;
\draw [shift={(240.44,119.67)}, rotate = 180] [color={rgb, 255:red, 0; green, 0; blue, 0 }  ][line width=0.75]    (10.93,-3.29) .. controls (6.95,-1.4) and (3.31,-0.3) .. (0,0) .. controls (3.31,0.3) and (6.95,1.4) .. (10.93,3.29)   ;
%Straight Lines [id:da5715443568775722] 
\draw    (240.44,119.67) -- (298.66,81.43) ;
\draw [shift={(300.33,80.33)}, rotate = 146.7] [color={rgb, 255:red, 0; green, 0; blue, 0 }  ][line width=0.75]    (10.93,-3.29) .. controls (6.95,-1.4) and (3.31,-0.3) .. (0,0) .. controls (3.31,0.3) and (6.95,1.4) .. (10.93,3.29)   ;
%Straight Lines [id:da7770854423418123] 
\draw    (240.44,119.67) -- (298.67,158.56) ;
\draw [shift={(300.33,159.67)}, rotate = 213.74] [color={rgb, 255:red, 0; green, 0; blue, 0 }  ][line width=0.75]    (10.93,-3.29) .. controls (6.95,-1.4) and (3.31,-0.3) .. (0,0) .. controls (3.31,0.3) and (6.95,1.4) .. (10.93,3.29)   ;
%Straight Lines [id:da054430206837292605] 
\draw    (180.33,119.67) -- (180.33,188) ;
\draw [shift={(180.33,190)}, rotate = 270] [color={rgb, 255:red, 0; green, 0; blue, 0 }  ][line width=0.75]    (10.93,-3.29) .. controls (6.95,-1.4) and (3.31,-0.3) .. (0,0) .. controls (3.31,0.3) and (6.95,1.4) .. (10.93,3.29)   ;
%Straight Lines [id:da5048436914641816] 
\draw    (30.22,159.67) -- (91.75,189.14) ;
\draw [shift={(93.56,190)}, rotate = 205.59] [color={rgb, 255:red, 0; green, 0; blue, 0 }  ][line width=0.75]    (10.93,-3.29) .. controls (6.95,-1.4) and (3.31,-0.3) .. (0,0) .. controls (3.31,0.3) and (6.95,1.4) .. (10.93,3.29)   ;
%Straight Lines [id:da7956678657628669] 
\draw    (29.56,220.5) -- (90.76,190.87) ;
\draw [shift={(92.56,190)}, rotate = 154.17] [color={rgb, 255:red, 0; green, 0; blue, 0 }  ][line width=0.75]    (10.93,-3.29) .. controls (6.95,-1.4) and (3.31,-0.3) .. (0,0) .. controls (3.31,0.3) and (6.95,1.4) .. (10.93,3.29)   ;
%Straight Lines [id:da3332475286582053] 
\draw    (92.56,190) -- (178.33,190) ;
\draw [shift={(180.33,190)}, rotate = 180] [color={rgb, 255:red, 0; green, 0; blue, 0 }  ][line width=0.75]    (10.93,-3.29) .. controls (6.95,-1.4) and (3.31,-0.3) .. (0,0) .. controls (3.31,0.3) and (6.95,1.4) .. (10.93,3.29)   ;
%Straight Lines [id:da3810245695382757] 
\draw    (180.33,190) -- (238.78,228.57) ;
\draw [shift={(240.44,229.67)}, rotate = 213.42] [color={rgb, 255:red, 0; green, 0; blue, 0 }  ][line width=0.75]    (10.93,-3.29) .. controls (6.95,-1.4) and (3.31,-0.3) .. (0,0) .. controls (3.31,0.3) and (6.95,1.4) .. (10.93,3.29)   ;
%Straight Lines [id:da8269806107726623] 
\draw    (240.44,229.67) -- (298.67,191.1) ;
\draw [shift={(300.33,190)}, rotate = 146.48] [color={rgb, 255:red, 0; green, 0; blue, 0 }  ][line width=0.75]    (10.93,-3.29) .. controls (6.95,-1.4) and (3.31,-0.3) .. (0,0) .. controls (3.31,0.3) and (6.95,1.4) .. (10.93,3.29)   ;
%Straight Lines [id:da8293040461117034] 
\draw    (240.44,229.67) -- (298.55,259.42) ;
\draw [shift={(300.33,260.33)}, rotate = 207.12] [color={rgb, 255:red, 0; green, 0; blue, 0 }  ][line width=0.75]    (10.93,-3.29) .. controls (6.95,-1.4) and (3.31,-0.3) .. (0,0) .. controls (3.31,0.3) and (6.95,1.4) .. (10.93,3.29)   ;
%Shape: Circle [id:dp7589589714070517] 
\draw  [draw opacity=0][fill={rgb, 255:red, 208; green, 2; blue, 27 }  ,fill opacity=1 ] (176.25,79.75) .. controls (176.25,77.49) and (178.08,75.67) .. (180.33,75.67) .. controls (182.59,75.67) and (184.42,77.49) .. (184.42,79.75) .. controls (184.42,82.01) and (182.59,83.83) .. (180.33,83.83) .. controls (178.08,83.83) and (176.25,82.01) .. (176.25,79.75) -- cycle ;
%Shape: Circle [id:dp6483957670164409] 
\draw  [draw opacity=0][fill={rgb, 255:red, 208; green, 2; blue, 27 }  ,fill opacity=1 ] (86.92,190) .. controls (86.92,187.74) and (88.74,185.92) .. (91,185.92) .. controls (93.26,185.92) and (95.08,187.74) .. (95.08,190) .. controls (95.08,192.26) and (93.26,194.08) .. (91,194.08) .. controls (88.74,194.08) and (86.92,192.26) .. (86.92,190) -- cycle ;
%Shape: Circle [id:dp04093894940371956] 
\draw  [draw opacity=0][fill={rgb, 255:red, 208; green, 2; blue, 27 }  ,fill opacity=1 ] (176.25,190) .. controls (176.25,187.74) and (178.08,185.92) .. (180.33,185.92) .. controls (182.59,185.92) and (184.42,187.74) .. (184.42,190) .. controls (184.42,192.26) and (182.59,194.08) .. (180.33,194.08) .. controls (178.08,194.08) and (176.25,192.26) .. (176.25,190) -- cycle ;
%Shape: Circle [id:dp9448149784544406] 
\draw  [draw opacity=0][fill={rgb, 255:red, 74; green, 144; blue, 226 }  ,fill opacity=1 ] (176.25,119.67) .. controls (176.25,117.41) and (178.08,115.58) .. (180.33,115.58) .. controls (182.59,115.58) and (184.42,117.41) .. (184.42,119.67) .. controls (184.42,121.92) and (182.59,123.75) .. (180.33,123.75) .. controls (178.08,123.75) and (176.25,121.92) .. (176.25,119.67) -- cycle ;
%Shape: Circle [id:dp8577908208543465] 
\draw  [draw opacity=0][fill={rgb, 255:red, 74; green, 144; blue, 226 }  ,fill opacity=1 ] (236.36,119.67) .. controls (236.36,117.41) and (238.19,115.58) .. (240.44,115.58) .. controls (242.7,115.58) and (244.53,117.41) .. (244.53,119.67) .. controls (244.53,121.92) and (242.7,123.75) .. (240.44,123.75) .. controls (238.19,123.75) and (236.36,121.92) .. (236.36,119.67) -- cycle ;
%Shape: Circle [id:dp8330847268561952] 
\draw  [draw opacity=0][fill={rgb, 255:red, 74; green, 144; blue, 226 }  ,fill opacity=1 ] (236.36,229.67) .. controls (236.36,227.41) and (238.19,225.58) .. (240.44,225.58) .. controls (242.7,225.58) and (244.53,227.41) .. (244.53,229.67) .. controls (244.53,231.92) and (242.7,233.75) .. (240.44,233.75) .. controls (238.19,233.75) and (236.36,231.92) .. (236.36,229.67) -- cycle ;

% Text Node
\draw (130.78,60.4) node [anchor=north west][inner sep=0.75pt]  [font=\scriptsize]  {$a_{1}$};
% Text Node
\draw (219.78,61.4) node [anchor=north west][inner sep=0.75pt]  [font=\scriptsize]  {$b_{1}$};
% Text Node
\draw (149.11,90.07) node [anchor=north west][inner sep=0.75pt]  [font=\scriptsize]  {$b_{1} a_{1}$};
% Text Node
\draw (249.78,91.07) node [anchor=north west][inner sep=0.75pt]  [font=\scriptsize]  {$a_{1}$};
% Text Node
\draw (248.78,137.4) node [anchor=north west][inner sep=0.75pt]  [font=\scriptsize]  {$b_{1}$};
% Text Node
\draw (195.78,105.07) node [anchor=north west][inner sep=0.75pt]  [font=\scriptsize]  {$a_{1} b_{1}$};
% Text Node
\draw (116.33,127.07) node [anchor=north west][inner sep=0.75pt]  [font=\scriptsize]  {$b_{1}^{-1} a_{1}^{-1} b_{1} a_{1}$};
% Text Node
\draw (135.44,144.07) node [anchor=north west][inner sep=0.75pt]  [font=\scriptsize]  {$=$};
% Text Node
\draw (116.33,153.07) node [anchor=north west][inner sep=0.75pt]  [font=\scriptsize]  {$a_{2}^{-1} b_{2}^{-1} a_{2} b_{2}$};
% Text Node
\draw (63.78,203.18) node [anchor=north west][inner sep=0.75pt]  [font=\scriptsize]  {$a_{2}$};
% Text Node
\draw (63.11,164.07) node [anchor=north west][inner sep=0.75pt]  [font=\scriptsize]  {$b_{2}$};
% Text Node
\draw (117.44,192.4) node [anchor=north west][inner sep=0.75pt]  [font=\scriptsize]  {$b_{2} a_{2}$};
% Text Node
\draw (254.11,245.73) node [anchor=north west][inner sep=0.75pt]  [font=\scriptsize]  {$b_{2}$};
% Text Node
\draw (254.78,197.73) node [anchor=north west][inner sep=0.75pt]  [font=\scriptsize]  {$a_{2}$};
% Text Node
\draw (184.33,207.4) node [anchor=north west][inner sep=0.75pt]  [font=\scriptsize]  {$a_{2} b_{2}$};

\end{tikzpicture}
    \caption{Trivalent resolution of genus 2 surface triangulation from Figure \ref{fig:g2triangulation}.}
    \label{fig:g2trivalentgp}
\end{figure}
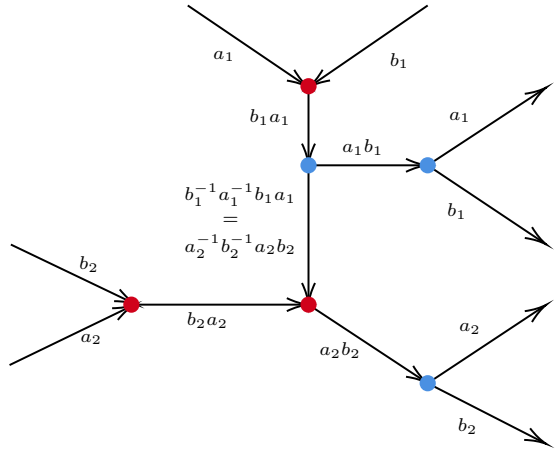

In our language, to compute a partition function, we first need to pick a 
triangulation of the worldsheet.  (Then, later, we will review how the results
are independent of the choice of triangulation.)
First, to set the stage, we briefly recall genus two computations in ordinary
orbifolds.  
One particular choice of triangulation for a $g=2$ surface is given in Figure 2 of \cite[figure 2]{Aspinwall:2000xv}, redrawn here as Figure~\ref{fig:g2triangulation}. It consists of an octagon whose edges are labeled by group elements of the group that is being gauged.
This gives rise to the (trivalent) dual graph shown in Figure~\ref{fig:g2trivalentgp}, where the multiplication and comultiplication morphisms are labeled by red and blue circles, respectively.

One of the important consistency checks that are satisfied in this perspective is that for the top part of the graph in Figure \ref{fig:g2trivalentgp}, which we can regard as a torus, to connect to the bottom part, another torus, it must hold that
\begin{equation}\label{eq:gen2comm}
    b_1^{-1}a_1^{-1}b_1a_1=a_2^{-1}b_2^{-1}a_2b_2,
\end{equation}
which is the familiar commutativity condition of a flat $G$-bundle on a genus $2$ surface. In particular, this means that the partition function will only sum over those simple objects $L_{a_1},L_{b_1},L_{a_2},L_{b_2}\in \text{ob}(\text{Vec}(G))$ whose group labels $a_1,b_1,a_2,b_2\in G$ satisfy Eq.(\ref{eq:gen2comm}).

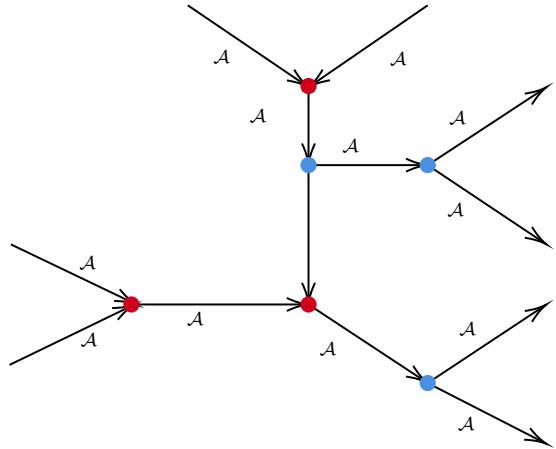
\begin{figure}
    \centering
\tikzset{every picture/.style={line width=0.75pt}} %set default line width to 0.75pt        
\begin{tikzpicture}[x=0.75pt,y=0.75pt,yscale=-1,xscale=1]

%Straight Lines [id:da09879049727609068] 
\draw    (119.44,39) -- (178.68,79.21) ;
\draw [shift={(180.33,80.33)}, rotate = 214.17] [color={rgb, 255:red, 0; green, 0; blue, 0 }  ][line width=0.75]    (10.93,-3.29) .. controls (6.95,-1.4) and (3.31,-0.3) .. (0,0) .. controls (3.31,0.3) and (6.95,1.4) .. (10.93,3.29)   ;
%Straight Lines [id:da022679162896319216] 
\draw    (240.44,39) -- (181.98,79.2) ;
\draw [shift={(180.33,80.33)}, rotate = 325.49] [color={rgb, 255:red, 0; green, 0; blue, 0 }  ][line width=0.75]    (10.93,-3.29) .. controls (6.95,-1.4) and (3.31,-0.3) .. (0,0) .. controls (3.31,0.3) and (6.95,1.4) .. (10.93,3.29)   ;
%Straight Lines [id:da6784132674816423] 
\draw    (180.33,80.33) -- (180.33,117.67) ;
\draw [shift={(180.33,119.67)}, rotate = 270] [color={rgb, 255:red, 0; green, 0; blue, 0 }  ][line width=0.75]    (10.93,-3.29) .. controls (6.95,-1.4) and (3.31,-0.3) .. (0,0) .. controls (3.31,0.3) and (6.95,1.4) .. (10.93,3.29)   ;
%Straight Lines [id:da641402190441402] 
\draw    (180.33,119.67) -- (238.44,119.67) ;
\draw [shift={(240.44,119.67)}, rotate = 180] [color={rgb, 255:red, 0; green, 0; blue, 0 }  ][line width=0.75]    (10.93,-3.29) .. controls (6.95,-1.4) and (3.31,-0.3) .. (0,0) .. controls (3.31,0.3) and (6.95,1.4) .. (10.93,3.29)   ;
%Straight Lines [id:da5715443568775722] 
\draw    (240.44,119.67) -- (298.66,81.43) ;
\draw [shift={(300.33,80.33)}, rotate = 146.7] [color={rgb, 255:red, 0; green, 0; blue, 0 }  ][line width=0.75]    (10.93,-3.29) .. controls (6.95,-1.4) and (3.31,-0.3) .. (0,0) .. controls (3.31,0.3) and (6.95,1.4) .. (10.93,3.29)   ;
%Straight Lines [id:da7770854423418123] 
\draw    (240.44,119.67) -- (298.67,158.56) ;
\draw [shift={(300.33,159.67)}, rotate = 213.74] [color={rgb, 255:red, 0; green, 0; blue, 0 }  ][line width=0.75]    (10.93,-3.29) .. controls (6.95,-1.4) and (3.31,-0.3) .. (0,0) .. controls (3.31,0.3) and (6.95,1.4) .. (10.93,3.29)   ;
%Straight Lines [id:da054430206837292605] 
\draw    (180.33,119.67) -- (180.33,188) ;
\draw [shift={(180.33,190)}, rotate = 270] [color={rgb, 255:red, 0; green, 0; blue, 0 }  ][line width=0.75]    (10.93,-3.29) .. controls (6.95,-1.4) and (3.31,-0.3) .. (0,0) .. controls (3.31,0.3) and (6.95,1.4) .. (10.93,3.29)   ;
%Straight Lines [id:da5048436914641816] 
\draw    (30.22,159.67) -- (91.75,189.14) ;
\draw [shift={(93.56,190)}, rotate = 205.59] [color={rgb, 255:red, 0; green, 0; blue, 0 }  ][line width=0.75]    (10.93,-3.29) .. controls (6.95,-1.4) and (3.31,-0.3) .. (0,0) .. controls (3.31,0.3) and (6.95,1.4) .. (10.93,3.29)   ;
%Straight Lines [id:da7956678657628669] 
\draw    (29.56,220.5) -- (90.76,190.87) ;
\draw [shift={(92.56,190)}, rotate = 154.17] [color={rgb, 255:red, 0; green, 0; blue, 0 }  ][line width=0.75]    (10.93,-3.29) .. controls (6.95,-1.4) and (3.31,-0.3) .. (0,0) .. controls (3.31,0.3) and (6.95,1.4) .. (10.93,3.29)   ;
%Straight Lines [id:da3332475286582053] 
\draw    (92.56,190) -- (178.33,190) ;
\draw [shift={(180.33,190)}, rotate = 180] [color={rgb, 255:red, 0; green, 0; blue, 0 }  ][line width=0.75]    (10.93,-3.29) .. controls (6.95,-1.4) and (3.31,-0.3) .. (0,0) .. controls (3.31,0.3) and (6.95,1.4) .. (10.93,3.29)   ;
%Straight Lines [id:da3810245695382757] 
\draw    (180.33,190) -- (238.78,228.57) ;
\draw [shift={(240.44,229.67)}, rotate = 213.42] [color={rgb, 255:red, 0; green, 0; blue, 0 }  ][line width=0.75]    (10.93,-3.29) .. controls (6.95,-1.4) and (3.31,-0.3) .. (0,0) .. controls (3.31,0.3) and (6.95,1.4) .. (10.93,3.29)   ;
%Straight Lines [id:da8269806107726623] 
\draw    (240.44,229.67) -- (298.67,191.1) ;
\draw [shift={(300.33,190)}, rotate = 146.48] [color={rgb, 255:red, 0; green, 0; blue, 0 }  ][line width=0.75]    (10.93,-3.29) .. controls (6.95,-1.4) and (3.31,-0.3) .. (0,0) .. controls (3.31,0.3) and (6.95,1.4) .. (10.93,3.29)   ;
%Straight Lines [id:da8293040461117034] 
\draw    (240.44,229.67) -- (298.55,259.42) ;
\draw [shift={(300.33,260.33)}, rotate = 207.12] [color={rgb, 255:red, 0; green, 0; blue, 0 }  ][line width=0.75]    (10.93,-3.29) .. controls (6.95,-1.4) and (3.31,-0.3) .. (0,0) .. controls (3.31,0.3) and (6.95,1.4) .. (10.93,3.29)   ;
%Shape: Circle [id:dp7589589714070517] 
\draw  [draw opacity=0][fill={rgb, 255:red, 208; green, 2; blue, 27 }  ,fill opacity=1 ] (176.25,79.75) .. controls (176.25,77.49) and (178.08,75.67) .. (180.33,75.67) .. controls (182.59,75.67) and (184.42,77.49) .. (184.42,79.75) .. controls (184.42,82.01) and (182.59,83.83) .. (180.33,83.83) .. controls (178.08,83.83) and (176.25,82.01) .. (176.25,79.75) -- cycle ;
%Shape: Circle [id:dp6483957670164409] 
\draw  [draw opacity=0][fill={rgb, 255:red, 208; green, 2; blue, 27 }  ,fill opacity=1 ] (86.92,190) .. controls (86.92,187.74) and (88.74,185.92) .. (91,185.92) .. controls (93.26,185.92) and (95.08,187.74) .. (95.08,190) .. controls (95.08,192.26) and (93.26,194.08) .. (91,194.08) .. controls (88.74,194.08) and (86.92,192.26) .. (86.92,190) -- cycle ;
%Shape: Circle [id:dp04093894940371956] 
\draw  [draw opacity=0][fill={rgb, 255:red, 208; green, 2; blue, 27 }  ,fill opacity=1 ] (176.25,190) .. controls (176.25,187.74) and (178.08,185.92) .. (180.33,185.92) .. controls (182.59,185.92) and (184.42,187.74) .. (184.42,190) .. controls (184.42,192.26) and (182.59,194.08) .. (180.33,194.08) .. controls (178.08,194.08) and (176.25,192.26) .. (176.25,190) -- cycle ;
%Shape: Circle [id:dp9448149784544406] 
\draw  [draw opacity=0][fill={rgb, 255:red, 74; green, 144; blue, 226 }  ,fill opacity=1 ] (176.25,119.67) .. controls (176.25,117.41) and (178.08,115.58) .. (180.33,115.58) .. controls (182.59,115.58) and (184.42,117.41) .. (184.42,119.67) .. controls (184.42,121.92) and (182.59,123.75) .. (180.33,123.75) .. controls (178.08,123.75) and (176.25,121.92) .. (176.25,119.67) -- cycle ;
%Shape: Circle [id:dp8577908208543465] 
\draw  [draw opacity=0][fill={rgb, 255:red, 74; green, 144; blue, 226 }  ,fill opacity=1 ] (236.36,119.67) .. controls (236.36,117.41) and (238.19,115.58) .. (240.44,115.58) .. controls (242.7,115.58) and (244.53,117.41) .. (244.53,119.67) .. controls (244.53,121.92) and (242.7,123.75) .. (240.44,123.75) .. controls (238.19,123.75) and (236.36,121.92) .. (236.36,119.67) -- cycle ;
%Shape: Circle [id:dp8330847268561952] 
\draw  [draw opacity=0][fill={rgb, 255:red, 74; green, 144; blue, 226 }  ,fill opacity=1 ] (236.36,229.67) .. controls (236.36,227.41) and (238.19,225.58) .. (240.44,225.58) .. controls (242.7,225.58) and (244.53,227.41) .. (244.53,229.67) .. controls (244.53,231.92) and (242.7,233.75) .. (240.44,233.75) .. controls (238.19,233.75) and (236.36,231.92) .. (236.36,229.67) -- cycle ;

% Text Node
\draw (130.78,60.4) node [anchor=north west][inner sep=0.75pt]  [font=\scriptsize]  {${\cal A}$};
% Text Node
\draw (219.78,61.4) node [anchor=north west][inner sep=0.75pt]  [font=\scriptsize]  {${\cal A}$};
% Text Node
\draw (149.11,90.07) node [anchor=north west][inner sep=0.75pt]  [font=\scriptsize]  {${\cal A}$};
% Text Node
\draw (249.78,91.07) node [anchor=north west][inner sep=0.75pt]  [font=\scriptsize]  {${\cal A}$};
% Text Node
\draw (248.78,137.4) node [anchor=north west][inner sep=0.75pt]  [font=\scriptsize]  {${\cal A}$};
% Text Node
\draw (195.78,105.07) node [anchor=north west][inner sep=0.75pt]  [font=\scriptsize]  {${\cal A}$};
% Text Node
\draw (63.78,203.18) node [anchor=north west][inner sep=0.75pt]  [font=\scriptsize]  {${\cal A}$};
% Text Node
\draw (63.11,164.07) node [anchor=north west][inner sep=0.75pt]  [font=\scriptsize]  {${\cal A}$};
% Text Node
\draw (117.44,192.4) node [anchor=north west][inner sep=0.75pt]  [font=\scriptsize]  {${\cal A}$};
% Text Node
\draw (254.11,245.73) node [anchor=north west][inner sep=0.75pt]  [font=\scriptsize]  {${\cal A}$};
% Text Node
\draw (254.78,197.73) node [anchor=north west][inner sep=0.75pt]  [font=\scriptsize]  {${\cal A}$};
% Text Node
\draw (184.33,207.4) node [anchor=north west][inner sep=0.75pt]  [font=\scriptsize]  {${\cal A}$};

\end{tikzpicture}
    \caption{Trivalent resolution of genus $2$ surface with Frobenius algebra object ${\cal A}$ defects. Red vertices indicate multiplications, and blue vertices indicate comultiplications.}
    \label{fig:g2trivalentfrobenius}
\end{figure}

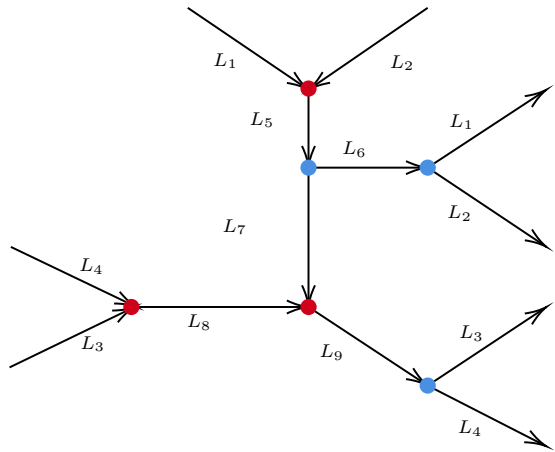
\begin{figure}
    \centering
\tikzset{every picture/.style={line width=0.75pt}} %set default line width to 0.75pt        

\begin{tikzpicture}[x=0.75pt,y=0.75pt,yscale=-1,xscale=1]

%Straight Lines [id:da09879049727609068] 
\draw    (119.44,39) -- (178.68,79.21) ;
\draw [shift={(180.33,80.33)}, rotate = 214.17] [color={rgb, 255:red, 0; green, 0; blue, 0 }  ][line width=0.75]    (10.93,-3.29) .. controls (6.95,-1.4) and (3.31,-0.3) .. (0,0) .. controls (3.31,0.3) and (6.95,1.4) .. (10.93,3.29)   ;
%Straight Lines [id:da022679162896319216] 
\draw    (240.44,39) -- (181.98,79.2) ;
\draw [shift={(180.33,80.33)}, rotate = 325.49] [color={rgb, 255:red, 0; green, 0; blue, 0 }  ][line width=0.75]    (10.93,-3.29) .. controls (6.95,-1.4) and (3.31,-0.3) .. (0,0) .. controls (3.31,0.3) and (6.95,1.4) .. (10.93,3.29)   ;
%Straight Lines [id:da6784132674816423] 
\draw    (180.33,80.33) -- (180.33,117.67) ;
\draw [shift={(180.33,119.67)}, rotate = 270] [color={rgb, 255:red, 0; green, 0; blue, 0 }  ][line width=0.75]    (10.93,-3.29) .. controls (6.95,-1.4) and (3.31,-0.3) .. (0,0) .. controls (3.31,0.3) and (6.95,1.4) .. (10.93,3.29)   ;
%Straight Lines [id:da641402190441402] 
\draw    (180.33,119.67) -- (238.44,119.67) ;
\draw [shift={(240.44,119.67)}, rotate = 180] [color={rgb, 255:red, 0; green, 0; blue, 0 }  ][line width=0.75]    (10.93,-3.29) .. controls (6.95,-1.4) and (3.31,-0.3) .. (0,0) .. controls (3.31,0.3) and (6.95,1.4) .. (10.93,3.29)   ;
%Straight Lines [id:da5715443568775722] 
\draw    (240.44,119.67) -- (298.66,81.43) ;
\draw [shift={(300.33,80.33)}, rotate = 146.7] [color={rgb, 255:red, 0; green, 0; blue, 0 }  ][line width=0.75]    (10.93,-3.29) .. controls (6.95,-1.4) and (3.31,-0.3) .. (0,0) .. controls (3.31,0.3) and (6.95,1.4) .. (10.93,3.29)   ;
%Straight Lines [id:da7770854423418123] 
\draw    (240.44,119.67) -- (298.67,158.56) ;
\draw [shift={(300.33,159.67)}, rotate = 213.74] [color={rgb, 255:red, 0; green, 0; blue, 0 }  ][line width=0.75]    (10.93,-3.29) .. controls (6.95,-1.4) and (3.31,-0.3) .. (0,0) .. controls (3.31,0.3) and (6.95,1.4) .. (10.93,3.29)   ;
%Straight Lines [id:da054430206837292605] 
\draw    (180.33,119.67) -- (180.33,188) ;
\draw [shift={(180.33,190)}, rotate = 270] [color={rgb, 255:red, 0; green, 0; blue, 0 }  ][line width=0.75]    (10.93,-3.29) .. controls (6.95,-1.4) and (3.31,-0.3) .. (0,0) .. controls (3.31,0.3) and (6.95,1.4) .. (10.93,3.29)   ;
%Straight Lines [id:da5048436914641816] 
\draw    (30.22,159.67) -- (91.75,189.14) ;
\draw [shift={(93.56,190)}, rotate = 205.59] [color={rgb, 255:red, 0; green, 0; blue, 0 }  ][line width=0.75]    (10.93,-3.29) .. controls (6.95,-1.4) and (3.31,-0.3) .. (0,0) .. controls (3.31,0.3) and (6.95,1.4) .. (10.93,3.29)   ;
%Straight Lines [id:da7956678657628669] 
\draw    (29.56,220.5) -- (90.76,190.87) ;
\draw [shift={(92.56,190)}, rotate = 154.17] [color={rgb, 255:red, 0; green, 0; blue, 0 }  ][line width=0.75]    (10.93,-3.29) .. controls (6.95,-1.4) and (3.31,-0.3) .. (0,0) .. controls (3.31,0.3) and (6.95,1.4) .. (10.93,3.29)   ;
%Straight Lines [id:da3332475286582053] 
\draw    (92.56,190) -- (178.33,190) ;
\draw [shift={(180.33,190)}, rotate = 180] [color={rgb, 255:red, 0; green, 0; blue, 0 }  ][line width=0.75]    (10.93,-3.29) .. controls (6.95,-1.4) and (3.31,-0.3) .. (0,0) .. controls (3.31,0.3) and (6.95,1.4) .. (10.93,3.29)   ;
%Straight Lines [id:da3810245695382757] 
\draw    (180.33,190) -- (238.78,228.57) ;
\draw [shift={(240.44,229.67)}, rotate = 213.42] [color={rgb, 255:red, 0; green, 0; blue, 0 }  ][line width=0.75]    (10.93,-3.29) .. controls (6.95,-1.4) and (3.31,-0.3) .. (0,0) .. controls (3.31,0.3) and (6.95,1.4) .. (10.93,3.29)   ;
%Straight Lines [id:da8269806107726623] 
\draw    (240.44,229.67) -- (298.67,191.1) ;
\draw [shift={(300.33,190)}, rotate = 146.48] [color={rgb, 255:red, 0; green, 0; blue, 0 }  ][line width=0.75]    (10.93,-3.29) .. controls (6.95,-1.4) and (3.31,-0.3) .. (0,0) .. controls (3.31,0.3) and (6.95,1.4) .. (10.93,3.29)   ;
%Straight Lines [id:da8293040461117034] 
\draw    (240.44,229.67) -- (298.55,259.42) ;
\draw [shift={(300.33,260.33)}, rotate = 207.12] [color={rgb, 255:red, 0; green, 0; blue, 0 }  ][line width=0.75]    (10.93,-3.29) .. controls (6.95,-1.4) and (3.31,-0.3) .. (0,0) .. controls (3.31,0.3) and (6.95,1.4) .. (10.93,3.29)   ;
%Shape: Circle [id:dp7589589714070517] 
\draw  [draw opacity=0][fill={rgb, 255:red, 208; green, 2; blue, 27 }  ,fill opacity=1 ] (176.25,79.75) .. controls (176.25,77.49) and (178.08,75.67) .. (180.33,75.67) .. controls (182.59,75.67) and (184.42,77.49) .. (184.42,79.75) .. controls (184.42,82.01) and (182.59,83.83) .. (180.33,83.83) .. controls (178.08,83.83) and (176.25,82.01) .. (176.25,79.75) -- cycle ;
%Shape: Circle [id:dp6483957670164409] 
\draw  [draw opacity=0][fill={rgb, 255:red, 208; green, 2; blue, 27 }  ,fill opacity=1 ] (86.92,190) .. controls (86.92,187.74) and (88.74,185.92) .. (91,185.92) .. controls (93.26,185.92) and (95.08,187.74) .. (95.08,190) .. controls (95.08,192.26) and (93.26,194.08) .. (91,194.08) .. controls (88.74,194.08) and (86.92,192.26) .. (86.92,190) -- cycle ;
%Shape: Circle [id:dp04093894940371956] 
\draw  [draw opacity=0][fill={rgb, 255:red, 208; green, 2; blue, 27 }  ,fill opacity=1 ] (176.25,190) .. controls (176.25,187.74) and (178.08,185.92) .. (180.33,185.92) .. controls (182.59,185.92) and (184.42,187.74) .. (184.42,190) .. controls (184.42,192.26) and (182.59,194.08) .. (180.33,194.08) .. controls (178.08,194.08) and (176.25,192.26) .. (176.25,190) -- cycle ;
%Shape: Circle [id:dp9448149784544406] 
\draw  [draw opacity=0][fill={rgb, 255:red, 74; green, 144; blue, 226 }  ,fill opacity=1 ] (176.25,119.67) .. controls (176.25,117.41) and (178.08,115.58) .. (180.33,115.58) .. controls (182.59,115.58) and (184.42,117.41) .. (184.42,119.67) .. controls (184.42,121.92) and (182.59,123.75) .. (180.33,123.75) .. controls (178.08,123.75) and (176.25,121.92) .. (176.25,119.67) -- cycle ;
%Shape: Circle [id:dp8577908208543465] 
\draw  [draw opacity=0][fill={rgb, 255:red, 74; green, 144; blue, 226 }  ,fill opacity=1 ] (236.36,119.67) .. controls (236.36,117.41) and (238.19,115.58) .. (240.44,115.58) .. controls (242.7,115.58) and (244.53,117.41) .. (244.53,119.67) .. controls (244.53,121.92) and (242.7,123.75) .. (240.44,123.75) .. controls (238.19,123.75) and (236.36,121.92) .. (236.36,119.67) -- cycle ;
%Shape: Circle [id:dp8330847268561952] 
\draw  [draw opacity=0][fill={rgb, 255:red, 74; green, 144; blue, 226 }  ,fill opacity=1 ] (236.36,229.67) .. controls (236.36,227.41) and (238.19,225.58) .. (240.44,225.58) .. controls (242.7,225.58) and (244.53,227.41) .. (244.53,229.67) .. controls (244.53,231.92) and (242.7,233.75) .. (240.44,233.75) .. controls (238.19,233.75) and (236.36,231.92) .. (236.36,229.67) -- cycle ;

% Text Node
\draw (130.78,60.4) node [anchor=north west][inner sep=0.75pt]  [font=\scriptsize]  {$L_1$};
% Text Node
\draw (219.78,61.4) node [anchor=north west][inner sep=0.75pt]  [font=\scriptsize]  {$L_2$};
% Text Node
\draw (149.11,90.07) node [anchor=north west][inner sep=0.75pt]  [font=\scriptsize]  {$L_5$};
% Text Node
\draw (249.78,91.07) node [anchor=north west][inner sep=0.75pt]  [font=\scriptsize]  {$L_1$};
% Text Node
\draw (248.78,137.4) node [anchor=north west][inner sep=0.75pt]  [font=\scriptsize]  {$L_2$};
% Text Node
\draw (195.78,105.07) node [anchor=north west][inner sep=0.75pt]  [font=\scriptsize]  {$L_6$};
% Text Node
\draw (116.33,127.07) node [anchor=north west][inner sep=0.75pt]  [font=\scriptsize]  {$ $};
% Text Node
\draw (135.44,144.07) node [anchor=north west][inner sep=0.75pt]  [font=\scriptsize]  {$L_7$};
% Text Node
\draw (116.33,153.07) node [anchor=north west][inner sep=0.75pt]  [font=\scriptsize]  {$ $};
% Text Node
\draw (63.78,203.18) node [anchor=north west][inner sep=0.75pt]  [font=\scriptsize]  {$L_3$};
% Text Node
\draw (63.11,164.07) node [anchor=north west][inner sep=0.75pt]  [font=\scriptsize]  {$L_4$};
% Text Node
\draw (117.44,192.4) node [anchor=north west][inner sep=0.75pt]  [font=\scriptsize]  {$L_8$};
% Text Node
\draw (254.11,245.73) node [anchor=north west][inner sep=0.75pt]  [font=\scriptsize]  {$L_4$};
% Text Node
\draw (254.78,197.73) node [anchor=north west][inner sep=0.75pt]  [font=\scriptsize]  {$L_3$};
% Text Node
\draw (184.33,207.4) node [anchor=north west][inner sep=0.75pt]  [font=\scriptsize]  {$L_9$};

\end{tikzpicture}
    \caption{Trivalent resolution of genus 2 surface triangulation from Figure \ref{fig:g2triangulation} in terms of simple lines. This diagram defines a genus 2 partial trace.}
    \label{fig:g2trivalentgpsimple}
\end{figure}

More generally, such a graph can be used to represent the $\mathcal{A}$-morphism that gives the coefficients the partition function of the $g=2$ surface. Given a Frobenius object ${\cal A}$, one such graph is  Figure~\ref{fig:g2trivalentfrobenius}. As in the genus $1$ case, it is often convenient to decompose this diagram into a sum of diagrams involving only simple lines, as in Figure \ref{fig:g2trivalentgpsimple}. From this diagram, we can read off the partition function in the \textit{multiplicity-free} case for a genus $2$ surface as:
\begin{equation}\label{eq:g2partitionfunction}
    Z_{g=2} = \sum_{L_1,\cdots,L_9} \mu_{L_1,L_2}^{L_5}\Delta_{L_5}^{L_6,L_7} \Delta_{L_6}^{L_2,L_1}\mu_{L_3,L_4}^{L_8}\mu_{L_7,L_8}^{L_9}\Delta_{L_9}^{L_4,L_3} Z_{L_1,L_2,L_3,L_4}^{L_5,L_6,L_7,L_8,L_9},
\end{equation}
where $Z_{L_1,L_2,L_3,L_4}^{L_5,L_6,L_7,L_8,L_9}$ is the partial trace involving $(L_1,\cdots,L_4)$ as external simple lines and $(L_5,\cdots,L_9)$ as internal simple lines, in the order specified in Figure \ref{fig:g2trivalentgpsimple}.

As discussed in \cite[section 5.1]{Fuchs:2002cm}, even though the gauging process of a Frobenius object uses a choice of worldsheet triangulation, the end result is independent of such a choice.
We outline this result below.  First, 
it is known that all triangulations of a given manifold can be related by a series of moves \cite{Alex30}, all of which are generated by two basic moves\cite[fig.~6]{Fukuma:1993hy}: the \textit{fusion} (F-) or $(2,2)$-move, and the \textit{bubble} move.

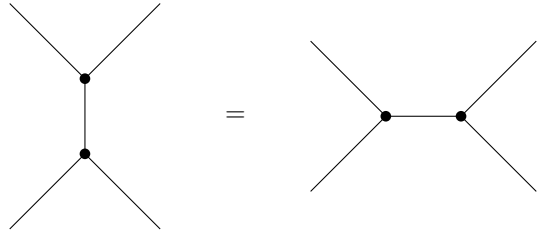
\begin{figure}[h]
    \centering
    \begin{tikzpicture}
        \draw (0,0) -- (1,1);
        \draw (2,0) -- (1,1);
        \draw (1,1) node [circle, fill, inner sep =0.5mm] {};
        \draw (1,1) -- (1,2);
        \draw (1,2) -- (0,3);
        \draw (1,2) -- (2,3);
        \draw (1,2) node [circle, fill, inner sep =0.5mm] {};
        \draw (3,1.5) node {$=$};
        \draw (4,0.5) -- (5,1.5);
        \draw (4,2.5) -- (5,1.5);
        \draw (5,1.5) node [circle, fill, inner sep=0.5mm] {};
        \draw (5,1.5) -- (6,1.5);
        \draw (6,1.5) -- (7,0.5);
        \draw (6,1.5) -- (7,2.5);
        \draw (6,1.5) node [circle, fill, inner sep =0.5mm] {};
    \end{tikzpicture}
    \caption{Shown is a fusion move.}
    \label{fig:fusion}
\end{figure}

Next, we outline these two moves and how both are realized by the axioms of a special symmetric Frobenius algebra\footnote{
So long as we draw diagrams with the Frobenius algebra ${\cal A}$ on legs, we can express everything in terms of axioms of the Frobenius algebra.  If we were to write the diagrams in components, then,
we would also need to take into account $F$ symbols / crossing kernels $\tilde{K}$.
}.
\begin{itemize}
    \item Fusion move = Frobenius:  In the dual graph, the fusion move is shown in figure~\ref{fig:fusion}.
    This corresponds to the Frobenius identities~(\ref{eq:Frobenius-identities}) for a special symmetric Frobenius algebra, namely
    \begin{equation}  
        \xymatrix{
        {\cal A} \otimes {\cal A} \ar[rr]^{\Delta_F \otimes \text{Id}_{\cal A}} \ar[d]_{\mu_*} 
        & & {\cal A} \otimes {\cal A} \otimes {\cal A} \ar[d]^{\text{Id}_{\cal A} \otimes \mu_*}
        \\
        {\cal A} \ar[rr]^{\Delta_F} & & {\cal A} \otimes {\cal A}
        }
    \end{equation}
    as discussed in appendix~\ref{app:Frobenius}.
    We have sketched the relation in figure~\ref{fig:fusion-as-Frobenius}.  Briefly, the diagram on the left in figure~\ref{fig:fusion-as-Frobenius} corresponds to the
    map
    \begin{equation}
        {\cal A} \otimes {\cal A} \: \stackrel{\mu_*}{\longrightarrow} \: {\cal A} \: \stackrel{\Delta_F}{\longrightarrow} \: {\cal A} \otimes {\cal A},
    \end{equation}
    and the diagram on the right in figure~\ref{fig:fusion-as-Frobenius} corresponds to the map
    \begin{equation}
        {\cal A} \otimes {\cal A} \: \stackrel{ \Delta_F \otimes {\rm Id}_{\cal A}}{\longrightarrow} \: 
        {\cal A} \otimes {\cal A} \otimes {\cal a} \: \stackrel{{\rm Id}_{\cal A} \otimes \mu_*}{\longrightarrow} \: {\cal A} \otimes {\cal A}.
    \end{equation}
    Equality of these two compositions is precisely the Frobenius identity above.

    \item Bubble move = special:  The bubble move is shown in figure~\ref{fig:bubble}.  Invariance under the bubble move is a consequence of the `special' condition~(\ref{eq:conds})
in the special symmetric Frobenius algebra, namely
\begin{equation}
    \mu_* \circ \Delta_F \: = \: {\rm Id}_{\cal A},
\end{equation}
as discussed in appendix~\ref{app:Frobenius}.
The relation between the diagram~\ref{fig:bubble} and the identity above should be clear from the labelling.

\end{itemize}
In this fashion we see that the partition function should be well-defined with respect to different choices of triangulation, and hence different descriptions as compositions of $\mu_*$ and $\Delta_F$.
An example is illustrated in figure~\ref{fig:simplify}.

\begin{figure}[h]
 \centering
    \begin{tikzpicture}
        \draw[->] (0,0) -- (0.5,0.5);  \draw (0.5,0.5) -- (1,1);
        \draw (0.1,0.5) node {${\cal A}$};
        \draw[->] (2,0) -- (1.5,0.5);  \draw (1.5,0.5) -- (1,1);
        \draw (1.9,0.5) node {${\cal A}$};
        \draw (1,1) node [circle, fill, inner sep =0.5mm] {};
        \draw (1.4,1) node {$\mu_*$};
        \draw[->] (1,1) -- (1,1.5);  \draw (1,1.5) -- (1,2);
        \draw[->] (1,2) -- (0.5,2.5);  \draw (0.5,2.5) -- (0,3);
        \draw (0.1,2.5) node {${\cal A}$};
        \draw[->] (1,2) -- (1.5,2.5); \draw (1.5,2.5) -- (2,3);
        \draw (1.9,2.5) node {${\cal A}$};
        \draw (1,2) node [circle, fill, inner sep =0.5mm] {};
        \draw (1.4,2) node {$\Delta_F$};
        \draw[thick,color=blue,->,opacity=0.5] (1,1) [partial ellipse=-135:90:0.2cm and 0.2cm];
        \draw[thick,color=blue,->,opacity=0.5] (1,2) [partial ellipse=-90:135:0.2cm and 0.2cm];
        \draw (3,1.5) node {$=$};
        \draw[->] (4,0) -- (4.5,0.5);  \draw (4.5,0.5) -- (5,1);
        \draw (4.1,0.5) node {${\cal A}$};
        \draw[->] (5,1) -- (4.5,2);  \draw (4.5,2) -- (4,3);
        \draw (4.4,2.8) node {${\cal A}$};  %(4.3,1.8)
        \draw (5,1) node [circle, fill, inner sep = 0.5mm] {};
        \draw (5.4,1) node {$\Delta_F$};
        \draw[->] (5,1) -- (5.5,1.5); \draw (5.5,1.5) -- (6,2);
        \draw (6.4,2.8) node {${\cal A}$};
        \draw[->] (7,0) -- (6.5,1); \draw (6.5,1) -- (6,2);
        \draw[->] (6,2) -- (6.5,2.5);  \draw (6.5,2.5) -- (7,3);
        \draw (6.4,0.5) node {${\cal A}$};
        \draw (6,2) node [circle, fill, inner sep = 0.5mm] {};
        \draw (6.4,2) node {$\mu_*$};
        \draw[thick,color=blue,->,opacity=0.5] (5,1) [partial ellipse=-135:135:0.2cm and 0.2cm];
        \draw[thick,color=blue,->,opacity=0.5] (6,2) [partial ellipse=-135:45:0.2cm and 0.2cm];
    \end{tikzpicture}
    \caption{Fusion move reinterpreted as Frobenius identity.  The other Frobenius identity gives a very similar diagram with some arrows reversed.}
    \label{fig:fusion-as-Frobenius}
\end{figure}
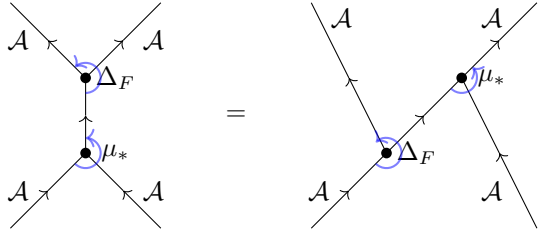

\begin{figure}[h]
    \centering
    \begin{tikzpicture}
        \draw[->] (0,0) -- (0.75,0);  \draw (0.75,0) -- (1.5,0);
        \draw (0.2,0.2) node {${\cal A}$};
        \draw[->] (2.5,0) -- (3.25,0);  \draw (3.25,0) -- (4,0);
        \draw (3.8,0.2) node {${\cal A}$};
        \draw[
            decoration = {markings, mark= at position 0.25 with {\arrow{<}}, 
                mark=at position 0.75 with {\arrow{>}}}, 
            postaction = {decorate}
            ] 
            (2,0) circle (0.5);
        \draw (1.5,0) node [circle, fill, inner sep=0.5mm] {};
        \draw (1.2,0.2) node {$\Delta_F$};
        \draw (2.5,0) node [circle, fill, inner sep=0.5mm] {};
        \draw (2.8,0.2) node {$\mu_*$};
        \draw[thick,color=blue,->,opacity=0.5] (1.5,0) [partial ellipse=-180:85:0.2cm and 0.2cm];
        \draw[thick,color=blue,->,opacity=0.5] (2.5,0) [partial ellipse=-265:0:0.2cm and 0.2cm];
        \draw (4.75,0) node {$=$};
        \draw[->] (5.5,0) -- (7.25,0);  \draw (7.25,0) -- (9,0);
        \draw (5.7,0.2) node {${\cal A}$};   \draw (8.8,0.2) node {${\cal A}$};
    \end{tikzpicture}
    \caption{Shown is a bubble move, with multiplication $\mu_*$ and
    comultiplication $\Delta_F$ inserted.}
    \label{fig:bubble}
\end{figure}
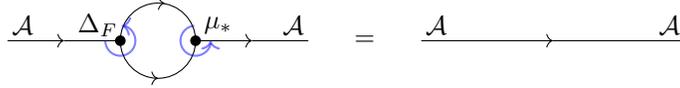

\begin{figure}[h]
\centering
\begin{tikzpicture}
    \draw[->] (0,4) -- (0.5,3.5);  \draw (0.5,3.5) -- (1,3);
    \draw (0.1,0.5) node {${\cal A}$};  \draw (0.1,3.5) node {${\cal A}$};
    \draw (1,3) node [circle, fill, inner sep=0.5mm] {};
    \draw (1.2,3.2) node {$\Delta_F$};
    \draw[thick,color=blue,->,opacity=0.5] (1,3) [partial ellipse=-225:-45:0.2cm and 0.2cm];
    \draw[->] (0,0) -- (0.5,0.5);  \draw (0.5,0.5) -- (1,1);
    \draw (1,1) node [circle, fill, inner sep=0.5mm] {};
    \draw[thick,color=blue,->,opacity=0.5] (1,1) [partial ellipse=-135:90:0.2cm and 0.2cm];
    \draw (1.2,0.8) node {$\mu_*$};
    \draw[->] (1,3) -- (1,2);  \draw (1,2) -- (1,1);
    \draw[->] (1,3) -- (1.5,2.5);  \draw (1.5,2.5) -- (2,2);
    \draw[->] (1,1) -- (1.5,1.5);  \draw (1.5,1.5) -- (2,2);
    \draw (2,2) node [circle, fill, inner sep=0.5mm] {};
    \draw[thick,color=blue,->,opacity=0.5] (2,2) [partial ellipse=-225:0:0.2cm and 0.2cm];
    \draw (2.2,2.2) node {$\mu_*$};
    \draw[->] (2,2) -- (2.5,2);  \draw (2.5,2) -- (3,2);
    \draw (3,2) node [circle, fill, inner sep=0.5mm] {};
    \draw[thick,color=blue,->,opacity=0.5] (3,2) [partial ellipse=-180:45:0.2cm and 0.2cm];
    \draw (2.8,1.8) node {$\Delta_F$};
    \draw[->] (3,2) -- (3.5,2.5);  \draw (3.5,2.5) -- (4,3);
    \draw[->] (3,2) -- (3.5,1.5); \draw (3.5,1.5) -- (4,1);
    \draw (3.9,1.5) node {${\cal A}$};  \draw (3.9,2.7) node {${\cal A}$};
    \draw (5,2) node {$=$};
    \draw[->] (6,3) -- (6.5,2.5);  \draw (6.5,2.5) -- (7,2);
    \draw[->] (6,1) -- (6.5,1.5);  \draw (6.5,1.5) -- (7,2);
    \draw (6.1,1.5) node {${\cal A}$};  \draw (6.1,2.5) node {${\cal A}$};
    \draw (7,2) node [circle, fill, inner sep=0.5mm] {};
    \draw (7.2,2.2) node {$\mu_*$};
    \draw[thick,color=blue,->,opacity=0.5] (7,2) [partial ellipse=-225:0:0.2cm and 0.2cm];
    \draw[->] (7,2) -- (7.5,2);  \draw (7.5,2) -- (8,2);
    \draw (8,2) node [circle, fill, inner sep=0.5mm] {};
    \draw (7.8,1.8) node {$\Delta_F$};
    \draw[thick,color=blue,->,opacity=0.5] (8,2) [partial ellipse=-180:85:0.2cm and 0.2cm];
    \draw[
            decoration = {markings, mark= at position 0.25 with {\arrow{<}}, 
                mark=at position 0.75 with {\arrow{>}}}, 
            postaction = {decorate}
            ]  (8.5,2) circle (0.5);
    \draw (9,2) node [circle, fill, inner sep=0.5mm] {};
    \draw (9.2,2.2) node {$\mu_*$};
    \draw[thick,color=blue,->,opacity=0.5] (9,2) [partial ellipse=-265:0:0.2cm and 0.2cm];
    \draw[->] (9,2) -- (9.5,2);  \draw (9.5,2) -- (10,2);
    \draw (10,2) node [circle, fill, inner sep=0.5mm] {};
    \draw (9.8,1.8) node {$\Delta_F$};
    \draw[thick,color=blue,->,opacity=0.5] (10,2) [partial ellipse=-180:85:0.2cm and 0.2cm];
    \draw[->] (10,2) -- (10.5,2.5);  \draw (10.5,2.5) -- (11,3);
    \draw[->] (10,2) -- (10.5,1.5);  \draw (10.5,1.5) -- (11,1);
    \draw (10.9,1.5) node {${\cal A}$};  \draw (10.9,2.5) node {${\cal A}$};
    \draw (12,2) node {$=$};
    \draw[->] (13,3) -- (13.5,2.5);  \draw (13.5,2.5) -- (14,2);
    \draw[->] (13,1) -- (13.5,1.5);  \draw (13.5,1.5) -- (14,2);
    \draw (13.1,1.5) node {${\cal A}$};  \draw (13.1,2.5) node {${\cal A}$};
    \draw (14,2) node [circle, fill, inner sep=0.5mm] {};
    \draw (14.2,2.2) node {$\mu_*$};
    \draw[thick,color=blue,->,opacity=0.5] (14,2) [partial ellipse=-225:0:0.2cm and 0.2cm];
    \draw[->] (14,2) -- (14.5,2); \draw (14.5,2) -- (15,2);
    \draw (15,2) node [circle, fill, inner sep=0.5mm] {};
    \draw (14.8,1.8) node {$\Delta_F$};
    \draw[thick,color=blue,->,opacity=0.5] (15,2) [partial ellipse=-180:45:0.2cm and 0.2cm];
    \draw[->] (15,2) -- (15.5,2.5);  \draw (15.5,2.5) -- (16,3);
    \draw[->] (15,2) -- (15.5,1.5);  \draw (15.5,1.5) -- (16,1);
    \draw (15.9,1.5) node {${\cal A}$};  \draw (15.9,2.5) node {${\cal A}$};
\end{tikzpicture}
\caption{Shown is an example of simplifying a diagram, using, from the left, a fusion move followed by a bubble move, resulting finally in a diagram given by $\Delta_F \circ \mu_*$.}
\label{fig:simplify}
\end{figure}
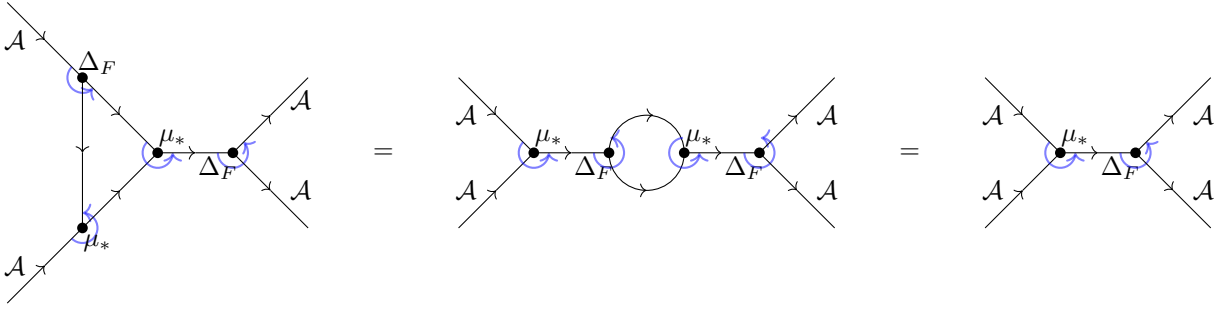

\subsection{Structure of state spaces for Rep$(G)$}
\label{sect:formal-state-spaces}

In this section we will utilize the existence of a noninvertible Rep$(G)$ symmetry in a $G$ orbifold, and the fact that
orbifolding by the quantum symmetry returns the original theory, to compare the organization of the state spaces arising
in orbifolds by a group $G$ and by a noninvertible symmetry Rep$(G)$.

Let us review the structure of the state space of local operators in a 2d theory ${\cal T}$ with an invertible global symmetry given by an abelian group $G$.  At the broadest level, there will be genuine local operators which can appear freestanding in correlation functions and twisted local operators which have support only on the end of TDLs.  That is, for each $g$ in $G$, there is a Hilbert space $\H_g$ of $g$-twisted operators living at the endpoint of a line labeled by $g$.  We can refine this description by noting that the symmetry $G$ naturally acts on the twist fields, which means that the states in $\H_g$ further break up into representations of $G$.  Accordingly, we will write $\H_{g,\hat{g}}$ for the Hilbert space of $g$-twisted states which transform in the $\hat{g}\in\hat{G}$ representation.

What happens to these states when we gauge $G$?  The resulting theory has a global quantum Rep$(G) = \hat{G}$ symmetry, which we can use to organize the result.  Those states that transform in the trivial representation of $G$ are gauge invariant, and become the genuine local (untwisted) states of the gauged theory.  More geometrically, if we think of gauging as flooding the worldsheet with a network of $G$ lines, the gauge-invariant operators that were previously bound to the end of those lines can move freely along that background network, making them genuinely local.  The states transforming in non-trivial representations of $G$ can be made gauge invariant by placing them at the end of $\hat{G}$ lines,
so these form the twisted sectors of the gauged theory.  Thus gauging has the effect of taking $\H_{g,\hat{g}}$ to $\H_{\hat{g},g}$.  The actions of $\gamma\in G$ and $\hat{\gamma}\in\hat{G}$ on a state $\mcO_{g,\hat{g}}$ in $H_{g,\hat{g}}$ are given by
\be
\label{localopaction}
\gamma\cdot\mcO_{g,\hat{g}} = \chi_{\hat{g}}(\gamma)\mcO_{g,\hat{g}}\hspace{1cm}\hat{\gamma}\cdot\mcO_{g,\hat{g}} = \chi_{\hat{\gamma}}(g)\mcO_{g,\hat{g}}.
\ee

When we allow $G$ to be non-abelian, some subtleties creep into the above story.  Consider the action of $G$ on an arbitrary twisted sector state.  This is given by the action of $G$ on itself, and the orbits for such an action are the conjugacy classes of $G$.  Twist fields in this case are thus labeled by conjugacy classes $[g]$ rather than individual group elements, as conjugate group elements give isomorphic Hilbert spaces.  For a given representative $g$ of $[g]$, the subgroup of $G$ that fixes $g$ is its stabilizer Stab$(g)$.\footnote{One sometimes sees this phrased in terms of centralizers, but for a group acting on itself by conjugation these are equivalent.}  Any other member of the conjugacy class has a stabilizer which is conjugate to Stab$(g)$, so if we only care about subgroups of $G$ up to conjugation (and conjugate subgroups do give equivalent symmetries), we can sensibly talk about Stab$([g])$ as the conjugacy class of subgroups which fixes $[g]$.  Thus, twist fields in this theory live in Hilbert spaces $\H_{[g],R}$ labeled by conjugacy classes of $G$ and representations $R$ of Stab$([g])$.

Let us proceed with an explicit example, with a global $S_3$ symmetry, where $S_3$ is the symmetric group on three objects.  We will use the same notation as in section~\ref{ssec:reps3}.  Here we have three conjugacy classes $[1]$, $[a]$ and $[b]$ with stabilizers $S_3$, $\Z_2$ and $\Z_3$.  Denoting respectively the trivial and non-trivial representations of $\Z_2$ by $+$ and $-$, and the representations of $\Z_3$ by $\omega^i$, the local operators in an $S_3$-symmetric theory break into
\be
\label{s3_states}
\begin{tabular}{l l l}
    $\H_{[1],1}$ & $\H_{[a],+}$ & $\H_{[b],\omega^0}$\\
    $\H_{[1],X}$ & $\H_{[a],-}$ & $\H_{[b],\omega^1}$\\
    $\H_{[1],Y}$ & & $\H_{[b],\omega^2}$
\end{tabular}
\ee
where each column is a twisted sector.  We can still leverage (\ref{localopaction}) to help us determine the fate of the genuine local operators under gauging $S_3$ or any of its subgroups.  $S_3$ has (up to conjugation) three non-trivial gaugeable subgroups: $\Z_2$, $\Z_3$ and all of $S_3$.  The algebra objects for gauging these can be written as $[1]+[a]$ for $\Z_2$, $[1]+2[b]$ for $\Z_3$ and $[1]+3[a]+2[b]$ for $S_3$.  Using (\ref{localopaction}), the genuine local operators are acted upon as
\begin{align}
    \frac{1}{2}([1]+[a])\cdot\mcO_{[1],1}&=\frac{1}{2}\left[\chi_{1}([1])+\chi_{1}([a])\right]\mcO_{[1],1}=\mcO_{[1],1}\\
    \frac{1}{2}([1]+[a])\cdot\mcO_{[1],X}&=\frac{1}{2}\left[\chi_{X}([1])+\chi_{X}([a])\right]\mcO_{[1],X}=0\\
    \frac{1}{2}([1]+[a])\cdot\mcO_{[1],Y}&=\frac{1}{2}\left[\chi_{Y}([1])+\chi_{Y}([a])\right]\mcO_{[1],Y}=\mcO_{[1],Y}\\
\end{align}
by $\Z_2$,
\begin{align}
    \frac{1}{3}([1]+2[b])\cdot\mcO_{[1],1}&=\frac{1}{3}\left[\chi_{1}([1])+2\chi_{1}([b])\right]\mcO_{[1],1}=\mcO_{[1],1}\\
    \frac{1}{3}([1]+2[b])\cdot\mcO_{[1],X}&=\frac{1}{3}\left[\chi_{X}([1])+2\chi_{X}([b])\right]\mcO_{[1],X}=\mcO_{[1],X}\\
    \frac{1}{3}([1]+2[b])\cdot\mcO_{[1],Y}&=\frac{1}{3}\left[\chi_{Y}([1])+2\chi_{Y}([b])\right]\mcO_{[1],Y}=0\\
\end{align}
by $\Z_3$ and 
\begin{align}
    \frac{1}{6}([1]+3[a]+2[b])\cdot\mcO_{[1],1}&=\frac{1}{6}\left[\chi_{1}([1])+3\chi_{1}([a])+2\chi_{1}([b])\right]\mcO_{[1],1}=\mcO_{[1],1}\\
    \frac{1}{6}([1]+3[a]+2[b])\cdot\mcO_{[1],X}&=\frac{1}{6}\left[\chi_{X}([1])+3\chi_{X}([a])+2\chi_{X}([b])\right]\mcO_{[1],X}=0\\
    \frac{1}{6}([1]+3[a]+2[b])\cdot\mcO_{[1],Y}&=\frac{1}{6}\left[\chi_{Y}([1])+3\chi_{Y}([a])+2\chi_{Y}([b])\right]\mcO_{[1],Y}=0\\
\end{align}
by $S_3$.  This gives us an idea of which genuine local operators are gauge-invariant and therefore will appear in the untwisted sector in an orbifold by any subgroup of $S_3$ above.  

Let us focus on the orbifold by the full $S_3$, $[{\cal T}/S_3]$, which will yield a theory with a global Rep$(S_3)$ quantum symmetry.  The untwisted states in (\ref{s3_states}) are already labeled by representations of $S_3$, so it is clear how they transform.  What about the twisted states, which are labeled by representations of subgroups?  These turn out to in general be $S_3$ multiplets, instead of singlets -- they transform under the $S_3$ representation induced by the representation of the stabilizier subgroup which labels them, which need not be irreducible.  In the example at hand, rewriting (\ref{s3_states}) in terms of induced representations gives
\be
\label{s3_states_induced}
\begin{tabular}{l l l}
    $\H_{[1],1}$ & $\H_{[a],1+Y}$ & $\H_{[b],1+X}$\\
    $\H_{[1],X}$ & $\H_{[a],X+Y}$ & $\H_{[b],Y}$\\
    $\H_{[1],Y}$ & & $\H_{[b],Y}$
\end{tabular}.
\ee
Concretely, then, we see that both $\H_{[a],+}$ and $\H_{[b],\omega^0}$ contain states which are $S_3$-invariant and therefore will end up in the untwisted sector of the $S_3$ orbifold.  The genuine local operators in $[{\cal T}/S_3]$, then, are made up of states $\mcO_{[1],1}$ arising from $\H_{[1],1}$, states $\mcO_{[a],1}$ arising from $\H_{[a],+}$ and states $\mcO_{[b],1}$ arising from $\H_{[b],\omega^0}$.  (\ref{localopaction}) gives us the action of Rep$(S_3)$ on these states:
\begin{align}
1\cdot\mcO_{[1],1}&=\chi_1([1])\mcO_{[1],1}=\mcO_{[1],1}\\
1\cdot\mcO_{[b],1}&=\chi_1([b])\mcO_{[b],1}=\mcO_{[b],1}\\
1\cdot\mcO_{[a],1}&=\chi_1([a])\mcO_{[a],1}=\mcO_{[a],1}\\
X\cdot\mcO_{[1],1}&=\chi_X([1])\mcO_{[1],1}=\mcO_{[1],1}\\
X\cdot\mcO_{[b],1}&=\chi_X([b])\mcO_{[b],1}=\mcO_{[b],1}\\
X\cdot\mcO_{[a],1}&=\chi_X([a])\mcO_{[a],1}=-\mcO_{[a],1}\\
Y\cdot\mcO_{[1],1}&=\chi_Y([1])\mcO_{[1],1}=2\mcO_{[1],1}\\
Y\cdot\mcO_{[b],1}&=\chi_Y([b])\mcO_{[b],1}=-\mcO_{[b],1}\\
Y\cdot\mcO_{[a],1}&=\chi_Y([a])\mcO_{[a],1}=0,
\end{align}
which agrees with the action given in \cite[section 5.2]{Bhardwaj:2023idu}.  The twisted sectors, in general, receive contributions from multiple ungauged sectors.  If we further gauge the regular representation of the Rep$(S_3)$ symmetry to recover the original $S_3$-symmetric theory, the resulting local operators should once again organize themselves by (\ref{s3_states}).

\subsection{Formal argument for algebra object as projector}  \label{sect:formal-proj}

Next, we consider gauging a noninvertible symmetry, and discuss how the algebra object ${\cal A}$ acts as a projector.

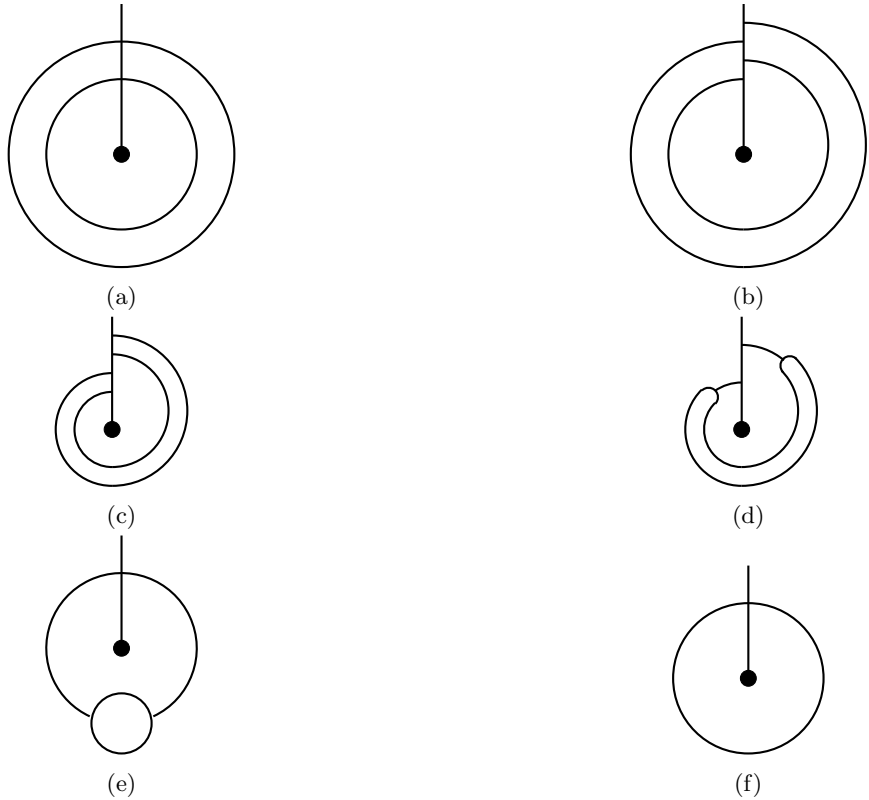
\begin{figure}
    \begin{subfigure}{0.5\textwidth}
    \centering
    \begin{tikzpicture}
        \draw[thick] (0,0) -- (0,2);
        \draw[thick] (0,0) [partial ellipse=0:360:1.0cm and 1.0cm];
        \draw[thick] (0,0) [partial ellipse=0:360:1.5cm and 1.5cm];
        \filldraw[black] (0,0) circle (3pt);
    \end{tikzpicture}
    \caption{}
    \label{statespace1}
    \end{subfigure}
    \begin{subfigure}{0.5\textwidth}
    \centering
    \begin{tikzpicture}
        \draw[thick] (0,0) -- (0,2);
        \draw[thick] (0,0) [partial ellipse=90:270:1.0cm and 1.0cm];
        \draw[thick] (0,0.125) [partial ellipse=-90:90:1.125cm and 1.125cm];
        \draw[thick] (0,0) [partial ellipse=90:270:1.5cm and 1.5cm];
        \draw[thick] (0,0.125) [partial ellipse=-90:90:1.625cm and 1.625cm];
        \filldraw[black] (0,0) circle (3pt);
    \end{tikzpicture}
    \caption{}
    \label{statespace2}
    \end{subfigure}
    \begin{subfigure}{0.5\textwidth}
    \centering
    \begin{tikzpicture}
        \draw[thick] (0,0) -- (0,1.5);
        \draw[thick] (0,0) [partial ellipse=90:270:0.5cm and 0.5cm];
        \draw[thick] (0,0.25) [partial ellipse=-90:90:0.75cm and 0.75cm];
        \draw[thick] (0,0) [partial ellipse=90:270:0.75cm and 0.75cm];
        \draw[thick] (0,0.25) [partial ellipse=-90:90:1.0cm and 1.0cm];
        \filldraw[black] (0,0) circle (3pt);
    \end{tikzpicture}
    \caption{}
    \label{statespace3}
    \end{subfigure}
    \begin{subfigure}{0.5\textwidth}
    \centering
    \begin{tikzpicture}
        \draw[thick] (0,0) -- (0,1.5);
        \draw[thick] (0,0) [partial ellipse=135:270:0.5cm and 0.5cm];
        \draw[thick] (0,0.25) [partial ellipse=-90:45:0.75cm and 0.75cm];
        \draw[thick] (-0.44,0.43) [partial ellipse=-45:135:0.125cm and 0.125cm];
        \draw[thick] (0,0) [partial ellipse=135:270:0.75cm and 0.75cm];
        \draw[thick] (0,0.25) [partial ellipse=-90:45:1.0cm and 1.0cm];
        \draw[thick] (0.64,0.85) [partial ellipse=45:225:0.125cm and 0.125cm];
        \filldraw[black] (0,0) circle (3pt);
        \draw[thick] (0,0) [partial ellipse=90:125:0.625cm and 0.625cm];
        \draw[thick] (0,0.25) [partial ellipse=90:50:0.875cm and 0.875cm];
    \end{tikzpicture}
    \caption{}
    \label{statespace4}
    \end{subfigure}
    \begin{subfigure}{0.5\textwidth}
    \centering
    \begin{tikzpicture}
        \draw[thick] (0,0) -- (0,1.5);
        \draw[thick] (0,0) [partial ellipse=-65:245:1.0cm and 1.0cm];
        \draw[thick] (0,-1) [partial ellipse=0:360:0.4cm and 0.4cm];
        \filldraw[black] (0,0) circle (3pt);
    \end{tikzpicture}
    \caption{}
    \label{statespace5}
    \end{subfigure}
    \begin{subfigure}{0.5\textwidth}
    \centering
    \begin{tikzpicture}
        \draw[thick] (0,0) -- (0,1.5);
        \draw[thick] (0,0) [partial ellipse=0:360:1.0cm and 1.0cm];
        \filldraw[black] (0,0) circle (3pt);
    \end{tikzpicture}
    \caption{}
    \label{statespace6}
    \end{subfigure}
    \caption{The algebra object $\mcA$ acts projectively on local states.}
    \label{statespace}
\end{figure}

As with invertible symmetries, the partition function obtained when gauging noninvertible symmetries will break into a sum of terms which describe the action of the algebra object on twist fields.  The properties of the Frobenius algebra, in particular the fact that the multiplication gives a map $\mcA\otimes\mcA\to\mcA$, ensure that $\mcA$ acts projectively on states.  Using the diagrammatic properties of the algebra object as given in e.g.~\cite[section 3]{Fuchs:2002cm}, we can follow \cite{Brunner:2013ota} in showing that $\mcA$ acts projectively on (twisted) local states -- this is shown in Figure~\ref{statespace}.

All of the lines drawn there are the algebra object $\mcA$, and in the middle we project $\mcA$ onto one of its constituent simple lines which then ends on a twist field.  The junctions contain multiplication or co-multiplication as necessary.  Arrows on lines are omitted for readability.  Moving from Figure~\ref{statespace1} to Figure~\ref{statespace2} we break up the two four-way junctions.  Using the equivalence of the configurations shown in Figure~\ref{frobassoc} allows us to deform Figure~\ref{statespace2} to Figure~\ref{statespace3}.  Applying two swap relations produces Figure~\ref{statespace4}, which we can deform to Figure~\ref{statespace5}.  Finally, the condition on composition of multiplication and co-mutiplication given in (\ref{eq:conds}) allows us to `pop' the bubble, leading to Figure~\ref{statespace6}.  In total, we find that the action of $\mcA\otimes\mcA$ on any local operator is equivalent to that of $\mcA$, as expected.

\subsection{Specialization to ordinary orbifolds}  \label{sect:ord-orbifold}

Previously in section~\ref{sect:genl-algebra} we introduced some abstract machinery to discuss gauging noninvertible symmetries, e.g.~Frobenius structures, Hopf algebras, and so forth.  In this section, we will demonstrate that that abstract machinery correctly reproduces standard results for ordinary orbifolds by (ordinary) groups.

Here, the fusion algebra is Vec$(G) = {\rm Rep}( {\mathbb C}[G]^*)$.
As this may be a bit obscure, let us take a moment to elaborate.
First, we describe ${\mathbb C}[G]^*$ as a vector space with basis elements $v_g$ for all $g \in G$, with multiplication
\begin{equation}
    v_g v_h \: = \: \delta_{g,h} v_g.
\end{equation}
The irreducible representations of ${\mathbb C}[G]^*$ are one-dimensional, and in one-to-one
correspondence with elements of $G$.  Let such an irreducible representation be denoted
$U_g$, with basis $\{ e_{Ug} \}$, and $G$-action defined linearly over
\begin{equation}
    v_h \cdot e_{Ug} \: = \: \delta_{g,h} e_{Ug}.
\end{equation}
The tensor product follows naturally from the group law:
\begin{equation}
    U_g \otimes U_h \: = \: U_{gh}.
\end{equation}
General representations, sums of irreducible representations, are $G$-graded vector spaces.
See also e.g.~\cite{se} for a (dual) description of $G$-graded vector spaces as comodules over ${\mathbb C}[G]$.

Now, in this case, we do not need to consider nontrivial junction defects, simply because
\begin{equation}
    {\rm Hom}(U_g \otimes U_h, U_k) \: = \: \left\{ \begin{array}{cl}
    {\mathbb C} & gh = k, \\
    0 & {\rm else}.
    \end{array} \right.
\end{equation}
This is certainly in agreement with standard orbifold constructions, which do not involve junction defects.

We next construct a special symmetric Frobenius algebra on ${\mathbb C}[G] = 
{\mathbb C}[G]^{**}$.  Proceeding as before, we inherit the product and unit structure from Vec($G$), namely
\begin{equation}  \label{eq:groupcase:mu}
    \mu_*(g,h) = gh, \: \: \: u_*(1) = 1.
\end{equation}
We then define a new coproduct $\Delta_F$ and counit $u^o_F$.
The integral element $\Lambda \in {\mathbb C}[G]^*$ given by
\begin{equation}
\Lambda = v_1,
\end{equation}
and the cointegral is
\begin{equation}
    \lambda = \frac{1}{|G|}\sum_{g\in G} g.
\end{equation}
We take $u^o_F: {\mathbb C}[G] \rightarrow {\mathbb C}$
to be given by $\Lambda^*:\C[G]\to\C$, that is,
\begin{equation}
    u^o_F(g) = |G| \, \delta_{1,g}.
\end{equation}
Next we work out the comultiplication. It corresponds to the dual of the map
\begin{eqnarray}
    v_g\otimes v_h\mapsto \sum_{k\in G}v_{gk}\otimes v_{k^{-1}}\otimes v_h\mapsto \sum_{k\in G}v_{gk}\otimes v_{k}\otimes v_h \mapsto v_{gh}\otimes 1
\end{eqnarray}
so that on the basis elements and normalizing this is
\begin{eqnarray}
    \Delta_F&:=& g\mapsto \frac{1}{|G|}\sum_{h\in G}gh\otimes h^{-1}
\end{eqnarray}
It is straightforward to check that this has the desired normalization,
$\mu_*\circ\Delta_F=\text{Id}_{{\cal H}^*}$, using the fact that $\mu_* = \Delta^*$ (from~(\ref{eq:defn:mu-star})), and in the present case, $\Delta = \mu^*$ (using the relation between
dual bialgebras, as discussed in section ~\ref{sect:genl-algebra}), hence $\mu_* = \Delta^* = \mu^{**} = \mu$.
Similarly, the coevaluation maps are dualized relative to Rep$(G)$:
\begin{equation}  \label{eq:coeval:vecg}
        \gamma_{\cal A}(1) \: = \: \sum_{h \in G} h \otimes v_h,
        \: \: \:
        \overline{\gamma}_{\cal A}(1) \: = \:  \sum_{h \in G} v_h \otimes h.
    \end{equation}
(Compare their expressions~(\ref{eq:coeval:repg}) for Rep$(G)$.)
Altogether, it is also straightforward to check that 
\begin{equation}
    ({\cal A} = {\cal H}^*, \mu_*, u_*, \Delta_F, u_F^o)
\end{equation}
satisfy the axioms to be a special symmetric Frobenius algebra.

Now, let us compute the genus one partition function~(\ref{eq:Z:genl1}), namely
\begin{equation}
    Z(\tau) \: = \: \sum_{g,h,k} \mu_{g,h}^k \Delta^{h,g}_k Z^k_{g,h}(\tau),
\end{equation}
where $\Delta$ indicates $\Delta_F$, the comultiplication in the Frobenius algebra.
(Since the lines are associated to group elements, we use $g, h, k$ in the remainder of this
section instead of $L$ to denote them.)
Expressing~(\ref{eq:groupcase:mu}) in components, we have\footnote{
In comparison to later computations, in which we will explicitly track intertwiners, here we take all intertwiners to be trivial.
}
\begin{equation}
    \mu_{g,h}^{k} \: = \: \left\{ \begin{array}{cl}
    1 & k = gh, \\
    0 & {\rm else},
    \end{array} \right.
\end{equation}
so we see that the only contributions to the sum are from $g, h \in G$ such that $gh = hg$
(otherwise one of the two factors $\mu_{g, h}^k$, $\Delta^{h, g}_k$ vanishes).
As a result, we can simplify the expression for the genus-one partition function and write
\begin{equation}
    Z \: = \: \frac{1}{|G|} \sum_{gh = hg} Z_{g,h}^{gh}, 
\end{equation}
which precisely matches the standard expression for genus one orbifold partition functions.

In the special case ${\cal H} = ({\mathbb C}[G])^*$, the Hopf subalgebras of ${\cal H}^*\cong\C[G]$ are all of the form $\C[H]\subset \C[G]$ for $H \subset G$ a subgroup (see e.g.~\cite[Prop. 2.1]{CRV14}, \cite{mo-subalg}). This recovers the familiar statement that the subcategories of $\text{Vec}(G)$ which we can gauge are precisely those of the form $\text{Vec}(H)$ for $H$ any subgroup of $G$.

So far we have discussed ordinary orbifolds, as special cases of the technology we have reviewed.  Next, we will describe how one can also describe orbifolds with discrete torsion, in the same
language.

Briefly, discrete torsion is encoded in a modification of the multiplication in the Frobenius
algebra.  (The starting Hopf algebra is unchanged.)
We modify the components of the multiplication $\mu_*$ in the Frobenius algebra to be given by
\begin{equation}
    \mu_{g,h}^{k} \: = \: \left\{ \begin{array}{cl}
    m_{g,h} & k = gh, \\
    0 & {\rm else},
    \end{array} \right.
\end{equation}
for some nonzero complex numbers $m_{g,h}$.
For now we will assume that the $G$ symmetry is non-anomalous, which means that we can take the associator to be trivial.\footnote{In the case of an anomalous group-like symmetry, the associator becomes a non-trivial 3-cocycle $\omega$, as described in the following section.  The condition on the discrete torsion 2-cochain $m$ is then that $dm=\omega$, rather than closure.}  The associativity constraint~(\ref{assoc}), namely
\be
\mu_* (\mu_*\otimes\text{Id}_{\cal A}) = \mu_* (\text{Id}_{\cal A}\otimes\mu_*),
\ee
reduces, in this basis, to
\be
m_{g,h} \, m_{gh,k} \:= \: m_{g,hk} \, m_{h,k}
\ee
which we recognize as the 2-cocycle condition\footnote{
Changing the 2-cocycles by 2-coboundaries changes the Frobenius algebra, but leaves the
partition function invariant.
} (and the normalization condition is indeed the usual cocycle normalization).  The unit axiom~(\ref{eq:id:unit}) requires
\begin{equation}
    m_{1,g} \: = \: m_{g,1} \: =  \: 1,
\end{equation}
which is a normalization condition that can always be imposed, without loss of
generality, on a group 2-cocycle.

The unit $u_*$ and co-unit $u^\circ_F$ in the Frobenius algebra take the form
\be
\label{unit_gl}
u_*:1\to \bigoplus_{g\in G}\delta_{1,g}g\hspace{0.5cm}u^\circ_F:g\to|G|\delta_{1,g}1.
\ee
The comultiplication $\Delta_F$ is given by
\begin{eqnarray}
    \Delta_F&:=& g\mapsto \frac{1}{|G|}\sum_{h\in G} \frac{1}{m_{gh, h^{-1}}} \,gh\otimes h^{-1}
\end{eqnarray}
which has components
\begin{equation}
    \Delta^{h,g}_k \: = \: \left\{ \begin{array}{cl}
    \frac{1}{|G|} \left( m_{h,g} \right)^{-1} & k = hg,
    \\
    0 & {\rm else}.
    \end{array} \right.
\end{equation}
The coevaluation maps are the same as~(\ref{eq:coeval:vecg}).
It is straightforward to check that with these definitions, $(\mu_*,u_*,\Delta_F, u_F^{\circ})$ satisfy the axioms to be a special symmetric Frobenius algebra.

Then, the resulting genus-one partition function is, from~(\ref{eq:Z:genl1}),
\begin{eqnarray}
    Z & = &
    \sum_{g,h,k} \mu_{g,h}^k \Delta^{h,g}_k Z_{g,h}, 
    \\
    & = & \frac{1}{|G|} \sum_{gh = hg} \frac{m_{g,h}}{m_{h,g}} \, Z_{g,h},
\end{eqnarray}
which is the standard result for genus-one partition functions of ordinary orbifolds with
discrete torsion.

As a solid consistency check, let us also check the genus 2 partition function, using our general technology,
in the case of Vec$(G)$ with discrete torsion, to compare to existing results for discrete torsion on
genus $2$ surfaces.  As discussed in subsection~\ref{ssec:genus2},
the genus 2 partition function is the sum over partial traces~(\ref{eq:g2partitionfunction}), namely
\begin{equation*}
    Z_{g=2} = \sum_{L_1,\cdots,L_9} \mu_{L_1,L_2}^{L_5}\Delta_{L_5}^{L_6,L_7} \Delta_{L_6}^{L_2,L_1}\mu_{L_3,L_4}^{L_8}\mu_{L_7,L_8}^{L_9}\Delta_{L_9}^{L_4,L_3} Z_{L_1,L_2,L_3,L_4}^{L_5,L_6,L_7,L_8,L_9},
\end{equation*} 
In the present case, all the internal simple lines are determined by the external lines, as in Figure \ref{fig:g2trivalentgp}. Since this renders the labels by internal lines on the partial traces redundant, we simply denote these as $Z_{b_1,a_1,b_2,a_2}$, much as the familiar $Z_{g,h}$ expression for genus 1 partial traces in group-like cases. Let $\gamma=b_1^{-1}a_1^{-1}b_1a_1$, for $a_{1,2}, b_{1,2} \in G$, then the partition function is
\begin{eqnarray}
   Z_{g=2} = \sum_{a_1,b_1,a_2,b_2} \frac{m_{b_1,a_1}}{m_{a_1b_1,\gamma}}\frac{m_{b_2,a_2}}{m_{a_1,b_1}}\frac{m_{\gamma,b_2a_2}}{m_{a_2,b_2}} \ Z_{b_1,a_1,b_2,a_2}.
\end{eqnarray}
We note that \begin{equation}
  \frac{m_{b_1,a_1}}{m_{a_1b_1,\gamma}}\frac{m_{b_2,a_2}}{m_{a_1,b_1}}\frac{m_{\gamma,b_2a_2}}{m_{a_2,b_2}}
\end{equation}
is precisely the discrete torsion phase for a partial trace of a genus $2$ surface \cite{Aspinwall:2000xv,Bantay:2000eq}.
Moreover, from the diagram we observe that for the partial trace coefficients to not vanish, the external lines need to satisfy the equation
\begin{equation}
    b_1^{-1}a_1^{-1}b_1a_1 = a_2^{-1}b_2^{-1}a_2b_2,
\end{equation}
which is the standard condition on genus-two orbifold contributions.

\section{Examples of partition function computations}  \label{sect:examples}

In this section we will compute genus-one partition functions explicitly in some examples of 
(multiplicity-free) fusion categories of the form Rep$({\cal H}$), for ${\cal H}$ a Hopf algebra.
In most of the examples, we will derive results expressed in terms of a general basis of fusion intertwiners.
To compute a partition function, one must specify a special symmetric Frobenius algebra, as already
discussed, and for each fusion category, we will list all choices and compute partition functions for each choice.
In every case, we will see that our general expression procedure yields a partition function that is a modular-invariant sum of partial traces, which is a strong check that the procedure is self-consistent.

\subsection{Rep$(S_3)$}\label{ssec:reps3}

First, we will consider Rep$(S_3)$, where $S_3$ is the symmetric group on three elements, which
can be presented as
\begin{equation}
    S_3 \: = \: \langle a, b \, | \, a^2 = 1 = b^3, \: a b a = b^2 \rangle.
\end{equation}

\subsubsection{Representation theory}

The group $S_3$ has three irreducible representations which we will label $1$, $X$, and $Y$.  The character table is:

\begin{center}
\begin{tabular}{c|c|c|c|}
& $[1]$ & $[a]$ & $[b]$ \\
\hline
$\chi_1$ & $1$ & $1$ & $1$ \\
\hline
$\chi_X$ & $1$ & $-1$ & $1$ \\
\hline
$\chi_Y$ & $2$ & $0$ & $-1$ \\
\hline
\end{tabular}
\end{center}

Specific realizations of these irreps are given by
\be
\rho_1(a)=\rho_1(b)=1,
\ee
\be
\rho_X(a)=-1,\qquad\rho_X(b)=1,
\ee
\be
\label{eq:rhoY}
\rho_Y(a)=\lp\begin{matrix} -1 & 0 \\ 0 & 1 \end{matrix}\rp,\qquad\rho_Y(b)=\lp\begin{matrix} -\hlf & -\frac{\sqrt{3}}{2} \\ \frac{\sqrt{3}}{2} & -\hlf \end{matrix}\rp.
\ee

The representations form a fusion algebra under tensor products, with
\be
X\otimes X\cong 1,\qquad X\otimes Y\cong Y,\qquad Y\otimes Y\cong 1\oplus X\oplus Y.
\ee
Additionally, all three of these irreducible representations are self-dual, meaning $1^\ast\cong 1$, $X^\ast\cong X$, and $Y^\ast\cong Y$.

In our fusion category $\text{Rep}(S_3)$ the objects will be labeled by representations of $S_3$, with the irreducible representations $1$, $X$, and $Y$ being the simple objects.  The spaces $\text{Hom}(R_1,R_2)$ will consist of intertwiners between the representations $R_1$ and $R_2$.  Explicitly, if $R_i$ consists of a vector space $V_i$ and a group homomorphism $\rho_i:S_3\rightarrow\GL(V_i)$, then each element of $\text{Hom}(R_1,R_2)$ is a linear map $\phi:V_1\rightarrow V_2$ satisfying
\begin{equation}
    \phi\circ\rho_1(g)=\rho_2(g)\circ\phi,\qquad\forall g\in S_3.
\end{equation}

\subsubsection{Computing the associator}

To do this computation, we first need to pick a basis of $\Hom(R_i\otimes R_j,R_k)$ for each fusion of 
irreducible representations $R_i$, $R_j$, $R_k$.  For Rep$(S_3)$, these spaces fortunately all have dimension zero or one, so the freedom is just a set of $\C^\ast$ parameters (corresponding to intertwiners) which we will leave arbitrary (except for fusions involving the trivial representation, for which there is a canonical choice).  We'll use $e$ as the basis vector\footnote{In doing so, we are using the fact that we have a fiber functor to define a basis on the underlying vector spaces. However, the maps herein described do exist in the fusion category as they are all equivariant.} for the trivial representation, $e_X$ as the basis vector for $X$, and $e_{Y1}$ and $e_{Y2}$ to be the basis vectors for $Y$ which are odd and even respectively under the action of $a$ (i.e.~so that $\rho_Y$ takes the form~(\ref{eq:rhoY})).  Letting $\beta_1, \cdots, \beta_6$ be the arbitrary parameters, we have the most general possible intertwiners
\begin{align}
e\otimes e\mapsto\ & e,\label{rs3int1}\\
e\otimes e_X\mapsto\ & e_X,\\
e\otimes e_{Y1}\mapsto\ & e_{Y1},\\
e\otimes e_{Y2}\mapsto\ & e_{Y2},\\
e_X \otimes e\mapsto\ & e_X,\\
e_X\otimes e_X\mapsto\ & \beta_1 \, e,\\
e_X\otimes e_{Y1}\mapsto\ & \beta_2 \, e_{Y2},  \label{eq:s3:int:xy1} \\
e_X\otimes e_{Y2}\mapsto\ & - \beta_2 \, e_{Y1}, \label{eq:s3:int:xy2} \\
e_{Y1}\otimes e\mapsto\ & e_{Y1},\\
e_{Y1}\otimes e_X\mapsto\ & \beta_3 \, e_{Y2},\\
e_{Y1}\otimes e_{Y1}\mapsto\ & \beta_4 \, e + \beta_5 \, e_{Y2},\\
e_{Y1}\otimes e_{Y2}\mapsto\ & \beta_6 \, e_X + \beta_5 \, e_{Y1},\\
e_{Y2}\otimes e\mapsto\ & e_{Y2},\\
e_{Y2}\otimes e_X\mapsto\ & - \beta_3 \, e_{Y1},\\
e_{Y2} \otimes e_{Y1}\mapsto\ & - \beta_6 \, e_X + \beta_5 \, e_{Y1},\\
e_{Y2}\otimes e_{Y2}\mapsto\ & \beta_4 \, e - \beta_5 \, e_{Y2}.\label{rs3int2}
\end{align}
Alternatively, in terms of explicit basis vectors $\lambda_{i,j}^k$ for $\text{Hom}(R_i\otimes R_j,R_k)$,
\begin{align}
    \lambda_{1,1}^1(e\otimes e)=\ & e,\\
    \lambda_{1,X}^X(e\otimes e_X)=\ & e_X,\\
    \lambda_{1,Y}^Y(e\otimes e_{Y1})=e_{Y1},\qquad & \lambda_{1,Y}^Y(e\otimes e_{Y2})=e_{Y2},\\
    \lambda_{X,1}^X(e_X\otimes e)=\ & e_X,\\
    \lambda_{X,X}^1(e_X\otimes e_X)=\ & \beta_1e,\\
    \lambda_{X,Y}^Y(e_X\otimes e_{Y1})=\beta_2e_{Y2},\qquad & \lambda_{X,Y}^Y(e_X\otimes e_{Y2})=-\beta_2e_{Y1},\\
    \lambda_{Y,1}^Y(e_{Y1}\otimes e)=e_{Y1},\qquad & \lambda_{Y,1}^Y(e_{Y2}\otimes e)=e_{Y2},\\
    \lambda_{Y,X}^Y(e_{Y1}\otimes e_X=\beta_3e_{Y2},\qquad & \lambda_{Y,X}^Y(e_{Y2}\otimes e_X)=-\beta_3e_{Y1},\\
    \lambda_{Y,Y}^1(e_{Y1}\otimes e_{Y1})=\beta_4e,\qquad & \lambda_{Y,Y}^1(e_{Y2}\otimes e_{Y2})=\beta_4e,\non\\
    \lambda_{Y,Y}^1(e_{Y1}\otimes e_{Y2})= & \lambda_{Y,Y}^1(e_{Y2}\otimes e_{Y2})=0,\\
    \lambda_{Y,Y}^X(e_{Y1}\otimes e_{Y2})=\beta_6e_X,\qquad & \lambda_{Y,Y}^X(e_{Y2}\otimes e_{Y1})=-\beta_6e_X,\non\\
    \lambda_{Y,Y}^X(e_{Y1}\otimes e_{Y1})= & \lambda_{Y,Y}^X(e_{Y2}\otimes e_{Y2})=0,\\
    \lambda_{Y,Y}^Y(e_{Y1}\otimes e_{Y1})=\beta_5e_{Y2},\qquad & \lambda_{Y,Y}^Y(e_{Y1}\otimes e_{Y2})=\beta_5e_{Y1},\non\\
    \lambda_{Y,Y}^Y(e_{Y2}\otimes e_{Y1})=\beta_5e_{Y1}, & \qquad\lambda_{Y,Y}^Y(e_{Y2}\otimes e_{Y2})=-\beta_5e_{Y2}.
\end{align}

Factors such as signs are determined by consistency with the group law.
To make this more explicit, consider the following example.
First, from the form of the representation $\rho_Y$,
\begin{equation}
    a \cdot e_{Y1} = - e_{Y1}, \: \: \:
a \cdot e_{Y2} = + e_{Y2},
\end{equation}
while
\begin{equation}
    a\cdot(e_X\otimes e_{Y1})=e_X\otimes e_{Y1},\: \: \: a\cdot(e_X\otimes e_{Y2})=-e_X\otimes e_{Y2}.
\end{equation}
Let $\lambda_{X,Y}^Y$ explicitly denote the intertwiner for the fusion $X\otimes Y\cong Y$.
Based on the action of $a$ above, the most general form of $\lambda_{X,Y}^Y$ is
\begin{equation}
\lambda_{X,Y}^Y\left( e_X \otimes e_{Y1}\right) \: = \: \beta_2 \, e_{Y2}, \: \: \:
\lambda_{X,Y}^Y\left( e_X \otimes e_{Y2}\right) \: = \: c_2 \, e_{Y1},
\end{equation}
where $\beta_2$ is as above and $c_2$ is another constant.
We then require
\begin{equation}
b \cdot \lambda_{X,Y}^Y\left( e_X \otimes e_{Y1} \right) \: = \: \lambda_{X,Y}^Y\left( e_X \otimes (b \cdot e_{Y1}) \right),
\end{equation}
and this fixes the value of $c_2$.
Explicitly, we are given
\begin{eqnarray}
b \cdot e_{Y1} & =  & - \frac{1}{2}\, e_{Y1} - \frac{ \sqrt{3} }{2} \, e_{Y2},
\\
b \cdot e_{Y2} & = & + \frac{ \sqrt{3} }{2}\, e_{Y1} - \frac{1}{2} \, e_{Y2},
\end{eqnarray}
and we compute
\begin{eqnarray}
b \cdot \lambda_{X,Y}^Y\left(e_X \otimes e_{Y1}\right) & = & \frac{\sqrt{3}}{2} \beta_2 \, e_{Y1} - \frac{1}{2} \beta_2 \, e_{Y2},
\\
\lambda_{X,Y}^Y\left( e_X \otimes (b \cdot e_{Y1})\right) & = & - \frac{1}{2} \beta_2 \, e_{Y2} - \frac{\sqrt{3}}{2} c_2 \, e_{Y1},
\end{eqnarray}
hence
\begin{equation}
c_2 \: = \: - \beta_2.
\end{equation}
Signs in other cases are determined similarly.

Next we compute the associators, which are elements $\al_{R_1,R_2,R_3}\in\Hom((R_1\otimes R_2)\otimes R_3,R_1\otimes(R_2\otimes R_3))$.  It turns out that as intertwiners, the associators in a representation category like this are canonical, acting simply as
\begin{equation}
    \al_{R_1,R_2,R_3}((v_1\otimes v_2)\otimes v_3)=v_1\otimes(v_2\otimes v_3),
\end{equation}
for all $v_i\in R_i$.  However, we would like to expand this simple map in terms of components related to the chosen bases for fusion intertwiners.  By taking fusions we can map a vector in $(R_1\otimes R_2)\otimes R_3$ to $R_4$, using some $R_5$ as an intermediate channel, by acting with the map $\lambda_{R_5,R_3}^{R_4}\circ(\lambda_{R_1,R_2}^{R_5}\otimes 1_{R_3})$.  On the other hand, we can similarly apply $\lambda_{R_1,R_6}^{R_4}\circ(1_{R_1}\otimes \lambda_{R_2,R_3}^{R_6})$ to a vector in $R_1\otimes(R_2\otimes R_3)$.  The sources of these two maps can be identified using the canonical associator map $\al_{R_1,R_2,R_3}$, so what we are looking for are the coefficients $\tilde{K}^{R_1,R_4}_{R_2,R_3}(R_5,R_6)$ relating these two, i.e.\footnote{The reason that dual representations appear in several of the indices is to match with~\cite{Chang:2018iay}, see footnote~\ref{fn:ChangCompare}.  Since in our examples we have $R\cong R^\ast$ for all irreducible representations $R$, we won't be bothered by taking the duals.}
\begin{equation}
        \lambda_{R_5,R_3}^{R_4}\circ(\lambda_{R_1,R_2}^{R_5}\otimes 1_{R_3})=\tilde{K}^{R_1^\ast,R_4}_{R_2^\ast,R_3^\ast}(R_5^\ast,R_6^\ast)\,\lambda_{R_1,R_6}^{R_4}\circ(1_{R_1}\otimes \lambda_{R_2,R_3}^{R_6})\circ\al_{R_1,R_2,R_3},
\end{equation}
as intertwiners in $\Hom((R_1\otimes R_2)\otimes R_3,R_4)$.  The quantities $\tilde{K}^{R_1,R_4}_{R_2,R_3}(R_5,R_6)$ here are just complex numbers\footnote{\label{fn:ChangCompare} This is a different perspective than in~\cite{Chang:2018iay}, where the $\tilde{K}^{R_1,R_4}_{R_2,R_3}(R_5,R_6)$ are viewed as maps, explicitly
\begin{equation}
    \tilde{K}^{R_1,R_4}_{R_2,R_3}(R_5,R_6):\Hom(R_1^\ast\otimes R_2^\ast,R_5^\ast)\otimes\Hom(R_5^\ast\otimes R_3^\ast,R_4)\rightarrow\Hom(R_2^\ast\otimes R_3^\ast,R_6^\ast)\otimes\Hom(R_1^\ast\otimes R_6^\ast,R_4).
\end{equation}
The quantities here are related by
\begin{equation}
    \left[\tilde{K}^{R_1,R_4}_{R_2,R_3}(R_5,R_6)_{there}\right](\lambda_{R_1^\ast,R_2^\ast}^{R_5^\ast}\otimes \lambda_{R_5^\ast,R_3^\ast}^{R_4})=\tilde{K}^{R_1,R_4}_{R_2,R_3}(R_5,R_6)_{here}\,\lambda_{R_2^\ast,R_3^\ast}^{R_6^\ast}\otimes \lambda_{R_1^\ast,R_6^\ast}^{R_4}.
\end{equation}
}, since we are restricting in this paper to the case where all the fusion and co-fusion $\Hom$ spaces have dimension at most one.

Since fusion involving the trivial representation was canonical (i.e.~$\lambda_{1,R}^R(e\otimes v)=v$ and $\lambda_{R,1}^R(v\otimes e)=v$ for all $v\in R$), we have
\be
\tilde{K}^{1,R_3}_{R_1,R_2}(R_1,R_3)=1,
\qquad
\tilde{K}^{R_1,R_3}_{1,R_2}(R_1,R_2)=1,
\qquad
\tilde{K}^{R_1,R_3}_{R_2,1}(R_3,R_2)=1.
\ee

So focusing now on the case with only non-trivial representations, we compute
\begin{align}
(e_X\otimes e_X)\otimes e_X\mapsto\ & \beta_1 \, e\otimes e_X\mapsto \beta_1 \, e_X,\non\\
e_X\otimes \left(e_X \otimes e_X\right) \mapsto\ & \beta_1 \, e_X\otimes e\mapsto \beta_1 \, e_X,\non\\
& \Rightarrow\tilde{K}^{X,X}_{X,X}(1,1)=1.
\end{align}
Here in the associator component $\tilde{K}^{R_1,R_4}_{R_2,R_3}(R_5,R_6)$, the representations $R_1$, $R_2$, and $R_3$ are the three appearing in the triple product, and $R_4$ is the representation for the final result.  The intermediate step when fusing the first and second vectors appears as $R_5$, while the intermediate step when fusing the second and third vectors appears as $R_6$.

Proceeding to other cases,
\begin{align}
\left(e_X\otimes e_X\right)\otimes e_{Y1}\mapsto\ & \beta_1 \, e\otimes e_{Y1}\mapsto \beta_1 \, e_{Y1},\non\\
e_X\otimes \left(e_X\otimes e_{Y1}\right)\mapsto\ & \beta_2 \, e_X\otimes e_{Y2}\mapsto - \beta_2^2 \, e_{Y1},\non\\
& \Rightarrow\tilde{K}^{X,Y}_{X,Y}(1,Y)=-\frac{\beta_1}{\beta_2^2}.
\end{align}
\begin{align}
\left(e_X\otimes e_X\right)\otimes e_{Y2}\mapsto\ & \beta_1 \, e\otimes e_{Y2}\mapsto \beta_1\, e_{Y2},\non\\
e_X\otimes \left(e_X\otimes e_{Y2}\right)\mapsto\ & - \beta_2 \, e_X\otimes e_{Y1}\mapsto - \beta_2^2 \, e_{Y2},\non\\
& \Rightarrow\tilde{K}^{X,Y}_{X,Y}(1,Y)=-\frac{\beta_1}{\beta_2^2}.
\end{align}
Note that we didn't really need to compute both of these; the second one didn't provide us with any new information.  Going forward we won't compute more than we need.
\begin{align}
\left(e_X\otimes e_{Y1}\right)\otimes e_X\mapsto\ & \beta_2 \, e_{Y2}\otimes e_X\mapsto - \beta_2 \beta_3 \, e_{Y1},\non\\
e_X\otimes\left(e_{Y1}\otimes e_X\right)\mapsto\ & \beta_3 \, e_X\otimes e_{Y2}\mapsto - \beta_2 \beta_3 \, e_{Y1},\non\\
& \Rightarrow\tilde{K}^{X,Y}_{Y,X}(Y,Y)=1.
\end{align}
\begin{align}
\left(e_X \otimes e_{Y1}\right)\otimes e_{Y1}\mapsto\ & \beta_2 \, e_{Y2}\otimes e_{Y1}\mapsto - \beta_2 \beta_6 \, e_X + \beta_2 \beta_5 \, e_{Y1},\non\\
e_X\otimes\left(e_{Y1}\otimes e_{Y1}\right)\mapsto\ & \beta_4 e_X\otimes e + \beta_5 e_X\otimes e_{Y2}\mapsto \beta_4 \, e_X - \beta_2 \beta_5 \, e_{Y1},\non\\
& \Rightarrow\tilde{K}^{X,X}_{Y,Y}(Y,1)=-\frac{\beta_2 \beta_6}{\beta_4},
\qquad
\tilde{K}^{X,Y}_{Y,Y}(Y,Y)=-1.
\end{align}
\begin{align}
\left(e_X \otimes e_{Y1}\right)\otimes e_{Y2}\mapsto\ & \beta_2 \, e_{Y2}\otimes e_{Y2}\mapsto \beta_2 \beta_4 \, e - \beta_2 \beta_5 \, e_{Y2},\non\\
e_X\otimes\left(e_{Y1}\otimes e_{Y2}\right)\mapsto\ & \beta_6 \, e_X\otimes e_X + \beta_5 \, e_X\otimes e_{Y1}\mapsto \beta_1 \beta_6 \, e + \beta_2 \beta_5 \, e_{Y2},\non\\
& \Rightarrow\tilde{K}^{X,1}_{Y,Y}(Y,X)=\frac{\beta_2 \beta_4}{\beta_1 \beta_6},
\qquad
\tilde{K}^{X,Y}_{Y,Y}(Y,Y)=-1.
\end{align}
\begin{align}
\left(e_{Y1} \otimes e_X\right)\otimes e_X\mapsto\ & \beta_3 \, e_{Y2}\otimes e_X\mapsto - \beta_3^2 \, e_{Y1},\non\\
e_{Y1}\otimes\left(e_X\otimes e_X\right)\mapsto\ & \beta_1 \, e_{Y1}\otimes e\mapsto \beta_1 \, e_{Y1},\non\\
& \Rightarrow\tilde{K}^{Y,Y}_{X,X}(Y,1)=-\frac{\beta_3^2}{\beta_1}.
\end{align}
\begin{align}
\left(e_{Y1}\otimes e_X\right)\otimes e_{Y1}\mapsto\ & \beta_3 \, e_{Y2}\otimes e_{Y1}\mapsto - \beta_3 \beta_6 \, e_X + \beta_3 \beta_5 \, e_{Y1},\non\\
e_{Y1}\otimes\left(e_X\otimes e_{Y1}\right)\mapsto\ & \beta_2 \, e_{Y1}\otimes e_{Y2}\mapsto \beta_2 \beta_6 \, e_X + \beta_2 \beta_5 \, e_{Y1},\non\\
& \Rightarrow\tilde{K}^{Y,X}_{X,Y}(Y,Y)=-\frac{\beta_3}{\beta_2},
\qquad
\tilde{K}^{Y,Y}_{X,Y}(Y,Y)=\frac{\beta_3}{\beta_2}.
\end{align}
\begin{align}
\left(e_{Y1}\otimes e_X\right)\otimes e_{Y2}\mapsto\ & \beta_3 \, e_{Y2}\otimes e_{Y2}\mapsto \beta_3 \beta_4 \, e - \beta_3 \beta_5 \, e_{Y2},\non\\
e_{Y1}\otimes\left(e_X\otimes e_{Y2}\right)\mapsto\ & - \beta_2 \, e_{Y1}\otimes e_{Y1}\mapsto - \beta_2 \beta_4 \, e - \beta_2 \beta_5 \, e_{Y2},\non\\
& \Rightarrow\tilde{K}^{Y,1}_{X,Y}(Y,Y)=-\frac{\beta_3}{\beta_2},
\qquad
\tilde{K}^{Y,Y}_{X,Y}(Y,Y)=\frac{\beta_3}{\beta_2}.
\end{align}
\begin{align}
\left(e_{Y1}\otimes e_{Y1}\right)\otimes e_X\mapsto\ & \beta_4 \, e\otimes e_X + \beta_5 \, e_{Y2}\otimes e_X \mapsto \beta_4 \, e_X - \beta_3 \beta_5 \, e_{Y1},\non\\
e_{Y1}\otimes \left(e_{Y1}\otimes e_X \right)\mapsto\ & \beta_3 \, e_{Y1}\otimes e_{Y2}\mapsto \beta_3 \beta_6 \, e_X + \beta_3 \beta_5 \, e_{Y1},\non\\
& \Rightarrow\tilde{K}^{Y,X}_{Y,X}(1,Y)=\frac{\beta_4}{\beta_3 \beta_6},
\qquad
\tilde{K}^{Y,Y}_{Y,X}(Y,Y)=-1.
\end{align}
\begin{align}
\left( e_{Y1}\otimes e_{Y2} \right)\otimes e_X\mapsto\ & \beta_6 \, e_X\otimes e_X + \beta_5 \, e_{Y1}\otimes e_X \mapsto \beta_1 \beta_6 \, e + \beta_3 \beta_5 e_{Y2},\non\\
e_{Y1} \otimes \left(e_{Y2} \otimes e_X \right)\mapsto\ & - \beta_3 \, e_{Y1} \otimes e_{Y1} \mapsto - \beta_3 \beta_4 \, e - \beta_3 \beta_5 \, e_{Y2},\non\\
& \Rightarrow\tilde{K}^{Y,1}_{Y,X}(X,Y)=-\frac{\beta_1 \beta_6}{\beta_3 \beta_4},
\qquad
\tilde{K}^{Y,Y}_{Y,X}(Y,Y)=-1.
\end{align}

Finally, we have the $Y^{\otimes 3}$ triple products.  Three of them are enough to determine the matrix
\be
\tilde{K}^{Y,Y}_{Y,Y}
=
\lp\begin{matrix}
\tilde{K}^{Y,Y}_{Y,Y}(1,1) & \tilde{K}^{Y,Y}_{Y,Y}(1,X) & \tilde{K}^{Y,Y}_{Y,Y}(1,Y)
\\ 
\tilde{K}^{Y,Y}_{Y,Y}(X,1) & \tilde{K}^{Y,Y}_{Y,Y}(X,X) & \tilde{K}^{Y,Y}_{Y,Y}(X,Y)
\\ 
\tilde{K}^{Y,Y}_{Y,Y}(Y,1) & \tilde{K}^{Y,Y}_{Y,Y}(Y,X) & \tilde{K}^{Y,Y}_{Y,Y}(Y,Y) \end{matrix}\rp.
\ee

To be systematic, let us choose bases for $(Y\otimes Y)\otimes Y$ and $Y\otimes(Y\otimes Y)$.   (Moving forward, we will start to leave the $\otimes$ symbol implicit.)
\begin{align}
E_L^Y=\ & \frac{1}{4 \beta_4 \beta_5}\lp \left(e_{Y1} e_{Y1}\right) e_{Y2} + \left(e_{Y1} e_{Y2} \right) e_{Y1} + \left( e_{Y2} e_{Y1} \right) e_{Y1} - \left(e_{Y2} e_{Y2} \right) e_{Y2} \rp
\\
& \mapsto
\frac{1}{2\beta_4}\lp e_{Y1} e_{Y1} + e_{Y2} e_{Y2} \rp\mapsto e,
\\
X_L^Y=\ & -\frac{1}{4 \beta_5 \beta_6}\lp \left( e_{Y1}e_{Y1} \right)e_{Y1} - \left( e_{Y1} e_{Y2} \right) e_{Y2} - \left(e_{Y2} e_{Y1} \right)e_{Y2} - \left(e_{Y2} e_{Y2} \right) e_{Y1}\rp
\\
& \mapsto\frac{1}{2\beta_6}\lp e_{Y} e_{Y2} - e_{Y2} e_{Y1} \rp\mapsto e_X,
\\
U_L^1=\ & \frac{1}{2\beta_4}\lp \left( e_{Y1} e_{Y1} \right) e_{Y1} + \left(e_{Y2} e_{Y2} \right) e_{Y1} \rp
\mapsto ee_{Y1} \mapsto e_{Y1},
\\
V_L^1=\ & \frac{1}{2\beta_4}\lp \left( e_{Y1} e_{Y1} \right) e_{Y2} + \left( e_{Y2} e_{Y2} \right) e_{Y2} \rp
\mapsto e e_{Y2}\mapsto e_{Y2},
\\
U_L^X=\ & -\frac{1}{2\beta_2 \beta_6}\lp \left( e_{Y1} e_{Y2} \right) e_{Y2} - \left(e_{Y2} e_{Y1} \right) e_{Y2} \rp
\mapsto -\frac{1}{2\beta_2}\lp  e_X e_{Y2} \rp\mapsto e_{Y1},
\\
V_L^X=\ & \frac{1}{2\beta_2 \beta_6}\lp \left( e_{Y1} e_{Y2} \right) e_{Y1} - \left(e_{Y2} e_{Y1} \right) e_{Y1} \rp
\mapsto\frac{1}{2\beta_2}\lp e_X e_{Y1} \rp\mapsto e_{Y2},
\\
U_L^Y=\ & \frac{1}{4\beta_5^2}\lp \left( e_{Y1} e_{Y1} \right) e_{Y1} + \left( e_{Y1} e_{Y2} \right) e_{Y2} + \left( e_{Y2} e_{Y1} \right) e_{Y2} - \left(e_{Y2} e_{Y2} \right) e_{Y1} \rp
\\
& \mapsto
\frac{1}{2\beta_5}\lp e_{Y1} e_{Y2} + e_{Y2} e_{Y1} \rp\mapsto e_{Y1},
\\
V_L^Y=\ & -\frac{1}{4\beta_5^2}\lp \left( e_{Y1} e_{Y1} \right) e_{Y2} - \left( e_{Y1} e_{Y2} \right) e_{Y1} - \left( e_{Y2} e_{Y1} \right) e_{Y1} - \left( e_{Y2} e_{Y2} \right) e_{Y2} \rp
\\
& \mapsto
\frac{1}{2\beta_5}\lp e_{Y1} e_{Y1} - e_{Y2} e_{Y2} \rp\mapsto e_{Y2}.
\end{align}
\begin{align}
E_R^Y=\ & \frac{1}{\beta_4 \beta_5}\lp  e_{Y1} \left( e_{Y1} e_{Y2} \right) + e_{Y1} \left( e_{Y2} e_{Y1} \right) + e_{Y2} \left( e_{Y1} e_{Y1} \right) - e_{Y2} \left( e_{Y2} e_{Y2} \right)\rp
\\
& \mapsto
\frac{1}{2\beta_4}\lp e_{Y1} e_{Y1} + e_{Y2} e_{Y2} \rp\mapsto e,
\\
X_R^Y=\ & \frac{1}{4 \beta_5 \beta_6}\lp  e_{Y1} \left( e_{Y1} e_{Y1} \right) - e_{Y1}  \left( e_{Y2} e_{Y2} \right) - e_{Y2} \left( e_{Y1} e_{Y2} \right) - e_{Y2} \left( e_{Y2} e_{Y1} \right)\rp
\\
& \mapsto
\frac{1}{2\beta_6}\lp e_{Y1} e_{Y2} - e_{Y2} e_{Y1} \rp\mapsto e_X,
\\
U_R^1=\ & \frac{1}{2\beta_4}\lp e_{Y1} \left( e_{Y1} e_{Y1} \right) + e_{Y1} \left( e_{Y2} e_{Y2} \right)\rp
\mapsto e_{Y1} e\mapsto e_{Y1},
\\
V_R^1=\ & \frac{1}{2\beta_4}\lp e_{Y2} \left( e_{Y1} e_{Y1} \right) + e_{Y2} \left( e_{Y2} e_{Y2} \right)\rp
\mapsto  e_{Y2} e\mapsto e_{Y2},
\\
U_R^X=\ & -\frac{1}{2\beta_3 \beta_6}\lp e_{Y2} \left( e_{Y1} e_{Y2} \right) - e_{Y2} \left( e_{Y2} e_{Y1} \right)\rp
\mapsto -\frac{1}{2\beta_3} e_{Y2} e_X \mapsto  e_{Y1},
\\
V_R^X=\ & \frac{1}{2\beta_3 \beta_6}\lp  e_{Y1} \left( e_{Y1} e_{Y2} \right) - e_{Y1} \left( e_{Y2} e_{Y1} \right)\rp
\mapsto
\frac{1}{2\beta_3} e_{Y1} e_X \mapsto e_{Y2},
\\
U_R^Y=\ & \frac{1}{4\beta_5^2}\lp e_{Y1} \left( e_{Y1} e_{Y1} \right) - e_{Y1} \left( e_{Y2} e_{Y2} \right) +  e_{Y2} \left( e_{Y1} e_{Y2} \right) + e_{Y2} \left( e_{Y2} e_{Y1} \right)\rp
\\
& \mapsto
\frac{1}{2\beta_5}\lp e_{Y1} e_{Y2} + e_{Y2} e_{Y1} \rp\mapsto e_{Y1} ,
\\
V_R^Y=\ & \frac{1}{4\beta_5^2}\lp e_{Y1} \left( e_{Y1} e_{Y2} \right) + e_{Y1} \left( e_{Y2} e_{Y1} \right) - e_{Y2} \left( e_{Y1} e_{Y1} \right) + e_{Y2} \left( e_{Y2} e_{Y2} \right)\rp
\\
& \mapsto
\frac{1}{2\beta_5}\lp e_{Y1} e_{Y1}  - e_{Y2} e_{Y2} \rp\mapsto e_{Y2}.
\end{align}
These combinations have been chosen so that both the final vector is one of our basis vectors, and the intermediate channel is through a single representation, indicated by the superscript.

Then to expand any triple product in this basis, we can compare with the fusions, for example
\be
\left( e_{Y1} e_{Y1} \right) e_{Y1}\mapsto \beta_4 e e_{Y1} + \beta_5 e_{Y2} e_{Y1} \mapsto - \beta_5 \beta_6 \, e_X + \lp \beta_4 + \beta_5^2 \rp e_{Y1},
\ee
tells us that
\be
\left( e_{Y1} e_{Y1} \right) e_{Y1} = - \beta_5 \beta_6 \, X_L^Y + \beta_4 U_L^1 + \beta_5^2 U_L^Y.
\ee
Similarly,
\be
 e_{Y1} \left( e_{Y1} e_{Y1} \right) = \beta_5 \beta_6 X_R^Y + \beta_4 U_R^1 + \beta_5^2 U_R^Y.
\ee
The $\al_{Y,Y,Y}$ associator should map these into each other.  Comparing the $X_L^Y$ and $X_R^Y$ terms, we conclude that
\be
\tilde{K}^{Y,X}_{Y,Y}(Y,Y)=-1.
\ee
Looking next at the $U_L$ and $U_R$ terms, we get a vector equation
\be
\tilde{K}^{Y,Y}_{Y,Y}\cdot\lp\begin{matrix} \beta_4 \\ 0 \\ \beta_5^2 \end{matrix}\rp=\lp\begin{matrix} \beta_4 \\ 0 \\ \beta_5^2 \end{matrix}\rp,
\ee
where $\tilde{K}^{Y,X}_{Y,Y}$ denotes a matrix with components $\tilde{K}^{Y,X}_{Y,Y}(R,S)$ for various irreducible
representations $R$, $S$.
This isn't enough to determine the matrix $\tilde{K}^{Y,Y}_{Y,Y}$, but we can continue.  We have
\begin{align}
\left( e_{Y1} e_{Y1} \right) e_{Y2}\mapsto\ & \beta_4 e e_{Y2} + \beta_5 e_{Y2} e_{Y2} \mapsto \beta_4 \beta_5 \, e + \lp \beta_4 - \beta_5^2\rp e_{Y2}
\non \\
& \quad\Rightarrow\quad
\left( e_{Y1} e_{Y1} \right) e_{Y2} = \beta_4 \beta_5 E_L^Y + \beta_4 V_L^1 - \beta_5^2 V_L^Y,
\\
e_{Y1} \left( e_{Y1} e_{Y2} \right)\mapsto\ & \beta_6 e_{Y1} e_X + \beta_5 e_{Y1} e_{Y1} \mapsto \beta_4 \beta_5 \, e + \lp \beta_3 \beta_6 + \beta_5^2 \rp  e_{Y2}
\non \\
& \quad\Rightarrow\quad 
e_{Y1} \left( e_{Y1} e_{Y2} \right) = \beta_4 \beta_5 E_R^Y + \beta_3 \beta_6 V_R^X + \beta_5^2 V_R^Y,
\end{align}
translating to
\be
\tilde{K}^{Y,1}_{Y,Y}(Y,Y)=1,
\qquad
\tilde{K}^{Y,Y}_{Y,Y}\cdot\lp\begin{matrix} 0 \\ \beta_3 \beta_6 \\ \beta_5^2 \end{matrix}\rp
=\lp\begin{matrix} \beta_4 \\ 0 \\ - \beta_5^2 \end{matrix}\rp.
\ee

Finally, also considering
\begin{align}
\left( e_{Y1} e_{Y2} \right) e_{Y1}\mapsto \beta_6 e_X e_{Y1} + \beta_5 e_{Y1} e_{Y1} &\mapsto \beta_4 \beta_5 \, e + \lp \beta_2 \beta_6 + \beta_5^2 \rp e_{Y2}
\non \\
& \quad\Rightarrow\quad 
\left(e_{Y1} e_{Y2} \right) e_{Y1} = \beta_4 \beta_5 E_L^Y + \beta_2 \beta_6 V_L^X + \beta_5^2V_L^Y,
\\
e_{Y1} \left( e_{Y2} e_{Y1} \right)\mapsto - \beta_6 e_{Y1} e_X + \beta_5 e_{Y1} e_{Y1} & \mapsto \beta_4 \beta_5 \, e + \lp - \beta_3 \beta_6 + \beta_5^2 \rp
\non \\
& \quad\Rightarrow\quad 
e_{Y1} \left( e_{Y2} e_{Y1} \right) = \beta_4 \beta_5 E_R^Y - \beta_3 \beta_6 V_R^X + \beta_5^2 V_R^Y,
\end{align}
leading to
\be
\tilde{K}^{Y,1}_{Y,Y}(Y,Y)=1,
\quad \tilde{K}^{Y,Y}_{Y,Y}\cdot\lp\begin{matrix} 0 \\ - \beta_3 \beta_6 \\ \beta_5^2 \end{matrix}\rp
=
\lp\begin{matrix} 0 \\ \beta_2 \beta_6 \\ \beta_5^2 \end{matrix}\rp.
\ee
These three vector equations uniquely determine the matrix $\tilde{K}^{Y,Y}_{Y,Y}$ to be
\be
\tilde{K}^{Y,Y}_{Y,Y} = \lp\begin{matrix} \hlf & \frac{\beta_4}{2 \beta_3 \beta_6} & \frac{\beta_4}{2 \beta_5^2} \\ -\frac{\beta_2 \beta_6}{2 \beta_4} & -\frac{\beta_2}{2 \beta_3} & \frac{\beta_2 \beta_6}{2 \beta_5^2} \\ \frac{\beta_5^2}{\beta_4} & -\frac{\beta_5^2}{\beta_3 \beta_6} & 0 \end{matrix}\rp.
\ee
We can verify this result by checking the other triple products,
\begin{align}
\left(e_{Y1} e_{Y2} \right) e_{Y2}\mapsto \beta_5 \beta_6 X_L^Y - \beta_2 \beta_6 U_L^X + \beta_5^2 U_L^Y,
\quad & 
e_{Y1} \left( e_{Y2} e_{Y2} \right)\mapsto - \beta_5 \beta_6 X_R^Y + \beta_4 U_R^1 - \beta_5^2 U_R^Y,\\
\left( e_{Y2} e_{Y1} \right) e_{Y1}\mapsto \beta_4 \beta_5 E_L^Y - \beta_2 \beta_6 V_L^X + \beta_5^2 V_L^Y,
\quad & 
e_{Y2} \left( e_{Y1} e_{Y1} \right)\mapsto \beta_4 \beta_5 E_R^Y + \beta_4 V_R^1 - \beta_5^2 V_R^Y,\\
\left( e_{Y2} e_{Y1} \right) e_{Y2}\mapsto \beta_5 \beta_6 X_L^Y + \beta_2 \beta_6 U_L^X + \beta_5^2 U_L^Y,
\quad & 
e_{Y2} \left( e_{Y1} e_{Y2} \right)\mapsto - \beta_5 \beta_6 X_R^Y - \beta_3 \beta_6 U_R^X + \beta_5^2 U_R^Y,\\
\left( e_{Y2} e_{Y2} \right) e_{Y1} \mapsto \beta_5 \beta_6 X_L^Y + \beta_4 U_L^1 - \beta_5^2 U_L^Y,
\quad & 
e_{Y2} \left( e_{Y2} e_{Y1} \right)\mapsto - \beta_5 \beta_6 X_R^Y + \beta_3 \beta_6 U_R^X + \beta_5^2U_R^Y,\\
\left( e_{Y2} e_{Y2} \right) e_{Y2} \mapsto - \beta_4 \beta_5 E_L^Y + \beta_4 V_L^1 + \beta_5^2 V_L^Y,
\quad & 
e_{Y2} \left( e_{Y2} e_{Y2} \right)\mapsto - \beta_4 \beta_5 E_R^Y + \beta_4 V_R^1 + \beta_5^2V_R^Y.
\end{align}

There are some other structures we can define on this fusion category, that will be useful once we start looking at Frobenius algebras.  Having established a standard basis for fusion, we then also have a set of associated evaluation morphisms $\overline{\e}_L:L\otimes L^\ast\rr\C$ and $\e_L:L^\ast\otimes L\rr\C$, by projecting the fusion on to the identity, e.g.~if we define $\pi:1\rightarrow\C$ by $\pi(e)=1$, then we have $\overline{\epsilon}_L=\pi\circ\lambda_{L,L^\ast}^1$, $\epsilon_L=\pi\circ\lambda_{L^\ast,L}^1$.  Thus,
\be
\overline{\e}_1(ee) = \e_1(ee)=1,
\qquad
\overline{\e}_X\left(e_X e_X\right) = \e_X\left(e_X e_X\right)=\beta_1,
\ee
\be
\overline{\e}_Y\left( e_{Y1} e_{Y1} \right) = \overline{\e}_Y\left( e_{Y2} e_{Y2} \right)
= \e_Y\left(e_{Y1} e_{Y1} \right) = \e_Y\left( e_{Y2} e_{Y2} \right) = \beta_4,
\ee
\be
%\quad
\overline{\e}_Y\left( e_{Y1} e_{Y2}\right) = \overline{\e}_Y\left( e_{Y2} e_{Y1} \right) = 
\e_Y\left( e_{Y1} e_{Y2} \right) = \e_Y\left( e_{Y2} e_{Y1} \right)=0.
\ee
Along with these we have canonically trivial pivotal structures~\cite{Bhardwaj:2017xup} $p_1$, $p_X$, and $p_Y$ which act as the identity intertwiner on $1$, $X$, and $Y$ respectively.

Next, we also have coevaluation morphisms $\overline{\gamma}_L:\C\rr L^\ast\otimes L$ and $\gamma_L:\C\rr L\otimes L^\ast$ which satisfy \cite[equ'n (3.14)]{Bhardwaj:2017xup}
\be
\lp \overline{\e}_L\otimes 1\rp\circ\al_{L,L^\ast,L}^{-1}\circ\lp 1\otimes \overline{\gamma}_L\rp=1,
\qquad
\lp 1\otimes\e_L\rp\circ\al_{L,L^\ast,L}\circ\lp\gamma_L\otimes 1\rp=1.
\ee

These fix the coevaluation morphisms to be
\be
\label{eq:StandardCoMult1}
\overline{\gamma}_1(1) = \gamma_1(1)=ee,
\quad
\overline{\gamma}_X(1) = \gamma_X(1) = \frac{1}{\beta_1} e_X e_X ,
\ee
and
\be
\label{eq:StandardCoMult2}
\overline{\gamma}_Y(1) = \gamma_Y(1) = \frac{1}{\beta_4}\lp  e_{Y1} e_{Y1} + e_{Y2} e_{Y2} \rp.
\ee

Finally, by combining coevaluation and fusion, we can establish a basis for the co-fusion intertwiners $\d_{R_1}^{R_2,R_3}\in\Hom(R_1,R_2\otimes R_3)$.  Explicitly, we'll construct these as\footnote{Alternatively, we could use
\begin{equation}
    \d_{R_1}^{R_2,R_3}(v_1)=\left[\left(1_{R_2}\otimes m_{R_2^\ast,R_1}^{R_3}\right)\circ\al_{R_2,R_2^\ast,R_1}\right](\g_{R_2}(1)\otimes v_1),\quad v_1\in R_1,
\end{equation}
resulting in a different basis.}
\begin{equation}
    \d_{R_1}^{R_2,R_3}(v_1)=\left[\left(\lambda_{R_1,R_3^\ast}^{R_2}\otimes 1_{R_3}\right)\circ\al_{R_1,R_3^\ast,R_3}^{-1}\right](v_1\otimes\overline{\g}_{R_3}(1)).
\end{equation}
If either of the outputs is the trivial representation, then we simply have $\d_R^{1,R}(v)=e\otimes v$ or $\d_R^{R,1}(v)=v\otimes e$.  For the others, we find (keeping $\otimes$ symbols implicit)
\begin{equation}
    \d_1^{X,X}(e)=\beta_1^{-1}e_Xe_X,\quad\d_1^{Y,Y}(e)=\beta_4^{-1}(e_{Y1}e_{Y1}+e_{Y2}e_{Y2}),\quad\d_X^{Y,Y}(e_X)=-\beta_2\beta_4^{-1}(e_{Y1}e_{Y2}-e_{Y2}e_{Y1}),\non
\end{equation}
\begin{equation}
    \d_Y^{X,Y}(e_{Y1})=\beta_6\beta_4^{-1}e_Xe_{Y2},\quad\d_Y^{X,Y}(e_{Y2})=-\beta_6\beta_4^{-1}e_Xe_{Y1},\non
\end{equation}
\begin{equation}
    \label{eq:RepS3deltaMaps}
    \d_Y^{Y,X}(e_{Y1})=\beta_3\beta_1^{-1}e_{Y2}e_X,\quad\d_Y^{Y,X}(e_{Y2})=-\beta_3\beta_1^{-1}e_{Y1}e_X,
\end{equation}
\begin{equation}
    \d_Y^{Y,Y}(e_{Y1})=\beta_5\beta_4^{-1}(e_{Y1}e_{Y2}+e_{Y2}e_{Y1}),\quad\d_Y^{Y,Y}(e_{Y2})=\beta_5\beta_4^{-1}(e_{Y1}e_{Y1}-e_{Y2}e_{Y2}).\non
\end{equation}

In all of the expressions above, we left the parameters $\beta_1, \cdots, \beta_6$ arbitrary.  In the group-like case, this corresponds to the freedom to choose a representative three-cocycle for our anomaly class in $H^3(G,\U(1))$ by shifting by a coboundary. 

In the next several subsections we will apply the computations above
to compute partition functions for various Frobenius algebras, following
the procedure described earlier in section~\ref{sect:genl-algebra}.

\subsubsection{Modular transformations}

Before computing the partition functions, let us take a moment to work
out the modular transformations, following section~\ref{sect:genl:modulartrans},
so that we can check that the partition functions we will derive momentarily,
are modular invariant.

Using equations~(\ref{eq:modtrans:T}), (\ref{eq:modtrans:S}), 
it is straightforward to compute that
\begin{align}
Z_{1,1}^1(\tau+1)=\ & Z_{1,1}^1(\tau),\\
Z_{1,X}^X(\tau+1)=\ & Z_{1,X}^X(\tau),\\
Z_{1,Y}^Y(\tau+1)=\ & Z_{1,Y}^Y(\tau),\\
Z_{X,1}^X(\tau+1)=\ & Z_{X,X}^1(\tau),\\
Z_{X,X}^1(\tau+1)=\ & Z_{X,1}^X(\tau),\\
Z_{X,Y}^Y(\tau+1)=\ & Z_{X,Y}^Y(\tau),\\
Z_{Y,1}^Y(\tau+1)=\ & Z_{Y,Y}^1(\tau),\\
Z_{Y,X}^Y(\tau+1)=\ & -\frac{\beta_3}{\beta_2} Z_{Y,Y}^X(\tau),\\
Z_{Y,Y}^1(\tau+1)=\ & \hlf Z_{Y,1}^Y(\tau) + \frac{\beta_4}{2 \beta_3 \beta_6} Z_{Y,X}^Y(\tau) + \frac{\beta_4}{2 \beta_5^2} Z_{Y,Y}^Y(\tau),\\
Z_{Y,Y}^X(\tau+1)=\ & -\frac{\beta_2 \beta_6}{2 \beta_4} Z_{Y,1}^Y(\tau) - \frac{\beta_2}{2 \beta_3} Z_{Y,X}^Y(\tau) + \frac{\beta_2 \beta_6}{2 \beta_5^2} Z_{Y,Y}^Y(\tau),\\
Z_{Y,Y}^Y(\tau+1)=\ & \frac{\beta_5^2}{\beta_4} Z_{Y,1}^Y(\tau) - \frac{\beta_5^2}{\beta_3 \beta_6} Z_{Y,X}^Y(\tau).
\end{align}
\begin{align}
Z_{1,1}^1(-1/\tau)=\ & Z_{1,1}^1(\tau),\\
Z_{1,X}^X(-1/\tau)=\ & Z_{X,1}^X(\tau),\\
Z_{1,Y}^Y(-1/\tau)=\ & Z_{Y,1}^Y(\tau),\\
Z_{X,1}^X(-1/\tau)=\ & Z_{1,X}^X(\tau),\\
Z_{X,X}^1(-1/\tau)=\ & Z_{X,X}^1(\tau),\\
Z_{X,Y}^Y(-1/\tau)=\ & \frac{\beta_2 \beta_4}{\beta_1 \beta_6} Z_{Y,X}^Y(\tau),\\
Z_{Y,1}^Y(-1/\tau)=\ & Z_{1,Y}^Y(\tau),\\
Z_{Y,X}^Y(-1/\tau)=\ & \frac{\beta_1 \beta_6}{\beta_2 \beta_4} Z_{X,Y}^Y(\tau),\\
Z_{Y,Y}^1(-1/\tau)=\ & \hlf Z_{Y,Y}^1(\tau) - \frac{\beta_4}{2 \beta_2 \beta_6} Z_{Y,Y}^X(\tau) + \frac{\beta_4}{2 \beta_5^2} Z_{Y,Y}^Y(\tau),\\
Z_{Y,Y}^X(-1/\tau)=\ & -\frac{\beta_2 \beta_6}{2 \beta_4} Z_{Y,Y}^1(\tau) + \hlf Z_{Y,Y}^X(\tau) + \frac{\beta_2 \beta_6}{2 \beta_5^2} Z_{Y,Y}^Y(\tau),\\
Z_{Y,Y}^Y(-1/\tau)=\ & \frac{\beta_5^2}{\beta_4} Z_{Y,Y}^1(\tau) + \frac{\beta_5^2}{\beta_2 \beta_6} Z_{Y,Y}^X(\tau).
\end{align}

Modular-invariant combinations of partial traces are
\be  \label{eq:reps3:modinvt1}
Z_{1,1}^1,\qquad Z_{1,X}^X+Z_{X,1}^X+Z_{X,X}^1,
\ee
\be  \label{eq:reps3:modinvt2}
Z_{1,Y}^Y+Z_{Y,1}^Y+Z_{Y,Y}^1+\frac{\beta_4}{2\beta_5^2}Z_{Y,Y}^Y,
\qquad 
Z_{X,Y}^Y+\frac{\beta_2 \beta_4}{\beta_1 \beta_6} Z_{Y,X}^Y - \frac{\beta_3 \beta_4}{\beta_1 \beta_6} Z_{Y,Y}^X - \frac{\beta_2 \beta_3 \beta_4}{2 \beta_1 \beta_5^2} Z_{Y,Y}^Y.
\ee

We will see these combinations reappear later when we compute partition functions corresponding to various Frobenius algebras.

\subsubsection{Subalgebras}
\label{sect:S3-subalg}

We will study Frobenius algebra structures indexed by Hopf subalgebras
${\mathbb C}[H] \subset {\mathbb C}[S_3]$, corresponding to (not necessarily normal) subgroups
$H \subset S_3$,
so
let us take a moment to enumerate
possible subgroups.  Recall that $S_3$ can be described as
\begin{equation}
    S_3 \: = \: \langle a, b \, | \, a^2 = 1 = b^3, \: a b a = b^2 \rangle.
\end{equation}

%**** Double-check ****

In terms of the generators above, the subgroups of $S_3$ are 
\begin{equation}
    \{ 1 \}, \: \: \:
    \langle a \rangle = {\mathbb Z}_2, \: \: \:
    \langle a b \rangle = {\mathbb Z}_2, \: \: \:
    \langle a b^2 \rangle = {\mathbb Z}_2, \: \: \:
    \langle b \rangle = {\mathbb Z}_3
\end{equation}
Now, $a$ is conjugate to $ab$ and $ab^2$, so there are really only three
distinct nontrivial subgroups for which we should compute corresponding Frobenius algebras
and partition functions:
\begin{equation}
    \langle b \rangle,
    \: \: \:
    \langle a \rangle, 
    \: \: \: 
    S_3.
\end{equation}
Indeed, in general, we will find that any two subgroups related by an automorphism will give rise to isomorphic Frobenius algebras.  For an inner automorphism the representation is the same, while for an outer automorphism representations may be exchanged (we'll see this latter possibility in the $D_4$ and $Q_8$ cases below).

Each of these subgroups will correspond to a Frobenius algebra.
In addition, since $S_3$ acts on $S_3/H$, if we construct a vector space with a basis labelled by the distinct cosets of $S_3/H$, that vector space with that $S_3$ action defines a representation of $S_3$.

We summarize the results in the table below:

\begin{center}
    \begin{tabular}{cccc}
    Coset & Subalgebra & Representation & Details in subsection \\ \hline
    $G/\langle b \rangle$ & ${\rm Span}[v_{K}, v_{a K}]$, $K = \langle b \rangle$ & $1+X$ & \ref{sect:S3:1pX}
\\
$G/\langle a \rangle$ & ${\rm Span}[v_{H}, v_{b H}, v_{b^2 H}]$, $H = \langle a \rangle$ & $1+Y$ & \ref{sect:S3:1pY}
\\
$G/1$ & ${\mathbb C}[G]^*$ & $1+X+2Y$ & \ref{sect:S3:regrep} \\
& & (the regular representation) 
\end{tabular}
\end{center}

\subsubsection{$1+X$ orbifold}
\label{sect:S3:1pX}

In this subsection, we consider the Frobenius algebra corresponding
to the subalgebra ${\mathbb C}[K]$ for $K = \langle b \rangle$.
To proceed further, we need to pick a basis for each case that makes the decomposition into irreps manifest.  We'll start with $1+X$.  The full representation has basis vectors $v_K$ and $v_{aK}$, and we can take a new basis
\be
e=u(1)=v_K + v_{aK},\qquad e_X=\frac{c}{\sqrt{2}}\lp v_K - v_{aK}\rp,
\ee
where the notation reflects the fact that the first element is a basis for the vector space
associated to $1$, and the second element is a basis for the vector space associated to
$X$.
Both vectors are invariant under the action of $b$ and are eigenvectors of the action of $a$, with $e$ having eigenvalue $+1$ and $x$ having eigenvalue $-1$.  Hence $e$ spans the $1$ sub-representation of $1+X$, while $x$ spans the $X$ sub-representation.  The constant $c$ can be any element of $\C^\ast$.

We also need to specify the multiplication and co-multiplication in this Frobenius algebra (as discussed in section~\ref{sect:genl-algebra}), which acts on the underlying basis vectors of the representation as
\begin{equation}
    \m_*(v_{gK} \otimes v_{hK})=\d_{gK,hK} \, v_{gK},\qquad\Delta_F(v_{gK})=v_{gK} \otimes v_{gK}.
\end{equation}
(This uses the fact that $\Delta_F$ is diagonal on the $v$ basis.)
Translating this to an action on our new basis, one easily obtains
\begin{align}
\m_*(e \otimes e)=\ & e,\\
\m_*\left(e \otimes e_X\right) = \m_*\left( e_X  \otimes e\right)=\ & e_X,\\
\m_*\left(e_X \otimes e_X \right) =\ & \frac{c^2}{2}e,
\end{align}
and
\begin{align}
    \Delta_F(e)=\ & \hlf e\otimes e + c^{-2}e_X \otimes e_X,\\
    \Delta_F(e_X)=\ & \hlf\lp e \otimes e_X + e_X \otimes e\rp.
\end{align}
Comparing with our standard bases (the $\lambda_{R_1,R_2}^{R_3}$'s (\ref{rs3int1}-\ref{rs3int2}) and $\d_{R_1}^{R_2,R_3}$'s \eqref{eq:RepS3deltaMaps}), we can represent the components of $\mu_*$, $\Delta_F$ as coefficients,
\be
\m_{1,1}^1=\m_{1,X}^X=\m_{X,1}^X=1,\qquad\m_{X,X}^1=\frac{c^2}{2\beta_1},
\ee
\be
\Delta_1^{1,1}=\Delta_X^{1,X}=\Delta_X^{X,1}=\hlf,\qquad\Delta_1^{X,X}=c^{-2}\beta_1.
\ee
Here for instance we used
\begin{equation}
    \m_*(e_X\otimes e_X)=\frac{c^2}{2}e=\left(\frac{c^2}{2\beta_1}\right)\lambda_{X,X}^1(e_X\otimes e_X),
\end{equation}
giving $\m_{X,X}^1=\frac{c^2}{2\beta_1}$, and
\begin{equation}
    \Delta_F(e)=\hlf e\otimes e+\frac{1}{c^2}e_X\otimes e_X=\hlf\delta_1^{1,1}(e)+\frac{\beta_1}{c^2}\delta_1^{X,X}(e),
\end{equation}
giving us $\Delta_1^{1,1}=\hlf$ and $\Delta_1^{X,X}=\frac{\beta_1}{c^2}$.

So now the recipe to compute the coefficients of partial traces is simply to assign to each $Z_{R_1,R_2}^{R_3}$ the coefficient $\m_{R_1,R_2}^{R_3}\Delta_{R_3}^{R_2,R_1}$.
\begin{align}
    Z_{1,1}^1:\ & \m_{1,1}^1\Delta_1^{1,1}=1\cdot\hlf=\hlf,\\
    Z_{1,X}^X:\ & \m_{1,X}^X\Delta_X^{X,1}=1\cdot\hlf=\hlf,\\
    Z_{X,1}^X:\ & \m_{X,1}^X\Delta_X^{1,X}=1\cdot\hlf=\hlf,\\
    Z_{X,X}^1:\ & \m_{X,X}^1\Delta_1^{X,X}=\frac{c^2}{2\beta_1}\cdot\frac{\beta_1}{c^2}=\hlf.
\end{align}
The full partition function is then simply
\be
\label{rs3_1+x_pf}
Z_{1+X}=\hlf\ls Z_{1,1}^1+Z_{1,X}^X+Z_{X,1}^X+Z_{X,X}^1\rs,
\ee
which is a sum of modular invariants in~(\ref{eq:reps3:modinvt1}), and also matches the standard form of a ${\mathbb Z}_2$ orbifold.

\subsubsection{$1+Y$ orbifold}
\label{sect:S3:1pY}

In this section we consider the Frobenius algebra and partition
function corresponding to the subalgebra ${\mathbb C}[H]$ for $H = \langle a \rangle$.
Here we have a basis $\{v_H,v_{bH},v_{b^2H}\}$ the complex vector space over $S_3/H$ and we identify
\be
e=v_H + v_{bH} + v_{b^2H},
\quad e_{Y1}=\frac{c}{\sqrt{2}}\lp v_{bH} - v_{b^2H}\rp,
\quad e_{Y2}=\frac{c}{\sqrt{6}}\lp - 2v_H + v_{bH} + v_{b^2H}\rp.
\ee
Then
\begin{align}
\m_*(e \otimes e)=\ & e,\\
\m_*\left(e \otimes e_{Y1}\right) = \m_*\left(e_{Y1} \otimes e\right)=\ & e_{Y1},\\
\m_*\left(e \otimes e_{Y2}\right) = \m_*\left(e_{Y2} \otimes e\right)=\ & e_{Y2},\\
\m_*\left( e_{Y1} \otimes e_{Y1}\right)=\ & \frac{c^2}{3}e+\frac{c}{\sqrt{6}} e_{Y2},\\
\m_*\left( e_{Y1} \otimes e_{Y2}\right) = \m_*\left( e_{Y2} \otimes e_{Y1} \right)=\ & \frac{c}{\sqrt{6}} e_{Y1},\\
\m_*\left( e_{Y2} \otimes e_{Y2} \right)=\ & \frac{c^2}{3}e-\frac{c}{\sqrt{6}} e_{Y2},
\end{align}
and (using the fact that $\Delta_F$ is diagonal on the $v$ basis)
\begin{align}
    \Delta_F(e)=\ & \frac{1}{3}e \otimes e + c^{-2}\lp e_{Y1} \otimes e_{Y1} + e_{Y2} \otimes e_{Y2}\rp,\\
    \Delta_F(e_{Y1})=\ & \frac{1}{3}\lp e \otimes e_{Y1} + e_{Y1} \otimes e\rp+\frac{1}{\sqrt{6}\,c}\lp e_{Y1} \otimes e_{Y2} + e_{Y2} \otimes e_{Y1}\rp,\\
    \Delta_F(e_{Y2})=\ & \frac{1}{3}\lp e \otimes e_{Y2} + e_{Y2}\otimes e\rp+\frac{1}{\sqrt{6}\,c}\lp e_{Y1} \otimes e_{Y1} - e_{Y2} \otimes e_{Y2}\rp.
\end{align}
So we have
\be
\m_{1,1}^1=\m_{1,Y}^Y=\m_{Y,1}^Y=1,\qquad\m_{Y,Y}^1=\frac{c^2}{3\beta_4},\qquad\m_{Y,Y}^Y=\frac{c}{\sqrt{6}\beta_5},
\ee
\be
\Delta_1^{1,1}=\Delta_Y^{1,Y}=\Delta_Y^{Y,1}=\frac{1}{3},\quad\Delta_1^{Y,Y}=\frac{\beta_4}{c^2},\quad\Delta_Y^{Y,Y}=\frac{\beta_4}{\sqrt{6}\,c\beta_5}.
\ee
Then for coefficients, we get
\begin{align}
Z_{1,1}^1:\ & \m_{1,1}^1\Delta_1^{1,1}=\frac{1}{3},\\
Z_{1,Y}^Y:\ & \m_{1,Y}^Y\Delta_Y^{Y,1}=\frac{1}{3},\\
Z_{Y,1}^Y:\ & \m_{Y,1}^Y\Delta_Y^{1,Y}=\frac{1}{3},\\
Z_{Y,Y}^1:\ & \m_{Y,Y}^1\Delta_1^{Y,Y}=\frac{1}{3},\\
Z_{Y,Y}^Y:\ & \m_{Y,Y}^Y\Delta_Y^{Y,Y}=\frac{\beta_4}{6\beta_5^2},
\end{align}
and the partition function is
\be
\label{rs3_1+y_pf}
Z_{1+Y}=\frac{1}{3}\ls Z_{1,1}^1+Z_{1,Y}^Y+Z_{Y,1}^Y+Z_{Y,Y}^1+\frac{\beta_4}{2\beta_5^2}Z_{Y,Y}^Y\rs,
\ee
which is a sum of modular invariants given in~(\ref{eq:reps3:modinvt1}),
(\ref{eq:reps3:modinvt2}).

\subsubsection{$1+X+2Y$ orbifold}
\label{sect:S3:regrep}

In this section, we consider the Frobenius algebra and partition function
corresponding to ${\mathbb C}[S_3]$.  Here, we have 
$A = 1 + X + 2Y$, corresponding to the regular representation of $S_3$.

This is slightly different since we have two copies of the $Y$ representation occurring in our algebra object, which we will denote $Y_1$ and $Y_2$ and keep separate until the end.  We start by picking a basis for the regular representation,
\begin{align}
e=\ & v_1+v_b+v_{b^2}+v_a+v_{ab}+v_{ab^2},\\
e_X =\ & \frac{c_x}{\sqrt{6}}\lp v_1+v_b+v_{b^2}-v_a-v_{ab}-v_{ab^2}\rp,\\
e_{Y_1 1}=\ & \frac{c_1}{2}\lp v_b-v_{b^2}-v_{ab}+v_{ab^2}\rp,\\
e_{Y_1 2} =\ & \frac{c_1}{2\sqrt{3}}\lp -2v_1+v_b+v_{b^2}-2v_a+v_{ab}+v_{ab^2}\rp,\\
e_{Y_2 1}=\ & \frac{c_2}{2\sqrt{3}}\lp -2v_1+v_b+v_{b^2}+2v_a-v_{ab}-v_{ab^2}\rp,\\
e_{Y_2 2}=\ & \frac{c_2}{2}\lp -v_b+v_{b^2}-v_{ab}+v_{ab^2}\rp.
\end{align}
Multiplication here is given by $\m_*(v_g \otimes v_h)=\d_{g,h} v_g$.  Note that $\m_*$ is symmetric by construction, so we do not have to compute both $\m_*(w_1 \otimes w_2)$ and $\m_*(w_2 \otimes w_1)$ separately.  Also, $\m_*(e \otimes w)=\m_*(w \otimes e)=w$ for all vectors $w$, so we omit those as well.  For the remaining ones, we have
\begin{align}
\m_*\left( e_X \otimes e_X \right)=\ & \frac{c_x^2}{6}e,\\
\m_*\left( e_X \otimes e_{Y_1 1} \right)=\ & -\frac{c_xc_1}{\sqrt{6}c_2} e_{Y_2 2},\\
\m_*\left( e_X \otimes e_{Y_1 2} \right)=\ & \frac{c_xc_1}{\sqrt{6}c_2} e_{Y_2 1},\\
\m_*\left( e_X \otimes e_{Y_2 1} \right)=\ & \frac{c_xc_2}{\sqrt{6}c_1} e_{Y_1 2},\\
\m_*\left( e_X \otimes e_{Y_2 2} \right)=\ & -\frac{c_xc_2}{\sqrt{6}c_1} e_{Y_1 1},
\end{align}
\begin{align}
\m_*\left( e_{Y_1 1} \otimes e_{Y_1 1} \right)=\ & \frac{c_1^2}{6}e + \frac{c_1}{2\sqrt{3}} e_{Y_1 2},\\
\m_*\left( e_{Y_1 1} \otimes e_{Y_1 2} \right)=\ & \frac{c_1}{2\sqrt{3}} e_{Y_1 1},\\
\m_*\left( e_{Y_1 2} \otimes e_{Y_1 2} \right)=\ & \frac{c_1^2}{6}e - \frac{c_1}{2\sqrt{3}} e_{Y_1 2},\\
\m_*\left( e_{Y_1 1} \otimes e_{Y_2 1} \right)=\ & -\frac{c_1}{2\sqrt{3}} e_{Y_2 2},\\
\m_*\left( e_{Y_1 1} \otimes e_{Y_2 2} \right)=\ & -\frac{c_1c_2}{\sqrt{6}c_x} e_X - \frac{c_1}{2\sqrt{3}} e_{Y_2 1},\\
\m_*\left( e_{Y_1 2} \otimes e_{Y_2 1} \right)=\ & \frac{c_1c_2}{\sqrt{6}c_x} e_X - \frac{c_1}{2\sqrt{3}} e_{Y_2 1},\\
\m_*\left( e_{Y_1 2} \otimes e_{Y_2 2} \right)=\ & \frac{c_1}{2\sqrt{3}} e_{Y_2 2},
\end{align}
\begin{align}
\m_*\left( e_{Y_2 1} \otimes e_{Y_2 1} \right)=\ & \frac{c_2^2}{6} e - \frac{c_2^2}{2\sqrt{3}c_1} e_{Y_1 2},\\
\m_*\left( e_{Y_2 1} \otimes e_{Y_2 2} \right)=\ & -\frac{c_2^2}{2\sqrt{3}c_1} e_{Y_1 1},\\
\m_*\left( e_{Y_2 2} \otimes e_{Y_2 2} \right)=\ & \frac{c_2^2}{6} e + \frac{c_2^2}{2\sqrt{3}c_1} e_{Y_1 2}.
\end{align}

Using the fact that $\Delta_F$ is diagonal on the $v$ basis, we find
\begin{align}
    \Delta_F(e)=\ & \frac{1}{6}e\otimes e + \frac{1}{c_x^2} e_X \otimes e_X + \frac{1}{c_1^2}\lp e_{Y_1 1} \otimes e_{Y_1 1}  + e_{Y_1 2} \otimes e_{Y_1 2}\rp + \frac{1}{c_2^2}\lp e_{Y_2 1} \otimes e_{Y_2 1} + e_{Y_2 2} \otimes e_{Y_2 2}\rp,\\
    \Delta_F(e_X)=\ & \frac{1}{6}\lp e \otimes e_X + e_X \otimes e\rp+\frac{c_x}{\sqrt{6}\,c_1c_2}\lp -e_{Y_11} \otimes e_{Y_22} + e_{Y_12} \otimes e_{Y_21} + e_{Y_21} \otimes e_{Y_12} - e_{Y_22} \otimes e_{Y_11}\rp,\\
    \Delta_F(e_{Y_11})=\ & \frac{1}{6}\lp e \otimes e_{Y_11} + e_{Y_11} \otimes e\rp - \frac{c_1}{\sqrt{6}\,c_xc_2}\lp e_X \otimes e_{Y_22} + e_{Y_22} \otimes e_X\rp\non\\
    & \quad +\frac{1}{2\sqrt{3}\,c_1}\lp e_{Y_11} \otimes e_{Y_12} + e_{Y_12} \otimes e_{Y_11}\rp-\frac{c_1}{2\sqrt{3}\,c_2^2}\lp e_{Y_21} \otimes e_{Y_22} + e_{Y_22} \otimes e_{Y_21}\rp,\\
    \Delta_F(e_{Y_12})=\ & \frac{1}{6}\lp e \otimes e_{Y_12} + e_{Y_12} \otimes e\rp+\frac{c_1}{\sqrt{6}\,c_xc_2}\lp e_X \otimes e_{Y_21} + e_{Y_21} \otimes e_X\rp\non\\
    & \quad +\frac{1}{2\sqrt{3}\,c_1}\lp e_{Y_11} \otimes e_{Y_11} - e_{Y_12} \otimes e_{Y_12}\rp-\frac{c_1}{2\sqrt{3}\,c_2^2}\lp e_{Y_21} \otimes e_{Y_21} - e_{Y_22} \otimes e_{Y_22}\rp,\\
    \Delta_F(e_{Y_21})=\ & \frac{1}{6}\lp e \otimes e_{Y_21} + e_{Y_21} \otimes e\rp+\frac{c_2}{\sqrt{6}\,c_xc_1}\lp e_X \otimes e_{Y_12} + e_{Y_12} \otimes e_X\rp\non\\
    & \quad -\frac{1}{2\sqrt{3}\,c_1}\lp e_{Y_11} \otimes e_{Y_22} + e_{Y_12} \otimes e_{Y_21} + e_{Y_21} \otimes e_{Y_12} + e_{Y_22} \otimes e_{Y_11}\rp,\\
    \Delta_F(e_{Y_22})=\ & \frac{1}{6}\lp e \otimes e_{Y_22} + e_{Y_22} \otimes e\rp-\frac{c_2}{\sqrt{6}\,c_xc_1}\lp e_X \otimes e_{Y_11} + e_{Y_11} \otimes e_X\rp\non\\
    & \quad -\frac{1}{2\sqrt{3}\,c_1}\lp e_{Y_11} \otimes e_{Y_21} - e_{Y_12} \otimes e_{Y_22} + e_{Y_21} \otimes e_{Y_11} - e_{Y_22} \otimes e_{Y_12}\rp.
\end{align}
This gives non-vanishing coefficients (apart from $\m_{1,R}^R=\m_{R,1}^R=1$, $\Delta_R^{1,R}=\Delta_R^{R,1}=\frac{1}{6}$)
\be
\m_{X,X}^1=\frac{c_x^2}{6\beta_1},\ \m_{X,Y_1}^{Y_2}=-\frac{c_xc_1}{\sqrt{6}c_2 \beta_2},\ \m_{X,Y_2}^{Y_1}=\frac{c_xc_2}{\sqrt{6}c_1 \beta_2},\ \m_{Y_1,X}^{Y_2}=-\frac{c_xc_1}{\sqrt{6}c_2 \beta_3},\ \m_{Y_2,X}^{Y_1}=\frac{c_xc_2}{\sqrt{6}c_1 \beta_3},\non
\ee
\be
\m_{Y_1,Y_1}^1=\frac{c_1^2}{6 \beta_4},\quad\m_{Y_2,Y_2}^1=\frac{c_2^2}{6 \beta_4},\quad\m_{Y_1,Y_2}^X=-\frac{c_1c_2}{\sqrt{6}c_x \beta_6},\quad\m_{Y_2,Y_1}^X=\frac{c_1c_2}{\sqrt{6}c_x \beta_6},\non
\ee
\be
\m_{Y_1,Y_1}^{Y_1}=\frac{c_1}{2\sqrt{3} \beta_5},\quad\m_{Y_1,Y_2}^{Y_2}=-\frac{c_1}{2\sqrt{3} \beta_5},\quad\m_{Y_2,Y_1}^{Y_2}=-\frac{c_1}{2\sqrt{3} \beta_5},\quad\m_{Y_2,Y_2}^{Y_1}=-\frac{c_2^2}{2\sqrt{3}c_1 \beta_5},
\ee
\be
\Delta_1^{X,X}=\frac{\beta_1}{c_x^2},\quad\Delta_1^{Y_1,Y_1}=\frac{\beta_4}{c_1^2},\quad\Delta_1^{Y_2,Y_2}=\frac{\beta_4}{c_2^2},\quad\Delta_X^{Y_1,Y_2}=\frac{c_x\beta_4}{\sqrt{6}\,c_1c_2\beta_2},\quad\Delta_X^{Y_2,Y_1}=-\frac{c_x\beta_4}{\sqrt{6}\,c_1c_2\beta_2},\non
\ee
\be
\Delta_{Y_1}^{X,Y_2}=-\frac{c_1\beta_4}{\sqrt{6}\,c_xc_2\beta_6},\quad\Delta_{Y_1}^{Y_1,Y_1}=\frac{\beta_4}{2\sqrt{3}\,c_1\beta_5},\quad\Delta_{Y_1}^{Y_2,X}=-\frac{c_1\beta_1}{\sqrt{6}\,c_xc_2\beta_3},\quad\Delta_{Y_1}^{Y_2,Y_2}=-\frac{c_1\beta_4}{2\sqrt{3}\,c_2^2\beta_5},
\ee
\be
\Delta_{Y_2}^{X,Y_1}=\frac{c_2\beta_4}{\sqrt{6}\,c_xc_1\beta_6},\quad\Delta_{Y_2}^{Y_1,X}=\frac{c_2\beta_1}{\sqrt{6}\,c_xc_1\beta_3},\quad\Delta_{Y_2}^{Y_1,Y_2}=-\frac{\beta_4}{2\sqrt{3}\,c_1\beta_5},\quad\Delta_{Y_2}^{Y_2,Y_1}=-\frac{\beta_4}{2\sqrt{3}\,c_1\beta_5}.\non
\ee

Now we put it together, keeping in mind that $Y_1$ and $Y_2$ both just become the simple line $Y$ when we talk about partial traces.
\begin{align}
Z_{1,1}^1:\ & \m_{1,1}^1\Delta_1^{1,1}=\frac{1}{6},\\
Z_{1,X}^X:\ & \m_{1,X}^X\Delta_X^{X,1}=\frac{1}{6},\\
Z_{1,Y}^Y:\ & \m_{1,Y_1}^{Y_1}\Delta_{Y_1}^{Y_1,1}+\m_{1,Y_2}^{Y_2}\Delta_{Y_2}^{Y_2,1}=\frac{1}{3},\\
%Z_{1,Y_2}^{Y_2}:\ & \m_{1,Y_2}^{Y_2}\Delta_{Y_2}^{Y_2,1}=\frac{1}{6},\\
Z_{X,1}^X:\ & \m_{X,1}^X\Delta_X^{1,X}=\frac{1}{6},\\
Z_{X,X}^1:\ & \m_{X,X}^1\Delta_1^{X,X}=\frac{1}{6},
\end{align}
\begin{align}
Z_{X,Y}^Y:\ & \m_{X,Y_1}^{Y_2}\Delta_{Y_2}^{Y_1,X}+\m_{X,Y_2}^{Y_1}\Delta_{Y_1}^{Y_2,X}=-\frac{\beta_1}{3\beta_2 \beta_3},\\
%Z_{X,Y_2}^{Y_1}:\ & \m_{X,Y_2}^{Y_1}\Delta_{Y_1}^{Y_2,X}=-\frac{\beta_1}{6 \beta_2 \beta_3},\\
Z_{Y,1}^Y:\ & \m_{Y_1,1}^{Y_1}\Delta_{Y_1}^{1,Y_1}+\m_{Y_2,1}^{Y_2}\Delta_{Y_2}^{1,Y_2}=\frac{1}{3},\\
Z_{Y,X}^Y:\ & \m_{Y_1,X}^{Y_2}\Delta_{Y_2}^{X,Y_1}+\m_{Y_2,X}^{Y_1}\Delta_{Y_1}^{X,Y_2}=-\frac{\beta_4}{3 \beta_3 \beta_6},\\
Z_{Y,Y}^1:\ & \m_{Y_1,Y_1}^{1}\Delta_{1}^{Y_1,Y_1}+\m_{Y_2,Y_2}^{1}\Delta_{1}^{Y_2,Y_2}=\frac{1}{3},\\
Z_{Y,Y}^X:\ & \m_{Y_1,Y_2}^X\Delta_X^{Y_2,Y_1}+\m_{Y_2,Y_1}^X\Delta_X^{Y_1,Y_2}=\frac{\beta_4}{3\beta_2\beta_6},\\
Z_{Y,Y}^Y:\ & \m_{Y_1,Y_1}^{Y_1}\Delta_{Y_1}^{Y_1,Y_1}+\m_{Y_1,Y_2}^{Y_2}\Delta_{Y_2}^{Y_2,Y_1}+\m_{Y_2,Y_1}^{Y_2}\Delta_{Y_2}^{Y_1,Y_2}+\m_{Y_2,Y_2}^{Y_1}\Delta_{Y_1}^{Y_2,Y_2}=\frac{\beta_4}{3 \beta_5^2}.
\end{align}
This gives the partition function
\begin{multline}
Z_{1+X+2Y}=\frac{1}{6}\ls Z_{1,1}^1+Z_{1,X}^X+2Z_{1,Y}^Y+Z_{X,1}^X+Z_{X,X}^1-\frac{2 \beta_1}{\beta_2 \beta_3}Z_{X,Y}^Y\right.\non\\
\left. +2Z_{Y,1}^Y-\frac{2\beta_4}{\beta_3 \beta_6}Z_{Y,X}^Y+2Z_{Y,Y}^1+\frac{2 \beta_4}{\beta_2 \beta_6}Z_{Y,Y}^X+\frac{2 \beta_4}{\beta_5^2}Z_{Y,Y}^Y\rs.
\end{multline}
This can also be written as
\begin{eqnarray}
\label{rs3_rr_pf}
    Z_{1+X+2Y} & = & \frac{1}{6} \biggl[  Z_{1,1}^1 \: + \: \left( Z_{1,X}^X + Z_{X,1}^X + Z^1_{X,X} \right) \: + \: 2 \left( Z_{1,Y}^Y + Z_{Y,1}^Y + Z_{Y,Y}^1 +  \frac{\beta_4}{2 \beta_5^2} Z_{Y,Y}^Y \right)
    \nonumber \\
    & & \hspace*{0.5in}
    \: - \:
    \frac{2 \beta_1}{\beta_2 \beta_3} \left(  Z^Y_{X,Y} +
    \frac{\beta_2 \beta_4}{\beta_1 \beta_6} Z^Y_{Y,X} - \frac{\beta_3 \beta_4}{\beta_1 \beta_6} Z_{Y,Y}^X - \frac{\beta_2 \beta_3 \beta_4}{2 \beta_1 \beta_5^2} Z_{Y,Y}^Y \right) \biggr]
\end{eqnarray}
in which form it is explicitly a sum of the modular invariants listed
in equations~(\ref{eq:reps3:modinvt1}), (\ref{eq:reps3:modinvt2}).

\subsection{Rep$(D_4)$}\label{ssec:repd4}

\subsubsection{Overview of $D_4$, $Q_8$, ${\cal H}_8$}

In Sections \ref{ssec:repd4}-\ref{ssec:reph8} we describe the gauging of $\text{Rep}(D_4)$, $\text{Rep}(Q_8)$, and $\text{Rep}({\cal H}_8)$, respectively,
where $D_4$ is\footnote{
The eight-element dihedral group is also often denoted $D_8$.  Our notation was chosen to be consistent with previous works on decomposition,
see e.g.~\cite{Hellerman:2006zs,Sharpe:2022ene}.
} the eight-element dihedral group,
$Q_8$ is the eight-element group of unit quaternions, and
${\cal H}_8$ is a Hopf algebra. These correspond to the three $\Z_2\times\Z_2$ Tambara-Yamagami categories that admit a fiber functor\footnote{For a 3D TFT approach to the $\mathbb{Z}_2\times \mathbb{Z}_2$ Tambara-Yamagami fusion category, see, e.g., \cite{Kaidi:2023maf,Yu:2023nyn}.}. Given that they are closely related, we first describe the general aspects of such categories and specialize afterwards.

A $G$ Tambara-Yamagami category \cite{ty} for $G$ a finite group can be thought of as a categorical extension of the fusion category $\text{Vec}(G)$ by $\Z_2$. For any such category $TY(G)$, the simple objects are labeled by elements $g\in G$, with an additional element $m$. The fusion rules are
\begin{eqnarray}
    g\otimes h \cong gh, & g\otimes m\cong m\otimes g\cong m, & m\otimes m\cong \bigoplus_{g\in G}g
\end{eqnarray}
The fusion rules do not uniquely characterize a fusion category, one also needs to provide associators, or $F$-symbols. For a fixed $G$, there may exist different choices of $F$-symbols that lead to inequivalent fusion categories. The equivalence classes of $TY(G)$ categories are characterized by two pieces of information: a square root, conventionally denoted $\tau$ (not to be confused with the modular parameter), of the reciprocal of the order of the group, 
\begin{equation}
    \tau^2=\frac{1}{|G|},
\end{equation}
and a class of bicharacter $\chi:G\times G\to \Bbbk^{\times}$ (here $\Bbbk=\C$). These two are used to compute the $F$-symbols.

We are interested in the case $G=\Z_2\times\Z_2=\{1,a,b,c\}$, where $ab=c$ (cyclically). (See appendix~\ref{app:ty} for a summary of $F$ symbols and modular transformations in these fusion categories.) Here, there are two choices of square roots $\tau=\pm 1/2,$ %$\frac{1}{2}$ 
and two choices of classes of characters: the trivial one, whose only nontrivial values are $\chi_1(a,b)=\chi_1(a,c)=\chi_1(b,c)=-1$ (symmetrically), and the nontrivial class generated by $\chi_c(a,a)=\chi_c(b,b)=-1$, for which in particular $\chi_c(a,b)=1$. This gives rise to four inequivalent classes of fusion categories $TY(\Z_2\times\Z_2)$. If we let $n=2\tau$, the different choices lead to the following fusion categories
\begin{center}
    \begin{tabular}{c|c|c}
     & $\chi_1$ & $\chi_c$\\ \hline
     $n=1$ & Rep($D_4$) & Rep(${\cal H}_8$)\\
     $n=-1$ & Rep($Q_8)$ & -
\end{tabular}
\end{center}
where the fourth category is omitted as it is not a representation category.

These $TY(\Z_2\times\Z_2)$ categories have the following associators\footnote{
For certain choices of associator maps.  Later in this section we will compute $F$ symbols (crossing kernels) for more general intertwiners.
}
\begin{eqnarray}
F^{q,p,pq}_{pqr,r,qr}=F^{q,p,pq}_{m,m,m}=F^{p,m,m}_{m,q,pq}=F^{m,p,m}_{pq,m,q}=F^{m,m,q}_{pq,p,m}=1;\\ F^{m,p,m}_{m,q,m}=F^{p,m,m}_{q,m,m}=\chi(p,q); \\F^{m,m,p}_{m,m,q}=\tfrac{1}{2}n\chi(p,q).
\end{eqnarray}
for $p,q,r\in\{1,a,b,c\}$.
The reader should note that for all simple objects $L$ in these categories,
$\overline{L} = L$, so we will often omit the duals in computations.

Over the next several sections, we will discuss Frobenius algebras constructed in each of the fusion categories Rep$(D_4)$, Rep$(Q_8)$, and Rep$({\cal H}_8)$.  For the first two, we will compute $F$ symbols (crossing kernels)
for more general intertwiner maps than described above.  For Rep$({\cal H}_8)$, we will use only the $F$ symbols above.

In this section we will focus on Rep$(D_4)$, where $D_4$ can be presented as
\begin{equation}
    \left\langle x,y|x^4=y^2=(xy)^2=1\right\rangle .
\end{equation}
In $D_4$ there are five conjugacy classes, $[1]=\{1\}$, $[x]=\{x,x^3\}$, $[x^2]=\{x^2\}$, $[y]=\{y,x^2y\}$, and $[xy]=\{xy,x^3y\}$.

\subsubsection{Representation theory}

The group $D_4$ has five irreducible representations which we'll label $1$, $a$, $b$, $c$, and $m$, and character table
\begin{center}
\begin{tabular}{c|c|c|c|c|c|}
& $[1]$ & $[x^2]$ & $[x]$ & $[y]$ & $[xy]$ \\
\hline
$\chi_1$ & $1$ & $1$ & $1$ & $1$ & $1$ \\
\hline
$\chi_a$ & $1$ & $1$ & $1$ & $-1$ & $-1$ \\
\hline
$\chi_b$ & $1$ & $1$ & $-1$ & $1$ & $-1$ \\
\hline
$\chi_c$ & $1$ & $1$ & $-1$ & $-1$ & $1$ \\
\hline
$\chi_m$ & $2$ & $-2$ & $0$ & $0$ & $0$ \\
\hline
\end{tabular}
\end{center}
From the character table we can determine the fusion rules as
\be
a\otimes a\cong b\otimes b\cong c\otimes c\cong 1,\qquad a\otimes b\cong c,\qquad a\otimes c\cong b,\qquad b\otimes c\cong a,\non
\ee
\be
a\otimes m\cong b\otimes m\cong c\otimes m\cong m,\qquad m\otimes m\cong 1\oplus a\oplus b\oplus c.
\ee

The one-dimensional irreps $1, a, b, c$ are given explicitly by their characters
as,
\be
\rho_1(x)=\rho_1(y)=1,\qquad\rho_a(x)=1,\ \rho_a(y)=-1,\non
\ee
\be
\rho_b(x)=-1,\ \rho_b(y)=1,\qquad\rho_c(x)=-1,\ \rho_c(y)=-1,
\ee
and the two-dimensional irrep $m$ is given by
\be
\rho_m(x)=\lp\begin{matrix} i & 0 \\ 0 & -i \end{matrix}\rp,
\qquad\rho_m(y)=\lp\begin{matrix} 0 & 1 \\ 1 & 0 \end{matrix}\rp,
\ee

\subsubsection{Cosets}

We have the following subgroups, and the corresponding coset representations:
\begin{itemize}
\item $H=D_4$, $D_4/H=\{H\}$, which transforms simply as the trivial irrep $1$.
\item $H=\langle x\rangle\cong\Z_4$, $D_4/H=\{H,yH\}$.  The corresponding representation has
\be
\rho(x)=\lp\begin{matrix} 1 & 0 \\ 0 & 1 \end{matrix}\rp,\qquad\rho(y)=\lp\begin{matrix} 0 & 1 \\ 1 & 0 \end{matrix}\rp,
\ee
and this decomposes into $1+a$.
\item $H=\langle x^2,y\rangle\cong(\Z_2)^2$, $D_4/H=\{H,xH\}$.  The representation is
\be
\rho(x)=\lp\begin{matrix} 0 & 1 \\ 1 & 0 \end{matrix}\rp,\qquad\rho(y)=\lp\begin{matrix} 1 & 0 \\ 0 & 1 \end{matrix}\rp,
\ee
corresponding to $1+b$.
\item $H=\langle x^2,xy\rangle\cong(\Z_2)^2$, $D_4/H=\{H,xH\}$.  The representation is
\be
\rho(x)=\lp\begin{matrix} 0 & 1 \\ 1 & 0 \end{matrix}\rp,\qquad\rho(y)=\lp\begin{matrix} 0 & 1 \\ 1 & 0 \end{matrix}\rp,
\ee
corresponding to $1+c$.  Note that the two $(\Z_2)^2$ subgroups (and their cosets) are exchanged under an outer automorphism of $D_4$.
\item $H=\langle x^2\rangle\cong\Z_2$, $D_4/H=\{H,xH,yH,xyH\}$.  We have
\be
\rho(x)=\lp\begin{matrix}0 & 1 & 0 & 0 \\ 1 & 0 & 0 & 0 \\ 0 & 0 & 0 & 1 \\ 0 & 0 & 1 & 0 \end{matrix}\rp,\qquad\rho(y)=\lp\begin{matrix} 0 & 0 & 1 & 0 \\ 0 & 0 & 0 & 1 \\ 1 & 0 & 0 & 0 \\ 0 & 1 & 0 & 0 \end{matrix}\rp,
\ee
for $1+a+b+c$.
\item $H=\langle y\rangle\cong\Z_2$, $D_4/H=\{H,xH,x^2H,x^3H\}$.
\be
\rho(x)=\lp\begin{matrix} 0 & 1 & 0 & 0 \\ 0 & 0 & 1 & 0 \\ 0 & 0 & 0 & 1 \\ 1 & 0 & 0 & 0 \end{matrix}\rp,\qquad\rho(y)=\lp\begin{matrix} 1 & 0 & 0 & 0 \\ 0 & 0 & 0 & 1 \\ 0 & 0 & 1 & 0 \\ 0 & 1 & 0 & 0 \end{matrix}\rp,
\ee
giving $1+b+m$.
\item $H=\langle xy\rangle\cong\Z_2$, $D_4/H=\{H,xH,x^2H,x^3H\}$.
\be
\rho(x)=\lp\begin{matrix} 0 & 1 & 0 & 0 \\ 0 & 0 & 1 & 0 \\ 0 & 0 & 0 & 1 \\ 1 & 0 & 0 & 0 \end{matrix}\rp,\qquad\rho(y)=\lp\begin{matrix} 0 & 0 & 0 & 1 \\ 0 & 0 & 1 & 0 \\ 0 & 1 & 0 & 0 \\ 1 & 0 & 0 & 0 \end{matrix}\rp,
\ee
resulting in $1+c+m$.
\item $H=\langle x^2y\rangle\cong\Z_2$, $D_4/H=\{H,xH,x^2H,x^3H\}$.
\be
\rho(x)=\lp\begin{matrix} 0 & 1 & 0 & 0 \\ 0 & 0 & 1 & 0 \\ 0 & 0 & 0 & 1 \\ 1 & 0 & 0 & 0 \end{matrix}\rp,\qquad\rho(y)=\lp\begin{matrix} 0 & 0 & 1 & 0 \\ 0 & 1 & 0 & 0 \\ 1 & 0 & 0 & 0 \\ 0 & 0 & 0 & 1 \end{matrix}\rp,
\ee
so $1+b+m$.
\item $H=\langle x^3y\rangle\cong\Z_2$, $D_4/H=\{H,xH,x^2H,x^3H\}$.
\be
\rho(x)=\lp\begin{matrix} 0 & 1 & 0 & 0 \\ 0 & 0 & 1 & 0 \\ 0 & 0 & 0 & 1 \\ 1 & 0 & 0 & 0 \end{matrix}\rp,\qquad\rho(y)=\lp\begin{matrix} 0 & 1 & 0 & 0 \\ 1 & 0 & 0 & 0 \\ 0 & 0 & 0 & 1 \\ 0 & 0 & 1 & 0 \end{matrix}\rp,
\ee
resulting in $1+c+m$.
\item Finally, $H=\{1\}$, $D_4/H\cong D_4=\{1,x,x^2,x^3,y,xy,x^2y,x^3y\}$ corresponding to the regular representation $1+a+b+c+2m$.
\end{itemize}

The two cases corresponding to $1+b+m$ are conjugate subgroups, as are the pair corresponding to $1+c+m$.  Furthermore, an outer automorphism exchanges these two cases.  So there are six physically distinct options for $D_4$: $1$, $1+a$, $1+b$, $1+a+b+c$, $1+b+m$, and $1+a+b+c+2m$.

Next, we will compute associators, crossing kernels, and modular transformations,
and compute modular-invariant partition functions for each of the six options above.

\subsubsection{Computing the associator}

In this section we will describe the computation of the associator and crossing kernels with
general coefficients in $D_4$.

Taking basis vectors $e$, $e_a$, $e_b$, $e_c$, and $e_{m1}$, $e_{m2}$ for $1$, $a$, $b$, $c$, and $m$ respectively (such that the irreps take the form from the previous subsection), we parameterize the most general basis for fusion intertwiners.  In the expressions below, we omit the tensor product symbol $\otimes$ to make expressions more compact, and we also omit trivial cases such as $e \otimes e_a \mapsto e_a$.
\begin{align}
e_a e_a\mapsto\ & \beta_1 \, e,  \label{eq:intertwinerd4} \\
e_a e_b \mapsto\ & \beta_2\, e_c,\\
e_a e_c \mapsto\ & \beta_3 \, e_b,\\
e_a e_{m1}\mapsto\ & \beta_4 \, e_{m1},\\
e_a e_{m2}\mapsto\ & - \beta_4 \, e_{m2},
\end{align}
\begin{align}
e_b e_a \mapsto\ & \beta_5 \, e_c, \\
e_b e_b\mapsto\ & \beta_6 \, e,  \label{eq:intertwinerd4:bb}\\
e_b e_c\mapsto\ & \beta_7 \, e_a,\\
e_b e_{m1}\mapsto\ & \beta_8 \, e_{m2},\\
e_b e_{m2}\mapsto\ & \beta_8 \, e_{m1},
\end{align}
\begin{align}
e_c e_a\mapsto\ & \beta_9 \, e_b,\\
e_c e_b\mapsto\ & \beta_{10} \, e_a,\\
e_c e_c\mapsto\ & \beta_{11} \, e, \\
e_c e_{m1}\mapsto\ & \beta_{12} \, e_{m2},\\
e_c e_{m2}\mapsto\ & - \beta_{12} \, e_{m1},
\end{align}
\begin{align}
e_{m1} e_a\mapsto\ & \beta_{13} \, e_{m1},\\
e_{m1} e_b\mapsto\ & \beta_{14} \, e_{m2} ,\\
e_{m1} e_c\mapsto\ & \beta_{15} e_{m2},\\
e_{m1} e_{m1}\mapsto\ & \beta_{16} \, e_b + \beta_{17} \, e_c,\\
e_{m1} e_{m2} \mapsto\ & \beta_{18} \, e + \beta_{19} \, e_a,\\
e_{m2} e_a\mapsto\ & - \beta_{13} \, e_{m2},\\
e_{m2} e_b\mapsto\ & \beta_{14} \, e_{m1},\\
e_{m2} e_c\mapsto\ & - \beta_{15} \, e_{m1},\\
e_{m2} e_{m1}\mapsto\ & \beta_{18} \, e - \beta_{19} \, e_a,\\
e_{m2} e_{m2}\mapsto\ & \beta_{16} \, e_b - \beta_{17} \, e_c.\label{eq:intertwinerd4-2}
\end{align}

This also gives rise to associated evaluation and coevaluation maps $\e_R$, $\overline{\e}_R$, $\g_R$, and $\overline{\g}_R$, and then to an associated basis for co-fusion homomorphisms which can be read off from
\begin{align}
    e\mapsto\ & ee+\beta_1^{-1}e_ae_a+\beta_6^{-1}e_be_b+\beta_{11}^{-1}e_ce_c+\beta_{18}^{-1}\lp e_{m1}e_{m2}+e_{m2}e_{m1}\rp,\\
    e_a\mapsto\ & ee_a+e_ae+\beta_3\beta_{11}^{-1}e_be_c+\beta_2\beta_6^{-1}e_ce_b+\beta_4\beta_{18}^{-1}\lp e_{m1}e_{m2}-e_{m2}e_{m1}\rp,\\
    e_b\mapsto\ & ee_b+e_be+\beta_7\beta_{11}^{-1}e_ae_c+\beta_5\beta_1^{-1}e_ce_a+\beta_8\beta_{18}^{-1}\lp e_{m1}e_{m1}+e_{m2}e_{m2}\rp,\\
    e_c\mapsto\ & ee_c+e_ce+\beta_{10}\beta_6^{-1}e_ae_b+\beta_9\beta_1^{-1}e_be_a-\beta_{12}\beta_{18}^{-1}\lp e_{m1}e_{m1}-e_{m2}e_{m2}\rp,\\
    e_{m1}\mapsto\ & ee_{m1}+e_{m1}e+\beta_{19}\beta_{18}^{-1}e_ae_{m1}+\beta_{13}\beta_1^{-1}e_{m1}e_a\non\\
    & \qquad +\beta_{16}\beta_{18}^{-1}e_be_{m2}+\beta_{14}\beta_6^{-1}e_{m2}e_b+\beta_{17}\beta_{18}^{-1}e_ce_{m2}+\beta_{15}\beta_{11}^{-1}e_{m2}e_c,\\
    e_{m2}\mapsto\ & ee_{m2}+e_{m2}e-\beta_{19}\beta_{18}^{-1}e_ae_{m2}-\beta_{13}\beta_1^{-1}e_{m2}e_a\non\\
    & \qquad +\beta_{16}\beta_{18}^{-1}e_be_{m1}+\beta_{14}\beta_6^{-1}e_{m1}e_b-\beta_{17}\beta_{18}^{-1}e_ce_{m1}-\beta_{15}\beta_{11}^{-1}e_{m1}e_c.
\end{align}

Now we compute the crossing kernels.  
\begin{align}
\left( e_a e_a \right) e_a = \beta_1 e e_a = \beta_1 \, e_a,\ 
& e_a \left( e_a e_a \right) = \beta_1 \, e_a e = \beta_1 \, e_a,\ \Rightarrow\ & \tilde{K}^{a,a}_{a,a}(1,1)=1,
\\
\left( e_a e_a \right) e_b = \beta_1 \, e e_b = \beta_1 \, e_b,\ & e_a \left( e_a e_b \right) = \beta_2 \, e_a e_c = \beta_2 \beta_3 \, e_b,\ \Rightarrow\ & \tilde{K}^{a,b}_{a,b}(1,c) = \frac{\beta_1}{\beta_2 \beta_3},
\\
\left( e_a e_a \right) e_c = \beta_1 \, e e_c = \beta_1 \, e_c,\ & e_a \left( e_a e_c \right) = \beta_3 \, e_a e_b = \beta_2 \beta_3 \, e_c,\ \Rightarrow\ & \tilde{K}^{a,c}_{a,c}(1,b) = \frac{\beta_1}{\beta_2 \beta_3},
\end{align}
\begin{align}
\left( e_a e_a \right) e_{m1} = \beta_1 \, e e_{m1} = \beta_1 \, e_{m1},\ & e_a \left( e_a e_{m1}\right) = \beta_4 e_a e_{m1} = \beta_4^2 \, e_{m1},\ \Rightarrow\ 
& \tilde{K}^{a,m}_{a,m}(1,m) = \frac{\beta_1}{\beta_4^2},
\\
\left( e_a e_b \right) e_a = \beta_2 \, e_c e_a = \beta_2 \beta_9 \, e_b,\ & e_a \left( e_b e_a \right) = \beta_5 \, e_a e_c = \beta_3 \beta_5 \, e_b,\ \Rightarrow\ & \tilde{K}^{a,b}_{b,a}(c,c) = \frac{\beta_2 \beta_9}{\beta_3 \beta_5},
\\
\left( e_a e_b \right) e_b = \beta_2 \, e_c e_b = \beta_2 \beta_{10} \, e_a,\ & e_a \left( e_b e_b \right) = \beta_6 \, e_a e = \beta_6 e_a,\ \Rightarrow\ 
& \tilde{K}^{a,a}_{b,b}(c,1) = \frac{\beta_2 \beta_{10}}{\beta_6},
\end{align}
\begin{align}
\left( e_a e_b \right) e_c = \beta_2 \, e_c e_c = \beta_2 \beta_{11} \, e,\ & e_a \left( e_b e_c \right) = \beta_7 \, e_a e_a = \beta_1 \beta_7 \, e,\ \Rightarrow\ 
& \tilde{K}^{a,1}_{b,c}(c,a) = \frac{\beta_2 \beta_{11}}{\beta_1 \beta_7},
\\
\left( e_a e_b \right) e_{m1} = \beta_2 \, e_c e_{m1} = \beta_2 \beta_{12} \, e_{m2},\ 
& e_a\left( e_b e_{m1} \right) = \beta_8 \, e_a e_{m2} = -\beta_4 \beta_8 \, e_{m2},\ \Rightarrow\ & 
\tilde{K}^{a,m}_{b,m}(c,m) = - \frac{\beta_2 \beta_{12}}{\beta_4 \beta_8},
\end{align}
\begin{align}
\left( e_a e_c \right) e_a = \beta_3 \, e_b e_a = \beta_3 \beta_5 \, e_c,\ & e_a \left( e_c e_a \right) = \beta_9 \, e_a e_b = \beta_2 \beta_9 \, e_c,\ \Rightarrow\ & \tilde{K}^{a,c}_{c,a}(b,b) = \frac{\beta_3 \beta_5}{\beta_2 \beta_9},
\\
\left( e_a e_c \right) e_b = \beta_3 \, e_b e_b = \beta_3 \beta_6 \, e,\ & e_a \left(e_c e_b \right) = \beta_{10} \, e_a e_a = \beta_1 \beta_{10} \, e,\ \Rightarrow\ & \tilde{K}^{a,1}_{c,b}(b,a) = \frac{\beta_3 \beta_6}{\beta_1 \beta_{10}},
\\
\left( e_a e_c \right) e_c = \beta_3 \, e_b e_c = \beta_3 \beta_7 \, e_a,\ & e_a \left( e_c e_c \right) = \beta_{11} \, e_a e = \beta_{11} \, e_a,\ \Rightarrow\ & \tilde{K}^{a,a}_{c,c}(b,1) = \frac{\beta_3 \beta_7}{\beta_{11}},
\end{align}
\begin{align}
\left( e_a e_c \right) e_{m1} = \beta_3 \, e_b e_{m1} = \beta_3 \beta_8 \, e_{m2},\ & e_a \left( e_c e_{m1} \right) = \beta_{12} \, e_a e_{m2} = - \beta_4 \beta_{12} \, e_{m2},\ \Rightarrow\ & \tilde{K}^{a,m}_{c,m}(b,m) = - \frac{\beta_3 \beta_8}{\beta_4 \beta_{12}},
\\
\left( e_a e_{m1} \right) e_a = \beta_4 \, e_{m1} e_a = \beta_4 \beta_{13} e_{m1},\ & e_a \left( e_{m1} e_a \right) = \beta_{13} \, e_a e_{m1} = \beta_4 \beta_{13} e_{m1},\ \Rightarrow\ & \tilde{K}^{a,m}_{m,a}(m,m)=1,
\\
\left( e_a e_{m1} \right) e_b = \beta_4 \, e_{m1} e_b = \beta_4 \beta_{14} \, e_{m2},\ & e_a \left( e_{m1} e_b \right) = \beta_{14} \, e_a e_{m2} = - \beta_4 \beta_{14} \, e_{m2},\ \Rightarrow\ & \tilde{K}^{a,m}_{m,b}(m,m)=-1,
\\
\left( e_a e_{m1} \right) e_c = \beta_4 \, e_{m1} e_c = \beta_4 \beta_{15} \, e_{m2},\ & e_a \left( e_{m1} e_c \right) = \beta_{15} \, e_a e_{m2} = - \beta_4 \beta_{15} \, e_{m2},\ \Rightarrow\ & \tilde{K}^{a,m}_{m,c}(m,m)=-1,
\end{align}
\begin{multline}
\left( e_a e_{m1} \right) e_{m1} = \beta_4 \, e_{m1} e_{m1}  = \beta_4 \beta_{16} \, e_b + \beta_4 \beta_{17} \, e_c,\quad e_a \left( e_{m1} e_{m1} \right) = \beta_{16} e_a e_b + \beta_{17} \, e_a e_c = \beta_3 \beta_{17} \, e_b + \beta_2 \beta_{16} \, e_c,\non
\\
\Rightarrow\quad\tilde{K}^{a,b}_{m,m}(m,c) = \frac{\beta_4 \beta_{16}}{\beta_3 \beta_{17}},\quad\tilde{K}^{a,c}_{m,m}(m,b) = \frac{\beta_4 \beta_{17}}{\beta_2 \beta_{16}},
\end{multline}
\begin{multline}
\left( e_a e_{m1} \right) e_{m2} = \beta_4 \, e_{m1} e_{m2}  = \beta_4 \beta_{18} \, e + \beta_4 \beta_{19} \, e_a,\quad 
e_a\left( e_{m1} e_{m2} \right) = \beta_{18} \, e_a e + \beta_{19} \, e_a e_a = \beta_1 \beta_{19} \, e + \beta_{18} \, e_a,\non\\
\Rightarrow\quad\tilde{K}^{a,1}_{m,m}(m,a) = \frac{\beta_4 \beta_{18}}{\beta_1 \beta_{19}},\quad
\tilde{K}^{a,a}_{m,m}(m,1) = \frac{\beta_4 \beta_{19}}{\beta_{18}},
\end{multline}
\begin{align}
\left( e_b e_a \right) e_a = \beta_5 \, e_c e_a = \beta_5 \beta_9 \, e_b,\ & e_b \left( e_a e_a \right) = \beta_1 \, e_b e = \beta_1 \, e_b,\ \Rightarrow\ & \tilde{K}^{b,b}_{a,a}(c,1) = \frac{\beta_5 \beta_9}{\beta_1},
\\
\left( e_b e_a \right) e_b = \beta_5 \, e_c e_b = \beta_5 \beta_{10} \, e_a,\ & e_b \left( e_a e_b \right) = \beta_2 \, e_b e_c = \beta_2 \beta_7 \, e_a,\ \Rightarrow\ & \tilde{K}^{b,a}_{a,b}(c,c) = \frac{\beta_5 \beta_{10}}{\beta_2 \beta_7},
\\
\left( e_b e_a \right) e_c = \beta_5 \, e_c e_c = \beta_5 \beta_{11} \, e,\ & e_b \left( e_a e_c \right) = \beta_3 \, e_b e_b = \beta_3 \beta_6 \, e,\ \Rightarrow\ & \tilde{K}^{b,1}_{a,c}(c,b) = \frac{\beta_5 \beta_{11}}{\beta_3 \beta_6},
\end{align}
\begin{align}
\left( e_b e_a \right) e_{m1} = \beta_5 \, e_c e_{m1} = \beta_5 \beta_{12} \, e_{m2},\ & 
e_b\left( e_a e_{m1} \right) = \beta_4 \, e_b e_{m1} = \beta_4 \beta_8 \, e_{m2},\ \Rightarrow\ & \tilde{K}^{b,m}_{a,y}(c,y) = \frac{\beta_5 \beta_{12}}{\beta_4 \beta_8},
\\
\left( e_b e_b \right) e_a = \beta_6 \, e e_a = \beta_6 \, e_a,\ & e_b \left( e_b e_a \right) = \beta_5 \, e_b e_c = \beta_5 \beta_7 \, e_a,\ \Rightarrow\ & \tilde{K}^{b,a}_{b,a}(1,c) = \frac{\beta_6}{\beta_5 \beta_7},
\\
\left( e_b e_b \right) e_b = \beta_6 \, e e_b = \beta_6 \, e_b,\ & e_b \left( e_b e_b \right) = \beta_6 \, e_b e = \beta_6 \, e_b,\ \Rightarrow\ 
& \tilde{K}^{b,b}_{b,b}(1,1)=1,
\end{align}
\begin{align}
\left( e_b e_b \right) e_c = \beta_6 \, e e_c = \beta_6 \, e_c,\ & e_b \left( e_b e_c \right) = \beta_7 \, e_b e_a = \beta_5 \beta_7 \, e_c,\ \Rightarrow\ & \tilde{K}^{b,c}_{b,c}(1,a) = \frac{\beta_6}{\beta_5 \beta_7},
\\
\left( e_b e_b \right) e_{m1} = \beta_6 \, e e_{m1} = \beta_6 \, e_{m1},\ & e_b \left( e_b e_{m1}\right) = \beta_8 \, e_b e_{m2} = \beta_8^2 \, e_{m1},\ \Rightarrow\ & \tilde{K}^{b,m}_{b,m}(1,m) = \frac{\beta_6}{\beta_8^2},
\\  
\left( e_b e_c \right) e_a = \beta_7 \, e_a e_a = \beta_1 \beta_7 \, e,\ & 
e_b \left( e_c e_a \right) = \beta_9 \, e_b e_b = \beta_6 \beta_9 \, e,\ \Rightarrow\ & 
\tilde{K}^{b,1}_{c,a}(a,b)=\frac{\beta_1 \beta_7}{\beta_6 \beta_9},
\end{align}
\begin{align}
\left( e_b e_c \right) e_b = \beta_7 \, e_a e_b = \beta_2 \beta_7 \, e_c,\ & 
e_b \left( e_c e_b\right) = \beta_{10} \, e_b e_a = \beta_5 \beta_{10} \, e_c,\ \Rightarrow\ & 
\tilde{K}^{b,c}_{c,b}(a,a) = \frac{\beta_2 \beta_7}{\beta_5 \beta_{10}},
\\
\left( e_b e_c \right) e_{m1} = \beta_7 \, e_a e_{m1} = \beta_4 \beta_7 \, e_{m1},\ 
& e_b \left( e_c e_{m1} \right) = \beta_{12} \, e_b e_{m2} = \beta_8 \beta_{12} \, e_{m1},\ \Rightarrow\ 
& \tilde{K}^{b,m}_{c,m}(a,m) = \frac{\beta_4 \beta_7}{\beta_8 \beta_{12}},
\\
\left( e_b e_{m1} \right) e_a = \beta_8 \, e_{m2} e_a = - \beta_8 \beta_{13} e_{m2},\ 
& e_b \left( e_{m1} e_a \right) = \beta_{13} \, e_b e_{m1} = \beta_8 \beta_{13} \, e_{m2},\ \Rightarrow\ 
& \tilde{K}^{b,m}_{m,a}(m,m)=-1,
\end{align}
\begin{align}
\left( e_b e_{m1} \right)e_b = \beta_8 \, e_{m2} e_b = \beta_8 \beta_{14} \, e_{m1},\ 
& e_b \left( e_{m1} e_b \right) = \beta_{14} \, e_b e_{m2} = \beta_8 \beta_{14} \, e_{m1},\ \Rightarrow\ 
& \tilde{K}^{b,m}_{m,b}(m,m)=1,
\\
\left( e_b e_{m1} \right) e_c = \beta_8 \, e_{m2} e_c = - \beta_8 \beta_{15} \, e_{m1},\ 
& e_b \left( e_{m1} e_c \right) = \beta_{15} \, e_b e_{m2} = \beta_8 \beta_{15} e_{m1},\ \Rightarrow\ 
& \tilde{K}^{b,m}_{m,c}(m,m)=-1,
\end{align}
\begin{multline}
\left( e_b e_{m1} \right) e_{m1} = \beta_8 \, e_{m2} e_{m1} = \beta_8 \beta_{18} \, e - \beta_8 \beta_{19} \, e_a,
\quad 
e_b \left( e_{m1} e_{m1} \right) = \beta_{16} \, e_b e_b + \beta_{17} \, e_b e_c = \beta_6 \beta_{16} \, e + \beta_9 \beta_{17} \, e_a,\non
\\
\Rightarrow\quad\tilde{K}^{b,1}_{m,m}(m,b) = \frac{\beta_8 \beta_{18}}{\beta_6 \beta_{16}},
\quad\tilde{K}^{b,a}_{m,m}(m,c) = - \frac{\beta_8 \beta_{19}}{\beta_9 \beta_{17}},
\end{multline}
\begin{multline}
\left( e_b e_{m1} \right) e_{m2} = \beta_8 \, e_{m2} e_{m2} = \beta_8 \beta_{16} \, e_b - \beta_8 \beta_{17} \, e_c,
\quad e_b \left( e_{m1} e_{m2} \right) = \beta_{18} \, e_b e + \beta_{19} \, e_b e_a = \beta_{18} \, b + \beta_5 \beta_{19} \, c,\non
\\
\Rightarrow\quad\tilde{K}^{b,b}_{m,m}(m,1) = \frac{\beta_{18}}{\beta_8 \beta_{16}},
\quad\tilde{K}^{b,c}_{m,m}(m,a) = - \frac{\beta_8 \beta_{17}}{\beta_5 \beta_{19}},
\end{multline}
\begin{align}
\left( e_c e_a \right) e_a = \beta_9 \, e_b e_a = \beta_5 \beta_9 \, e_c,\ 
& e_c\left( e_a e_a \right) = \beta_1 \, e_c e = \beta_1 \, e_c,\ \Rightarrow\ 
& \tilde{K}^{c,c}_{a,a}(b,1) = \frac{\beta_5 \beta_9}{\beta_1},
\\
\left( e_c e_a\right) e_b = \beta_9 \, e_b e_b = \beta_6 \beta_9 \, e,\ & 
e_c\left( e_a e_b \right) = \beta_2 \, e_c e_c = \beta_2 \beta_{11} \, e,\ \Rightarrow\ 
& \tilde{K}^{c,1}_{a,b}(b,c) = \frac{\beta_6 \beta_9}{\beta_2 \beta_{11}},
\\
\left( e_c e_a \right) e_c = \beta_9 \, e_b e_c = \beta_7 \beta_9 \, e_a,\ 
& e_c \left( e_a e_c \right) = \beta_3 \, e_c e_b = \beta_3 \beta_{10} \, e_a,\ \Rightarrow\ 
& \tilde{K}^{c,a}_{a,c}(b,b) = \frac{\beta_7 \beta_9}{\beta_3 \beta_{10}},
\end{align}
\begin{align}
\left( e_c e_a \right) e_{m1} = \beta_9 \, e_b e_{m1} = \beta_8 \beta_9 \, e_{m2},\ 
& e_c\left( e_a e_{m1} \right) = \beta_4 \, e_c e_{m1} = \beta_4 \beta_{12} \, e_{m2},\ \Rightarrow\ 
& \tilde{K}^{c,m}_{a,m}(b,m) = \frac{\beta_8 \beta_9}{\beta_4 \beta_{12}},
\\
\left( e_c e_b \right) e_a = \beta_{10} \, e_a e_a = \beta_1 \beta_{10} \, e,\ 
& e_c \left( e_b e_a \right) = \beta_5 \, e_c e_c = \beta_5 \beta_{11} \, e,\ \Rightarrow\ 
& \tilde{K}^{c,1}_{b,a}(a,c) = \frac{\beta_1 \beta_{10}}{\beta_5 \beta_{11}},
\\
\left( e_c e_b \right) e_b = \beta_{10} \, e_a e_b = \beta_2 \beta_{10} \, e_c,\ 
& e_c \left( e_b e_b \right) = \beta_6 \, e_c e = \beta_6 \, e_c,\ \Rightarrow\ 
& \tilde{K}^{c,c}_{b,b}(a,1) = \frac{\beta_2 \beta_{10}}{\beta_6},
\end{align}
\begin{align}
\left( e_c e_b \right) e_c = \beta_{10} \, e_a e_c = \beta_3 \beta_{10} \, e_b,\ 
& e_c \left( e_b e_c \right) = \beta_7 \, e_c e_a = \beta_7 \beta_9 \, e_b,\ \Rightarrow\ 
& \tilde{K}^{c,b}_{b,c}(a,a) = \frac{\beta_3 \beta_{10}}{\beta_7 \beta_9},
\\
\left( e_c e_b \right) e_{m1} = \beta_{10} \, e_a e_{m1} = \beta_4 \beta_{10} \, e_{m1}\,\ 
& e_c\left( e_b e_{m1} \right) = \beta_8 \, e_c e_{m2} = - \beta_8 \beta_{12} \, e_{m1},\ \Rightarrow\ 
& \tilde{K}^{c,m}_{b,m}(a,m) = - \frac{\beta_4 \beta_{10}}{\beta_8 \beta_{12}},
\\
\left( e_c e_c \right) e_a = \beta_{11} \, e e_a = \beta_{11} \, e_a,\ 
& e_c\left( e_c e_a \right) = \beta_9 \, e_c e_b = \beta_9 \beta_{10} \, e_a,\ \Rightarrow\ 
& \tilde{K}^{c,a}_{c,a}(1,b) = \frac{\beta_{11}}{\beta_9 \beta_{10}},
\end{align}
\begin{align}
\left( e_c e_c \right) e_b = \beta_{11} \, e e_b = \beta_{11} \, e_b,\ 
& e_c\left( e_c e_b \right) = \beta_{10} \, e_c e_a = \beta_9 \beta_{10}\, e_b,\ \Rightarrow\ 
& \tilde{K}^{c,b}_{c,b}(1,a) = \frac{\beta_{11}}{\beta_9 \beta_{10}},
\\
\left( e_c e_c \right) e_c = \beta_{11} \, e e_c = \beta_{11} \, c,\ 
& e_c\left( e_c e_c \right) = \beta_{11} \, e_c e = \beta_{11} \, e_c,\ \Rightarrow\ 
& \tilde{K}^{c,c}_{c,c}(1,1)=1,
\\
\left( e_c e_c \right) e_{m1} = \beta_{11} \, e e_{m1} = \beta_{11} \, e_{m1},\ 
& e_c \left( e_c e_{m1} \right) = \beta_{12} e_c e_{m2} = - \beta_{12}^2 \, e_{m1},\ \Rightarrow\ 
& \tilde{K}^{c,m}_{c,m}(1,m) = - \frac{\beta_{11}}{\beta_{12}^2},
\end{align}
\begin{align}
\left( e_c e_{m1} \right) e_a = \beta_{12} \, e_{m2} e_a = - \beta_{12} \beta_{13} \, e_{m2},\ 
& e_c\left( e_{m1} e_a\right) = \beta_{13} \, e_c e_{m1} = \beta_{12} \beta_{13} e_{m2},\ \Rightarrow\ 
& \tilde{K}^{c,m}_{m,a}(m,m)=-1,
\\
\left( e_c e_{m1} \right) e_b = \beta_{12} e_{m2} \, e_b = \beta_{12} \beta_{14} \, e_{m1},\ 
& e_c \left( e_{m1} e_b) \right) = \beta_{14} \, e_c e_{m2} = - \beta_{12} \beta_{14} \, e_{m1},\ \Rightarrow\ 
& \tilde{K}^{c,m}_{m,b}(m,m)=-1,
\\  
\left( e_c e_{m1}\right) e_c = \beta_{12} e_{m2} e_c = - \beta_{12}\beta_{15} \, e_{m1},\ 
& e_c \left( e_{m1} e_c \right) = \beta_{15} e_c e_{m2} = - \beta_{12}\beta_{15} e_{m1},\ \Rightarrow\ 
& \tilde{K}^{c,m}_{m,c}(m,m)=1,
\end{align}
\begin{multline}
\left( e_c e_{m1} \right) e_{m1} = \beta_{12} \, e_{m2} e_{m1} = \beta_{12} \beta_{18} \, e - \beta_{12} \beta_{19} \, e_a,
\quad 
e_c\left( e_{m1} e_{m1} \right) = \beta_{16} \, e_c e_b + \beta_{17} \, e_c e_c = \beta_{11} \beta_{17} \, e + \beta_{10} \beta_{16} \, e_a,\non
\\
\Rightarrow\quad\tilde{K}^{c,1}_{m,m}(m,c) = \frac{\beta_{12} \beta_{18}}{\beta_{11}\beta_{17}},
\quad\tilde{K}^{c,a}_{m,m}(m,b) = - \frac{\beta_{12} \beta_{19}}{\beta_{10}\beta_{16}},
\end{multline}
\begin{multline} 
\left( e_c e_{m1} \right) e_{m2} =  \beta_{12} \, e_{m2} e_{m2}  = \beta_{12} \beta_{16} \, e_b - \beta_{12}\beta_{17} \, e_c,\quad
e_c \left( e_{m1} e_{m2}\right) = \beta_{18}\, e_c e + \beta_{19} \, e_c e_a = \beta_9\beta_{19} \, e_b + \beta_{18} \, e_c,\non\\
\Rightarrow\quad\tilde{K}^{c,b}_{m,m}(m,a) = \frac{\beta_{12}\beta_{16}}{\beta_9\beta_{19}},
\quad\tilde{K}^{c,c}_{m,m}(m,1) = - \frac{\beta_{12}\beta_{17}}{\beta_{18}},
\end{multline}
\begin{align}
\left( e_{m1} e_a \right) e_a = \beta_{13} \, e_{m1} e_a = \beta_{13}^2 \, e_{m1},\ 
& e_{m1}\left(e_a e_a\right) = \beta_1 \, e_{m1} e = \beta_1 \, e_{m1},\ \Rightarrow\ 
& \tilde{K}^{m,m}_{a,a}(m,1) = \frac{\beta_{13}^2}{\beta_1},
\\
\left( e_{m1} e_a\right) e_b = \beta_{13} \, e_{m1} e_b = \beta_{13} \beta_{14} \, e_{m2},\ 
& e_{m1}\left( e_a e_b \right) = \beta_2 \, e_{m1} e_c = \beta_2 \beta_{15} \, e_{m2},\ \Rightarrow\ 
& \tilde{K}^{m,m}_{a,b}(m,c) = \frac{\beta_{13}\beta_{14}}{\beta_2 \beta_{15}},
\\
\left( e_{m1} e_a \right) e_c = \beta_{13} \, e_{m1} e_c = \beta_{13} \beta_{15} \, e_{m2},\ 
& e_{m1} \left( e_a e_c \right) = \beta_3 \, e_{m1} e_b = \beta_3 \beta_{14} \, e_{m2},\ \Rightarrow\ 
& \tilde{K}^{m,m}_{a,c}(m,b) = \frac{\beta_{13}\beta_{15}}{\beta_3 \beta_{14}},
\end{align}
\begin{multline}
\left( e_{m1} e_a \right) e_{m1} = \beta_{13} \, e_{m1} e_{m1} = \beta_{13}\beta_{16} \, e_b + \beta_{13}\beta_{17} \, e_c,\quad e_{m1}\left( e_a e_{m1}\right) = \beta_4 \, e_{m1} e_{m1} = \beta_4 \beta_{16} \, e_b + \beta_4 \beta_{17} \, e_c,\non\\
\Rightarrow\quad\tilde{K}^{m,b}_{a,m}(m,m) = \frac{\beta_{13}}{\beta_4},
\quad\tilde{K}^{m,c}_{a,m}(m,m) = \frac{\beta_{13}}{\beta_4},
\end{multline}
\begin{multline}
\left( e_{m1} e_a \right) e_{m2} = \beta_{13} \, e_{m1} e_{m2} = \beta_{13} \beta_{18} \, e + \beta_{13} \beta_{19} \, e_a,
\quad e_{m1} \left( e_a e_{m2} \right) = - \beta_4 \, e_{m1} e_{m2} = - \beta_4 \beta_{18} \, e - \beta_4 \beta_{19} \, e_a,\non\\
\Rightarrow\quad\tilde{K}^{m,1}_{a,m}(m,m) = - \frac{\beta_{13}}{\beta_4},
\quad\tilde{K}^{m,a}_{a,m}(m,m) = - \frac{\beta_{13}}{\beta_4},
\end{multline}
\begin{align}  
\left( e_{m1} e_b \right) e_a = \beta_{14} e_{m2} e_a = - \beta_{13} \beta_{14} e_{m2},\ 
& e_{m1}\left( e_b e_a \right) = \beta_5 \, e_{m1} e_c = \beta_5 \beta_{15} \, e_{m2},\ \Rightarrow\ 
& \tilde{K}^{m,m}_{b,a}(m,c) = - \frac{\beta_{13}\beta_{14}}{\beta_5 \beta_{15}},
\\
\left( e_{m1} e_b \right) e_b = \beta_{14} e_{m2} e_b = \beta_{14}^2 \, e_{m1},\ 
& e_{m1} \left( e_b e_b \right) = \beta_6 \, e_{m1} e = \beta_6 \, e_{m1},\ \Rightarrow\ 
& \tilde{K}^{m,m}_{b,b}(m,1) = \frac{\beta_{14}^2}{\beta_6},
\\
\left( e_{m1} e_b \right) e_c = \beta_{14} \, e_{m2} e_c = - \beta_{14}\beta_{15} \, e_{m1},\ 
& e_{m1} \left( e_b e_c \right) = \beta_7 \, e_{m1} e_a = \beta_7 \beta_{13} \, e_{m1},\ \Rightarrow\ 
& \tilde{K}^{m,m}_{b,c}(m,a) = - \frac{\beta_{14}\beta_{15}}{\beta_7 \beta_{13}},
\end{align}
\begin{multline}
\left( e_{m1} e_b \right) e_{m1} = \beta_{14} \, e_{m2} e_{m1} = \beta_{14} \beta_{18} \, e - \beta_{14} \beta_{19} \, e_a,
\quad e_{m1} \left( e_b e_{m1} \right) = \beta_8 \, e_{m1} e_{m2} = \beta_8 \beta_{18} \,e + \beta_8 \beta_{19} \, e_a,\non
\\
\Rightarrow\quad\tilde{K}^{m,1}_{b,m}(m,m) = \frac{\beta_{14}}{\beta_8},
\quad\tilde{K}^{m,a}_{b,m}(m,m) = - \frac{\beta_{14}}{\beta_8},
\end{multline}
\begin{multline}
\left( e_{m1} e_b \right) e_{m2} = \beta_{14} \ ,e_{m2} e_{m2} = \beta_{14} \beta_{16} \, e_b - \beta_{14} \beta_{17} \, e_c,
\quad e_{m1} \left( e_b e_{m2}\right) = \beta_8 e_{m1} e{m1} = \beta_8 \beta_{16} \, e_b + \beta_8 \beta_{17} \, e_c,\non
\\
\Rightarrow\quad\tilde{K}^{m,b}_{b,m}(m,m) = \frac{\beta_{14}}{\beta_8},
\quad\tilde{K}^{m,c}_{b,m}(m,m) = - \frac{\beta_{14}}{\beta_8},
\end{multline}
\begin{align}
\left( e_{m1} e_c \right) e_a = \beta_{15} \, e_{m2} e_a = - \beta_{13}\beta_{15} \, e_{m2},\ 
& e_{m1}\left( e_c e_a \right) = \beta_9 \, e_{m1} e_b = \beta_9 \beta_{14} \, e_{m2},\ \Rightarrow\ 
& \tilde{K}^{m,m}_{c,a}(m,b) = - \frac{\beta_{13}\beta_{15}}{\beta_9 \beta_{14}},
\\
\left( e_{m1} e_c \right) e_b = \beta_{15} \, e_{m2} e_b = \beta_{14} \beta_{15} \, e_{m1},\ 
& e_{m1} \left( e_c e_b\right) = \beta_{10} \, e_{m1} e_a = \beta_{10}\beta_{13}\, e_{m1},\ \Rightarrow\ 
& \tilde{K}^{m,m}_{c,b}(m,a) = \frac{\beta_{14}\beta_{15}}{\beta_{10}\beta_{13}},
\\
\left( e_{m1} e_c \right) e_c = \beta_{15} \, e_{m2} e_c = - \beta_{15}^2 e_{m1},\ 
& e_{m1} \left( e_c e_c\right) = \beta_{11} \, e_{m1} e = \beta_{11} \, e_{m1},\ \Rightarrow\ 
& \tilde{K}^{m,m}_{c,c}(m,1) = - \frac{\beta_{15}^2}{\beta_{11}},
\end{align}
\begin{multline}
\left( e_{m1} e_c \right) e_{m1} = \beta_{15} \, e_{m2} e_{m1} = \beta_{15}\beta_{18} \, e - \beta_{15} \beta_{19} \, e_a,
\quad e_{m1}\left( e_c e_{m1}\right) = \beta_{12} \, e_{m1} e_{m2} = \beta_{12}\beta_{18} \,e + \beta_{12}\beta_{19} \, e_a,\non
\\
\Rightarrow\quad\tilde{K}^{m,1}_{c,m}(m,m) = \frac{\beta_{15}}{\beta_{12}},
\quad\tilde{K}^{m,a}_{c,m}(m,m) = - \frac{\beta_{15}}{\beta_{12}},
\end{multline}
\begin{multline}
\left( e_{m1} e_c \right) e_{m2} = \beta_{15} \, e_{m2} e_{m2} = \beta_{15} \beta_{16} \, e_b - \beta_{15} \beta_{17} \, e_c,
\quad e_{m1}\left( e_c e_{m2}\right) = - \beta_{12} \, e_{m1} e_{m1} = - \beta_{12}\beta_{16} \, e_b - \beta_{12}\beta_{17} \, e_c,\non
\\
\Rightarrow\quad\tilde{K}^{m,b}_{c,m}(m,m) = - \frac{\beta_{15}}{\beta_{12}},
\quad\tilde{K}^{m,c}_{c,m}(m,m) = \frac{\beta_{15}}{\beta_{12}},
\end{multline}
\begin{multline}
\left( e_{m1} e_{m1} \right) e_a = \beta_{16} \, e_b e_a + \beta_{17} \, e_c e_a = \beta_9 \beta_{17}\, e_b + \beta_5 \beta_{16} \, e_c,
\quad e_{m1}\left( e_{m1} e_a \right) = \beta_{13} \, e_{m1} e_{m1} = \beta_{13} \beta_{16} \, e_b + \beta_{13} \beta_{17} \, e_c,\non
\\
\Rightarrow\quad\tilde{K}^{m,b}_{m,a}(c,m) = \frac{\beta_9\beta_{17}}{\beta_{13}\beta_{16}},
\quad\tilde{K}^{m,c}_{m,a}(b,m) = \frac{\beta_5\beta_{16}}{\beta_{13}\beta_{17}},
\end{multline}
\begin{multline}
\left( e_{m1} e_{m1} \right) e_b = \beta_{16} \, e_b e_b + \beta_{17} \, e_c e_b = \beta_6 \beta_{16}\, e + \beta_{10} \beta_{17} \, e_a,
\quad e_{m1}\left( e_{m1} e_b \right) = \beta_{14} \, e_{m1} e_{m2} = \beta_{14} \beta_{18} \, e + \beta_{14} \beta_{19} \, e_a,\non
\\
\Rightarrow\quad\tilde{K}^{m,1}_{m,b}(b,m) = \frac{\beta_6\beta_{16}}{\beta_{14}\beta_{18}},
\quad\tilde{K}^{m,a}_{m,b}(c,m) = \frac{\beta_{10}\beta_{17}}{\beta_{14}\beta_{19}},
\end{multline}
\begin{multline}
\left( e_{m1} e_{m1} \right) e_c = \beta_{16} \, e_b e_c + \beta_{17} \, e_c e_c = \beta_{11} \beta_{17}\, e + \beta_7 \beta_{16} \, e_a,
\quad e_{m1}\left( e_{m1} e_c \right) = \beta_{15} \, e_{m1} e_{m2} = \beta_{15} \beta_{18} \, e + \beta_{15} \beta_{19} \, e_a,\non
\\
\Rightarrow\quad\tilde{K}^{m,1}_{m,c}(c,m) = \frac{\beta_{11}\beta_{17}}{\beta_{15}\beta_{18}},
\quad\tilde{K}^{m,a}_{m,c}(b,m) = \frac{\beta_7\beta_{16}}{\beta_{15}\beta_{19}},
\end{multline}
\begin{multline}
\left( e_{m1} e_{m2} \right) e_a = \beta_{18}\, e e_a + \beta_{19} \, e_a e_a = \beta_1 \beta_{19} \, e + \beta_{18}\, e_a,
\quad e_{m1} \left( e_{m2} e_a \right) = - \beta_{13} \, e_{m1} e_{m2} = - \beta_{13} \beta_{18} \, e - \beta_{13} \beta_{19} \, e_a,\non
\\
\Rightarrow\quad\tilde{K}^{m,1}_{m,a}(a,m) = - \frac{\beta_1 \beta_{19}}{\beta_{13}\beta_{18}},
\quad\tilde{K}^{m,a}_{m,a}(1,m) = - \frac{\beta_{18}}{\beta_{13}\beta_{19}},
\end{multline}
\begin{multline}
\left( e_{m1} e_{m2} \right) e_b = \beta_{18} \, e e_b + \beta_{19} \, e_a e_b = \beta_{18} \, e_b + \beta_2\beta_{19} \, e_c,
\quad e_{m1}\left( e_{m2} e_b \right) = \beta_{14} \, e_{m1} e_{m1} = \beta_{14} \beta_{16} \, e_b + \beta_{14} \beta_{17} \, e_c,\non
\\
\Rightarrow\quad\tilde{K}^{m,b}_{m,b}(1,m) = \frac{\beta_{18}}{\beta_{14}\beta_{16}},
\quad\tilde{K}^{m,c}_{m,b}(a,m) = \frac{\beta_2 \beta_{19}}{\beta_{14}\beta_{17}},
\end{multline}
\begin{multline}
\left( e_{m1} e_{m2} \right) e_c = \beta_{18} \, e e_c + \beta_{19} \, e_a e_c = \beta_3\beta_{19} \, e_b + \beta_{18} \, e_c,
\quad e_{m1} \left( e_{m2} e_c \right)=-\beta_{15}\, e_{m1} e_{m1} = -\beta_{15} \beta_{16} e_b - \beta_{15} \beta_{17} e_c,\non\\
\Rightarrow\quad\tilde{K}^{m,b}_{m,c}(a,m) = - \frac{\beta_3\beta_{19}}{\beta_{15}\beta_{16}},
\quad\tilde{K}^{m,c}_{m,c}(1,m) = - \frac{\beta_{18}}{\beta_{15}\beta_{17}},
\end{multline}
\begin{eqnarray}
\left( e_{m1} e_{m1} \right) e_{m1} = \beta_{16} \, e_b e_{m1} + \beta_{17} \, e_c e_{m1}
& = & 
\lp \beta_8 \beta_{16} + \beta_{12} \beta_{17}\rp e_{m2},
\\ %\quad 
e_{m1} \left( e_{m1} e_{m1} \right) = \beta_{16} \, e_{m1} e_b + \beta_{17} \, e_{m1} e_c 
& = &
\lp \beta_{14}\beta_{16} + \beta_{15} \beta_{17}\rp e_{m2},\non
\\
\left( e_{m1} e_{m1} \right) e_{m2} = \beta_{16} \, e_b e_{m2} + \beta_{17} \, e_c e_{m2}
& = & 
\lp \beta_8 \beta_{16} - \beta_{12} \beta_{17}\rp e_{m1},
\\ %\quad 
e_{m1} \left( e_{m1} e_{m2} \right) = \beta_{18} \, e_{m1} e + \beta_{19} \, e_{m1} e_a
& = & 
\lp \beta_{18} + \beta_{13} \beta_{19}\rp e_{m1},\non
\\
\left( e_{m1} e_{m2} \right) e_{m1} = \beta_{18} \, e e_{m1} + \beta_{19} \, e_a e_{m1} 
& = &
\lp \beta_{18} + \beta_4 \beta_{19}\rp e_{m1}, 
\\ %\quad 
e_{m1} \left( e_{m2} e_{m1} \right) = \beta_{18} \, e_{m1} e - \beta_{19} \, e_{m1} e_a 
& = &
\lp \beta_{18} - \beta_{13} \beta_{19}\rp e_{m1},\non
\\
\left( e_{m1} e_{m2} \right) e_{m2} = \beta_{18} \, e e_{m2} + \beta_{19} \, e_a e_{m2}
& = &
\lp \beta_{18} - \beta_4 \beta_{19}\rp e_{m2},
\\ %\quad 
e_{m1} \left( e_{m2} e_{m2} \right) = \beta_{16} \, e_{m1} e_b - \beta_{17} \, e_{m1} e_c 
& = & 
\lp \beta_{14} \beta_{16} - \beta_{15} \beta_{17}\rp e_{m2},\non
\end{eqnarray}
\begin{equation}
\lp\begin{matrix} \tilde{K}^{m,m}_{m,m}(1,1) & \tilde{K}^{m,m}_{m,m}(1,a) & \tilde{K}^{m,m}_{m,m}(1,b) & \tilde{K}^{m,m}_{m,m}(1,c) \\ \tilde{K}^{m,m}_{m,m}(a,1) & \tilde{K}^{m,m}_{m,m}(a,a) & \tilde{K}^{m,m}_{m,m}(a,b) & \tilde{K}^{m,m}_{m,m}(a,c) \\ \tilde{K}^{m,m}_{m,m}(b,1) & \tilde{K}^{m,m}_{m,m}(b,a) & \tilde{K}^{m,m}_{m,m}(b,b) & \tilde{K}^{m,m}_{m,m}(b,c) \\ \tilde{K}^{m,m}_{m,m}(c,1) & \tilde{K}^{m,m}_{m,m}(c,a) & \tilde{K}^{m,m}_{m,m}(c,b) & \tilde{K}^{m,m}_{m,m}(c,c) \end{matrix}\rp
=
\hlf\lp\begin{matrix} 
1 & -\frac{\beta_{18}}{\beta_{13}\beta_{19}} & \frac{\beta_{18}}{\beta_{14}\beta_{16}} & -\frac{\beta_{18}}{\beta_{15}\beta_{17}} \\ \frac{\beta_4\beta_{19}}{\beta_{18}} & -\frac{\beta_4}{\beta_{13}} & -\frac{\beta_4\beta_{19}}{\beta_{14}\beta_{16}} & \frac{\beta_4\beta_{19}}{\beta_{15}\beta_{17}} \\ 
\frac{\beta_8\beta_{16}}{\beta_{18}} & \frac{\beta_8\beta_{16}}{\beta_{13}\beta_{19}} & \frac{\beta_8}{\beta_{14}} & \frac{\beta_8\beta_{16}}{\beta_{15}\beta_{17}} \\
-\frac{\beta_{12}\beta_{17}}{\beta_{18}} & -\frac{\beta_{12}\beta_{17}}{\beta_{13}\beta_{19}} & \frac{\beta_{12}\beta_{17}}{\beta_{14}\beta_{16}} & \frac{\beta_{12}}{\beta_{15}}
\end{matrix}\rp.
\end{equation}

\subsubsection{Modular transformations}
\label{sect:d4:modtrans}

For $D_4$, equations \eqref{eq:modtrans:T} and \eqref{eq:modtrans:S} give us (here $\m$ represents any of the irreps)
\begin{align}
    Z_{1,m}^m(\tau+1)=\ & Z_{1,m}^m(\tau),\\
    Z_{m,1}^m(\tau+1)=\ & Z_{m,m}^1(\tau),\\
    Z_{a,a}^1(\tau+1)=\ & Z_{a,1}^a(\tau),
    \end{align}
    \begin{align}
    Z_{a,b}^c(\tau+1)=\ & \frac{\beta_2\beta_9}{\beta_3\beta_5}Z_{a,c}^b(\tau),\\
    Z_{a,c}^b(\tau+1)=\ & \frac{\beta_3\beta_5}{\beta_2\beta_9}Z_{a,b}^c(\tau),\\
    Z_{a,m}^m(\tau+1)=\ & Z_{a,m}^m(\tau),\\
    Z_{b,a}^c(\tau+1)=\ & \frac{\beta_5\beta_{10}}{\beta_2\beta_7}Z_{b,c}^a(\tau),\\
    Z_{b,b}^1(\tau+1)=\ & Z_{b,1}^b(\tau),\\
    Z_{b,c}^a(\tau+1)=\ & \frac{\beta_2\beta_7}{\beta_5\beta_{10}}Z_{b,a}^c(\tau),\\
    Z_{b,m}^m(\tau+1)=\ & Z_{b,m}^m(\tau),\\
    Z_{c,a}^b(\tau+1)=\ & \frac{\beta_7\beta_9}{\beta_3\beta_{10}}Z_{c,b}^a(\tau),\\
    Z_{c,b}^a(\tau+1)=\ & \frac{\beta_3\beta_{10}}{\beta_7\beta_9}Z_{c,a}^b(\tau),
    \end{align}
    \begin{align}
    Z_{c,c}^1(\tau+1)=\ & Z_{c,1}^c(\tau),\\
    Z_{c,m}^m(\tau+1)=\ & Z_{c,m}^m(\tau),\\
    Z_{m,a}^m(\tau+1)=\ & -\frac{\beta_{13}}{\beta_4}Z_{m,m}^a(\tau),\\
    Z_{m,b}^m(\tau+1)=\ & \frac{\beta_{14}}{\beta_8}Z_{m,m}^b(\tau),\\
    Z_{m,c}^m(\tau+1)=\ & \frac{\beta_{15}}{\beta_{12}}Z_{m,m}^c(\tau),
    \end{align}
    \begin{align}
    Z_{m,m}^1(\tau+1)=\ & \hlf\lp Z_{m,1}^m(\tau)-\frac{\beta_{18}}{\beta_{13}\beta_{19}}Z_{m,a}^m(\tau)+\frac{\beta_{18}}{\beta_{14}\beta_{16}}Z_{m,b}^m(\tau)-\frac{\beta_{18}}{\beta_{15}\beta_{17}}Z_{m,c}^m(\tau)\rp,\\
    Z_{m,m}^a(\tau+1)=\ & \hlf\lp\frac{\beta_4\beta_{19}}{\beta_{18}}Z_{m,1}^m(\tau)-\frac{\beta_4}{\beta_{13}}Z_{m,a}^m(\tau)-\frac{\beta_4\beta_{19}}{\beta_{14}\beta_{16}}Z_{m,b}^m(\tau)+\frac{\beta_4\beta_{19}}{\beta_{15}\beta_{17}}Z_{m,c}^m(\tau)\rp,\\
    Z_{m,m}^b(\tau+1)=\ & \hlf\lp\frac{\beta_8\beta_{16}}{\beta_{18}}Z_{m,1}^m(\tau)+\frac{\beta_8\beta_{16}}{\beta_{13}\beta_{19}}Z_{m,a}^m(\tau)+\frac{\beta_8}{\beta_{14}}Z_{m,b}^m(\tau)+\frac{\beta_8\beta_{16}}{\beta_{15}\beta_{17}}Z_{m,c}^m(\tau)\rp,\\
    Z_{m,m}^c(\tau+1)=\ & \hlf\lp -\frac{\beta_{12}\beta_{17}}{\beta_{18}}Z_{m,1}^m(\tau)-\frac{\beta_{12}\beta_{17}}{\beta_{13}\beta_{19}}Z_{m,a}^m(\tau)+\frac{\beta_{12}\beta_{17}}{\beta_{14}\beta_{16}}Z_{m,b}^m(\tau)+\frac{\beta_{12}}{\beta_{15}}Z_{m,c}^m(\tau)\rp,
\end{align}
\begin{align}
    Z_{1,m}^m(-1/\tau)=\ & Z_{m,1}^m(\tau),\\
    Z_{m,1}^m(-1/\tau)=\ & Z_{1,m}^m(\tau),\\
    Z_{a,a}^1(-1/\tau)=\ & Z_{a,a}^1(\tau),
    \end{align}
    \begin{align}
    Z_{a,b}^c(-1/\tau)=\ & \frac{\beta_2\beta_6\beta_9}{\beta_1\beta_5\beta_{10}}Z_{b,a}^c(\tau),\\
    Z_{a,c}^b(-1/\tau)=\ & \frac{\beta_3\beta_5\beta_{11}}{\beta_1\beta_7\beta_9}Z_{c,a}^b(\tau),\\
    Z_{a,m}^m(-1/\tau)=\ & \frac{\beta_4\beta_{18}}{\beta_1\beta_{19}}Z_{m,a}^m(\tau),\\
    Z_{b,a}^c(-1/\tau)=\ & \frac{\beta_1\beta_5\beta_{10}}{\beta_2\beta_6\beta_9}Z_{a,b}^c(\tau),\\
    Z_{b,b}^1(-1/\tau)=\ & Z_{b,b}^1(\tau),
    \end{align}
    \begin{align}
    Z_{b,c}^a(-1/\tau)=\ & \frac{\beta_2\beta_7\beta_{11}}{\beta_3\beta_6\beta_{10}}Z_{c,b}^a(\tau),\\
    Z_{b,m}^m(-1/\tau)=\ & \frac{\beta_8\beta_{18}}{\beta_6\beta_{16}}Z_{m,b}^m(\tau),\\
    Z_{c,a}^b(-1/\tau)=\ & \frac{\beta_1\beta_7\beta_9}{\beta_3\beta_5\beta_{11}}Z_{a,c}^b(\tau),\\
    Z_{c,b}^a(-1/\tau)=\ & \frac{\beta_3\beta_6\beta_{10}}{\beta_2\beta_7\beta_{11}}Z_{b,c}^a(\tau),
    \end{align}
    \begin{align}
    Z_{c,c}^1(-1/\tau)=\ & Z_{c,c}^1(\tau),\\
    Z_{c,m}^m(-1/\tau)=\ & \frac{\beta_{12}\beta_{18}}{\beta_{11}\beta_{17}}Z_{m,c}^m(\tau),\\
    Z_{m,a}^m(-1/\tau)=\ & \frac{\beta_1\beta_{19}}{\beta_4\beta_{18}}Z_{a,m}^m(\tau),\\
    Z_{m,b}^m(-1/\tau)=\ & \frac{\beta_6\beta_{16}}{\beta_8\beta_{18}}Z_{b,m}^m(\tau),\\
    Z_{m,c}^m(-1/\tau)=\ & \frac{\beta_{11}\beta_{17}}{\beta_{12}\beta_{18}}Z_{c,m}^m(\tau),
    \end{align}
    \begin{align}
    Z_{m,m}^1(-1/\tau)=\ & \hlf\lp Z_{m,m}^1(\tau)+\frac{\beta_{18}}{\beta_4\beta_{19}}Z_{m,m}^a(\tau)+\frac{\beta_{18}}{\beta_8\beta_{16}}Z_{m,m}^b(\tau)-\frac{\beta_{18}}{\beta_{12}\beta_{17}}Z_{m,m}^c(\tau)\rp, \\
    Z_{m,m}^a(-1/\tau)=\ & \hlf\lp\frac{\beta_4\beta_{19}}{\beta_{18}}Z_{m,m}^1(\tau)+Z_{m,m}^a(\tau)-\frac{\beta_4\beta_{19}}{\beta_8\beta_{16}}Z_{m,m}^b(\tau)+\frac{\beta_4\beta_{19}}{\beta_{12}\beta_{17}}Z_{m,m}^c(\tau)\rp,\\
    Z_{m,m}^b(-1/\tau)=\ & \hlf\lp\frac{\beta_8\beta_{16}}{\beta_{18}}Z_{m,m}^1(\tau)-\frac{\beta_8\beta_{16}}{\beta_4\beta_{19}}Z_{m,m}^a(\tau)+Z_{m,m}^b(\tau)+\frac{\beta_8\beta_{16}}{\beta_{12}\beta_{17}}Z_{m,m}^c(\tau)\rp,\\
    Z_{m,m}^c(-1/\tau)=\ & \hlf\lp -\frac{\beta_{12}\beta_{17}}{\beta_{18}}Z_{m,m}^1(\tau)+\frac{\beta_{12}\beta_{17}}{\beta_4\beta_{19}}Z_{m,m}^a(\tau)+\frac{\beta_{12}\beta_{17}}{\beta_8\beta_{16}}Z_{m,m}^b(\tau)+Z_{m,m}^c(\tau)\rp,
\end{align}

We find that the following combinations of partial traces are modular invariant:
\begin{align}
    & Z_{1,1}^1,\\
    & Z_{1,a}^a+Z_{a,1}^a+Z_{a,a}^1,\\
    & Z_{1,b}^b+Z_{b,1}^b+Z_{b,b}^1,\\
    & Z_{1,c}^c+Z_{c,1}^c+Z_{c,c}^1,\\
    & Z_{a,b}^c+\frac{\beta_2\beta_9}{\beta_3\beta_5}Z_{a,c}^b+\frac{\beta_2\beta_6\beta_9}{\beta_1\beta_5\beta_{10}}Z_{b,a}^c+\frac{\beta_6\beta_9}{\beta_1\beta_7}Z_{b,c}^a+\frac{\beta_2\beta_{11}}{\beta_1\beta_7}Z_{c,a}^b+\frac{\beta_2\beta_9\beta_{11}}{\beta_1\beta_3\beta_{10}}Z_{c,b}^a,\\
    & Z_{1,m}^m-\frac{\beta_1}{\beta_4\beta_{13}}Z_{a,m}^m+Z_{m,1}^m-\frac{\beta_{18}}{\beta_{13}\beta_{19}}Z_{m,a}^m+Z_{m,m}^1+\frac{\beta_{18}}{\beta_4\beta_{19}}Z_{m,m}^a,\\
    & Z_{1,m}^m+\frac{\beta_6}{\beta_8\beta_{14}}Z_{b,m}^m+Z_{m,1}^m+\frac{\beta_{18}}{\beta_{14}\beta_{16}}Z_{m,b}^m+Z_{m,m}^1+\frac{\beta_{18}}{\beta_8\beta_{16}}Z_{m,m}^b,\\
    & Z_{1,m}^m-\frac{\beta_{11}}{\beta_{12}\beta_{15}}Z_{c,m}^m+Z_{m,1}^m-\frac{\beta_{18}}{\beta_{15}\beta_{17}}Z_{m,c}^m+Z_{m,m}^1-\frac{\beta_{18}}{\beta_{12}\beta_{17}}Z_{m,m}^c.
\end{align}

In the next several subsections, we will compute genus-one partition functions from our general
formula~(\ref{eq:Z:genl1}), corresponding to each of the
physically-different cosets, and we will see explicitly that the resulting partition functions 
are linear combinations of the modular-invariant combinations above, and hence are modular-invariant.

\subsubsection{$H=\mathbb{Z}_4$: $1+a$ orbifold}

We pick the Frobenius subalgebra $1+a$, corresponding to $H = {\mathbb Z}_4$. 
From the coset $D_4/H=\{ H, yH \}$, following the general analysis in section~\ref{sect:genl-algebra},
we define
\begin{eqnarray}
    e_1&=& v_H+v_{yH},
    \\
    e_a&=& v_H-v_{yH},
\end{eqnarray}
for $e_1$ the basis element of $1$ and $e_a$ the basis element of $a$. Note that $e_1$ and $e_a$ are both invariant under the group action of $x$. Moreover, only $e_1$ is invariant under the group action of $y$, while $e_a$ is an eigenvector with eigenvalue $-1$.

As in the $\Rep(S_3)$ case, we compute the action of the product and co-product on these basis vectors.  As usual, the multiplications involving $e_1$ are trivial, so we have
\begin{equation}
    \m_{1,1}^1=\m_{1,a}^a=\m_{a,1}^a=1,
\end{equation}
while
\begin{equation}
    \m_*(e_a\otimes e_a)=v_H+v_{yH}=e_1=\beta_1^{-1}\lambda_{a,a}^1(e_ae_a)\quad\Rightarrow\quad\m_{a,a}^1=\beta_1^{-1},
\end{equation}
where the coefficient is obtained by comparing the multiplication above with our basis fusion homomorphism which sends $e_a\otimes e_a\mapsto\beta_1e$.

Similarly we have co-products
\begin{align}
    \Delta_F(e_1)=\ & v_Hv_H+v_{yH}v_{yH}=\hlf\lp e_1e_1+e_ae_a\rp,\\
    \Delta_F(e_a)=\ & v_Hv_H-v_{yH}v_{yH}=\hlf\lp e_1e_a+e_ae_1\rp,
\end{align}
so
\begin{equation}
    \Delta_1^{1,1}=\Delta_a^{1,a}=\Delta_a^{a,1}=\hlf,\qquad\Delta_1^{a,a}=\hlf\beta_1.
\end{equation}
Then including each partial trace $Z_{R_1,R_2}^{R_3}$ with a coefficient $\m_{R_1,R_2}^{R_3}\Delta_{R_3}^{R_2,R_1}$, we find the partition function
\begin{eqnarray}
    Z_{1+a}&=& \mu_{1,1}^1 \Delta_1^{1,1} Z_{1,1}^1 + \mu_{1,a}^a\Delta_a^{a,1} Z_{1,a}^a+\mu_{a,1}^a\Delta_a^{1,a} Z_{a,1}^a+\mu_{a,a}^1\Delta_1^{a,a} Z_{a,a}^1 ,
    \\
    &=& \frac{1}{2}\left(Z_{1,1}^1 + Z_{1,a}^a + Z_{a,1}^a + Z^1_{a,a}\right),
    \label{d4_orb_1a}
\end{eqnarray}
which is a linear combination of two of the modular-invariant combinations given in
section~\ref{sect:d4:modtrans}, demonstrating that the expression above is modular-invariant.

Note that $a\otimes a=1$, i.e.~$a$ corresponds to a $\mathbb{Z}_2$ line defect. The above partition function is thus for a $\mathbb{Z}_2$ orbifold from gauging the $\mathbb{Z}_2$ subalgebra generated by $\{1, a\}\subset \text{Rep}(D_4)$. The resulting $\mathbb{Z}_2$ quantum symmetry is given by the quotient group $D_4/\mathbb{Z}_4\cong \mathbb{Z}_2$.

\subsubsection{$H=\mathbb{Z}_2\times \mathbb{Z}_2$: $1+b$ orbifold}

Consider the Frobenius subalgebra $1+b$. From the coset $D_4/H=\{ H, xH \}$, we define basis vectors as:
\begin{eqnarray}
    e_1 &=& v_H+v_{xH},\\
    e_b &=& v_H-v_{xH}.
\end{eqnarray}
It is straightforward to check that the (co)product computation is exactly the as that of the $1+a$ orbifold. 
For completeness, we sketch the details here.
First, using
\begin{equation}
    \m(e_1\otimes v)=\m(v\otimes e_1)=v,\qquad\m(e_b\otimes e_b)=e_1,
\end{equation}
and
\begin{align}
    \Delta(e_1)=\ & \hlf\lp e_1e_1+e_be_b\rp,\\
    \Delta(e_b)=\ & \hlf\lp e_1e_b+e_be_1\rp,
\end{align}
and comparing with our established basis, we obtain
\begin{equation}
    \m_{1,1}^1=\m_{1,b}^b=\m_{b,1}^b=1,\qquad\m_{b,b}^1=\beta_6^{-1},
\end{equation}
\begin{equation}
    \Delta_1^{1,1}=\Delta_b^{1,b}=\Delta_b^{b,1}=\hlf,\qquad\Delta_1^{b,b}=\hlf\beta_6.
\end{equation}

The partition function is then
\begin{equation}
    Z_{1+b} = \sum_{i,j,k} \mu_{i,j}^k \Delta_k^{j,i} Z_{i,j}^k = \frac{1}{2}\left( Z_{1,1}^1+Z_{1,b}^b+Z_{b,1}^b+Z_{b,b}^1 \right),
    \label{d4_orb_1b}
\end{equation}
which is a linear combination of modular-invariant expressions,
hence itself modular invariant.
It is again a $\mathbb{Z}_2$ orbifold theory as a result of gauging the $\mathbb{Z}_2$ subalgebra generated by $\{ 1,b \}\subset \text{Rep}(D_4)$. The resulting quantum symmetry is again $\mathbb{Z}_2$, but this time given by a different quotient $D_4/(\mathbb{Z}_2\times \mathbb{Z}_2)\cong \mathbb{Z}_2$.

\subsubsection{$H=\mathbb{Z}_2$ normal: $1+a+b+c$ orbifold}

Consider the Frobenius subalgebra $\mathcal{A}=1+a+b+c$. From the coset $D_4/H=\{ H, xH, yH, xyH \}$, we define basis vectors as 
\begin{equation}
\begin{split}
     e_1=v_H+v_{xH}+v_{yH}+v_{xyH},\\
     e_a=v_H+v_{xH}-v_{yH}-v_{xyH},\\
     e_b=v_H-v_{xH}+v_{yH}-v_{xyH},\\
     e_c=v_H-v_{xH}-v_{yH}+v_{xyH}.
\end{split}
\end{equation}

The multiplication is 
\begin{equation}
    \begin{split}
        &\mu_*\left(e_1 \otimes e_1\right) = \mu_*\left(e_a \otimes e_a\right) = \mu_*\left(e_b \otimes e_b\right) = \mu_*\left(e_c \otimes e_c\right) = e_1,
        \\
        &\mu_*\left(e_1 \otimes e_a\right) = \mu_*\left(e_a \otimes e_1\right) = e_a, \: \: \:
        \mu_*\left( e_1 e_b \right) = \mu_*\left(e_b \otimes e_1\right) = e_b,
        \\
        &\mu_*\left(e_1 \otimes e_c\right) = \mu_*\left(e_c \otimes e_1\right) = e_c, \: \: \:
        \mu_*\left(e_a \otimes e_b\right) = \mu_*\left(e_b \otimes e_a\right) = e_c,
        \\
        & \mu_*\left(e_a \otimes e_c\right) = \mu_*\left(e_c \otimes e_a\right) = e_b, \: \: \:
        \mu_*\left(e_b \otimes e_c\right) = \mu_*\left(e_c \otimes e_b\right) = e_a,
    \end{split}
\end{equation}
while the comultiplication reads
\begin{equation}
    \begin{split}
        \Delta_F\left(e_1\right) = \frac{1}{4}\left( e_1 \otimes e_1 + e_a \otimes e_a + e_b \otimes e_b + e_c \otimes e_c \right),\\
        \Delta_F\left(e_a\right) = \frac{1}{4}\left( e_1 \otimes e_a + e_a \otimes e_1 + e_b \otimes e_c + e_c \otimes e_b \right),\\
        \Delta_F\left(e_b\right) = \frac{1}{4}\left( e_1 \otimes e_b + e_b \otimes e_1 + e_a \otimes e_c + e_c \otimes e_a \right),\\
        \Delta_F\left(e_c\right) = \frac{1}{4}\left( e_1 \otimes e_c + e_c \otimes e_1 + e_a \otimes e_b + e_b \otimes e_a \right),
    \end{split}
\end{equation}
which was derived using the fact that $\Delta_F$ is diagonal on the $v$ basis.

Combining coefficients in the (co)multiplication and the those in the intertwiner yields the 
components
\be
\mu_{1,1}^1 = 1, 
\quad\m_{a,a}^1 = \beta_1^{-1},
\quad\m_{a,b}^c = \beta_2^{-1},
\quad\m_{a,c}^b = \beta_3^{-1},
\ee
\be
\m_{b,a}^c = \beta_5^{-1},
\quad\m_{b,b}^1 = \beta_6^{-1},
\quad\m_{b,c}^a = \beta_7^{-1},
\ee
\be
\m_{c,a}^b = \beta_9^{-1},
\quad\m_{c,b}^a = \beta_{10}^{-1},
\quad\m_{c,c}^1 = \beta_{11}^{-1},
\ee
\begin{equation}
    \Delta_1^{1,1} = \frac{1}{4},
    \quad\Delta_1^{a,a} = \frac{\beta_1}{4},
    \quad\Delta_c^{a,b} = \frac{\beta_6}{4\beta_{10}},
    \quad\Delta_b^{a,c} = \frac{\beta_{11}}{4\beta_7},
\end{equation}
\begin{equation}
    \Delta_c^{b,a} = \frac{\beta_1}{4\beta_9},
    \quad\Delta_1^{b,b} = \frac{\beta_6}{4},
    \quad\Delta_a^{b,c} = \frac{\beta_{11}}{4\beta_3},
\end{equation}
\begin{equation}
    \Delta_b^{c,a} = \frac{\beta_1}{4\beta_5},
    \quad\Delta_a^{c,b} = \frac{\beta_6}{4\beta_2},
    \quad\Delta_1^{c,c} = \frac{\beta_{11}}{4}.
\end{equation}

Plugging into the general formula~(\ref{eq:Z:genl1}) for one-loop partition functions, we find
\begin{align}
\label{d4_orb_1abc}
Z_{1+a+b+c} = \ & \frac{1}{4}\ls Z_{1,1}^1 + Z_{1,a}^a + Z_{1,b}^b + Z_{1,c}^c + Z_{a,1}^a + Z_{a,a}^1 + \frac{\beta_1}{\beta_2\beta_9} Z_{a,b}^c + \frac{\beta_1}{\beta_3\beta_5}Z_{a,c}^b \right.\non
\\
& \qquad\left. +Z_{b,1}^b + \frac{\beta_6}{\beta_5\beta_{10}}Z_{b,a}^c + Z_{b,b}^1 +  \frac{\beta_6}{\beta_2\beta_7} Z_{b,c}^a+Z_{c,1}^c + \frac{\beta_{11}}{\beta_7\beta_9} Z_{c,a}^b + \frac{\beta_{11}}{\beta_3\beta_{10}} Z_{c,b}^a + Z_{c,c}^1\rs.
\end{align}

This is a $\mathbb{Z}_2\times \mathbb{Z}_2$ orbifold theory, the result of gauging the $\mathbb{Z}_2\times \mathbb{Z}_2$ subalgebra generated by $\{ 1, a, b, c \}\subset \text{Rep}(D_4)$. The resulting quantum symmetry corresponds to the coset  $D_4/\mathbb{Z}_2 \cong \mathbb{Z}_2\times \mathbb{Z}_2$.

In passing, the reader might ask, since this is essentially the same as a
${\mathbb Z}_2 \times {\mathbb Z}_2$ orbifold, why factors of $\beta$ appear here.  Technically, this is because the associator in ${\mathbb Z}_2 \times {\mathbb Z}_2$ is a pullback from the associator of Rep($D_4$), and although this will be anomaly-free, in the language
of group cocycles, this formally corresponds to a trivial element of $H^3(
{\mathbb Z}_2 \times {\mathbb Z}_2,U(1))$ that is represented by a cocycle that is cohomologically trivial but not identically 1.
This will also arise in other examples.

It will be useful to write the partition function~(\ref{d4_orb_1abc}) in the form 
\begin{eqnarray}
    Z_{1+a+b+c} & = & \frac{1}{4} \biggl[ 
     Z_{1,1}^1 + Z_{1,a}^a + Z_{1,b}^b + Z_{1,c}^c + Z_{a,1}^a + Z_{a,a}^1 + Z_{b,1}^b + Z_{b,b}^1 + Z_{c,1}^c + Z_{c,c}^1
      \\
     & & \qquad 
     + \frac{\beta_1}{\beta_2\beta_9} \left( Z_{a,b}^c+\frac{\beta_2\beta_9}{\beta_3\beta_5}Z_{a,c}^b+\frac{\beta_2\beta_6\beta_9}{\beta_1\beta_5\beta_{10}}Z_{b,a}^c+\frac{\beta_6\beta_9}{\beta_1\beta_7}Z_{b,c}^a+\frac{\beta_2\beta_{11}}{\beta_1\beta_7}Z_{c,a}^b+\frac{\beta_2\beta_9\beta_{11}}{\beta_1\beta_3\beta_{10}}Z_{c,b}^a
     \right) \biggr].
     \nonumber
\end{eqnarray}
In this expression, the first and second lines are separately modular-invariant, as can be seen from
our results earlier in section~\ref{sect:d4:modtrans}.  

One implication is that the partition function~(\ref{d4_orb_1abc}) is modular-invariant.
A second implication is that, at least naively, we can construct a second modular-invariant partition function, through an analogue of discrete torsion.
Let $\omega(g,h)$ be cocycles representing the nontrivial element of 
$H^2({\mathbb Z}_2 \times {\mathbb Z}_2, U(1)) = 
{\mathbb Z}_2$, and define
\begin{equation}
    \epsilon(g,h) \: = \: \frac{ \omega(g,h) }{ \omega(h,g) }.
\end{equation}
In an ordinary ${\mathbb Z}_2 \times {\mathbb Z}_2$ orbifold, the phases $\epsilon(g,h)$ are the discrete torsion
phase factors multiplying $Z_{g,h}$.  Now, it can be shown that 
\begin{equation}
    \epsilon(g,h) \: = \: \left\{ \begin{array}{cl}
    -1 & g \neq h, g \neq 1, \mbox{ and } h \neq 1,
    \\
    +1 & g = h \mbox{ or } g = 1 \mbox{ or } h = 1.
    \end{array} \right.
\end{equation}
With this in mind, we can define a new, explicitly modular-invariant partition function,
given by
\begin{eqnarray}
    Z_{1+a+b+c} & = & \frac{1}{4} \biggl[ 
     Z_{1,1}^1 + Z_{1,a}^a + Z_{1,b}^b + Z_{1,c}^c + Z_{a,1}^a + Z_{a,a}^1 + Z_{b,1}^b + Z_{b,b}^1 + Z_{c,1}^c + Z_{c,c}^1
      \\
     & & \qquad 
     - \frac{\beta_1}{\beta_2\beta_9} \left( Z_{a,b}^c+\frac{\beta_2\beta_9}{\beta_3\beta_5}Z_{a,c}^b+\frac{\beta_2\beta_6\beta_9}{\beta_1\beta_5\beta_{10}}Z_{b,a}^c+\frac{\beta_6\beta_9}{\beta_1\beta_7}Z_{b,c}^a+\frac{\beta_2\beta_{11}}{\beta_1\beta_7}Z_{c,a}^b+\frac{\beta_2\beta_9\beta_{11}}{\beta_1\beta_3\beta_{10}}Z_{c,b}^a
     \right) \biggr].
     \nonumber
\end{eqnarray}
This expression is nearly the same as the previous partition function, except that the second line of terms are subtracted, rather than added.
Because each line is separately modular-invariant, this expression is manifestly modular-invariant.
For suitable intertwiners, this reduces precisely to the partition function of a ${\mathbb Z}_2 \times {\mathbb Z}_2$ orbifold with discrete torsion.

Naively, one would expect that the minus sign on the second line (arising from discrete torsion) could be absorbed into the intertwiners.  However, physics is a bit more subtle:  the definition of the partial traces $Z_{i,j}^k$ depends upon the intertwiners, so changing the intertwiners also changes the partial traces.  Later in section~\ref{sect:gauging-reph8-ising} we will discuss a close analogue of this choice in a Rep$({\cal H}_8)$ gauging.  In that section, we will give explicit expressions for the partial traces, and we will see explicitly that the two choices of discrete torsion, the two choices of sign, are physically distinct.
We will see analogous examples in other sections.

This example gives an initial demonstration that there exists some analogue of discrete torsion when gauging noninvertible symmetries.
We leave a first-principles understanding for future work.
Furthermore, as previously discussed, we have not attempted to check e.g.~multiloop factorization in such noninvertible analogues of discrete torsion, hence it is possible that some choices may not be physically sensible.  We leave this also for future work.

\subsubsection{$H=\Z_2$ non-normal: $1+b+m$ orbifold}

Now, we consider the Frobenius subalgebra ${\cal A}=1+b+m$. From the coset $D_4/H=\{ H, xH, x^2H, x^3H \}$, we define the basis vectors as
\begin{eqnarray}
    e_1 &=& v_H+v_{xH}+v_{x^2H}+v_{x^3H},\\
    e_b &=& v_H-v_{xH}+ v_{x^2H}- v_{x^3H}.\\
    e_{m1} &=& v_H-iv_{xH} -v_{x^2H} +iv_{x^3H},\\
    e_{m2} &=& v_H+iv_{xH}-v_{x^2H}-iv_{x^3H}.
\end{eqnarray}

The non-trivial multiplications are
\begin{equation}
    \m(e_be_b)=e_1,\quad\m(e_be_{m1})=e_{m2},\quad\m(e_be_{m2})=e_{m1},\quad\m(e_{m1}e_{m1})=e_b,\quad\m(e_{m1}e_{m2})=e_1,\quad\m(e_{m2}e_{m2})=e_b,
\end{equation}
while the comultiplication reads
\begin{eqnarray}
    \Delta_F\left(e_1 \right) &=& \frac{1}{4}\left(e_1 \otimes e_1 + e_b \otimes e_b + e_{m1} \otimes e_{m2} + e_{m2} \otimes e_{m1}\right),\\
    \Delta_F\left(e_b \right) &=& \frac{1}{4}\left(e_1 \otimes e_b + e_b \otimes e_1 + e_{m1} \otimes e_{m1} + e_{m2} \otimes e_{m2} \right),\\
    \Delta_F\left(e_{m1}\right) &=& \frac{1}{4}\left(e_1 \otimes e_{m1} + e_{m1} \otimes e_1 + e_b \otimes e_{m2} + e_{m2} \otimes e_b\right),\\
    \Delta_F\left(e_{m2}\right) &=& \frac{1}{4}\left(e_1 \otimes e_{m2} + e_{m2} \otimes e_1 + e_b \otimes e_{m1} + e_{m1} \otimes e_b\right),
\end{eqnarray}
which was derived using the fact that $\Delta_F$ is diagonal on the $v$ basis.
Combining coefficients in the (co)multiplication and the those in the intertwiners, we derive the components below:
\be
\m_{b,b}^1 = \beta_6^{-1}, \quad\m_{b,m}^m = \beta_8^{-1}, \quad\m_{m,b}^m = \beta_{14}^{-1}, \quad\m_{m,m}^1 = \beta_{18}^{-1},\quad\m_{m,m}^b = \beta_{16}^{-1},
\ee
\be
\Delta_1^{b,b}=\frac{\beta_6}{4},\quad\Delta_1^{m,m}=\frac{\beta_{18}}{4},\quad\Delta_b^{m,m}=\frac{\beta_{18}}{4\beta_8},\quad\Delta_m^{b,m}=\frac{\beta_{18}}{4\beta_{16}},\quad\Delta_m^{m,b}=\frac{\beta_6}{4\beta_{14}},
\ee
thus
\begin{align}
\label{d4_orb_1bm}
Z_{1+b+m} = \ & \frac{1}{4}\ls Z_{1,1}^1+Z_{1,b}^b+Z_{1,m}^m+Z_{b,1}^b+Z_{b,b}^1+\frac{\beta_6}{\beta_8\beta_{14}}Z_{b,m}^m\right.\\
& \qquad\left. +Z_{m,1}^m+\frac{\beta_{18}}{\beta_{14}\beta_{16}}Z_{m,b}^m+Z_{m,m}^1+\frac{\beta_{18}}{\beta_8\beta_{16}}Z_{m,m}^b\rs
\nonumber 
\end{align}

We can write this as
\begin{eqnarray}
    Z_{1+b+m} & = & \frac{1}{4} \biggl[
    Z_{1,1}^1 + Z_{1,b}^b + Z_{b,1}^b + Z_{b,b}^1
    \\
    & & \qquad 
    + Z_{1,m}^m + \frac{\beta_6}{\beta_8\beta_{14}}Z_{b,m}^m +Z_{m,1}^m+\frac{\beta_{18}}{\beta_{14}\beta_{16}}Z_{m,b}^m+Z_{m,m}^1+\frac{\beta_{18}}{\beta_8\beta_{16}}Z_{m,m}^b
    \biggr].
    \nonumber 
\end{eqnarray}
In this expression, each line is separately modular-invariant, using the results
of section~\ref{sect:d4:modtrans}, hence we see that $Z_{1+b+m}$ is modular-invariant.

\subsubsection{$H=1$: $1+a+b+c+2m$ orbifold}

We now consider gauging the regular representation, i.e. $\mathcal{A}=1+a+b+c+2m=1+a+b+c+m_1+m_2$ (where $m_1\cong m_2\cong m$). The basis vectors for all objects can be defined as
\be
e=v_1+v_x+v_{x^2}+v_{x^3}+v_y+v_{xy}+v_{x^2y}+v_{x^3y},\quad e_a=v_1+v_x+v_{x^2}+v_{x^3}-v_y-v_{xy}-v_{x^2y}-v_{x^3y},
\ee
\be
e_b=v_1-v_x+v_{x^2}-v_{x^3}+v_y-v_{xy}+v_{x^2y}-v_{x^3y},\quad e_c=v_1-v_x+v_{x^2}-v_{x^3}-v_y+v_{xy}-v_{x^2y}+v_{x^3y},
\ee
\be
e_{m_11}=v_1+iv_x-v_{x^2}-iv_{x^3}+v_y+iv_{xy}-v_{x^2y}-iv_{x^3y},\quad e_{m_12}=v_1-iv_x-v_{x^2}+iv_{x^3}+v_y-iv_{xy}-v_{x^2y}+iv_{x^3y},
\ee
\be
e_{m_21}=v_1+iv_x-v_{x^2}-iv_{x^3}-v_y-iv_{xy}+v_{x^2y}+iv_{x^3y},\quad e_{m_22}=-v_1+iv_x+v_{x^2}-iv_{x^3}+v_y-iv_{xy}-v_{x^2y}+iv_{x^3y}.
\ee
The multiplication coefficients are $\m_{1,R}^R=\m_{R,1}^R=1$ along with
\be
\m_{a,a}^1=\beta_1^{-1},\quad\m_{a,b}^c=\beta_2^{-1},\quad\m_{a,c}^b=\beta_3^{-1},\quad\m_{a,m_1}^{m_2}=\beta_4^{-1},\quad\m_{a,m_2}^{m_1}=\beta_4^{-1},\quad\m_{b,a}^c=\beta_5^{-1},\quad\m_{b,b}^1=\beta_6^{-1},
\ee
\be
\m_{b,c}^a=\beta_7^{-1},\quad\m_{b,m_1}^{m_1}=\beta_8^{-1},\quad\m_{b,m_2}^{m_2}=-\beta_8^{-1},\quad\m_{c,a}^b=\beta_9^{-1},\quad\m_{c,b}^a=\beta_{10}^{-1},\quad\m_{c,c}^1=\beta_{11}^{-1},
\ee
\be
\m_{c,m_1}^{m_2}=-\beta_{12}^{-1},\quad\m_{c,m_2}^{m_1}=\beta_{12}^{-1},\quad\m_{m_1,a}^{m_2}=\beta_{13}^{-1},\quad\m_{m_1,b}^{m_1}=\beta_{14}^{-1},\quad\m_{m_1,c}^{m_2}=-\beta_{15}^{-1},\quad\m_{m_1,m_1}^1=\beta_{18}^{-1},
\ee
\be
\m_{m_1,m_1}^b=\beta_{16}^{-1},\quad\m_{m_1,m_2}^a=-\beta_{19}^{-1},\quad\m_{m_1,m_2}^c=\beta_{17}^{-1},\quad\m_{m_2,a}^{m_1}=\beta_{13}^{-1},\quad\m_{m_2,b}^{m_2}=-\beta_{14}^{-1},\quad\m_{m_2,c}^{m_1}=\beta_{15}^{-1},
\ee
\be
\m_{m_2,m_1}^a=\beta_{19}^{-1},\quad\m_{m_2,m_1}^c=\beta_{17}^{-1},\quad\m_{m_2,m_2}^1=-\beta_{18}^{-1},\quad\m_{m_2,m_2}^b=\beta_{16}^{-1},
\ee
Some multiplication coefficients are zero (e.g.~$\m_{a,m_1}^{m_1}$) and we have omitted these.

The co-multiplication coefficients are $\Delta_R^{1,R}=\Delta_R^{R,1}=\frac{1}{8}$, and
\be
\Delta_1^{a,a}=\frac{\beta_1}{8},\quad\Delta_1^{b,b}=\frac{\beta_6}{8},\quad\Delta_1^{c,c}=\frac{\beta_{11}}{8},\quad\Delta_1^{m_1,m_1}=\frac{\beta_{18}}{8},\quad\Delta_1^{m_2,m_2}=-\frac{\beta_{18}}{8},\quad\Delta_a^{b,c}=\frac{\beta_{11}}{8\beta_3},\quad\Delta_a^{c,b}=\frac{\beta_6}{8\beta_2},
\ee
\be
\Delta_a^{m_1,m_2}=-\frac{\beta_{18}}{8\beta_4},\quad\Delta_a^{m_2,m_1}=\frac{\beta_{18}}{8\beta_4},\quad\Delta_b^{a,c}=\frac{\beta_{11}}{8\beta_7},\quad\Delta_b^{c,a}=\frac{\beta_1}{8\beta_5},\quad\Delta_b^{m_1,m_1}=\frac{\beta_{18}}{8\beta_8},\quad\Delta_b^{m_2,m_2}=\frac{\beta_{18}}{8\beta_8},
\ee
\be
\Delta_c^{a,b}=\frac{\beta_6}{8\beta_{10}},\quad\Delta_c^{b,a}=\frac{\beta_1}{8\beta_9},\quad\Delta_c^{m_1,m_2}=-\frac{\beta_{18}}{8\beta_{12}},\quad\Delta_c^{m_2,m_1}=-\frac{\beta_{18}}{8\beta_{12}},\quad\Delta_{m_1}^{a,m_2}=\frac{\beta_{18}}{8\beta_{19}},\quad\Delta_{m_1}^{b,m_1}=\frac{\beta_{18}}{8\beta_{16}},
\ee
\be
\Delta_{m_1}^{c,m_2}=-\frac{\beta_{18}}{8\beta_{17}},\quad\Delta_{m_1}^{m_1,b}=\frac{\beta_6}{8\beta_{14}},\quad\Delta_{m_1}^{m_2,a}=\frac{\beta_1}{8\beta_{13}},\quad\Delta_{m_1}^{m_2,c}=-\frac{\beta_{11}}{8\beta_{15}},\quad\Delta_{m_2}^{a,m_1}=\frac{\beta_{18}}{8\beta_{19}},\quad\Delta_{m_2}^{b,m_2}=-\frac{\beta_{18}}{8\beta_{16}},
\ee
\be
\Delta_{m_2}^{c,m_1}=\frac{\beta_{18}}{8\beta_{17}},\quad\Delta_{m_2}^{m_1,a}=\frac{\beta_1}{8\beta_{13}},\quad\Delta_{m_2}^{m_1,c}=\frac{\beta_{11}}{8\beta_{15}},\quad\Delta_{m_2}^{m_2,b}=-\frac{\beta_6}{8\beta_{14}}.
\ee

The partition function is then
\begin{align}
\label{d4_orb_1abc2m}
    Z_{1+a+b+c+2m}=\ & \frac{1}{8}\ls Z_{1,1}^1+Z_{1,a}^a+Z_{1,b}^b+Z_{1,c}^c+2Z_{1,m}^m+Z_{a,1}^a+Z_{a,a}^1+\frac{\beta_1}{\beta_2\beta_9}Z_{a,b}^c+\frac{\beta_1}{\beta_3\beta_5}Z_{a,c}^b\right.\non\\
    & \qquad\left. +\frac{2\beta_1}{\beta_4\beta_{13}}Z_{a,m}^m+Z_{b,1}^b+\frac{\beta_6}{\beta_5\beta_{10}}Z_{b,a}^c+Z_{b,b}^1+\frac{\beta_6}{\beta_2\beta_7}Z_{b,c}^a+\frac{2\beta_6}{\beta_8\beta_{14}}Z_{b,m}^m\right.\non\\
    & \qquad\left. +Z_{c,1}^c+\frac{\beta_{11}}{\beta_7\beta_9}Z_{c,a}^b + \frac{\beta_{11}}{\beta_3\beta_{10}}Z_{c,b}^a + Z_{c,c}^1-\frac{2\beta_{11}}{\beta_{12}\beta_{15}}Z_{c,m}^m\right.\non\\
    & \qquad\left. +2Z_{m,1}^m+\frac{2\beta_{18}}{\beta_{13}\beta_{19}}Z_{m,a}^m+\frac{2\beta_{18}}{\beta_{14}\beta_{16}}Z_{m,b}^m-\frac{2\beta_{18}}{\beta_{15}\beta_{17}}Z_{m,c}^m\right.\non\\
    & \qquad\left. +2Z_{m,m}^1-\frac{2\beta_{18}}{\beta_4\beta_{19}}Z_{m,m}^a +\frac{2\beta_{18}}{\beta_8\beta_{16}}Z_{m,m}^b-\frac{2\beta_{18}}{\beta_{12}\beta_{17}}Z_{m,m}^c\rs.
\end{align}

It will be useful to rewrite this expression as follows:
\begin{eqnarray}
    Z_{1+a+b+c+2m} & = & \frac{1}{8} \biggl[
    Z_{1,1}^1 + Z_{1,a}^a + + Z_{a,1}^a + Z_{a,a}^1 + Z_{1,b}^b + Z_{b,1}^b +Z_{b,b}^1 + Z_{1,c}^c +Z_{c,1}^c + Z_{c,c}^1 
    \\
    & & \qquad
     + \frac{\beta_1}{\beta_2\beta_9} \left(
     Z_{a,b}^c + \frac{\beta_2\beta_9}{\beta_3\beta_5} Z_{a,c}^b + \frac{\beta_2\beta_6\beta_9}{\beta_1\beta_5\beta_{10}} Z_{b,a}^c + \frac{\beta_6\beta_9}{\beta_1\beta_7} Z_{b,c}^a + \frac{\beta_2\beta_{11}}{\beta_1\beta_7} Z_{c,a}^b + \frac{\beta_2\beta_9\beta_{11}}{\beta_1\beta_3\beta_{10}} Z_{c,b}^a
     \right)
    \nonumber \\
     & & \qquad
     - 2 \left( Z_{1,m}^m  - \frac{\beta_1}{\beta_4\beta_{13}}Z_{a,m}^m + Z_{m,1}^m - \frac{\beta_{18}}{\beta_{13}\beta_{19}} Z_{m,a}^m + Z_{m,m}^1 + \frac{\beta_{18}}{\beta_4\beta_{19}} Z_{m,m}^a
     \right)
     \nonumber \\
     & & \qquad
     + 2 \left( Z_{1,m}^m + \frac{\beta_6}{\beta_8\beta_{14}}Z_{b,m}^m + Z_{m,1}^m + \frac{\beta_{18}}{\beta_{14}\beta_{16}} Z_{m,b}^m + Z_{m,m}^1 + \frac{\beta_{18}}{\beta_8\beta_{16}}Z_{m,m}^b
     \right)
     \nonumber \\
     & & \qquad  
     + 2 \left( Z_{1,m}^1  - \frac{\beta_{11}}{\beta_{12}\beta_{15}}Z_{c,m}^m + Z_{m,1}^m - \frac{\beta_{18}}{\beta_{15}\beta_{17}} Z_{m,c}^m + Z_{m,m}^1 - \frac{\beta_{18}}{\beta_{12}\beta_{17}} Z_{m,m}^c
     \right) \biggr]
     \nonumber
\end{eqnarray}
In the expression above, each line is separately modular-invariant, using the results
of section~\ref{sect:d4:modtrans}, hence $Z_{1+a+b+c+2m}$ is modular-invariant.

At least naively, we can also turn on discrete torsion in the
${\mathbb Z}_2 \times {\mathbb Z}_2$ subalgebra,
by adding phases, in exactly the same fashion as discussed for the $1+a+b+c$ Frobenius algebra.
In the present case, this gives the modular-invariant partition function
\begin{eqnarray}
    Z_{1+a+b+c+2m, {\rm dt}} & = & \frac{1}{8} \biggl[
    Z_{1,1}^1 + Z_{1,a}^a + + Z_{a,1}^a + Z_{a,a}^1 + Z_{1,b}^b + Z_{b,1}^b +Z_{b,b}^1 + Z_{1,c}^c +Z_{c,1}^c + Z_{c,c}^1 
    \\
    & & \qquad
     - \frac{\beta_1}{\beta_2\beta_9} \left(
     Z_{a,b}^c + \frac{\beta_2\beta_9}{\beta_3\beta_5} Z_{a,c}^b + \frac{\beta_2\beta_6\beta_9}{\beta_1\beta_5\beta_{10}} Z_{b,a}^c + \frac{\beta_6\beta_9}{\beta_1\beta_7} Z_{b,c}^a + \frac{\beta_2\beta_{11}}{\beta_1\beta_7} Z_{c,a}^b + \frac{\beta_2\beta_9\beta_{11}}{\beta_1\beta_3\beta_{10}} Z_{c,b}^a
     \right)
    \nonumber \\
     & & \qquad
     - 2 \left( Z_{1,m}^m  - \frac{\beta_1}{\beta_4\beta_{13}}Z_{a,m}^m + Z_{m,1}^m - \frac{\beta_{18}}{\beta_{13}\beta_{19}} Z_{m,a}^m + Z_{m,m}^1 + \frac{\beta_{18}}{\beta_4\beta_{19}} Z_{m,m}^a
     \right)
     \nonumber \\
     & & \qquad
     + 2 \left( Z_{1,m}^m + \frac{\beta_6}{\beta_8\beta_{14}}Z_{b,m}^m + Z_{m,1}^m + \frac{\beta_{18}}{\beta_{14}\beta_{16}} Z_{m,b}^m + Z_{m,m}^1 + \frac{\beta_{18}}{\beta_8\beta_{16}}Z_{m,m}^b
     \right)
     \nonumber \\
     & & \qquad 
     + 2 \left( Z_{1,m}^1  - \frac{\beta_{11}}{\beta_{12}\beta_{15}}Z_{c,m}^m + Z_{m,1}^m - \frac{\beta_{18}}{\beta_{15}\beta_{17}} Z_{m,c}^m + Z_{m,m}^1 - \frac{\beta_{18}}{\beta_{12}\beta_{17}} Z_{m,m}^c
     \right) \biggr].
     \nonumber 
\end{eqnarray}
which differs from the first partition function by a sign flip on the second line.

As in the $1+a+b+c$ example, naively it appears as if such a sign choice could be absorbed into the intertwiners $\beta$, but changing the $\beta$'s also changes the partial traces $Z_{i,j}^k$, and as the later Rep$({\cal H}_8)$ example in section~\ref{sect:gauging-reph8-ising} demonstrates, the choice of discrete torsion in this context can be meaningful.

Of course, one could multiply the separate modular-invariant rows by any phase to get a new modular-invariant result, but to be a well-defined degree of freedom, one must impose, for example, multiloop factorization, which we have not attempted to check.  We will not attempt to give a first-principles derivation or classification in this paper.

\subsection{$\text{Rep}(Q_8)$}\label{ssec:repq8}

In this section, we study gauging Frobenius algebras derived from
Rep$(Q_8)$, where $Q_8$ denotes the eight-element group of unit
quaternions
\begin{equation}
    \{ \pm 1, \pi i, \pi j, \pi k \}.
\end{equation}
We will present this group as
\begin{equation}
    \left\langle x,y|x^2=y^2=(xy)^2, x^4=1\right\rangle.
\end{equation}
In $Q_8$ we have five conjugacy classes, $[1]=\{1\}$, $[x]=\{x,x^3\}$, $[x^2]=\{x^2\}$, $[y]=\{y,y^3\}$, and $[xy]=\{xy,x^3y\}$, the same number as for $D_4$.

\subsubsection{Representation theory}

Just as the group $D_4$, the group $Q_8$ has five irreducible representations which we'll label $1$, $a$, $b$, $c$, and $m$,
which have the same character table as for $D_4$:
\begin{center}
\begin{tabular}{c|c|c|c|c|c|}
& $[1]$ & $[x^2]$ & $[x]$ & $[y]$ & $[xy]$ \\
\hline
$\chi_1$ & $1$ & $1$ & $1$ & $1$ & $1$ \\
\hline
$\chi_a$ & $1$ & $1$ & $1$ & $-1$ & $-1$ \\
\hline
$\chi_b$ & $1$ & $1$ & $-1$ & $1$ & $-1$ \\
\hline
$\chi_c$ & $1$ & $1$ & $-1$ & $-1$ & $1$ \\
\hline
$\chi_m$ & $2$ & $-2$ & $0$ & $0$ & $0$ \\
\hline
\end{tabular}
\end{center}
The fusion rules are the same as for Rep$(D_4)$, namely
\be
a\otimes a\cong b\otimes b\cong c\otimes c\cong 1,\qquad a\otimes b\cong c,\qquad a\otimes c\cong b,\qquad b\otimes c\cong a,\non
\ee
\be
a\otimes m\cong b\otimes m\cong c\otimes m\cong m,\qquad m\otimes m\cong 1\oplus a\oplus b\oplus c.
\ee

The one-dimensional irreducible representations are given explicitly by
their characters, and are the same as for $D_4$:
\be
\rho_1(x)=\rho_1(y)=1,\qquad\rho_a(x)=1,\ \rho_a(y)=-1,\non
\ee
\be
\rho_b(x)=-1,\ \rho_b(y)=1,\qquad\rho_c(x)=-1,\ \rho_c(y)=-1.
\ee
The two-dimensional irreducible representation $m$ is different from that
of $D_4$, and here is described by
\be
\rho'_m(x)=\lp\begin{matrix} i & 0 \\ 0 & -i \end{matrix}\rp,\qquad\rho'_m(y)=\lp\begin{matrix} 0 & 1 \\ -1 & 0 \end{matrix}\rp.
\ee

\subsubsection{Cosets}

We have
\begin{itemize}
\item $H=Q_8$, $Q_8/H=\{H\}$, giving $1$.
\item $H=\langle x\rangle\cong\Z_4$, $Q_8/H=\{H,yH\}$.
\be
\rho(x)=\lp\begin{matrix} 1 & 0 \\ 0 & 1 \end{matrix}\rp,\qquad\rho(y)=\lp\begin{matrix} 0 & 1 \\ 1 & 0 \end{matrix}\rp,
\ee
for $1+a$.
\item $H=\langle y\rangle\cong\Z_4$, $Q_8/H=\{H,xH\}$.
\be
\rho(x)=\lp\begin{matrix} 0 & 1 \\ 1 & 0 \end{matrix}\rp,\qquad\rho(y)=\lp\begin{matrix} 1 & 0 \\ 0 & 1 \end{matrix}\rp,
\ee
for $1+b$.
\item $H=\langle xy\rangle\cong\Z_4$, $Q_8/H=\{H,xH\}$.
\be
\rho(x)=\lp\begin{matrix} 0 & 1 \\ 1 & 0 \end{matrix}\rp,\qquad\rho(y)=\lp\begin{matrix} 0 & 1 \\ 1 & 0 \end{matrix}\rp,
\ee
for $1+c$.
\item $H=\langle x^2\rangle\cong\Z_2$, $Q_8/H=\{H,xH,yH,xyH\}$.
\be
\rho(x)=\lp\begin{matrix} 0 & 1 & 0 & 0 \\ 1 & 0 & 0 & 0 \\ 0 & 0 & 0 & 1 \\ 0 & 0 & 1 & 0 \end{matrix}\rp,\qquad\rho(y)=\lp\begin{matrix} 0 & 0 & 1 & 0 \\ 0 & 0 & 0 & 1 \\ 1 & 0 & 0 & 0 \\ 0 & 1 & 0 & 0 \end{matrix}\rp,
\ee
for $1+a+b+c$.
\item $H=\{1\}$, $Q_8/H\cong Q_8=\{1,x,x^2,x^3,y,xy,x^2y,x^3y\}$ corresponding to the regular representation $1+a+b+c+2m$.
\end{itemize}

In this case all three of the $\Z_4$ subgroups are permuted by the $S_3$ outer automorphism group, so the physically distinct options for $Q_8$ are only $1$, $1+a$, $1+a+b+c$, and $1+a+b+c+2m$.
We will compute partition functions for each of these cases later in this section.

\subsubsection{Computing the associator}

The results for associators and crossing kernels for $Q_8$ will be very similar to those of
$D_4$.

Taking basis vectors $e$, $e_a$, $e_b$, $e_c$, and $e_{m1}$, $e_{m2}$ for $1$, $a$, $b$, $c$, and $m$ respectively (such that the irreps take the form from the previous subsection), we parameterize the most general fusion intertwiners.  

For $Q_8$ we have
\begin{align}
\label{q8inter1}
e_a e_a\mapsto\ & \beta_1' \, e,\\
e_a e_b\mapsto\ & \beta_2' \, e_c,\\
e_a e_c\mapsto\ & \beta_3' \, e_b,\\
e_a e_{m1}\mapsto\ & \beta_4' \, e_{m1},\\
e_a e_{m2}\mapsto\ & - \beta_4' \, e_{m2},\\
e_b e_a\mapsto\ & \beta_5' \, e_c,\\
e_b e_b\mapsto\ & \beta_6' \, e,\\
e_b e_c\mapsto\ & \beta_7' \, e_a,\\
e_b e_{m1}\mapsto\ & \beta_8' \, e_{m2},\\
\color{red} e_b e_{m2}\mapsto\ & \color{red} - \beta_8' \, e_{m1},\\
e_c e_a\mapsto\ & \beta_9' \, e_b,\\
e_c e_b\mapsto\ & \beta_{10}' \, e_a,\\
e_c e_c\mapsto\ & \beta_{11}' \, e,\\
e_c e_{m1}\mapsto\ & \beta_{12}' \, e_{m2},\\
\color{red} e_c e_{m2}\mapsto\ & \color{red} \beta_{12}' \, e_{m1},\\
e_{m1} e_a\mapsto\ & \beta_{13}' \, e_{m1},\\
e_{m1} e_b\mapsto\ & \beta_{14}' \, e_{m2},\\
e_{m1} e_c\mapsto\ & \beta_{15}' \, e_{m2},\\
e_{m1} e_{m1}\mapsto\ & \beta_{16}' \, e_b + \beta_{17}' \, e_c,\\
e_{m1} e_{m2}\mapsto\ & \beta_{18}' \, e + \beta_{19}' \, e_a,\\
e_{m2} e_a\mapsto\ & - \beta_{13}' \, e_{m2},\\
\color{red} e_{m2} e_b\mapsto\ & \color{red} - \beta_{14}' \, e_{m1},\\
\color{red} e_{m2} e_c\mapsto\ & \color{red} \beta_{15}' \, e_{m1},\\
\color{red} e_{m2}  e_{m1}\mapsto\ & \color{red} - \beta_{18}' \, e + \beta_{19}' \, e_a,\\
e_{m2} e_{m2}\mapsto\ & \beta_{16}' \, e_b - \beta_{17}' \, e_c.
\label{q8inter2}
\end{align}
These differ by a few key signs from those for $D_4$, but only in the lines colored red.

We get evaluation maps $\epsilon_L$ and $\overline{\epsilon}_L$ by projecting onto the identity line as usual, and then we can determine the coevaluation maps $\g_R$ and $\overline{\g}_R$.  Here we have to be a bit careful when dealing with the $m$ line.  We have
\begin{equation}
    \epsilon_m(e_{m1}e_{m1})=\epsilon_m(e_{m2}e_{m2})=\overline{\epsilon}_m(e_{m1}e_{m1})=\overline{\e}_m(e_{m2}e_{m2})=0,\non
\end{equation}
\begin{equation}
    \epsilon_m(e_{m1}e_{m2})=\overline{\e}_m(e_{m1}e_{m2})=\beta_{18}',\qquad\epsilon_m(e_{m2}e_{m1})=\overline{\epsilon}_m(e_{m2}e_{m1})=-\beta_{18}',
\end{equation}
The coevaluation maps pick up a relative sign,
\begin{equation}
    \g_m(1)=\overline{\g}_m(1)=-\beta_{18}^{\prime\,-1}\left(e_{m1}e_{m2}-e_{m2}e_{m1}\right).
\end{equation}
Then using the coevaluation map along with fusion, we can produce a basis for co-fusions,
\begin{align}
    e\mapsto\ & ee+\beta_1^{\prime\,-1}e_ae_a+\beta_6^{\prime\,-1}e_be_b+\beta_{11}^{\prime\,-1}e_ce_c-\beta_{18}^{\prime\,-1}\left(e_{m1}e_{m2}-e_{m2}e_{m1}\right),\\
    e_a\mapsto\ & ee_a+e_ae+\beta_3'\beta_{11}^{\prime\,-1}e_be_c+\beta_2'\beta_6^{\prime\,-1}e_ce_b-\beta_4'\beta_{18}^{\prime\,-1}\left(e_{m1}e_{m2}+e_{m2}e_{m1}\right),\\
    e_b\mapsto\ & ee_b+e_be+\beta_7'\beta_{11}^{\prime\,-1}e_ae_c+\beta_5'\beta_1^{\prime\,-1}e_ce_a-\beta_8'\beta_{18}^{\prime\,-1}\left(e_{m1}e_{m1}+e_{m2}e_{m2}\right),\\
    e_c\mapsto\ & ee_c+e_ce+\beta_{10}'\beta_6^{\prime\,-1}e_ae_b+\beta_9'\beta_1^{\prime\,-1}e_be_a+\beta_{12}'\beta_{18}^{\prime\,-1}\left(e_{m1}e_{m1}-e_{m2}e_{m2}\right),\\
    e_{m1}\mapsto\ & ee_{m1}+e_{m1}e+\beta_{19}'\beta_{18}^{\prime\,-1}e_ae_{m1}+\beta_{13}'\beta_1^{\prime\,-1}e_{m1}e_a\non\\
    & \qquad -\beta_{16}'\beta_{18}^{\prime\,-1}e_be_{m2}+\beta_{14}'\beta_6^{\prime\,-1}e_{m2}e_b-\beta_{17}'\beta_{18}^{\prime\,-1}e_ce_{m2}+\beta_{15}'\beta_{11}^{\prime\,-1}e_{m2}e_c,\\
    e_{m2}\mapsto\ & ee_{m2}+e_{m2}e-\beta_{19}'\beta_{18}^{\prime\,-1}e_ae_{m2}-\beta_{13}'\beta_1^{\prime\,-1}e_{m2}e_a\non\\
    & \qquad +\beta_{16}'\beta_{18}^{\prime\,-1}e_be_{m1}-\beta_{14}'\beta_6^{\prime\,-1}e_{m1}e_b-\beta_{17}'\beta_{18}^{\prime\,-1}e_ce_{m1}+\beta_{15}'\beta_{11}^{\prime\,-1}e_{m1}e_c.
\end{align}

Repeating the crossing kernel computation for $Q_8$ just changes some signs relative to $D_4$.  We list below only the components where there are sign changes from $D_4$ results:
\begin{align}
\left( e_b e_b \right) e_{m1} = \beta_6' \, e e_{m1} = \beta_6' e_{m1},\ 
& e_b \left( e_b e_{m1}\right) = \beta_8' \, e_b e_{m2} = - (\beta_8')^2 e_{m1},\ \Rightarrow\ 
& \tilde{K}^{b,m}_{b,m}(1,m) = -\frac{\beta_6'}{(\beta_8')^2},
\\
\left( e_b e_c \right) e_{m1} = \beta_7' \, e_a e_{m1} = \beta_4' \beta_7' e_{m1},\ 
& e_b \left( e_c e_{m1}\right) = \beta_{12}' \, e_b e_{m2} = - \beta_8' \beta_{12}' e_{m1},\ \Rightarrow\ 
& \tilde{K}^{b,m}_{c,m}(a,m) = - \frac{\beta_4' \beta_7'}{\beta_8' \beta_{12}'},
\\
\left( e_b e_{m1}\right) e_b = \beta_8' \, e_{m2} e_b = - \beta_8' \beta_{14}' \, e_{m1},\ 
& e_b \left( e_{m1} e_b \right) = \beta_{14}' \, e_b e_{m2} = - \beta_8' \beta_{14}' e_{m1},\ \Rightarrow\ 
& \tilde{K}^{b,m}_{m,b}(m,m)=1,
\\
\left( e_b e_{m1} \right) e_c = \beta_8' \, e_{m2} e_c = \beta_8' \beta_{15}' e_{m1},\ 
& e_b\left( e_{m1} e_c\right) = \beta_{15}' \, e_b e_{m2} = - \beta_8' \beta_{15}' e_{m1},\ \Rightarrow\ 
& \tilde{K}^{b,m}_{m,c}(m,m)=-1,
\end{align}
\begin{multline}
\left( e_b e_{m1}\right) e_{m1} = \beta_8' \, e_{m2} e_{m1} = - \beta_8' \beta_{18}' \, e + \beta_8' \beta_{19}' \, e_a,
\quad e_b\left( e_{m1} e_{m1} \right) = \beta_{16}' \, e_b e_b + \beta_{17}' \, e_b e_c = \beta_6' \beta_{16}' \, e + \beta_9' \beta_{17}' e_a,\non
\\
\Rightarrow\quad\tilde{K}^{b,1}_{m,m}(m,b) = - \frac{\beta_8' \beta_{18}'}{\beta_6' \beta_{16}'},
\quad\tilde{K}^{b,a}_{m,m}(m,c) = \frac{\beta_8' \beta_{19}'}{\beta_9' \beta_{17}'},
\end{multline}
\begin{align}
\left( e_c e_b \right) e_{m1} = \beta_{10}' \, e_a e_{m1} = \beta_4' \beta_{10}' \, e_{m1},\ 
& e_c\left( e_b e_{m1}\right) = \beta_8' \, e_c e_{m2} = \beta_8' \beta_{12}' e_{m1},\ \Rightarrow\ 
& \tilde{K}^{c,m}_{b,m}(a,m) = \frac{\beta_4' \beta_{10}'}{\beta_8' \beta_{12}'},
\\
\left( e_c e_c \right) e_{m1} = \beta_{11}' \, e e_{m1} = \beta_{11}' \, e_{m1},\ 
& e_c\left(e_c e_{m1} \right) = \beta_{12}' \, e_c e_{m2} = (\beta_{12}')^2 e_{m1},\ \Rightarrow\ 
& \tilde{K}^{c,m}_{c,m}(1,m) = \frac{\beta_{11}'}{(\beta_{12}')^2},
\\
\left( e_c e_{m1} \right) e_b = \beta_{12}' \, e_{m2} e_b = - \beta_{12}' \beta_{14}' \, e_{m1},\ 
& e_c\left( e_{m1} e_b\right) = \beta_{14}' \, e_c e_{m2} = \beta_{12}' \beta_{14}' e_{m1},\ \Rightarrow\ 
& \tilde{K}^{c,m}_{m,b}(m,m)=-1,
\\
\left( e_c e_{m1} \right) e_c = \beta_{12}' \, e_{m2} e_c = \beta_{12}' \beta_{15}' \, e_{m1},\ 
& e_c \left( e_{m1} e_c \right) = \beta_{15}' \, e_c e_{m2} = \beta_{12}' \beta_{15}' e_{m1},\ \Rightarrow\ 
& \tilde{K}^{c,m}_{m,c}(m,m)=1,
\end{align}
\begin{multline}
\left( e_c e_{m1} \right) e_{m1} = \beta_{12}' \, e_{m2} e_{m1} = - \beta_{12}' \beta_{18}' \, e + \beta_{12}' \beta_{19}' \, e_a,
\quad e_c\left( e_{m1} e_{m1} \right) = \beta_{16}' \, e_c e_b + \beta_{17}' \, e_c e_c = \beta_{11}' \beta_{17}' \, e + \beta_{10}' \beta_{16}' \, e_a,\non
\\
\Rightarrow\quad\tilde{K}^{c,1}_{m,m}(m,c) = - \frac{\beta_{12}'\beta_{18}'}{\beta_{11}'\beta_{17}'},
\quad\tilde{K}^{c,a}_{m,m}(m,b) = \frac{\beta_{12}' \beta_{19}'}{\beta_{10}' \beta_{16}'},
\end{multline}
\begin{align}
\left( e_{m1} e_b\right) e_b = \beta_{14}' e_{m2} e_b = - (\beta_{14}')^2 \, e_{m1},\ 
& e_{m1} \left( e_b e_b\right) = \beta_6' \, e_{m1} e = \beta_6' \, e_{m1},\ \Rightarrow\ 
& \tilde{K}^{m,m}_{b,b}(m,1) = - \frac{(\beta_{14}')^2}{\beta_6'},
\\
\left( e_{m1} e_b\right) e_c = \beta_{14}' \, e_{m2} e_c = \beta_{14}' \beta_{15}' \, e_{m1},\ 
& e_{m1} \left( e_b e_c \right) = \beta_7' \, e_{m1} e_a = \beta_7' \beta_{13}' \, e_{m1},\ \Rightarrow\ 
& \tilde{K}^{m,m}_{b,c}(m,a) = \frac{\beta_{14}'\beta_{15}'}{\beta_7'\beta_{13}'},
\end{align}
\begin{multline}
\left( e_{m1} e_b\right) e_{m1} = \beta_{14}' \, e_{m2} e_{m1} = - \beta_{14}' \beta_{18}' \, e + \beta_{14}' \beta_{19}' \, e_a,
\quad e_{m1} \left( e_b e_{m1}\right) = \beta_8' \, e_{m1} e_{m2} = \beta_8' \beta_{18}' \, e + \beta_8' \beta_{19}'a,\non
\\
\Rightarrow\quad\tilde{K}^{m,1}_{b,m}(m,m) = - \frac{\beta_{14}'}{\beta_8'},
\quad\tilde{K}^{m,a}_{b,m}(m,m) = \frac{\beta_{14}'}{\beta_8'},
\end{multline}
\begin{multline}
\left( e_{m1} e_b\right) e_{m2} = \beta_{14}' \, e_{m2} e_{m2} = \beta_{14}' \beta_{16}' \, e_b - \beta_{14}' \beta_{17}' \, e_c,
\quad e_{m1} \left( e_b e_{m2}\right) = - \beta_8' \, e_{m1} e_{m1} = - \beta_8' \beta_{16}' \, e_b - \beta_8' \beta_{17}' \, e_c,\non
\\
\Rightarrow\quad\tilde{K}^{m,b}_{b,m}(m,m) = - \frac{\beta_{14}'}{\beta_8'},
\quad\tilde{K}^{m,c}_{b,m}(m,m) = \frac{\beta_{14}'}{\beta_8'},
\end{multline}
\begin{align}
\left( e_{m1} e_c\right) e_b = \beta_{15}' \, e_{m2} e_b = - \beta_{14}' \beta_{15}' \, e_{m1},\ 
& e_{m1} \left( e_c e_b \right) = \beta_{10}' \, e_{m1} e_a = \beta_{10}' \beta_{13}' \, e_{m1},\ \Rightarrow\ 
& \tilde{K}^{m,m}_{c,b}(m,a) = - \frac{\beta_{14}'\beta_{15}'}{\beta_{10}'\beta_{13}'},
\\
\left( e_{m1} e_c \right) e_c = \beta_{15}' \, e_{m2} e_c = (\beta_{15}')^2 \, e_{m1},\ 
& e_{m1} \left(e_c e_c \right) = \beta_{11}' \, e_{m1} e = \beta_{11}' \, e_{m1},\ \Rightarrow\ 
& \tilde{K}^{m,m}_{c,c}(m,1) = \frac{(\beta_{15}')^2}{\beta_{11}'},
\end{align}
\begin{multline}
\left( e_{m1} e_c \right) e_{m1} = \beta_{15}' \, e_{m2} e_{m1} = - \beta_{15}' \beta_{18}' \, e + \beta_{15}' \beta_{19}' e_a,
\quad e_{m1}  \left( e_c e_{m1}\right) = \beta_{12}' \, e_{m1} e_{m2} = \beta_{12}' \beta_{18}' \, e + \beta_{12}' \beta_{19}' \, e_a,\non
\\
\Rightarrow\quad\tilde{K}^{m,1}_{c,m}(m,m) = - \frac{\beta_{15}'}{\beta_{12}'},
\quad\tilde{K}^{m,a}_{c,m}(m,m) = \frac{\beta_{15}'}{\beta_{12}'},
\end{multline}
\begin{multline}
\left( e_{m1} e_c \right) e_{m2} = \beta_{15}' \, e_{m2} e_{m2} = \beta_{15}' \beta_{16}' \, e_b - \beta_{15}' \beta_{17}' \, e_c,
\quad e_{m1} \left( e_c e_{m2} \right) = \beta_{12}' \, e_{m1} e_{m1} = \beta_{12}' \beta_{16}' \, e_b + \beta_{12}' \beta_{17}' \, e_c,\non
\\
\Rightarrow\quad\tilde{K}^{m,b}_{c,m}(m,m) = \frac{\beta_{12}'}{\beta_{15}'},
\quad\tilde{K}^{m,c}_{c,m}(m,m) = - \frac{\beta_{12}'}{\beta_{15}'},
\end{multline}
\begin{multline}
\left( e_{m1} e_{m2} \right) e_b = \beta_{18}' \, e e_b + \beta_{19}' \, e_a e_b = \beta_{18}' \, e_b + \beta_2'\beta_{19}' \, e_c,
\quad e_{m1} \left( e_{m2} e_b \right) = - \beta_{14}' \, e_{m1} e_{m1} = - \beta_{14}' \beta_{16}' \, e_b - \beta_{14}' \beta_{17}' \, e_c,\non
\\
\Rightarrow\quad\tilde{K}^{m,b}_{m,b}(1,m) = - \frac{\beta_{18}'}{\beta_{14}'\beta_{16}'},
\quad\tilde{K}^{m,c}_{m,b}(a,m) = - \frac{\beta_2' \beta_{19}'}{\beta_{14}' \beta_{17}'},
\end{multline}
\begin{multline}
\left( e_{m1} e_{m1} \right) e_c = \beta_{18}' \, e e_c + \beta_{19}' \, e_a e_c = \beta_3' \beta_{19}' \, e_b + \beta_{18}' \, e_c,
\quad e_{m1} \left( e_{m2} e_c \right) = \beta_{15}' \, e_{m1} e_{m1} = \beta_{15}' \beta_{16}' \, e_b + \beta_{15}' \beta_{17}' \, e_c,\non
\\
\Rightarrow\quad\tilde{K}^{m,b}_{m,c}(a,m) = \frac{\beta_3' \beta_{19}'}{\beta_{15}' \beta_{16}'},
\quad\tilde{K}^{m,c}_{m,c}(1,m) = \frac{\beta_{18}'}{\beta_{15}'\beta_{17}'},
\end{multline}
\begin{eqnarray}
\left( e_{m1} e_{m1} \right) e_{m1} = \beta_{16}' \, e_b e_{m1} + \beta_{17}' \, e_c e_{m1} 
& = &
\lp \beta_8'\beta_{16}' + \beta_{12}' \beta_{17}'\rp e_{m2},
\non \\ %\quad 
e_{m1} \left( e_{m1} e_{m1} \right) = \beta_{16}' \, e_{m1} e_b + \beta_{17}' \, e_{m1} e_c  & = & 
\lp \beta_{14}' \beta_{16}' + \beta_{15}' \beta_{17}'\rp v,\non
\\
\left( e_{m1} e_{m1} \right) e_{m2} = \beta_{16}' \, e_b e_{m2} + \beta_{17}' \, e_c e_{m2} & = & -
\lp \beta_8' \beta_{16}' - \beta_{12}' \beta_{17}'\rp e_{m1},
\non \\ %\quad 
e_{m1} \left( e_{m1} e_{m2} \right) = \beta_{18}' \, e_{m1} e + \beta_{19}' \, e_{m1} e_a 
& = & 
\lp \beta_{18}' + \beta_{13}' \beta_{19}'\rp e_{m1},\non
\\
\left( e_{m1} e_{m2} \right) e_{m1} = \beta_{18}' \, e e_{m1} + \beta_{19}' \, e_a e_{m1}
& = &
\lp \beta_{18}' + \beta_4'\beta_{19}'\rp e_{m1},
\non \\ %\quad 
e_{m1} \left( e_{m2} e_{m1} \right) = - \beta_{18}' \, e_{m1} e + \beta_{19}' e_{m1} e_a
& = & -
\lp \beta_{18}' - \beta_{13}' \beta_{19}'\rp e_{m1},\non
\\
\left( e_{m1} e_{m2} \right) e_{m2} = \beta_{18}' \, e e_{m2} + \beta_{19}' \, e_a e_{m2} 
& = &
\lp \beta_{18}' - \beta_4' \beta_{19}'\rp e_{m2},
\non \\ %\quad 
e_{m1} \left( e_{m2} e_{m2} \right) = \beta_{16}' \, e_{m1} e_b - \beta_{17}' \, e_{m1} e_c
& = &
\lp \beta_{14}' \beta_{16}' - \beta_{15}' \beta_{17}'\rp e_{m2},\non
\end{eqnarray}
\begin{multline}
\lp\begin{matrix} \tilde{K}^{m,m}_{m,m}(1,1) & \tilde{K}^{m,m}_{m,m}(1,a) & \tilde{K}^{m,m}_{m,m}(1,b) & \tilde{K}^{m,m}_{m,m}(1,c) \\ \tilde{K}^{m,m}_{m,m}(a,1) & \tilde{K}^{m,m}_{m,m}(a,a) & \tilde{K}^{m,m}_{m,m}(a,b) & \tilde{K}^{m,m}_{m,m}(a,c) \\ \tilde{K}^{m,m}_{m,m}(b,1) & \tilde{K}^{m,m}_{m,m}(b,a) & \tilde{K}^{m,m}_{m,m}(b,b) & \tilde{K}^{m,m}_{m,m}(b,c) \\ \tilde{K}^{m,m}_{m,m}(c,1) & \tilde{K}^{m,m}_{m,m}(c,a) & \tilde{K}^{m,m}_{m,m}(c,b) & \tilde{K}^{m,m}_{m,m}(c,c) \end{matrix}\rp\non
\\
=
\hlf\lp\begin{matrix} -1 & \frac{\beta_{18}'}{\beta_{13}' \beta_{19}'} & \frac{\beta_{18}'}{\beta_{14}' \beta_{16}'} & -\frac{\beta_{18}'}{\beta_{15}' \beta_{17}'} \\
-\frac{\beta_4' \beta_{19}'}{\beta_{18}'} & \frac{\beta_4'}{\beta_{13}'} & -\frac{\beta_4' \beta_{19}'}{\beta_{14}' \beta_{16}'} & \frac{\beta_4' \beta_{19}'}{\beta_{15}' \beta_{17}'} \\
-\frac{\beta_8' \beta_{16}'}{\beta_{18}'} & -\frac{\beta_8' \beta_{16}'}{\beta_{13}' \beta_{19}'} & \frac{\beta_8'}{\beta_{14}'} & \frac{\beta_8' \beta_{16}'}{\beta_{15}' \beta_{17}'} \\
\frac{\beta_{12}' \beta_{17}'}{\beta_{18}'} & \frac{\beta_{12}' \beta_{17}'}{\beta_{13}' \beta_{19}'} & \frac{\beta_{12}' \beta_{17}'}{\beta_{14}' \beta_{16}'} & \frac{\beta_{12}'}{\beta_{15}'}
\end{matrix}\rp.
\end{multline}

\subsubsection{Modular transformations}
\label{sect:q8:modtrans}

For $Q_8$ a few of the modular transformations pick up signs relative to those for $D_4$, specifically
\begin{align}
    Z_{m,b}^m(\tau+1)=\ & -\frac{\beta_{14}'}{\beta_8'}Z_{m,m}^b(\tau),\\
    Z_{m,c}^m(\tau+1)=\ & -\frac{\beta_{15}'}{\beta_{12}'}Z_{m,m}^c(\tau),
\end{align}
\begin{align}
    Z_{m,m}^1(\tau+1)=\ & \hlf\lp -Z_{m,1}^m(\tau)+\frac{\beta_{18}'}{\beta_{13}'\beta_{19}'}Z_{m,a}^m(\tau)+\frac{\beta_{18}'}{\beta_{14}'\beta_{16}'}Z_{m,b}^m(\tau)-\frac{\beta_{18}'}{\beta_{15}'\beta_{17}'}Z_{m,c}^m(\tau)\rp,\\
    Z_{m,m}^a(\tau+1)=\ & \hlf\lp -\frac{\beta_4'\beta_{19}'}{\beta_{18}'}Z_{m,1}^m(\tau)+\frac{\beta_4'}{\beta_{13}'}Z_{m,a}^m(\tau)-\frac{\beta_4'\beta_{19}'}{\beta_{14}'\beta_{16}'}Z_{m,b}^m(\tau)+\frac{\beta_4'\beta_{19}'}{\beta_{15}'\beta_{17}'}Z_{m,c}^m(\tau)\rp,\\
    Z_{m,m}^b(\tau+1)=\ & \hlf\lp -\frac{\beta_8'\beta_{16}'}{\beta_{18}'}Z_{m,1}^m(\tau)-\frac{\beta_8'\beta_{16}'}{\beta_{13}'\beta_{19}'}Z_{m,a}^m(\tau)+\frac{\beta_8'}{\beta_{14}'}Z_{m,b}^m(\tau)+\frac{\beta_8'\beta_{16}'}{\beta_{15}'\beta_{17}'}Z_{m,c}^m(\tau)\rp,\\
    Z_{m,m}^c(\tau+1)=\ & \hlf\lp\frac{\beta_{12}'\beta_{17}'}{\beta_{18}'}Z_{m,1}^m(\tau)+\frac{\beta_{12}'\beta_{17}'}{\beta_{13}'\beta_{19}'}Z_{m,a}^m(\tau)+\frac{\beta_{12}'\beta_{17}'}{\beta_{14}'\beta_{16}'}Z_{m,b}^m(\tau)+\frac{\beta_{12}'}{\beta_{15}'}Z_{m,c}^m(\tau)\rp,
\end{align}
\begin{align}
    Z_{b,m}^m(-1/\tau)=\ & -\frac{\beta_8'\beta_{18}'}{\beta_6'\beta_{16}'}Z_{m,b}^m(\tau),\\
    Z_{c,m}^m(-1/\tau)=\ & -\frac{\beta_{12}'\beta_{18}'}{\beta_{11}'\beta_{17}'}Z_{m,c}^m(\tau),\\
    Z_{m,b}^m(-1/\tau)=\ & -\frac{\beta_6'\beta_{16}'}{\beta_8'\beta_{!8}'}Z_{b,m}^m(\tau),\\
    Z_{m,c}^m(-1/\tau)=\ & -\frac{\beta_{11}'\beta_{17}'}{\beta_{12}'\beta_{18}'}Z_{c,m}^m(\tau),
\end{align}
\begin{align}
    Z_{m,m}^1(-1/\tau)=\ & \hlf\lp -Z_{m,m}^1(\tau)-\frac{\beta_{18}'}{\beta_4'\beta_{19}'}Z_{m,m}^a(\tau)-\frac{\beta_{18}'}{\beta_8'\beta_{16}'}Z_{m,m}^b(\tau)+\frac{\beta_{18}'}{\beta_{12}'\beta_{17}'}Z_{m,m}^c(\tau)\rp,\\
    Z_{m,m}^a(-1/\tau)=\ & \hlf\lp -\frac{\beta_4'\beta_{19}'}{\beta_{18}'}Z_{m,m}^1(\tau)-Z_{m,m}^a(\tau)+\frac{\beta_4'\beta_{19}'}{\beta_8'\beta_{16}'}Z_{m,m}^b(\tau)-\frac{\beta_4'\beta_{19}'}{\beta_{12}'\beta_{17}'}Z_{m,m}^c(\tau)\rp,\\
    Z_{m,m}^b(-1/\tau)=\ & \hlf\lp -\frac{\beta_8'\beta_{16}'}{\beta_{18}'}Z_{m,m}^1(\tau)+\frac{\beta_8'\beta_{16}'}{\beta_4'\beta_{19}'}Z_{m,m}^a(\tau)-Z_{m,m}^b(\tau)-\frac{\beta_8'\beta_{16}'}{\beta_{12}'\beta_{17}'}Z_{m,m}^c(\tau)\rp,\\
    Z_{m,m}^c(-1/\tau)=\ & \hlf\lp\frac{\beta_{12}'\beta_{17}'}{\beta_{18}'}Z_{m,m}^1(\tau)-\frac{\beta_{12}'\beta_{17}'}{\beta_4'\beta_{19}'}Z_{m,m}^a(\tau)-\frac{\beta_{12}'\beta_{17}'}{\beta_8'\beta_{16}'}Z_{m,m}^b(\tau)-Z_{m,m}^c(\tau)\rp.
\end{align}
The rest of the modular transformations are the same as for $D_4$ but with $\beta'$ instead of $\beta$.

There are then fewer modular invariant combinations than existed in the $D_4$ case, which we list
below:
\begin{align}
    & Z_{1,1}^1,\\
    & Z_{1,a}^a+Z_{a,1}^a+Z_{a,a}^1,\\
    & Z_{1,b}^b+Z_{b,1}^b+Z_{b,b}^1,\\
    & Z_{1,c}^c+Z_{c,1}^c+Z_{c,c}^1,\\
    & Z_{a,b}^c+\frac{\beta_2'\beta_9'}{\beta_3'\beta_5'}Z_{a,c}^b+\frac{\beta_2'\beta_6'\beta_9'}{\beta_1'\beta_5'\beta_{10}'}Z_{b,a}^c+\frac{\beta_6'\beta_9'}{\beta_1'\beta_7'}Z_{b,c}^a+\frac{\beta_2'\beta_{11}'}{\beta_1'\beta_7'}Z_{c,a}^b+\frac{\beta_2'\beta_9'\beta_{11}'}{\beta_1'\beta_3'\beta_{10}'}Z_{c,b}^a,\\
    & Z_{1,m}^m+\frac{\beta_1'}{\beta_4'\beta_{13}'}Z_{a,m}^m-\frac{\beta_6'}{\beta_8'\beta_{14}'}Z_{b,m}^m+\frac{\beta_{11}'}{\beta_{12}'\beta_{15}'}Z_{c,m}^m+Z_{m,1}^m+\frac{\beta_{18}'}{\beta_{13}'\beta_{19}'}Z_{m,a}^m+\frac{\beta_{18}'}{\beta_{14}'\beta_{16}'}Z_{m,b}^m-\frac{\beta_{18}'}{\beta_{15}'\beta_{17}'}Z_{m,c}^m\non\\
    & \qquad +Z_{m,m}^1-\frac{\beta_{18}'}{\beta_4'\beta_{19}'}Z_{m,m}^a-\frac{\beta_{18}'}{\beta_8'\beta_{16}'}Z_{m,m}^b+\frac{\beta_{18}'}{\beta_{12}'\beta_{17}'}Z_{m,m}^c.
\end{align}

\subsubsection{$H=\Z_4$: $1+a$ orbifold}
We have basis vectors for the trivial and $a$-representation:
\be
    e_1=v_H+v_{yH},\qquad e_a=v_H-v_{yH},
    \ee
The multiplication with $e_1$ acts as identity, and the multiplication of $e_a$ with itself is
\be
\mu_*(e_a\otimes e_a) = e_1,
\ee
as expected from the fact that $a\otimes a\cong 1$. We can compute the comultiplication as
\begin{eqnarray}
    \Delta_F(e_1) &=& v_Hv_H+ v_{yH}v_{yH} = \tfrac{1}{2}(e_1e_1+e_ae_a),\\
    \Delta_F(e_a) &=& v_Hv_H-v_{yH}v_{yH} = \tfrac{1}{2}(e_1e_a+e_ae_1).
\end{eqnarray}
Expanding this in the (co-)fusion basis gives the coefficients
    \be
     \m_{1,1}^1=\m_{1,a}^a=\m_{a,1}^a=1,\quad \m_{a,a}^1=\beta_1^{\prime\,-1},
    \ee
    \be
   \Delta_{1,1}^1=\Delta^{1,a}_a=\Delta^{a,1}_a=\tfrac{1}{2},\quad \Delta_{1}^{a,a} = \frac{\beta'_1}{2}.
    \ee
    As we have seen, these coefficients can be used to compute the partition function as $Z=\sum_{i,j,k}\mu_{i,j}^k\Delta_k^{j,i} Z_{i,j}^k$. Then the partition function is
    \be   \label{eq:q8-orb-1a}
    Z_{1+a}=\hlf\ls Z_{1,1}^1+Z_{1,a}^a+Z_{a,1}^a+Z_{a,a}^1\rs.
    \ee
This expression for the partition function is a linear combination of the modular invariants computed in section~\ref{sect:q8:modtrans}, hence $Z_{1+a}$ is modular-invariant.

\subsubsection{$H=\Z_2$: $1+a+b+c$ orbifold}
We first define basis vectors for the representations $\{1,a,b,c\}$
    \be
    e_1=v_H+v_{xH}+v_{yH}+v_{xyH},\qquad e_a=v_H+v_{xH}-v_{yH}-v_{xyH},
    \ee
       \be
    e_b=v_H-v_{xH}+v_{yH}-v_{xyH},\qquad e_c=v_H-v_{xH}-v_{yH}+v_{xyH},
    \ee
The multiplication is
\begin{equation}
    \begin{split}
        &\mu_*\left(e_1 \otimes e_1\right) = \mu_*\left(e_a \otimes e_a\right) = \mu_*\left(e_b \otimes e_b\right) = \mu_*\left(e_c \otimes e_c\right) = e_1,
        \\
        &\mu_*\left(e_1 \otimes e_a\right) = \mu_*\left(e_a \otimes e_1\right) = e_a, \: \: \:
        \mu_*\left( e_1 e_b \right) = \mu_*\left(e_b \otimes e_1\right) = e_b,
        \\
        &\mu_*\left(e_1 \otimes e_c\right) = \mu_*\left(e_c \otimes e_1\right) = e_c, \: \: \:
        \mu_*\left(e_a \otimes e_b\right) = \mu_*\left(e_b \otimes e_a\right) = e_c,
        \\
        & \mu_*\left(e_a \otimes e_c\right) = \mu_*\left(e_c \otimes e_a\right) = e_b, \: \: \:
        \mu_*\left(e_b \otimes e_c\right) = \mu_*\left(e_c \otimes e_b\right) = e_a,
    \end{split}
\end{equation}
while the comultiplication reads
\begin{equation}
    \begin{split}
        \Delta_F\left(e_1\right) = \frac{1}{4}\left( e_1 \otimes e_1 + e_a \otimes e_a + e_b \otimes e_b + e_c \otimes e_c \right),\\
        \Delta_F\left(e_a\right) = \frac{1}{4}\left( e_1 \otimes e_a + e_a \otimes e_1 + e_b \otimes e_c + e_c \otimes e_b \right),\\
        \Delta_F\left(e_b\right) = \frac{1}{4}\left( e_1 \otimes e_b + e_b \otimes e_1 + e_a \otimes e_c + e_c \otimes e_a \right),\\
        \Delta_F\left(e_c\right) = \frac{1}{4}\left( e_1 \otimes e_c + e_c \otimes e_1 + e_a \otimes e_b + e_b \otimes e_a \right),
    \end{split}
\end{equation}
which was derived using the fact that $\Delta_F$ is diagonal on the $v$ basis. We can then expand this in the (co-)fusion basis to compute the relevant coefficients as
    \be
\m_{1,1}^1=1,\quad \m_{a,a}^1=\beta_1^{\prime\,-1},\quad\m_{a,b}^c=\beta_2^{\prime\,-1},\quad\m_{a,c}^b=\beta_3^{\prime\,-1},
    \ee
    \be
    \m_{b,a}^c=\beta_5^{\prime\,-1},\quad\m_{b,b}^1=\beta_6^{\prime\,-1},\quad\m_{b,c}^a=\beta_7^{\prime\,-1},
    \ee
    \be
    \m_{c,a}^b=\beta_9^{\prime\,-1},\quad\m_{c,b}^a=\beta_{10}^{\prime\,-1},\quad\m_{c,c}^1=\beta_{11}^{\prime\,-1},
    \ee
    \be
    \Delta_{1}^{1,1}=\tfrac{1}{4},\quad \Delta_{1}^{a,a}=\frac{\beta_1'}{4},\quad \Delta_{c}^{b,a}=\frac{\beta_1'}{4\beta_9'},\quad \Delta_{b}^{c,a}=\frac{\beta_1'}{4\beta_5'},
    \ee
    \be
    \Delta_{c}^{a,b}=\frac{\beta_6'}{4\beta_{10}'},\quad \Delta_{1}^{b,b}=\frac{\beta_6'}{4},\quad \Delta_{a}^{c,b}=\frac{\beta_6'}{4\beta_2'},
    \ee
    \be
    \Delta_{b}^{a,c}=\frac{\beta_{11}'}{4\beta_7'},\quad \Delta_{a}^{b,c}=\frac{\beta_{11}'}{\beta_3'},\quad \Delta_{1}^{c,c}=\frac{\beta_{11}'}{4},
    \ee
   thus giving the partition function
    \begin{align}  \label{eq:q8-orb-1abc}
    Z_{1+a+b+c} = \ & \frac{1}{4}\ls Z_{1,1}^1+Z_{1,a}^a+Z_{1,b}^b+Z_{1,c}^c+Z_{a,1}^a+Z_{a,a}^1+\frac{\beta_1'}{\beta_2'\beta_9'}Z_{a,b}^c+\frac{\beta_1'}{\beta_3'\beta_5'}Z_{a,c}^b \right. \\
    & \qquad\left. +Z_{b,1}^b+\frac{\beta_6'}{\beta_5'\beta_{10}'}Z_{b,a}^c+Z_{b,b}^1+\frac{\beta_6'}{\beta_2'\beta_7'}Z_{b,c}^a+Z_{c,1}^c+\frac{\beta_{11}'}{\beta_7'\beta_9'}Z_{c,a}^b+\frac{\beta_{11}'}{\beta_3'\beta_{10}'}Z_{c,b}^a+Z_{c,c}^1\rs.
    \nonumber
    \end{align}

We can rewrite the partition function as
\begin{eqnarray}
    Z_{1+a+b+c} & = & \frac{1}{4} \biggl[ 
    Z_{1,1}^1 + Z_{1,a}^a + Z_{a,1}^a + Z_{a,a}^1 + Z_{1,b}^b + Z_{b,1}^b + Z_{b,b}^1 + Z_{1,c}^c + Z_{c,1}^c + Z_{c,c}^1 
    \\
    & & \qquad
    + \frac{\beta_1'}{\beta_2'\beta_9'} \left( 
    Z_{a,b}^c+\frac{\beta_2'\beta_9'}{\beta_3'\beta_5'}Z_{a,c}^b+\frac{\beta_2'\beta_6'\beta_9'}{\beta_1'\beta_5'\beta_{10}'}Z_{b,a}^c+\frac{\beta_6'\beta_9'}{\beta_1'\beta_7'}Z_{b,c}^a+\frac{\beta_2'\beta_{11}'}{\beta_1'\beta_7'}Z_{c,a}^b+\frac{\beta_2'\beta_9'\beta_{11}'}{\beta_1'\beta_3'\beta_{10}'}Z_{c,b}^a
    \right) \biggr].
    \nonumber 
\end{eqnarray}
Each line above is separately modular invariant, using the results
computed in section~\ref{sect:q8:modtrans}, and so $Z_{1+a+b+c}$ is manifestly modular-invariant.

Just as in the Rep$(D_4)$ cases, we can formally turn on an analogue of discrete torsion, which results in the modular-invariant partition function
\begin{eqnarray}
    Z_{1+a+b+c} & = & \frac{1}{4} \biggl[ 
    Z_{1,1}^1 + Z_{1,a}^a + Z_{a,1}^a + Z_{a,a}^1 + Z_{1,b}^b + Z_{b,1}^b + Z_{b,b}^1 + Z_{1,c}^c + Z_{c,1}^c + Z_{c,c}^1 
    \\
    & & \qquad
    - \frac{\beta_1'}{\beta_2'\beta_9'} \left( 
    Z_{a,b}^c+\frac{\beta_2'\beta_9'}{\beta_3'\beta_5'}Z_{a,c}^b+\frac{\beta_2'\beta_6'\beta_9'}{\beta_1'\beta_5'\beta_{10}'}Z_{b,a}^c+\frac{\beta_6'\beta_9'}{\beta_1'\beta_7'}Z_{b,c}^a+\frac{\beta_2'\beta_{11}'}{\beta_1'\beta_7'}Z_{c,a}^b+\frac{\beta_2'\beta_9'\beta_{11}'}{\beta_1'\beta_3'\beta_{10}'}Z_{c,b}^a
    \right) \biggr].
    \nonumber 
\end{eqnarray}
This differs from the previous partition function by a sign on the second line.
As before, one cannot merely absorb the sign into the $\beta'$'s, as this also changes the partial traces $Z_{i,j}^k$.  Also as before, we do not claim to have checked multiloop factorization, so it is entirely possible that some choices of discrete torsion do not yield physically-sensible theories.  We leave a detailed analysis and first-principles understanding of such phases for future work.

\subsubsection{$H=1$: $1+a+b+c+2m$ orbifold}

Finally, we gauge the regular representation $1+a+b+c+2m=1+a+b+c+m_1+m_2$. We choose basis vectors for these irreps, where the basis vectors $e_{m_ij}$ carry two indices $1\leq i,j\leq 2$ to denote the $i$th basis vector of the $j$th copy of $m$ appearing in the regular representation.
\be
e=v_1+v_x+v_{x^2}+v_{x^3}+v_y+v_{xy}+v_{x^2y}+v_{x^3y},\quad e_a=v_1+v_x+v_{x^2}+v_{x^3}-v_y-v_{xy}-v_{x^2y}-v_{x^3y},
\ee
\be
e_b=v_1-v_x+v_{x^2}-v_{x^3}+v_y-v_{xy}+v_{x^2y}-v_{x^3y},\quad e_c=v_1-v_x+v_{x^2}-v_{x^3}-v_y+v_{xy}-v_{x^2y}+v_{x^3y},
\ee
\be
e_{m_11}=v_1+iv_x-v_{x^2}-iv_{x^3}+v_y+iv_{xy}-v_{x^2y}-iv_{x^3y},\quad e_{m_12}=v_1-iv_x-v_{x^2}+iv_{x^3}-v_y+iv_{xy}+v_{x^2y}-iv_{x^3y},
\ee
\be
e_{m_21}=v_1+iv_x-v_{x^2}-iv_{x^3}-v_y-iv_{xy}+v_{x^2y}+iv_{x^3y},\quad e_{m_22}=-v_1+iv_x+v_{x^2}-iv_{x^3}-v_y+iv_{xy}+v_{x^2y}-iv_{x^3y}.
\ee
By first computing the (co-)multiplications and then expanding in terms of the (co-)fusion basis as before, we compute the coefficients as
\be
\mu_{1,R}^R = \mu_{R,1}^R=1,\quad \Delta_R^{1,R}=\Delta_R^{1,R}=\tfrac{1}{8},
\ee
\be
\m_{a,a}^1=\beta_1^{\prime\,-1},\quad\m_{a,b}^c=\beta_2^{\prime\,-1},\quad\m_{a,c}^b=\beta_3^{\prime\,-1},\quad\m_{a,m_1}^{m_2}=\beta_4^{\prime\,-1},\quad\m_{a,m_2}^{m_1}=\beta_4^{\prime\,-1},\quad\m_{b,a}^c=\beta_5^{\prime\,-1},\quad\m_{b,b}^1=\beta_6^{\prime\,-1},
\ee
\be
\m_{b,c}^a=\beta_7^{\prime\,-1},\quad\m_{b,m_1}^{m_2}=-\beta_8^{\prime\,-1},\quad\m_{b,m_2}^{m_1}=\beta_8^{\prime\,-1},\quad\m_{c,a}^b=\beta_9^{\prime\,-1},\quad\m_{c,b}^a=\beta_{10}^{\prime\,-1},\quad\m_{c,c}^1=\beta_{11}^{\prime\,-1},
\ee
\be
\m_{c,m_1}^{m_1}=\beta_{12}^{\prime\,-1},\quad\m_{c,m_2}^{m_2}=-\beta_{12}^{\prime\,-1},\quad\m_{m_1,a}^{m_2}=\beta_{13}^{\prime\,-1},\quad\m_{m_1,b}^{m_2}=-\beta_{14}^{\prime\,-1},\quad\m_{m_1,c}^{m_1}=\beta_{15}^{\prime\,-1},\quad\m_{m_1,m_1}^a=\beta_{19}^{\prime\,-1},
\ee
\be
\m_{m_1,m_1}^b=\beta_{16}^{\prime\,-1},\quad\m_{m_1,m_2}^1=-\beta_{18}^{\prime\,-1},\quad\m_{m_1,m_2}^c=\beta_{17}^{\prime\,-1},\quad\m_{m_2,a}^{m_1}=\beta_{13}^{\prime\,-1},\quad\m_{m_2,b}^{m_1}=\beta_{14}^{\prime\,-1},\quad\m_{m_2,c}^{m_2}=-\beta_{15}^{\prime\,-1},
\ee
\be
\m_{m_2,m_1}^1=\beta_{18}^{\prime\,-1},\quad\m_{m_2,m_1}^c=\beta_{17}^{\prime\,-1},\quad\m_{m_2,m_2}^a=-\beta_{19}^{\prime\,-1},\quad\m_{m_2,m_2}^b=\beta_{16}^{\prime\,-1},
\ee
\be
\Delta_{1}^{a,a}=\frac{\beta_1'}{8},\quad \Delta_{c}^{b,a}=\frac{\beta_1'}{8\beta_9'},\quad \Delta_{b}^{c,a}=\frac{\beta_1'}{8\beta_5'},\quad \Delta_{m_2}^{m_1,a}=\frac{\beta_1'}{8\beta_{13}'},\quad \Delta_{m_1}^{m_2,a}=\frac{\beta_1'}{8\beta_{13}'},\quad \Delta_{c}^{a,b}=\frac{\beta_6'}{8\beta_{10}'},\quad \Delta_{1}^{b,b}=\frac{\beta_6'}{8}
\ee
\be
\Delta_{a}^{c,b}=\frac{\beta_6'}{8\beta_2'},\quad \Delta_{m_2}^{m_1,b}=\frac{\beta_6'}{8\beta_{14}'},\quad \Delta_{m_1}^{m_2,b}=-\frac{\beta_6'}{8\beta_{14}'},\quad \Delta_{b}^{a,c}=\frac{\beta_{11}'}{8\beta_7'},\quad \Delta_{a}^{b,c}=\frac{\beta_{11}'}{8\beta_3'},\quad \Delta_1^{c,c}=\frac{\beta_{11}'}{8,}
\ee
\be
\Delta_{m_1}^{m_1,c}=\frac{\beta_{11}'}{8\beta_{15}'},\quad \Delta_{m_2}^{m_2,c}=-\frac{\beta_{11}'}{8\beta_{15}'},\quad \Delta_{m_2}^{a,m_1} = \frac{\beta_{18}'}{8\beta_{19}'}, \quad \Delta_{m_2}^{b,m_1} =-\frac{\beta_{18}'}{8\beta_{16}'},\quad \Delta_{m_1}^{c,m_1}=-\frac{\beta_{18}'}{8\beta_{17}'},\quad \Delta_{a}^{m_1,m_1}=-\frac{\beta_{18}'}{8\beta_4'},
\ee
\be
\Delta_{b}^{m_1,m_1}=-\frac{\beta_{18}'}{8\beta_8},\quad \Delta_{1}^{m_2,m_1}=-\frac{\beta_{18}'}{8},\quad \Delta_{c}^{m_2,m_1}=\frac{\beta_{18}'}{8\beta_{12}'},\quad \Delta_{m_1}^{a,m_2}=\frac{\beta_{18}'}{8\beta_{19}'},\quad \Delta_{m_1}^{b,m_1}=\frac{\beta_{18}'}{8\beta_{16}'},\quad \Delta_{m_2}^{c,m_2}=\frac{\beta_{18}'}{8\beta_{17}'},
\ee
\be
\Delta_1^{m_1,m_2}=\frac{\beta_{18}'}{8},\quad \Delta_c^{m_1,m_2}=\frac{\beta_{18}'}{8\beta_{12}'},\quad \Delta_a^{m_2,m_2}=\frac{\beta_{18}'}{8\beta_{4}'},\quad \Delta_b^{m_2,m_2}=-\frac{\beta_{18}'}{8\beta_{8}'}.
\ee
Some multiplication coefficients are zero (e.g.~$\m_{a,m_1}^{m_1}$) and we have omitted these.  Note also that a difference in this example is that the identity occurs in the multiplication of $m_1$ with $m_2$, not in multiplication of $m_1$ with $m_1$ or $m_2$ with $m_2$.
The partition function is then
\begin{align}
%\label{d4_orb_1abc2m} 
\label{eq:q8-orb-1abc2m}
    Z_{1+a+b+c+2m}=\ & \frac{1}{8}\ls Z_{1,1}^1+Z_{1,a}^a+Z_{1,b}^b+Z_{1,c}^c+2Z_{1,m}^m+Z_{a,1}^a+Z_{a,a}^1+\frac{\beta_1'}{\beta_2'\beta_9'}Z_{a,b}^c+\frac{\beta_1'}{\beta_3'\beta_5'}Z_{a,c}^b\right.\non\\
    & \qquad\left. +\frac{2\beta_1'}{\beta_4'\beta_{13}'}Z_{a,m}^m+Z_{b,1}^b+\frac{\beta_6'}{\beta_5'\beta_{10}'}Z_{b,a}^c+Z_{b,b}^1+\frac{\beta_6'}{\beta_2'\beta_7'}Z_{b,c}^a-\frac{2\beta_6'}{\beta_8'\beta_{14}'}Z_{b,m}^m\right.\non\\
    & \qquad\left. +Z_{c,1}^c+\frac{\beta_{11}'}{\beta_7'\beta_9'}Z_{c,a}^b+\frac{\beta_{11}'}{\beta_3'\beta_{10}'}Z_{c,b}^a+Z_{c,c}^1+\frac{2\beta_{11}'}{\beta_{12}'\beta_{15}'}Z_{c,m}^m\right.\non\\
    & \qquad\left. +2Z_{m,1}^m+\frac{2\beta_{18}'}{\beta_{13}'\beta_{19}'}Z_{m,a}^m+\frac{2\beta_{18}'}{\beta_{14}'\beta_{16}'}Z_{m,b}^m-\frac{2\beta_{18}'}{\beta_{15}'\beta_{17}'}Z_{m,c}^m\right.\non\\
    & \qquad\left. +2Z_{m,m}^1-\frac{2\beta_{18}'}{\beta_4'\beta_{19}'}Z_{m,m}^a -\frac{2\beta_{18}'}{\beta_8'\beta_{16}'}Z_{m,m}^b+\frac{2\beta_{18}'}{\beta_{12}'\beta_{17}'}Z_{m,m}^c\rs.
\end{align}

To check modular invariance, we rewrite the partition function above in the form
\begin{eqnarray}
    Z_{1+a+b+c+2m} & = & \frac{1}{8} \biggl[
    \left( Z_{1,1}^1 + Z_{1,a}^a + Z_{a,1}^a + Z_{a,a}^1 + Z_{1,b}^b + Z_{b,1}^b + Z_{b,b}^1 + Z_{1,c}^c + Z_{c,1}^c + Z_{c,c}^1 \right)
    \\
    & & \qquad
    + \frac{\beta_1'}{\beta_2'\beta_9'} \left(
     Z_{a,b}^c + \frac{\beta_2'\beta_9'}{\beta_3'\beta_5'} Z_{a,c}^b + \frac{\beta_2'\beta_6'\beta_9'}{\beta_1'\beta_5'\beta_{10}'} Z_{b,a}^c + \frac{\beta_6'\beta_9'}{\beta_1'\beta_7'} Z_{b,c}^a + \frac{\beta_2'\beta_{11}'}{\beta_1'\beta_7'} Z_{c,a}^b + \frac{\beta_2'\beta_9'\beta_{11}'}{\beta_1'\beta_3'\beta_{10}'} Z_{c,b}^a
    \right)
    \nonumber \\
     & & \qquad 
     + 2 \biggl( 
     Z_{1,m}^m + \frac{\beta_1'}{\beta_4'\beta_{13}'} Z_{a,m}^m - \frac{\beta_6'}{\beta_8'\beta_{14}'} Z_{b,m}^m 
     + \frac{\beta_{11}'}{\beta_{12}'\beta_{15}'}Z_{c,m}^m + Z_{m,1}^m + \frac{\beta_{18}'}{\beta_{13}'\beta_{19}'}Z_{m,a}^m
     \nonumber \\
     & & \qquad \qquad \qquad 
     + \frac{\beta_{18}'}{\beta_{14}'\beta_{16}'} Z_{m,b}^m - \frac{\beta_{18}'}{\beta_{15}'\beta_{17}'}Z_{m,c}^m + Z_{m,m}^1  - \frac{\beta_{18}'}{\beta_4'\beta_{19}'}Z_{m,m}^a - \frac{\beta_{18}'}{\beta_8'\beta_{16}'}Z_{m,m}^b 
     \nonumber \\
     & & \qquad \qquad \qquad \qquad 
     + \frac{\beta_{18}'}{\beta_{12}'\beta_{17}'}Z_{m,m}^c
     \biggr) \biggr].
     \nonumber
\end{eqnarray}
Each of the three quantities enclosed in parantheses above is separately modular-invariant, from the
results of section~\ref{sect:q8:modtrans}, hence $Z_{1+a+b+c+2m}$ is explicitly modular-invariant.

As in the Rep$(D_4)$ case, we can turn on an analogue of discrete torsion in the ${\mathbb Z}_2 \times {\mathbb Z}_2$ subalgebra, essentially by multiplying the second line of modular invariants by $-1$, resulting in the modular-invariant partition function
\begin{eqnarray}
    Z_{1+a+b+c+2m, {\rm dt}} & = & \frac{1}{8} \biggl[
    \left( Z_{1,1}^1 + Z_{1,a}^a + Z_{a,1}^a + Z_{a,a}^1 + Z_{1,b}^b + Z_{b,1}^b + Z_{b,b}^1 + Z_{1,c}^c + Z_{c,1}^c + Z_{c,c}^1 \right)
    \\
    & & \qquad
    - \frac{\beta_1'}{\beta_2'\beta_9'} \left(
     Z_{a,b}^c + \frac{\beta_2'\beta_9'}{\beta_3'\beta_5'} Z_{a,c}^b + \frac{\beta_2'\beta_6'\beta_9'}{\beta_1'\beta_5'\beta_{10}'} Z_{b,a}^c + \frac{\beta_6'\beta_9'}{\beta_1'\beta_7'} Z_{b,c}^a + \frac{\beta_2'\beta_{11}'}{\beta_1'\beta_7'} Z_{c,a}^b + \frac{\beta_2'\beta_9'\beta_{11}'}{\beta_1'\beta_3'\beta_{10}'} Z_{c,b}^a
    \right)
    \nonumber \\
     & & \qquad 
     + 2 \biggl( 
     Z_{1,m}^m + \frac{\beta_1'}{\beta_4'\beta_{13}'} Z_{a,m}^m - \frac{\beta_6'}{\beta_8'\beta_{14}'} Z_{b,m}^m 
     + \frac{\beta_{11}'}{\beta_{12}'\beta_{15}'}Z_{c,m}^m + Z_{m,1}^m + \frac{\beta_{18}'}{\beta_{13}'\beta_{19}'}Z_{m,a}^m
     \nonumber \\
     & & \qquad \qquad \qquad 
     + \frac{\beta_{18}'}{\beta_{14}'\beta_{16}'} Z_{m,b}^m - \frac{\beta_{18}'}{\beta_{15}'\beta_{17}'}Z_{m,c}^m + Z_{m,m}^1  - \frac{\beta_{18}'}{\beta_4'\beta_{19}'}Z_{m,m}^a - \frac{\beta_{18}'}{\beta_8'\beta_{16}'}Z_{m,m}^b 
     \nonumber \\
     & & \qquad \qquad \qquad \qquad 
     + \frac{\beta_{18}'}{\beta_{12}'\beta_{17}'}Z_{m,m}^c
     \biggr) \biggr].
     \nonumber
\end{eqnarray}
As before, such a sign cannot be simply absorbed into the $\beta'$'s without also modifying the partial traces $Z_{i,j}^k$.  A closely related choice will also appear in the Rep$({\cal H}_8)$
analysis in section~\ref{sect:gauging-reph8-ising}, which will make it clear that there really does exist a physically-distinct discrete-torsion-like degree of freedom here.  As noted earlier, we have not checked e.g.~multiloop factorization, so not all choices of discrete torsion may be sensible.  We leave a detailed first-principles analysis for the future.

\subsection{$\text{Rep}({\cal H}_8)$}\label{ssec:reph8}

In this subsection, we gauge a $\text{Rep}({\cal H}_8)$ symmetry, the remaining $TY(\Z_2\times\Z_2)$ category, where ${\cal H}_8$ is the eight-dimensional Kac-Paljutkin Hopf algebra \cite{kp66}. 
Unlike the previous examples, in this example we will not compute the most intertwiners and so forth, but rather
will use existing relations for crossing kernels.  (See appendix~\ref{app:ty} for a summary of existing results and implied modular transformations.)

The Hopf algebra ${\cal H}_8$ has generators $\{x,y,z\}$ satisfying the relations
\begin{eqnarray}
    x^2=y^2=z^2=1;& xz=zx;& zy=yz; \\ xyz=yx & &
\end{eqnarray}
so that ${\cal H}_8$ is spanned by basis elements
\begin{equation}\label{eq:h8span}
    H_8=\text{Span}(1,x,y,z,xy,xz,yz,yx).
\end{equation}
The comultiplication is
\begin{eqnarray}
    \Delta(x)&=& xe_0\otimes x+xe_1\otimes y=\frac{1}{2}(x\otimes x+xz\otimes x+x\otimes y-xz\otimes y)\\
    \Delta(y) &=& ye_1\otimes x + ye_0\otimes y=\frac{1}{2}(y\otimes x-yz\otimes x+y\otimes y+yz\otimes y)\\
    \Delta(z) &=& z\otimes z\\
    \Delta(xy)&=&xye_0\otimes xy+xye_1\otimes yx=\frac{1}{2}(xy\otimes xy+yx\otimes xy+xy\otimes yx-yx\otimes yx)\\
    \Delta(xz)&=&xe_0\otimes xz-xe_1\otimes yz=\frac{1}{2}(x\otimes xz+xz\otimes xz-x\otimes yz+xz\otimes yz)\\
    \Delta(yz)&=&ye_0\otimes yz-ye_1\otimes xz=\frac{1}{2}(y\otimes yz+yz\otimes yz-y\otimes xz+yz\otimes xz)\\
    \Delta(yx)&=&yxe_0\otimes yx+yxe_1\otimes xy=\frac{1}{2}(yx\otimes yx+xy\otimes yx+yx\otimes xy-xy\otimes xy)
\end{eqnarray}
where the elements $e_0,e_1$ are two central orthogonal idempotents defined as
\begin{eqnarray}
    e_0 &=& \frac{1}{2}(1+z)\\
    e_1 &=& \frac{1}{2}(1-z)
\end{eqnarray}
and counit $u^o(x)=u^o(y)=u^o(z)=1$. There is also an antipode map defined as
\begin{eqnarray}
    S(x)&=&xe_0+ye_1=\frac{1}{2}(x+xz+y-yz), \\
    S(y)&=&xe_1+ye_0=\frac{1}{2}(x-xz+y+yz), \\
    S(z)&=&z, \\
    S(xy)&=&yx, \\
    S(xz)&=& \frac{1}{2}(xz+yz+x-y), \\
    S(yz)&=&\frac{1}{2}(-x+y+yz+xz), \\
    S(yx)&=& xy.
\end{eqnarray}

We now want to endow ${\cal H}_8^*$ with the structure of a symmetric special Frobenius algebra. In principle, we use a multiplication $\mu_*$ and unit $u_*$ inherited from ${\cal H}_8^*$, and then add a comultiplication $\Delta_F$ (not directly inherited from the Hopf algebra) and counit $u_F^o$ (also not directly inherited).  It is important to note that ${\cal H}_8$ is a self-dual algebra \cite{Alaoui03}, and a basis is given by linear functions $v_a(b)=\delta_{a,b}$ where $a,b$ are elements of the basis from Equation~(\ref{eq:h8span}). 

First, we describe the algebra structure $(\mu_*, u_*)$ on ${\cal H}_8^*$ with such a basis. It is inherited from the coalgebra structure on ${\cal H}_8$. The unit $u_*:\C\to {\cal H}_8^*$ is
\begin{eqnarray}
    u_*(1)&=& v_1+v_x+v_y+v_z+v_{xy}+v_{xz}+v_{yz}+v_{yx}.
\end{eqnarray}
The Frobenius product $\mu_*(v_a \otimes v_b) = %v_a\otimes v_b=
\Delta^*(v_a\otimes v_b)$:
\begin{eqnarray}
    \mu_*(v_1 \otimes v_b) &=& \delta_{1,b}v_1,
    \\
    \mu_*(v_x \otimes v_b) &=& \frac{1}{2}\left(\delta_{x,b}v_x+\delta_{y,b}v_x+\delta_{xz,b}v_{xz}-\delta_{yz,b}v_{xz}\right),
    \\
    \mu_*(v_y \otimes v_b) &=& \frac{1}{2}\left(\delta_{x,b}v_y+\delta_{y,b}v_y+\delta_{yz,b}v_{yz}-\delta_{xz,b}v_{yz}\right),
    \\
    \mu_*(v_z \otimes v_b) &=& \delta_{z,b}v_z,
    \\
    \mu_*(v_{xy} \otimes v_b) &=& \frac{1}{2}\left(\delta_{xy,b}v_{xy}+\delta_{yx,b}v_{xy}+\delta_{yx,b}v_{yx}-\delta_{xy,b}v_{yx}\right),
    \\
    \mu_*(v_{xz} \otimes v_b) &=& \frac{1}{2}\left(\delta_{xz,b}v_{xz}+\delta_{yz,b}v_{xz}+\delta_{x,b}v_x-\delta_{y,b}v_x\right),
    \\
    \mu_*(v_{yz} \otimes v_b)&=&\frac{1}{2}\left(-\delta_{x,b}v_y+\delta_{y,b}v_y+\delta_{yz,b}v_{yz}+\delta_{xz,b}v_{yz}\right),
    \\
    \mu_*(v_{yx} \otimes v_b) &=& \frac{1}{2}\left(\delta_{xy,b}v_{xy}-\delta_{yx,b}v_{xy}+\delta_{yx,b}v_{yx}+\delta_{xy,b}v_{yx}\right).
\end{eqnarray}

As described in Section~\ref{sec:general}, any finite-dimensional Hopf algebra ${\cal H}$ can be endowed with a Frobenius algebra structure $(\mu_*, u_*, \Delta_F, u_F^o)$ by using integral elements of ${\cal H}$ and ${\cal H}^*$. An integral element of ${\cal H}_8$ is
\cite{Burciu17}
\begin{eqnarray}
    \Lambda &=& 1+x+y+z+xy+xz+yz+yx,
\end{eqnarray}
which serves as the counit via evaluation. An integral element of ${\cal H}_8^*$ is simply
\begin{eqnarray}
    \lambda &=& v_1.
\end{eqnarray}
Then, in terms of the present basis, the counit is
\begin{eqnarray}
    u^o_F: & {\cal H}^*\to \C\\
    & v_a\mapsto v_a(\Lambda)=1
\end{eqnarray}

The comultiplication is computed as the dual map~(\ref{eq:DeltaF-defn}), namely
\begin{eqnarray}
    \Delta_F:= (\text{Id}_{\cal H}\otimes (\lambda\circ\mu)\circ(\text{Id}_{\cal H}\otimes S\otimes\text{Id}_{\cal H})\circ(\Delta\circ\text{Id}_{\cal H}))^*.
\end{eqnarray}
For the dual basis, this comultiplication is
\begin{eqnarray}
    \Delta_F(v_1) &=& v_1\otimes v_1, \\
    \Delta_F(v_x) &=& \frac{1}{2}(v_x\otimes v_x+v_x\otimes v_y -v_{xz}\otimes v_y+v_{xz}\otimes v_x), \\
    \Delta_F(v_y) &=& \frac{1}{2}(v_y\otimes v_x+v_y\otimes v_y-v_{yz}\otimes v_x+v_{yz}\otimes v_y), \\
    \Delta_F(v_z) &=& v_z\otimes v_z\\
    \Delta_F(v_{xy}) &=& \frac{1}{2}(v_{xy}\otimes v_{xy}+v_{xy}\otimes v_{yx}+v_{yx}\otimes v_{xy}-v_{yx}\otimes v_{yx}), \\
    \Delta_F(v_{xz})&=&\frac{1}{2}(v_x\otimes v_{xz}-v_x\otimes v_{yz}+v_{xz}\otimes v_{xz}+v_{xz}\otimes v_{yz}), \\
    \Delta_F(v_{yz}) &=& \frac{1}{2}(v_y\otimes v_{yz}-v_y\otimes v_{xz}+v_{yz}\otimes v_{yz}+v_{yz}\otimes v_{xz}), \\
    \Delta_F(v_{yx}) &=& \frac{1}{2}(-v_{xy}\otimes v_{xy}+v_{xy}\otimes v_{yx}+v_{yx}\otimes v_{xy}+v_{yx}\otimes v_{yx}),
\end{eqnarray}
reflecting the fact that $\Delta_F$ is diagonal on the $v$ basis.
One can readily check that
\begin{eqnarray}
    \mu_*\circ\Delta_F = \text{Id}_{{\cal H}^*},\\
    u_*^o\circ u_* = \text{dim}({\cal H}_8^*)\ \text{Id}_{\C}=8\ \text{Id}_{\C}.
\end{eqnarray}

The Hopf algebra ${\cal H}_8$ has five irreducible representations \cite{Burciu17}, four of which are one-dimensional.
The trivial representation $1$ has actions
\begin{eqnarray}
    \rho_1(x)v=\rho_1(y)v=\rho_1(z)v=v.
\end{eqnarray}
The other three one-dimensional irreps are labeled as $a,b,c$ and form a Klein group. The actions are defined as
\begin{eqnarray}
    \rho_a(x)v=-v; &\rho_a(y)v=v;& \rho_a(z)v=v, \\
    \rho_b(x)v=v; &\rho_b(y)v=-v;& \rho_b(z)v=v, \\
    \rho_c(x)v=-v; &\rho_c(y)v=-v;& \rho_c(z)v=v.
\end{eqnarray}
There is also a two-dimensional irreducible representation, which we denote as $m$. For $v_1, v_2$ the generators of the underlying vector space, the actions are defined as follows
\begin{eqnarray}
    \rho_a(x)v_1=v_1; &\rho_a(y)v_1=v_2;& \rho_a(z)v_1=-v_1, \\
    \rho_a(x)v_2=-v_2; &\rho_a(y)v_2=v_1;& \rho_a(z)v_2=-v_2.
\end{eqnarray}

Gauging the whole category corresponds to gauging by the
regular representation $R= 1+a+b+c+2m$. As before, we choose a suitable basis in ${\cal H}_8^*$:
\begin{eqnarray}
    e_1 &=& v_1+v_z+v_x+v_y+v_{xy}+v_{xz}+v_{yz}+v_{yx},\\
    e_a &=& v_1 + v_z - v_x + v_y - v_{xy} - v_{xz}+ v_{yz}- v_{yx},\\
    e_b &=& v_1 + v_z + v_x - v_y - v_{xy} + v_{xz}- v_{yz}- v_{yx},\\
    e_c &=& v_1 + v_z - v_x - v_y + v_{xy} - v_{xz}- v_{yz}+ v_{yx}, \\
    e_{m_11} &=& v_1-v_z+v_x+v_y+v_{xy}-v_{yx}-v_{xz}-v_{yz},\\
    e_{m_12} &=& v_1-v_z-v_x+v_y-v_{xy}+v_{yx}+v_{xz}-v_{yz},\\
    e_{m_21} &=& v_1-v_z+v_x-v_y-v_{xy}+v_{yx}-v_{xz}+v_{yz},\\
    e_{m_22} &=& -v_1+v_z+v_x+v_y-v_{xy}+v_{yx}-v_{xz}-v_{yz}.
\end{eqnarray}

As expected, the products of $\{e_a,e_b,e_c\}$ are cyclic. The product table is
\begin{center}
    \begin{tabular}{c|c|c|c|c|c|c|c}
     &  $e_a$& $e_b$& $e_c$ & $e_{m_11}$ & $e_{m_12}$ & $e_{m_21} $& $e_{m_22}$ \\ \hline
   $e_a$  & $e_1$ & $e_c$ & $e_b$ & $e_{m_12}$ & $e_{m_11}$ & $-e_{m_22}$ & $-e_{m_21}$\\
    $e_b$ & $e_c$ & $e_1$ & $e_a$ & $e_{m_21}$ & $-e_{m_22}$ & $e_{m_11}$ & $-e_{m_12}$\\
    $e_c$ & $e_b$ & $e_a$& $e_1$& $-e_{m_22}$ & $e_{m_21}$ & $e_{m_12}$ & $-e_{m_11}$\\
   $e_{m_11}$  & $e_{m_21}$ & $e_{m_12}$ &$-e_{m_22}$ & $e_1$ & $e_b$ & $e_a$ & $-e_c$\\
    $e_{m_12}$ & $-e_{m_22}$ & $e_{m_11}$ & $e_{m_21}$& $e_a$ & $e_c$ & $e_1$ & $-e_b$\\
    $e_{m_21}$ & $e_{m_11}$ & $-e_{m_22}$ &$e_{m_12}$ & $e_b$ & $e_1$ & $e_c$ & $-e_a$\\
    $e_{m_22}$ & $-e_{m_12}$ & $-e_{m_21}$ &$-e_{m_11}$ & $-e_c$ & $-e_a$& $-e_b$& $e_1$
\end{tabular}
\end{center}
where the row elements multiply from the left and the column elements multiply from the right. Notice that, unlike the other examples considered so far, this algebra is non-commutative. For example, from the table above we have
\begin{eqnarray*}
    \mu_*(e_a \otimes e_{m_11}) &=& e_{m_12}\\
    \mu_*(e_{m_11} \otimes e_a) &=& e_{m_21}.
\end{eqnarray*}
As described below, this becomes important when computing the partial traces appearing in the partition function of the regular representation, since, just as for group-like orbifolds, this non-commutativity does not allow a consistent insertion of $a$ and $m$ defects on the nontrivial $2$-torus cycles.

We can also compute the coproduct of the unit as
\begin{eqnarray}
    \Delta_F(u(1))&=&\frac{1}{8}\big(e_1\otimes e_1 + e_a\otimes e_a + e_b\otimes e_b + e_c\otimes e_c  + e_{m_11}\otimes e_{m_11} + e_{m_12}\otimes e_{m_21}\\ 
    & &  \qquad \qquad  + e_{m_21}\otimes e_{m_12}+e_{m_22}\otimes e_{m_22}\big).
\end{eqnarray}

Next, we briefly discuss the intertwiners relating basis elements for the representations $1, a, b, c, m$,
which we will denote $e_1, e_a, e_b, e_c, e_{m1}, e_{m2}$.
The intertwiners are listed
\begin{center}
\begin{tabular}{c|c|c|c|c|c}
     & $e_a$ & $e_b$ & $e_c$ & $e_{m1}$ & $e_{m2}$ \\ \hline
    $e_a$ & $e_1$ & $e_c$ & $e_b$ & $e_{m1}$ & $- e_{m2}$\\
    $e_b$ & $e_c$ & $e_1$ & $e_a$ & $e_{m2}$ & $e_{m1}$ \\
    $e_c$ & $e_b$ & $e_a$ &  $e_1$ & $- e_{m2}$ & $e_{m1}$\\
    $e_{m1}$ & $e_{m1}$ & $e_{m2}$ & $e_{m2}$ & $e_1 + e_b$ & $e_a + e_c$ \\
    $e_{m2}$ & $- e_{m2}$ & $e_{m1}$ & $- e_{m1}$ & $e_a-e_c$& $e_1-e_b$ 
\end{tabular}
\end{center}
which are to be read in the same order as the product table, for example
\begin{eqnarray*}
    e_{m1} \otimes e_{m2} &=& e_a + e_c,\\
    e_{m2} \otimes e_{m1}  &=& e_a - e_c.
\end{eqnarray*}
For this example, we do not try to compute general intertwiners.

We now gauge the regular representation, the symmetric special Frobenius algebra corresponding to
$1+a+b+c+2m$.  

An important difference between this case and the other two gaugeable $\Z_2\times\Z_2$ TY fusion categories analyzed previously is that in this case this algebra object is non-commutative, hence on general grounds we do not expect all the partial traces that appeared previously to be admissible here.

This is indeed the case for the partial traces involving the noninvertible object $m$ together with $a$ or $b$. Notice for example that $e_a e_{m_11} = e_{m_12}$, so that we would expect a partial trace $Z_{a,m_1}^{m_1}$. However, one has that
\begin{eqnarray}
    a\otimes m_1\xrightarrow{\mu_*} m_1\xrightarrow{\Delta_F} m_2\otimes a\\
    e_a e_{m_11}\mapsto e_{m_12}\mapsto -e_{m_22}e_a
\end{eqnarray}
and similarly
\begin{eqnarray}
    b\otimes m_1\xrightarrow{\mu_*} m_2\xrightarrow{\Delta_F} m_2\otimes b\\
    e_b\otimes e_{m_11}\mapsto e_{m_21}\mapsto -e_{m_22}e_b,
\end{eqnarray}
which suggests that the partial traces involving $a$ or $b$ along with $m$ do not appear this time, in contrast with $\text{Rep}(D_4)$ and $\text{Rep}(Q_8)$. Indeed, one can proceed as before and compute the coefficients $\mu_{L_1,L_2}^{L_3},\Delta_{L_3}^{L_2,L_1}$:
\be
\mu_{a,b}^c=1 \ \text{(cyclic)},\quad \mu_{1,m_1}^{m_1}=1,\quad \mu_{1,m_2}^{m_2}=1,\quad \mu_{m_1,1}^{m_1}=1,\quad \mu_{m_2,1}^{m_2}=1, \mu_{m_1,m_1}^1=1,\quad \mu_{m_2,m_2}^1=1,
\ee
\be
\mu_{c,m_1}^{m_2}=1, \quad \mu_{c,m_2}^{m_1}=1, \quad \mu_{m_1,c}^{m_2}=-1,\quad \mu_{m_2,c}^{m_1}=1,\quad \mu_{m_1,m_2}^c=-1,\quad \mu_{m_2,m_1}^c=1,
\ee
\be
\Delta_{a}^{b,c}=\tfrac{1}{8} \ \text{(cyclic)},\quad \Delta_{m_1}^{m_1,1}=\tfrac{1}{8}, \quad \Delta_{m_2}^{m_2,1}=\tfrac{1}{8},\quad \Delta_{m_1}^{1,m_1}=\tfrac{1}{8},\quad \Delta_{m_2}^{1,m_2}=\tfrac{1}{8},\Delta_1^{m_1,m_1}=\tfrac{1}{8}, \quad \Delta_1^{m_2,m_2}=\tfrac{1}{8},
\ee
\be
\Delta_{m_2}^{m_1,c}=\tfrac{1}{8}, \quad \Delta_{m_1}^{m_2,c}=\tfrac{1}{8}, \quad \Delta_{m_2}^{c,m_1}=-\tfrac{1}{8}, \quad \Delta_{m_1}^{c,m_2}=\tfrac{1}{8},\quad \Delta_c^{m_2,m_1}=-\tfrac{1}{8},\quad \Delta_c^{m_1,m_2}=\tfrac{1}{8},
\ee
where all others vanish. In particular, this confirms that the partial traces $Z_{a,m}^m$, $Z_{m,a}^m$, $Z_{m,b}^m$, $Z^m_{b,m}$, $Z_{m,m}^a$, and $Z_{m,m}^b$ do not appear in the partition function.

With these coefficients, we get the partition function
\begin{eqnarray}  \label{eq:h8-z}
    Z&=& \frac{1}{8}\left(\left(\sum_{g,h\in\{1,a,b,c\}}Z_{g,h}^{gh}\right) + 2\left(Z_{1,m}^m+Z_{m,1}^m+Z_{c,m}^m+Z_{m,c}^m+Z_{m,m}^1+Z_{m,m}^c\right)\right).
\end{eqnarray}

This result of the torus partition function for gauging the whole $\text{Rep}(\mathcal{H}_8)$ symmetry appeared in \cite[Eq. 3.15]{CLS23} while the present paper was in preparation.

In principle one can also construct Frobenius algebras associated with other Hopf ideals,
which correspond to \cite[appendix D.2]{buerschaper-thesis}
\begin{equation}
    1, \: \: \:
    1+a, \: \: \:
    1+b, \: \: \:
    1+c, \: \: \:
    1+a+b+c,
\end{equation}
just as we did for Rep$(S_3)$, Rep($D_4$), and Rep($Q_8$) in previous sections.
However, for reasons of brevity, we defer such discussions in Rep(${\cal H}_8$) to future work.

In passing, just as we have seen in previous examples, it is also possible to create a new theory by turning on
an analogue of discrete torsion, which specifically will be determined by the discrete torsion in an
ordinary ${\mathbb Z}_2 \times {\mathbb Z}_2$ orbifold.  If we let $\omega$ denote a cocycle representing the
nontrivial element of $H^2({\mathbb Z}_2 \times {\mathbb Z}_2,U(1)) = {\mathbb Z}_2$, 
and define
\begin{equation}
    \epsilon(g,h) \: = \: \frac{\omega(g,h)}{\omega(h,g)},
\end{equation}
then the second theory
has partition function
\begin{equation} \label{eq:h8-z-dt}
Z \: = \: \frac{1}{8}\left(\left(\sum_{g,h\in\{1,a,b,c\}} \epsilon(g,h) Z_{g,h}^{gh}\right) + 2\left(Z_{1,m}^m+Z_{m,1}^m+Z_{c,m}^m+Z_{m,c}^m+Z_{m,m}^1+Z_{m,m}^c\right)\right).
\end{equation}
Later in section~\ref{sect:gauging-reph8-ising}, we will see that these two gauged Rep$({\cal H}_8)$ theories, with and without this analogue of discrete torsion, are physically distinct.  As noted earlier, we have not checked e.g.~multiloop factorization, so it is possible that some choices of discrete torsion in some theories may not be physically consistent.
Furthermore, and also as noted earlier, we do not have a first-principles understanding of analogues of discrete torsion in gauged noninvertible symmetries.  We hope to return to this topic in future work.

\subsection{Summary of results}
Here, for ease of reference, we will compile the partition functions computed in this section.  While most of these computations were done with arbitrary coefficients $\beta_i$ appearing in the intertwinters, here we will specialize to the choices which appear most often in the literature.

\subsubsection{Rep$(S_3)$ partition functions}

We can match conventions in the existing literature for Rep$(S_3)$ by choosing
\be
\beta_2 = \beta_4 = \beta_5 = 1,\quad \beta_1 = \beta_3 = \beta_6 =-1
\ee
for the coefficients appearing in (\ref{rs3int1})-(\ref{rs3int2}).  Then all components of the associator are trivial except for
\be
\label{eq:ScalarMinusOnes}
\tilde{K}^{X,Y}_{Y,Y}(Y,Y) = \tilde{K}^{Y,Y}_{X,Y}(Y,Y) = \tilde{K}^{Y,Y}_{Y,X}(Y,Y) = \tilde{K}^{Y,X}_{Y,Y}(Y,Y)=-1.
\ee
and
\be
\tilde{K}^{Y,Y}_{Y,Y} = \lp\begin{matrix} \hlf & \hlf & \hlf \\ \hlf & \hlf & -\hlf \\ 1 & -1 & 0 \end{matrix}\rp.
\ee
Comparing with \cite[(6.17)]{Chang:2018iay}, this matches their $\om=1$ result except for the components in (\ref{eq:ScalarMinusOnes}) which they don't mention as differing from one.  These signs do appear in \cite{Barter_2022}, in which the calculation is also done in a general gauge.

With such a choice, the Rep$(S_3)$ partition functions of section~\ref{ssec:reps3} are
\begin{align}
Z_{1+X}&= \frac{1}{2}\bigg[ Z_{1,1}^1+(Z_{1,X}^X+Z_{X,1}^X+Z_{X,X}^1)\bigg],\\
Z_{1+Y}&=\frac{1}{3}\ls
Z_{1,1}^1+\left(Z_{1,Y}^Y+Z_{Y,1}^Y+Z_{Y,Y}^1+\frac{1}{2}Z_{Y,Y}^Y\right)\rs,\\
Z_{1+X+2Y} & = \frac{1}{6} \biggl[  Z_{1,1}^1 \: + \: \left( Z_{1,X}^X + Z_{X,1}^X + Z^1_{X,X} \right) \: + \: 2 \left( Z_{1,Y}^Y + Z_{Y,1}^Y + Z_{Y,Y}^1 +  \frac{1}{2} Z_{Y,Y}^Y \right)
    \\\nonumber
    &  \hspace*{0.5in}
    \: - \:
    2 \left(  Z^Y_{X,Y} +
     Z^Y_{Y,X} + Z_{Y,Y}^X - \frac{1}{2} Z_{Y,Y}^Y \right) \biggr].
\end{align}

\subsubsection{Rep$(D_4)$ partition functions}

We will choose the intertwiner coefficients appearing in (\ref{eq:intertwinerd4})-(\ref{eq:intertwinerd4-2}) to be
\be
\beta_1=\beta_4^2,\quad \beta_3=\frac{\beta_4^2}{\beta_2},\quad \beta_5=-\beta_2,\quad \beta_6=\beta_4^2,\quad \beta_7=-\frac{\beta_4^2}{\beta_2},\quad \beta_8=\pm \beta_4,\quad \beta_9=-\frac{\beta_4^2}{\beta_2},\non
\ee
\be
\beta_{10}=\frac{\beta_4^2}{\beta_2},\quad \beta_{11}=-\frac{\beta_4^4}{\beta_2^2},\quad \beta_{12}=\mp\frac{\beta_4^2}{\beta_2},\quad \beta_{13}=-\beta_4,\quad \beta_{14}=\pm \beta_4,\quad \beta_{15}=\mp\frac{\beta_4^2}{\beta_2},\non
\ee
\be
\label{eq:StandardD4Coeffs}
\beta_{17}=\frac{\beta_2\beta_{16}}{\beta_4},\quad \beta_{18}=\pm \beta_4 \beta_{16},\quad \beta_{19}=\pm \beta_{16}.
\ee
Then all components of the associator are equal to $+1$ except for
\be
\tilde{K}^{m,i}_{j,m}(m,m) = \tilde{K}^{i,m}_{m,j}(m,m) = \chi(i,j),
\qquad\tilde{K}^{m,m}_{m,m}(i,j)=\hlf\chi(i,j),
\ee
where $i$ and $j$ run over $1,a,b,c$ and
\be
\chi(i,j)=\lp\begin{matrix} 1 & 1 & 1 & 1 \\ 1 & 1 & -1 & -1 \\ 1 & -1 & 1 & -1 \\ 1 & -1 & -1 & 1 \end{matrix}\rp,
\ee
is the non-trivial bi-character for $\Z_2^2$.  This matches the standard Tambara-Yamagami associator \cite{ty}.

The various Rep$(D_4)$ partition functions from section~\ref{ssec:repd4} are then
\begin{align}
Z_{1+a}&=\frac{1}{2}\bigg[Z_{1,1}^1 + (Z_{1,a}^a + Z_{a,1}^a + Z^1_{a,a})\bigg],\\
Z_{1+b}&=\frac{1}{2}\bigg[ Z_{1,1}^1+(Z_{1,b}^b+Z_{b,1}^b+Z_{b,b}^1) \bigg],\\
 Z_{1+a+b+c} & = \frac{1}{4} \biggl[ 
     Z_{1,1}^1 + (Z_{1,a}^a + Z_{a,1}^a + Z_{a,a}^1) + (Z_{1,b}^b + Z_{b,1}^b + Z_{b,b}^1) + (Z_{1,c}^c + Z_{c,1}^c + Z_{c,c}^1)
    \\\nonumber
     & \qquad 
     + \left( Z_{a,b}^c+Z_{a,c}^b+Z_{b,a}^c+Z_{b,c}^a+Z_{c,a}^b+Z_{c,b}^a
     \right) \biggr],\\
Z_{1+b+m} & = \frac{1}{4} \biggl[
    Z_{1,1}^1 + (Z_{1,b}^b + Z_{b,1}^b + Z_{b,b}^1)
    + (Z_{1,m}^m + +Z_{m,1}^m + Z_{m,m}^1 + Z_{b,m}^m +Z_{m,b}^m+Z_{m,m}^b)
    \biggr],\\
Z_{1+a+b+c+2m} & = \frac{1}{8} \biggl[
    Z_{1,1}^1 + (Z_{1,a}^a + + Z_{a,1}^a + Z_{a,a}^1) + (Z_{1,b}^b + Z_{b,1}^b +Z_{b,b}^1) + (Z_{1,c}^c +Z_{c,1}^c + Z_{c,c}^1)
    \\
    & \qquad
      + \left( Z_{a,b}^c+Z_{a,c}^b+Z_{b,a}^c+Z_{b,c}^a+Z_{c,a}^b+Z_{c,b}^a
     \right)
    \nonumber \\
     & \qquad
     - 2 \left( Z_{1,m}^m + Z_{m,1}^m + Z_{m,m}^1 + Z_{a,m}^m  + Z_{m,a}^m + Z_{m,m}^a
     \right)
     \nonumber \\
     & \qquad
     + 2 \left( Z_{1,m}^m + Z_{m,1}^m + Z_{m,m}^1 + Z_{b,m}^m + Z_{m,b}^m + Z_{m,m}^b
     \right)
     \nonumber \\\nonumber
     & \qquad  
     + 2 \left( Z_{1,m}^1 + Z_{m,1}^m + Z_{m,m}^1 + Z_{c,m}^m + Z_{m,c}^m + Z_{m,m}^c
     \right) \biggr].
\end{align}

\subsubsection{Rep$(Q_8)$ partition functions}

This time we can set (highlighting the differences from Rep($D_4$) in red) the coefficients of (\ref{q8inter1})-(\ref{q8inter2}) to
\be
\beta_1'=(\beta_4')^2,\quad \beta_3'=\frac{(\beta_4')^2}{\beta_2'},\quad \beta_5'=- \beta_2',\quad \beta_6'=(\beta_4')^2,\quad \beta_7' = -\frac{(\beta_4')^2}{\beta_2'},\quad {\color{red}\beta_8'=\pm i\beta_4'},\quad \beta_9'=-\frac{(\beta_4')^2}{a_2'},\non
\ee
\be
\beta_{10}' = \frac{(\beta_4')^2}{\beta_2'},\quad \beta_{11}' = -\frac{(\beta_4')^4}{(\beta_2')^2},\quad {\color{red}\beta_{12}'=\mp i\frac{(\beta_4')^2}{\beta_2'}},\quad \beta_{13}'=- \beta_4',\quad {\color{red}\beta_{14}'=\mp i \beta_4'},\quad {\color{red} \beta_{15}'=\pm i\frac{(\beta_4')^2}{\beta_2'}},\non
\ee
\be
\beta_{17}'=\frac{\beta_2' \beta_{16}'}{\beta_4'},\quad {\color{red}\beta_{18}'=\pm i \beta_4' \beta_{16}'},\quad {\color{red} \beta_{19}'=\pm i \beta_{16}'},
\ee
with the result that the only non-trivial components of the Rep$(Q_8)$ associator are
\be
\label{eq:Q8SimpleAssociator}
\tilde{K}^{m,i}_{j,m}(m,m) = \tilde{K}^{i,m}_{m,j}(m,m) = \chi(i,j),
\qquad\tilde{K}^{m,m}_{m,m}(i,j) = -\hlf\chi(i,j).
\ee

With such choices, the Rep$(Q_8)$ partition functions appearing in section~\ref{ssec:repq8} become
\begin{align}
Z_{1+a} &= \frac{1}{2}\bigg[ Z_{1,1}^1+Z_{1,a}^a+Z_{a,1}^a+Z_{a,a}^1\bigg],\\
Z_{1+a+b+c} & = \frac{1}{4} \biggl[ 
     Z_{1,1}^1 + (Z_{1,a}^a + Z_{a,1}^a + Z_{a,a}^1) + (Z_{1,b}^b + Z_{b,1}^b + Z_{b,b}^1) + (Z_{1,c}^c + Z_{c,1}^c + Z_{c,c}^1)
    \\\nonumber
     & \qquad 
     + \left( Z_{a,b}^c+Z_{a,c}^b+Z_{b,a}^c+Z_{b,c}^a+Z_{c,a}^b+Z_{c,b}^a
     \right) \biggr],\\
Z_{1+a+b+c+2m} &= \frac{1}{8}\bigg[ Z_{1,1}^1+(Z_{1,a}^a+Z_{a,1}^a+Z_{a,a}^1)+(Z_{1,b}^b+Z_{b,1}^b+Z_{b,b}^1)+(Z_{1,c}^c+Z_{c,1}^c+Z_{c,c}^1)\\
    & \qquad\left. +(Z_{a,b}^c+Z_{a,c}^b+Z_{b,a}^c+Z_{b,c}^a+Z_{c,a}^b+Z_{c,b}^a)\right.\non\\
    & \qquad\left. +2(Z_{1,m}^m+Z_{m,1}^m+Z_{m,m}^1-Z_{a,m}^m-Z_{m,a}^m-Z_{m,m}^a\right.\non\\\non
    & \qquad\left. -Z_{b,m}^m-Z_{m,b}^m-Z_{m,m}^b-Z_{c,m}^m-Z_{m,c}^m-Z_{m,m}^c)\right.\bigg].
\end{align}

\subsubsection{Rep$(\H_8)$ partition function}

In section~\ref{ssec:reph8} we computed only the partition function for the regular representation of Rep$(\H_8)$, with coefficients already chosen to match the usual Tambara-Yamagami setup.  The result was
\begin{align}
Z_{1+a+b+c+2m} & = \frac{1}{8} \biggl[
    Z_{1,1}^1 + (Z_{1,a}^a + + Z_{a,1}^a + Z_{a,a}^1) + (Z_{1,b}^b + Z_{b,1}^b +Z_{b,b}^1) + (Z_{1,c}^c +Z_{c,1}^c + Z_{c,c}^1)
    \\
    & \qquad
      + \left( Z_{a,b}^c+Z_{a,c}^b+Z_{b,a}^c+Z_{b,c}^a+Z_{c,a}^b+Z_{c,b}^a
     \right)
    \nonumber \\\non
     & \qquad  
     + 2 \left( Z_{1,m}^1 + Z_{m,1}^m + Z_{m,m}^1 + Z_{c,m}^m + Z_{m,c}^m + Z_{m,m}^c
     \right) \biggr].
\end{align}

\subsection{Existence of non-multiplicity-free examples}  \label{sect:rep-a4}

The examples we have studied in this paper all have the property that all spaces of junction operators are at most one dimensional (equivalently, the fusion categories are multiplicity-free).  Concretely, this means that in all  examples studied in this paper, for all irreducible representations $R$, $S$, $T$,
\begin{equation}
    \dim {\rm Hom}(R \otimes S, T) \: \in \: \{0, 1\}.
\end{equation}

In this section, we will observe that Rep$(A_4)$, where $A_4=(\mathbb{Z}_2\times \mathbb{Z}_2)\rtimes \mathbb{Z}_3$ the alternating group on four elements, is a non-multiplicity-free\footnote{
In fact, this seems to be the simplest such example.  The next largest finite nonabelian group beyond those we have
described so far is the ten-element dihedral group $D_5$.  However, this does not happen in that example,
see e.g.~\cite[p. 6]{lpr}.
} example.

In $A_4$, let $m$ denote the irreducible three-dimensional representation.
Then, the tensor product $m \otimes m \otimes m$ has multiple singlets.
This is an example to illustrate the necessity of keeping track of more than just simple objects in multiplications and comultiplications.
We can see this as follows.
\begin{equation}
    m \otimes m=1+a+b+2m, \: \: \:
    a\otimes a=b, \: \: \:
    b\otimes b=a, \: \: \:
    a \otimes b=1, \: \: \:
    a\otimes m=m, \: \: \:
    b \otimes m=m.
\end{equation}
As a result,
\begin{eqnarray}
    m \otimes m \otimes m & = & m \otimes (1 + a + b + 2m) \: = \: m + m + m + 2 (1 + a + b + 2m),
    \\
    & = & 2(1) + 2 a + 2 b + 7 m.
\end{eqnarray}
A presentation of $A_4$ in terms of four generators is \cite{GKKL11}
\begin{equation}
    A_4:=\langle w,x,y,z\ \vert \  w^2=x^2=y^2=z^2=1; (x_ix_j)^2=1\rangle
\end{equation}
where $x_i,x_j\in\{w,x,y,z\}$ and $x_i\neq x_j$.

We will not describe the partition function of a theory with a gauged Rep$(A_4)$ symmetry here, but instead will return
to non-multiplicity-free examples in our followup paper \cite{toappear}.

\section{Applications: Duality defects from gauging noninvertible symmetries}  \label{sect:apps}

In this section, we present several explicit examples of gauging noninvertible symmetries in $c=1$ CFTs as $\mathbb{Z}_2$ orbifold of the compact boson \cite{Dixon:1985jw, Hamidi:1986vh, Dixon:1986qv}. In~\cite{Becker:2017zai}, a large class of topological defects were written down.  This paper constructed more general conformal interfaces between $c=1$ theories using the folding trick (finding conformal boundaries in the tensor product theory) and computed their fusions.  Among these, the interfaces which were actually topological were identified, and by specializing to cases with the same theory on either side of the interface, one obtains a large set of topological defects.  For the case of defects in the free boson theories, at generic radius $R$ the only identified topological defects were invertible, generating the symmetry group $G_R$ defined below.  At non-generic radii (rational $R^2$) more possibilities appear, including noninvertible topological defects, but the constructive methods of~\cite{Becker:2017zai} are also not exhaustive, missing some defects.  For the orbifold branch at a generic radius, the identified topological defects included eight invertible defects, labeled $\mathcal{I}_{1,1}^{(+)OO}(\al_0;\beta_0;\epsilon)$, where $\al_0,\beta_0\in\{0,\pi\}$ and $\epsilon\in\{\pm 1\}$.  These defects generate a $D_4$ symmetry under fusion,
\begin{equation}
    \mathcal{I}_{1,1}^{(+)OO}(\al_0;\beta_0;\epsilon)\times\mathcal{I}_{1,1}^{(+)OO}(\al_0';\beta_0';\epsilon')=\mathcal{I}_{1,1}^{(+)OO}(\al_0+\al_0';\beta_0+\beta_0';(-1)^{\beta_0\al_0'/\pi^2}\epsilon\epsilon').
\end{equation}
Additionally, there is an infinite family of simple noninvertible defects 
\begin{equation}
    \mathcal{I}_{1,1}^{(+)OO}(\al,\beta), \qquad (\al,\beta)\notin\{(0,0),(0,\pi),(\pi,0),(\pi,\pi)\}
\end{equation}
with fusions
\begin{equation}
    \mathcal{I}_{1,1}^{(+)OO}(\al;\beta)\times\mathcal{I}_{1,1}^{(+)OO}(\al_0;\beta_0;\epsilon)=\mathcal{I}_{1,1}^{(+)OO}(\al_0;\beta_0;\epsilon)\times\mathcal{I}_{1,1}^{(+)OO}(\al;\beta)=\mathcal{I}_{1,1}^{(+)OO}(\al+\al_0;\beta+\beta_0),
\end{equation}
\begin{equation}
    \mathcal{I}_{1,1}^{(+)OO}(\al;\beta)\times\mathcal{I}_{1,1}^{(+)OO}(\al';\beta')=\mathcal{I}_{1,1}^{(+)OO}(\al+\al';\beta+\beta')\oplus\mathcal{I}_{1,1}^{(+)OO}(\al-\al';\beta-\beta'),
\end{equation}
where on the right-hand side of the second line when both arguments of either defect land in the set $\{0,\pi\}$, then that defect should be understood to decompose as
\begin{equation}
\mathcal{I}_{1,1}^{(+)OO}(\al_0;\beta_0)=\mathcal{I}_{1,1}^{(+)OO}(\al_0;\beta_0;+)\oplus\mathcal{I}_{1,1}^{(+)OO}(\al_0;\beta_0;-).
\end{equation}
Again, at non-generic radii more topological defects were identified, but the methods of~\cite{Becker:2017zai} also miss some defects.  

These structures were rediscovered in~\cite{Thorngren:2021yso} and embedded in a larger structure including also the missing defects, and was explored in great detail using more modern methods.

We will show that there exist self-dualities under gauging noninvertible symmetries. In particular, this self-duality can happen via gauging a full fusion category or a Frobenius subalgebra. We present and discuss both cases with concrete examples and build  duality defects with specific fusion rules via performing half-space gauging.

Our starting point is the 2D compact boson with radius $R$. Using $2\pi R$-periodic left- and right-moving fields $X_L$ and $X_R$, we define $U(1)$-valued fields as 
\begin{equation}
    \theta_m=\frac{X_L+X_R}{R},~ \theta_w=R(X_L-X_R)
\end{equation}
where subindices $m$ and $w$ denote momentum and winding, respectively. The celebrated (invertible) T-duality is given by 
\begin{equation}
\begin{split}
    R&\rightarrow \frac{1}{R},\\
    \theta_m &\leftrightarrow \theta_w,\\
    X_R&\rightarrow -X_R
\end{split}
\end{equation}
The global symmetry at generic $R$ reads
\begin{equation}
   G_R=( U(1)_m\times U(1)_w )\rtimes \mathbb{Z}_2^r
\end{equation}
where the two $U(1)$'s denote the obvious shifting for $\theta_m$ and $\theta_w$, while the $\mathbb{Z}_2$ extension arises from the reflection (sometimes also referred to as ``charge conjugation"):
\begin{equation}
   \mathbb{Z}_2^r: (\theta_m, \theta_w)\rightarrow (-\theta_m, -\theta_w).
\end{equation}

Let us denote the compact boson CFT at radius $R$ with its target space $[S^1_R]$. As systematically investigated in \cite{Becker:2017zai,Thorngren:2021yso}, a large class of noninvertible symmetries arise in the $\mathbb{Z}_2^r$ orbifold theory, which we denote as  $[S^1_R/\mathbb{Z}_2^r]$. Briefly speaking, one can focus on various subgroups of $U(1)_m\times U(1)_w$ and its associated $\mathbb{Z}_2^r$ extension to a non-abelian group $G$, then gauging the non-abelian group or its non-normal $\mathbb{Z}_2^r$ subgroup \cite{Bhardwaj:2017xup}, lead to noninvertible symmetries in the orbifold theory $[S^1_R/G]$ \footnote{
Although in this section we mainly focus on $c=1$ CFTs, and in particular, rational CFTs, there are also cases where one can build non-invertible symmetries for $c>1$ irrational CFTs, see e.g.~\cite{Brunner:2014lua, Bhardwaj:2017xup, Nagoya:2023zky}
}.

A natural question is: starting with a given noninvertible symmetry in $[S^1_R/G]$, what can we learn from gauging part of (i.e., Frobenius subalgebra) or the full noninvertible symmetry? Below, we will show that one can build new noninvertible  defects via gauging noninvertible symmetries.\footnote{For generic discussions on building topological interfaces via gauging noninvertible symmetries, we refer the reader to \cite{Wang:2023, Chang:202x}.}

\subsection{Gauging Frobenius subalgebra}

\subsubsection{Gauging $1+Y$ of Rep$(S_3)$ in $SU(2)_4/U(1)$}

Let us start with the Rep$(S_3)$ symmetry. This noninvertible symmetry can be realized via considering a $\mathbb{Z}_{3(m)}$ subgroup of $U(1)_m$:
\begin{equation}
    \mathbb{Z}_{3(m)}: \theta \rightarrow \theta + \frac{2\pi}{3}.
\end{equation}
Extended by $\mathbb{Z}_2^r$ symmetry, this gives rise to a 
\begin{equation}
    S_3=\mathbb{Z}_{3(m)}\rtimes \mathbb{Z}_2^r
\end{equation}
symmetry in the compact boson CFT $[S^1_R]$. A Rep$(S_3)$ noninvertible symmetry is then realized via gauging the $S_3$, leading to the orbifold theory $[S^1_R/S_3]$. Alternatively, one can think of this $S_3$ orbifolding as first gauging the $\mathbb{Z}_{3(m)}$ symmetry and then gauging the $\mathbb{Z}_2^r$ symmetry. Notice that the $\mathbb{Z}_{3(m)}$ symmetry is a 3-fold rotation symmetry of the target space circle $S^1$. Thus, gauging this symmetry is performing a quotient of the target space as $R\rightarrow R/3$. Therefore we have 
\begin{equation}\label{eq: c=1 gauging S3 via two steps}
    [S^1_R/S_3]=[S^1_{\frac{R}{3}}/\mathbb{Z}_2^r].
\end{equation}
In the notation of~\cite{Becker:2017zai}, the Rep$(S_3)$ symmetry is generated by
\begin{equation}
    1=\mathcal{I}_{1,1}^{(+)OO}(0;0;+),\quad X=\mathcal{I}_{1,1}^{(+)OO}(0;0;-),\quad Y=\mathcal{I}_{1,1}^{(+)OO}\left(\frac{2\pi}{3};0\right).
\end{equation}

 As we discussed in Section \ref{sect:S3:1pY},  for theories with Rep$(S_3)$ global symmetry, in addition to gauging the full Rep$(S_3)$ or its $\mathbb{Z}_2$ subgroup, it is also possible to gauge its Frobenius subalgebra, corresponding to the algebra objects $1+Y$. Furthermore, if the theory itself is an orbifold theory as $[\mathcal{T}/S_3]$, then gauging $1+Y$ ends up with the $\mathbb{Z}_2$ orbifold of the theory $[\mathcal{T}]$, namely
\begin{equation}
    [\mathcal{T}/\mathbb{Z}_2] =[\left( \mathcal{T}/S_3 \right)/(1+Y)].
\end{equation}
Coming back to our case of interest where $[\mathcal{T}]=[S^1_R]$ and  $\mathbb{Z}_2=\mathbb{Z}_2^r$, we obtain
\begin{equation}
    [S^1_R/\mathbb{Z}_2^r]=[(S^1_R/S_3)/(1+Y)]=[(S^1_{\frac{R}{3}}/\mathbb{Z}_2^r)/(1+Y)],
\end{equation}
where in the second step we use (\ref{eq: c=1 gauging S3 via two steps}). With the T-duality
\begin{equation}
    [S_R^1]=[S_\frac{1}{R}^1],
\end{equation}
we conclude at $R=\sqrt{3}$,
\begin{equation}
     [S^1_{\sqrt{3}}/\mathbb{Z}_2^r]=[(S^1_{\sqrt{3}}/\mathbb{Z}_2^r)/(1+Y)],
\end{equation}
namely the orbifold theory $[S^1_{\sqrt{3}}/\mathbb{Z}_2^r]$ is self-dual under gauging Frobenius subalgebra $1+Y$. In fact, at this special radius, the $c=1$ CFT is rational and enjoys a coset description as $SU(2)_4/U(1)$ with the diagonal modular invariant \cite{Gepner:1986hr, Thorngren:2021yso}. According to the above self-duality under gauging Frobenius algebra $1+Y$, we can build a duality defect in this theory via gauging $1+Y$ in half of the spacetime and imposing a Dirichlet boundary on the resulting interface, as in \cite{Choi:2021kmx}. Let us denote the resulting duality defect as $\mathcal{D}$, and its fusion can then be expressed as
\begin{equation}\label{eq: fl for subgauging reps3}
\begin{split}
     &\mathcal{D}\otimes \mathcal{D}=1\oplus Y,\\
    &\mathcal{D}\otimes Y=Y\otimes \mathcal{D}=2\mathcal{D}.
\end{split}
\end{equation}
From the fusion rule, one can see this is indeed a new noninvertible defect line, whose quantum dimension is $\sqrt{3}$. This non-integer quantum dimension implies the whole fusion category containing $\mathcal{D}$ is not gaugeable, though there could exist a gaugeable Frobenius algebra containing ${\cal D}$. The defect $\mathcal{D}$ does not appear among those studied in~\cite{Becker:2017zai}, which is not surprising since it only exists at a particular value of the radius.

\subsubsection{Gauging $1+b+m$ of Rep$(D_4)$ in $SU(2)_1/(\mathbb{Z}_2\times \mathbb{Z}_2)$} 

Similarly, one can consider Rep$(D_4)$ symmetry of the orbifold theory
\begin{equation}
    [S^1_R/D_4]=[S^1_{\frac{R}{4}}/\mathbb{Z}_2^r],
\end{equation}
and gauging its Frobenius subalgebra associated with the algebra objects $1+b+m$ as in Section \ref{ssec:repd4}. In the notation of~\cite{Becker:2017zai}, we identify
\begin{equation}
    1=\mathcal{I}_{1,1}^{(+)OO}(0;0;+),\quad b=\mathcal{I}_{1,1}^{(+)OO}(0;0;-),\quad m=\mathcal{I}_{1,1}^{(+)OO}\left(\frac{\pi}{2};0\right).
\end{equation}
We obtain the relation similar to the Rep$(S_3)$ case 
\begin{equation}
    [S^1_R/\mathbb{Z}_2^r]=[(S^1_R/D_4)/(1+b+m)]=[(S^1_{\frac{R}{4}}/\mathbb{Z}_2^r)/(1+b+m)].
\end{equation}
At $R=2$ , we have the self-duality 
\begin{equation}
    [S^1_2/\mathbb{Z}_2^r]=[(S^1_2/\mathbb{Z}_2^r)/(1+b+m)]
\end{equation}
for $[S^1_2/\mathbb{Z}_2^r]$ theory under gauging $1+b+m$ Frobenius subalgebra. At this special radius, the theory is rational and enjoys a coset description as $SU(2)_1/(\mathbb{Z}_2\times \mathbb{Z}_2)$ \cite{Thorngren:2021yso}, which also corresponds to the continuum limit of the 4-state Potts model \cite{Kadanoff:1978ve, Nienhuis:1984wm, DiFrancesco:1987gwq}. The resulting duality defect $\mathcal{D}$ via half-space gauging of this Frobenius subalgebra enjoys the fusion rule
\begin{equation}\label{eq: fl for subgauging repd4}
\begin{split}
    &\mathcal{D}\otimes \mathcal{D}=1\oplus b \oplus m,\\
    &\mathcal{D}\otimes b=b\otimes \mathcal{D}=\mathcal{D},\\
    &\mathcal{D}\otimes m=m\otimes \mathcal{D}=2\mathcal{D}
\end{split}
\end{equation}
which is another new noninvertible defect line, with quantum dimension 2 (though not among the defects identified in~\cite{Becker:2017zai}, which all exist for generic radii). This integer quantum dimension implies the possibility of further gauging the noninvertible symmetry category containing $\mathcal{D}$, which we leave for future work.

\subsection{Gauging Rep$({\cal H}_8)$ in (Ising $\otimes$ Ising) CFT}
\label{sect:gauging-reph8-ising}

At radius $R=\sqrt{2}$, the $\mathbb{Z}_2^r$ orbifold theory of $c=1$ compact boson is isomorphic to the tensor product of two Ising CFTs, which we denote as (Ising $\otimes$ Ising) CFT \cite{Thorngren:2021yso}:
\begin{equation}
    [S^1_{\sqrt{2}}/\mathbb{Z}_2^r]\cong [\text{Ising} \otimes \text{Ising}].
\end{equation}
The noninvertible symmetry this theory enjoys is the Rep$({\cal H}_8)$ symmetry. The simple objects of Rep$({\cal H}_8)$ are composed of topological lines in Ising CFT as 
\begin{equation}
\begin{split}
     &1=1_1\otimes 1_2, \\
     &a=\eta_1\otimes 1_2, ~b=1_1\otimes \eta_2, ~c=\eta_1\otimes \eta_2, \\
     &m=\mathcal{N}_1\otimes \mathcal{N}_2,
\end{split}
\end{equation}
where $1_i, \eta_i$ and $\mathcal{N}_i$ are the identity line, the $\mathbb{Z}_2$ defect and the Kramers-Wannier duality defect for the $i$-th Ising CFT, respectively. The fusion rules for Rep$({\cal H}_8)$ are determined by 
\begin{equation}
  \eta_i\otimes \eta_i=1_i, ~~\eta_i\otimes \mathcal{N}_i=\mathcal{N}_i\otimes \eta_i=\mathcal{N}_i, ~~\mathcal{N}_i\otimes \mathcal{N}_i=1_i\oplus \eta_i.
\end{equation}

It was argued in \cite{Choi:2023xjw} the Rep$({\cal H}_8)$ categorical symmetry in (Ising $\otimes$ Ising) CFT is gaugeable, due to the fact that it admits a boundary state which is invariant under the action of all topological defects. Here we review the proposal briefly. Start with a critial Ising CFT, with three Kramers-Wannier duality line defects inserted. 
\begin{figure}
    \centering
\includegraphics[width=12cm]{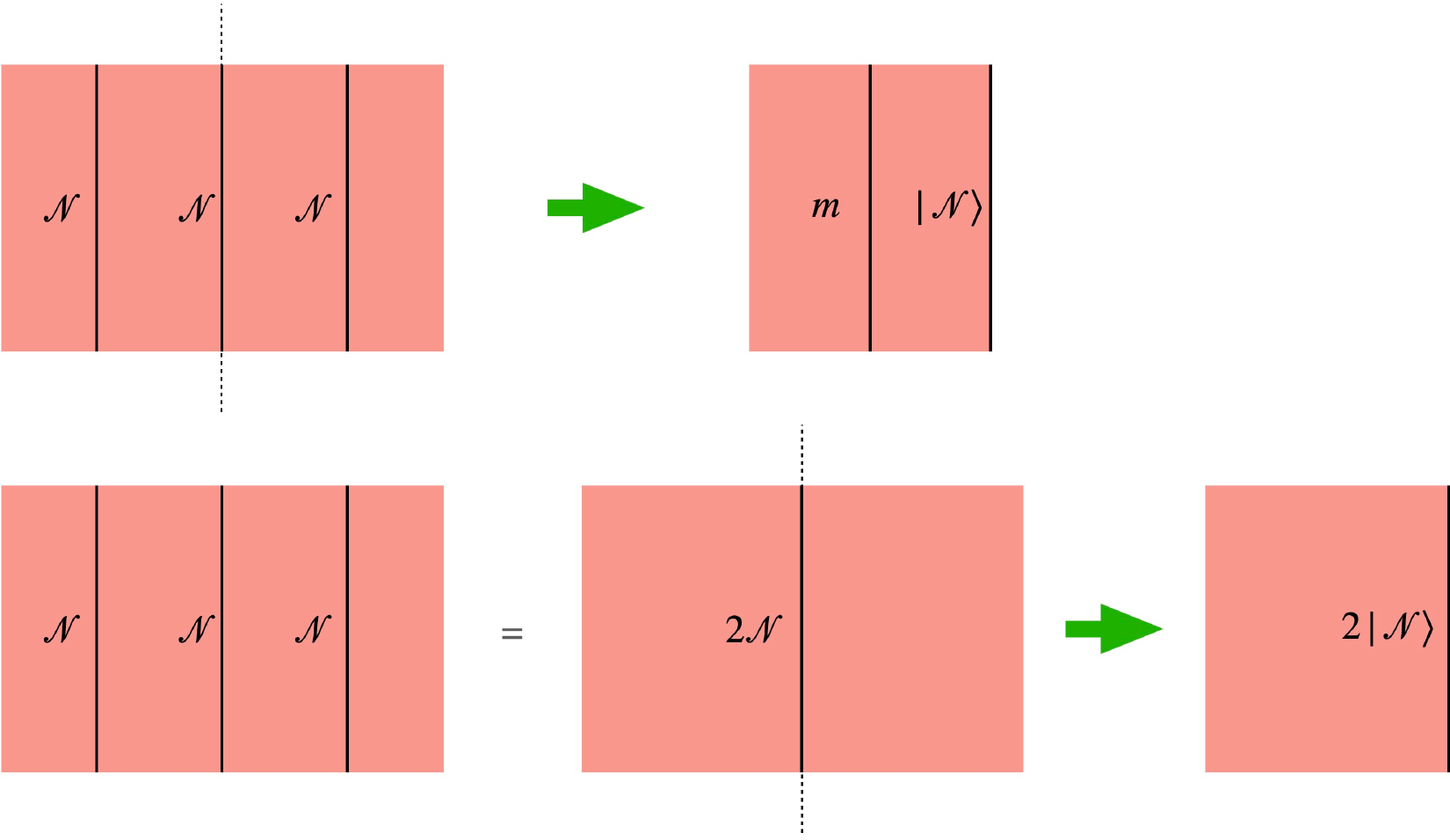}
    \caption{Top: Folding Ising CFT along the middle Kramers-Wannier line gives rise to $m$ line acting on the boundary state $|\mathcal{N}\rangle$ in the (Ising $\otimes$ Ising) CFT. Bottom: Fusing the three Kramers-Wannier lines and then folding the Ising CFT along the $2\mathcal{N}$ line give rise to boundary state $2|\mathcal{N}\rangle$ in the (Ising $\otimes$ Ising) CFT.}
    \label{fig: folding Ising}
\end{figure}
In order to build a (Ising $\otimes$ Ising) CFT, one can consider two alternative ways as follows
\begin{itemize}
    \item Fold the worldsheet along the middle defect, as shown in the top picture in Figure \ref{fig: folding Ising}. The middle defect after folding becomes a $\mathcal{N}$ interface between the (Ising $\otimes$ Ising) CFT and a trivial theory, thus realizing a boundary state denoted as $| \mathcal{N}\rangle$. Notice that the left and the right $\mathcal{N}$ defects fuse into the $m$ defect line after the folding. Therefore, the whole configuration is exactly the noninvertible line $m$ acting on the boundary state $| \mathcal{N} \rangle$ in the (Ising $\otimes$ Ising) CFT
    \begin{equation}
        m| \mathcal{N}\rangle.
    \end{equation}
    \item Fuse the left two $\mathcal{N}$ defects into $1\oplus \eta$, then fuse them with the rightmost $\mathcal{N}$ line $(1\oplus \eta)\otimes \mathcal{N}=2\mathcal{N}$. Fold the worldsheet along the resulting $2\mathcal{N}$ defect, as shown in the bottom picture in Figure \ref{fig: folding Ising}. Now, the folding manipulation gives rise to a $2| \mathcal{N}\rangle$ boundary state for the (Ising $\otimes$ Ising) CFT. 
\end{itemize}
The equivalence of these two folding steps give rise to 
\begin{equation}
    m|\mathcal{N}\rangle =2|\mathcal{N}\rangle,
\end{equation}
which is a strongly symmetric condition for the boundary state due to the quantum dimension $\langle m \rangle =2$. We then conclude the categorical symmetry Rep$({\cal H}_8)$ is gaugeable. 

\paragraph{Partition functions and self-duality via gauging Rep$(\mathcal{H}_8)$.} The folding method discussed above makes it possible to write a closed form for various twisted partition functions $Z_{L_1, L_2}^{L_3}$ for (Ising $\otimes$ Ising) CFT. Let us start with the simple cases where $L_1$ or $L_2$ is the identity line defect $1$. Recall in the Ising CFT, there are three primary operators whose characters are given by 
\begin{equation}
\begin{split}
    &\chi_0=\frac{1}{2}\left[ \left( \frac{\theta_3}{\eta} \right)^\frac{1}{2}+\left( \frac{\theta_4}{\eta} \right)^\frac{1}{2} \right], ~\chi_\frac{1}{2}=\frac{1}{2}\left[ \left( \frac{\theta_3}{\eta} \right)^\frac{1}{2}-\left( \frac{\theta_4}{\eta} \right)^\frac{1}{2} \right], ~\chi_{\frac{1}{16}}=\frac{1}{\sqrt{2}}\left( \frac{\theta_2}{\eta} \right)^{1/2},\\
    &\bar{\chi}_{h}\equiv \chi_h(\bar{\tau}), ~|\chi_{h}|^2\equiv \chi_{h}\bar{\chi}_{h}.
\end{split}
\end{equation}
Partition functions for the Ising CFT with various line insertions read (see, e.g., \cite{Lin:2019kpn, Lin:2019hks, Thorngren:2019iar})
\begin{equation}\label{eq: Ising2 twisted sector with invertible lines}
\begin{split}
    Z_{1,1}^{1}[\text{Ising}]&=|\chi_0|^2+|\chi_{\frac{1}{2}}|^2+|\chi_{\frac{1}{16}}|^2,\\
    Z_{1,\eta}^{\eta}[\text{Ising}]&=|\chi_0|^2+|\chi_{\frac{1}{2}}|^2-|\chi_{\frac{1}{16}}|^2,\\
    Z_{\eta,1}^{\eta}[\text{Ising}]&=\chi_0\bar{\chi}_\frac{1}{2}+\chi_\frac{1}{2}\bar{\chi}_0+|\chi_{\frac{1}{16}}|^2,\\
    Z_{\eta, \eta}^{1}[\text{Ising}]&=-\chi_0\bar{\chi}_\frac{1}{2}-\chi_\frac{1}{2}\bar{\chi}_0+|\chi_{\frac{1}{16}}|^2,\\
    Z_{1,\mathcal{N}}^\mathcal{N}[\text{Ising}]&=\sqrt{2}(|\chi_0|^2-|\chi_\frac{1}{2}|^2),\\
    Z_{\mathcal{N},1}^\mathcal{N}[\text{Ising}]&=\bar{\chi}_\frac{1}{16}(\chi_0+\chi_\frac{1}{2})+c.c.,\\
    Z_{\mathcal{N},\eta}^{\mathcal{N}}[\text{Ising}]&=i\bar{\chi}_{\frac{1}{16}}(\chi_0+\chi_\frac{1}{2})+c.c.
\end{split}
\end{equation}
We can then readily obtain various twisted partition functions for (Ising $\otimes$ Ising) CFT. With invertible lines inserted, we have
\begin{equation}
\begin{split}
    &Z_{1,1}^1[\text{Ising}\otimes\text{Ising}]=(Z_{1,1}^{1}[\text{Ising}])^2,\\
    &Z_{1,a}^a[\text{Ising}\otimes\text{Ising}]=Z_{1,b}^b[\text{Ising}\otimes\text{Ising}]=Z_{1,\eta}^{\eta}[\text{Ising}]\times Z_{1,1}^{1}[\text{Ising}],\\
    &Z_{1,c}^c[\text{Ising}\otimes\text{Ising}]=(Z_{1,\eta}^{\eta}[\text{Ising}])^2,\\
    &
Z_{a,1}^a[\text{Ising}\otimes\text{Ising}]=Z_{b,1}^b[\text{Ising}\otimes\text{Ising}]=Z_{\eta,1}^{\eta}[\text{Ising}]\times Z_{1,1}^{1}[\text{Ising}],\\
& Z_{c,1}^c[\text{Ising}\otimes\text{Ising}]=(Z_{\eta,1}^{\eta}[\text{Ising}])^2,\\
&Z_{a,a}^1[\text{Ising}\otimes\text{Ising}]=Z_{b,b}^1[\text{Ising}\otimes\text{Ising}]=Z_{\eta,\eta}^{1}[\text{Ising}]\times Z_{1,1}^{1}[\text{Ising}],\\
&Z_{a,b}^{ab}[\text{Ising}\otimes\text{Ising}]=Z_{b,a}^{ab}[\text{Ising}\otimes\text{Ising}]=Z_{\eta,1}^{\eta}[\text{Ising}]\times Z_{1,\eta}^{\eta}[\text{Ising}],\\
&Z_{a,c}^{b}[\text{Ising}\otimes\text{Ising}]=Z_{b,c}^{a}[\text{Ising}\otimes\text{Ising}]=Z_{\eta,\eta}^{1}[\text{Ising}]\times Z_{1,\eta}^{\eta}[\text{Ising}],\\
&Z_{c,a}^{b}[\text{Ising}\otimes\text{Ising}]=Z_{c,b}^{a}[\text{Ising}\otimes\text{Ising}]=Z_{\eta,\eta}^{1}[\text{Ising}]\times Z_{\eta,1}^{\eta}[\text{Ising}],\\
& Z_{c,c}^1[\text{Ising}\otimes\text{Ising}]=(Z_{\eta,\eta}^{\eta}[\text{Ising}])^2,
\end{split}
\end{equation}
Recall that in gauging Rep$(\mathcal{H}_8)$, not all twisted partition functions with noninvertible line $m$ insertions are included, but only 
\begin{equation}
    Z_{1,m}^m, Z_{m,1}^, Z_{c,m}^m, Z_{m,c}^m, Z_{m,m}^1, Z_{m,m}^c
\end{equation}
are present. Thus, we need to obtain their expressions, three of which can be derived from products of Ising CFT partition functions
\begin{equation}
\begin{split}
   & Z_{1,m}^m[\text{Ising}\otimes\text{Ising}]=(Z_{1,\mathcal{N}}^{\mathcal{N}}[\text{Ising}])^2,\\
&Z_{m,1}^m[\text{Ising}\otimes\text{Ising}]=(Z_{\mathcal{N},1}^{\mathcal{N}}[\text{Ising}])^2,\\
&Z_{m,c}^m[\text{Ising}\otimes\text{Ising}]=(Z_{\mathcal{N},\eta}^{\mathcal{N}}[\text{Ising}])^2,\\
\end{split}
\end{equation}
while the other three partition functions can in turn be derived from the Rep$(\mathcal{H}_8)$ modular transformation 
\begin{equation}
    \begin{split}
        &Z_{c,m}^m[\text{Ising}\otimes\text{Ising}](\tau)=Z_{m,c}^m[\text{Ising}\otimes\text{Ising}]\left(-\frac{1}{\tau}\right),\\
&Z_{m,m}^1[\text{Ising}\otimes\text{Ising}](\tau)=Z_{m,1}^m[\text{Ising}\otimes\text{Ising}](\tau+1),\\
&Z_{m,m}^c[\text{Ising}\otimes\text{Ising}](\tau)=Z_{m,c}^m[\text{Ising}\otimes\text{Ising}](\tau+1),
    \end{split}
\end{equation}
The sum over twisted partition functions present in gauging Rep$(\mathcal{H}_8)$ can then be computed as 
\begin{equation}\label{eq: ising2 non-invertible part}
\begin{split}
    \frac{2}{8}(Z_{1,m}^m+ Z_{m,1}^+ Z_{c,m}^m+ Z_{m,c}^m+ Z_{m,m}^1+ Z_{m,m}^c)[\text{Ising}\otimes\text{Ising}]\\
    =2|\bar{\chi}_\frac{1}{16}(\chi_0+\chi_\frac{1}{2})|^2+\frac{1}{2}(|\chi_0|^4+|\chi_\frac{1}{2}|^4-(\bar{\chi}_0\chi_{\frac{1}{2}})^2-(\bar{\chi}_\frac{1}{2}\chi_{0})^2)
\end{split}
\end{equation}

Recall that the (Ising $\otimes$ Ising) CFT is self-dual under gauging $\mathbb{Z}_2\times \mathbb{Z}_2$ symmetry without discrete torsion, whose half-space gauging shows the presence of the noninvertible duality defect $m$ \footnote{It is also possible to gauge $\mathbb{Z}_2\times \mathbb{Z}_2$ with discrete torsion turned on. This is equivalent to gauging the diagonal $\mathbb{Z}_2$ of the two Ising CFTs, leading to the compact boson at radius $R=\sqrt{2}$ \cite{Dijkgraaf:1989hb, Hsin:2020nts}.}. Thus, we conclude the sum over twisted partition function with invertible line insertions as 
\begin{equation}\label{eq: ising2 invertible part}
    \frac{1}{8}\sum_{g,h\in \{ 1,a,b,c \}}Z_{g,h}^{gh}[\text{Ising}\otimes\text{Ising}]=\frac{1}{2}Z_{1,1}^1[\text{Ising}\otimes\text{Ising}]=\frac{1}{2}(|\chi_0|^2+|\chi_\frac{1}{2}|^2+|\chi_{\frac{1}{16}}|^2)^2.
\end{equation}

The resulting partition function after gauging Rep$(\mathcal{H}_8)$, which can be computed straightforwardly using (\ref{eq: Ising2 twisted sector with invertible lines}), (\ref{eq: ising2 non-invertible part}), and (\ref{eq: ising2 invertible part}), interestingly, is the same as that of the (Ising $\otimes$ Ising) CFT (see also \cite{CLS23}):
\begin{equation}
\begin{split}
&Z[(\text{Ising}\otimes\text{Ising})/\text{Rep}(\mathcal{H}_8)]\\
&=\frac{1}{8}\left(\left(\sum_{g,h\in\{1,a,b,c\}} \epsilon(g,h) Z_{g,h}^{gh}\right) + 2\left(Z_{1,m}^m+Z_{m,1}^m+Z_{c,m}^m+Z_{m,c}^m+Z_{m,m}^1+Z_{m,m}^c\right)\right)\\
  &=(|\chi_0|^2+|\chi_\frac{1}{2}|^2+|\chi_{\frac{1}{16}}|^2)^2,
\end{split}
\end{equation}
The self-duality of (Ising $\otimes$ Ising) CFT under this gauging \footnote{Naively, one can also try summing over the twisted partition function without any minus signs in front of $Z_{g,h}^{gh}$, as in (\ref{eq:h8-z}). However, the resulting partition function will be ill-defined, implying it is not an admitted gauging. In fact, in the case of gauging Rep$(\mathcal{H}_8)$, the expression of summing over twisted partition functions is dependent on whether the theory is self-dual under gauging $\mathbb{Z}_2\times \mathbb{Z}_2$ with/without discrete torsion. See, e.g. \cite{CLS23}. We thank Yifan Wang for the discussions on this point.} implies a new duality defect $\mathcal{D}$, which can be built by performing a half-space gauging. The associated fusion rule is similar to (\ref{eq: fl for subgauging reps3}) and (\ref{eq: fl for subgauging repd4})
\begin{equation}\label{eq: fl for gauging reph8}
\begin{split}
    &\mathcal{D}\times \mathcal{D}=1+a+b+c+2m,\\
    &\mathcal{D}\times m=m\times \mathcal{D}=2\mathcal{D},\\
    &\mathcal{D}\times g=g\times\mathcal{D}=\mathcal{D}, g\in \{ a,b,c \},
\end{split}
\end{equation}
from which one reads the quantum dimension of $\mathcal{D}$ is $\sqrt{8}$. This non-integer quantum dimension implies the whole fusion category containing $\mathcal{D}$ is not gaugeable\footnote{While finishing this paper, the reference \cite{CLS23} appeared, which includes a systematic study of this new duality defect in (Ising $\otimes$ Ising) CFT. We refer the reader to that reference for more details.}, though there might exist a gaugeable Frobenius algebra containing ${\cal D}$.

\begin{figure}
    \centering
    \includegraphics[width=12cm]{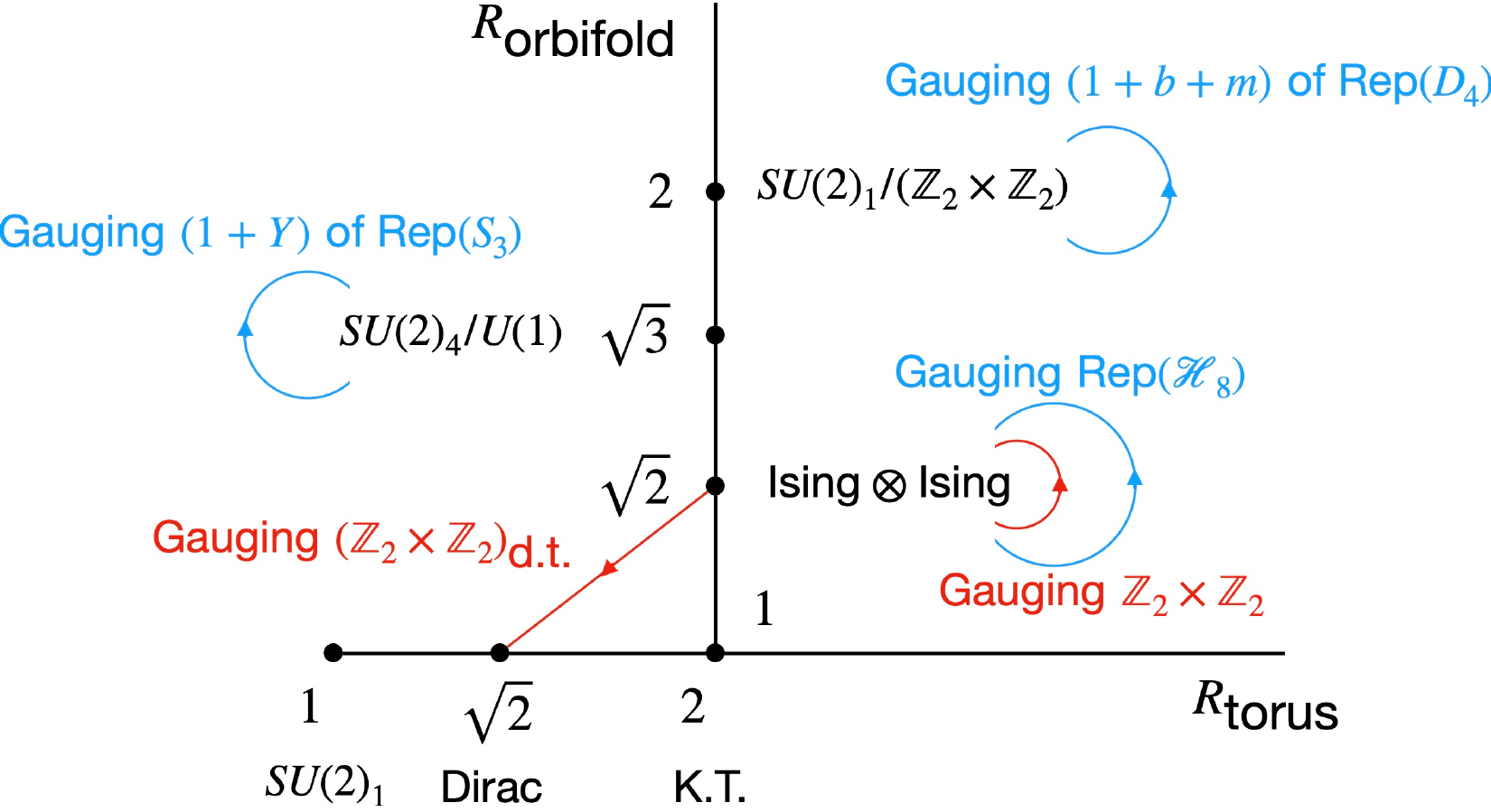}
    \caption{Self-dualities for several $c=1$ CFTs from  gauging. The horizontal and vertical axes denote the compact boson's radius and its $\mathbb{Z}_2^r$ orbifold theory, respectively. These two branches meet at $[S_2^1]=[S_1^1/\mathbb{Z}_2^r]$, which corresponds to the Kosterlitz-Thouless transition point of the XY-model \cite{kadanoff1979multicritical}. Loops colored in blue denote self-dualities under gauging noninvertible symmetries, whose associated duality defects enjoy the fusion rules presented in (\ref{eq: fl for subgauging reps3}), (\ref{eq: fl for subgauging repd4}) and (\ref{eq: fl for gauging reph8}). Lines colored in red denote the previously known cases that (Ising $\otimes$ Ising) CFT is self-dual under gauging $\mathbb{Z}_2\times \mathbb{Z}_2$ without discrete torsion and transformed to the diagonal bosonization of the Dirac fermion with discrete torsion. }
    \label{fig: sel-duality in moduli space}
\end{figure}

We conclude this section by embedding the new self-dualities we find via gauging the noninvertible symmetries in the moduli space of $c=1$ CFTs \cite{Dijkgraaf:1987vp} (see also \cite{Thorngren:2021yso}), as shown in Figure \ref{fig: sel-duality in moduli space}. 

\section{Decomposition}  \label{sect:decomp}

Briefly, decomposition is the statement that a local quantum field theory in $d$ dimensions with a global $(d-1)$-form symmetry is equivalent to a disjoint union of local quantum field theories,
see e.g.~\cite{Hellerman:2006zs,Sharpe:2022ene}.  In two dimensions,
standard examples involving gauge theories in which a subgroup of the (zero-form) gauge group acts trivially.  The resulting theory has a global one-form symmetry, and so is equivalent to a disjoint union.

In this section, we will discuss simple prototypical examples of decomposition arising when gauging a trivially-acting noninvertible zero-form symmetry.  We begin by defining what it means for a noninvertible symmetry group to act trivially, state a conjecture for the form of the result, and then compute some examples.

\subsection{Definition of trivially-acting noninvertible 0-form symmetry}

For an element of an ordinary, 
invertible 0-form symmetry to act trivially means, for example,
that it leaves invariant all local operators.  If $g$ denotes a
symmetry operator, then, 
\begin{equation} 
g \cdot {\cal O} \: = \: {\cal O}.
\end{equation}
In terms of line operators, we can describe the action of $g$ on
a local operator ${\cal O}$ as (see e.g.~\cite[section 1]{Lin:2022dhv})
where we imagine the line collapsing onto the local operator ${\cal O}$
to form the local operator ${\cal O}'$, and where, for a trivial action,
${\cal O}' = {\cal O}$.
This is illustrated schematically in figures~\ref{fig:act1}, \ref{fig:act2}.

\begin{figure}[h]
\begin{subfigure}{0.5\textwidth}
\centering
    \begin{tikzpicture}
			\draw[thick,->] (0,0) [partial ellipse=270:90:0.75cm and 0.75cm];
			\draw[thick,->] (0,0) [partial ellipse=270:-90:0.75cm and 0.75cm];
			\filldraw[black] (0,0) circle (2pt);
			\node at (0,0.3) {$\mathcal{O}$};
			\node at (-0.75,0.75) {$k$};
		\end{tikzpicture}
		\caption{}
		\label{fig:act1}
	\end{subfigure}
 \begin{subfigure}{0.5\textwidth}
\centering
    \begin{tikzpicture}
    \filldraw[black] (0,0) circle (2pt);
    \node at (0,0.3) {$\mathcal{O}'$};
    \end{tikzpicture}
    \caption{}
    \label{fig:act2}
    \end{subfigure}
\end{figure}

For an invertible 0-form symmetry, associated to some group $G$,
abstractly one can always declare that the
entire group acts trivially.  Indeed, gauging such trivially-acting
symmetry groups is at the heart of both pure gauge theories
as well as Dijkgraaf-Witten theory \cite{Dijkgraaf:1989pz}.  

For noninvertible symmetries, this is often not possible.
For a simple example, consider Rep($S_3$).
Explicitly, there are three irreducible representations:
\begin{itemize}
\item the trivial representation, which we denote $1$,
\item the sign representation, a one-dimensional representation we denote
$X$,
\item a two-dimensional representation we denote $Y$.
\end{itemize}

The nontrivial products
are as follows:
\begin{equation}
X^2 \: = \: 1, \: \: \: X \otimes Y \: = \: Y, \: \: \:
Y^2 \: = \: 1 + X + Y.
\end{equation}

In this example, we see that the best we can hope for is for
the subgroup $\{1, X\}$ to act trivially on local operators.
The action of $Y$ is then constrained to obey
\begin{equation}
Y^2 \cdot {\cal O} \: = \: (1 + X + Y) \cdot {\cal O},
\end{equation}
so if $1$ and $X$ both act trivially, in the sense that
e.g.~$X \cdot {\cal O} = {\cal O}$, then
\begin{equation}
Y^2 \cdot {\cal O} \: = \: (2 + Y) \cdot {\cal O}.
\end{equation}
This will be satisfied if $Y \cdot {\cal O} = 2 {\cal O}$,
but not if $Y$ acts trivially.

In passing, the reader should note that $\{ 1, X \}$ generate an
invertible 0-form symmetry, namely ${\mathbb Z}_2$.  Only when we extend
to the full noninvertible case do we find that not all of the line
operators can act trivially.

With this example in mind, we define a line $L$ in a fusion category to act trivially if, for any local
operator ${\cal O}$,
\be   \label{eq:defn-triv-action}
L\cdot\mcO = \sum_{i,j=1}^{|L|}\delta_{i,j}\mcO=|L| \, \mcO.
\ee
We motivate this definition below.

Assume we have a 2d QFT with symmetry given by a fusion category $\mcC$.  We also assume that the there exists a fiber functor $F:\mcC\to {\rm Vec}$ to the category ${\rm Vec}$ of vector spaces, such that $F(L)= |L|1$, (meaning technically that the direct sum of the monoidal unit in ${\rm Vec}$ $|L|$ times, which clearly requires $|L|$ to be a nonnegative natural number). (As remarked earlier, gaugeable noninvertible symmetry categories will always admit a fiber functor.)  One way of understanding~(\ref{eq:defn-triv-action}) is via the fiber functor, which maps $L$ to $|L|$ times the identity.

When that symmetry acts trivially, $\dim(\Hom_{\rm Vec}(F(1),F(L)))=\dim(\Hom_{\rm Vec}(F(L),F(1)))=|L|$ for each simple object $L$ of $\mcC$, where $|L|$ is the quantum dimension of $L$.  Physically this means that there are (non-trivial) topological operators that sit at two-way junctions between lines.  These operators can be used to `unwrap' TDLs.  This was used in \cite{Robbins:2022wlr} in the group-like case.  For instance \cite[Figure 6]{Robbins:2022wlr}, reproduced here as Figure~\ref{fig:trivaction}, illustrates this visually.

\begin{figure}[h]
	\begin{subfigure}{0.5\textwidth}
		\centering
		\begin{tikzpicture}
			\draw[thick,->] (0,0) [partial ellipse=270:90:0.75cm and 0.75cm];
			\draw[thick,->] (0,0) [partial ellipse=270:-90:0.75cm and 0.75cm];
			\filldraw[black] (0,0) circle (2pt);
			\node at (0,0.3) {$\mathcal{O}$};
			\node at (-0.75,0.75) {$k$};
		\end{tikzpicture}
		\caption{}
		\label{fig:trivaction1}
	\end{subfigure}
	\begin{subfigure}{0.5\textwidth}
		\centering
		\begin{tikzpicture}
			\draw[thick,->] (0,0) [partial ellipse=270:90:0.75cm and 0.75cm];
			\draw[thick,->] (0,0) [partial ellipse=270:-90:0.75cm and 0.75cm];
			\filldraw[black] (0,0) circle (2pt);
			\node at (0,0.3) {$\mathcal{O}$};
			\node at (-0.75,0.75) {$k$};
			\filldraw[black] (0.75,0) circle (2pt);
			\node at (1.1,0) {$\ell_1$};
		\end{tikzpicture}
		\caption{}
		\label{fig:trivaction2}
	\end{subfigure}
	\begin{subfigure}{0.5\textwidth}
		\centering
		\begin{tikzpicture}
			\draw[thick] (0,0) [partial ellipse=180:45:0.75cm and 0.75cm];
			\draw[thick,->] (0,0) [partial ellipse=-45:-180:0.75cm and 0.75cm];
			\draw[thick,dashed,->] (0,0) [partial ellipse=45:0:0.75cm and 0.75cm];
			\draw[thick,dashed] (0,0) [partial ellipse=0:-45:0.75cm and 0.75cm];
			\filldraw[black] (0,0) circle (2pt);
			\node at (0,0.3) {$\mathcal{O}$};
			\node at (-0.75,0.75) {$k$};
			\filldraw[black] (0.54,0.54) circle (2pt);
			\filldraw[black] (0.54,-0.54) circle (2pt);
			\node at (0.9,0.9) {$\ell_{k^{-1}}$};
			\node at (1,0) {$1$};
			\node at (0.85,-0.9) {$\ell_k$};
		\end{tikzpicture}
		\caption{}
		\label{fig:trivaction3}
	\end{subfigure}
	\begin{subfigure}{0.5\textwidth}
		\centering
		\begin{tikzpicture}
			\draw[thick] (0,0) [partial ellipse=180:135:0.75cm and 0.75cm];
			\draw[thick,->] (0,0) [partial ellipse=-135:-180:0.75cm and 0.75cm];
			\draw[thick,dashed,->] (0,0) [partial ellipse=135:0:0.75cm and 0.75cm];
			\draw[thick,dashed] (0,0) [partial ellipse=0:-135:0.75cm and 0.75cm];
			\filldraw[black] (0,0) circle (2pt);
			\node at (0,0.3) {$\mathcal{O}$};
			\node at (-1,0) {$k$};
			\filldraw[black] (-0.54,0.54) circle (2pt);
			\filldraw[black] (-0.54,-0.54) circle (2pt);
			\node at (-0.85,0.85) {$\ell_{k^{-1}}$};
			\node at (1,0) {$1$};
			\node at (-0.85,-0.9) {$\ell_k$};
		\end{tikzpicture}
		\caption{}
		\label{fig:trivaction4}
	\end{subfigure}
	\begin{subfigure}{0.5\textwidth}
		\centering
		\begin{tikzpicture}
			\draw[thick,dashed,->] (0,0) [partial ellipse=270:90:0.75cm and 0.75cm];
			\draw[thick,dashed,->] (0,0) [partial ellipse=270:-90:0.75cm and 0.75cm];
			\filldraw[black] (0,0) circle (2pt);
			\node at (0,0.3) {$\mathcal{O}$};
			\filldraw[black] (-0.75,0) circle (2pt);
			\node at (-1.1,0) {$\ell_1$};
		\end{tikzpicture}
		\caption{}
		\label{fig:trivaction5}
	\end{subfigure}
	\caption{}
	\label{fig:trivaction}
\end{figure}

We can extend this calculation to noninvertible lines.  Let us again take a simple object $L$.  Pick a basis $\ell^i_L$ for $\Hom_{\rm Vec}(F(1),F(L))$ and $\overline{\ell}^i_L$ for $\Hom_{\rm Vec}(F(L),F(1))$.\footnote{In the group-like case we are able to identify $\bar{\ell}_g$ with $\ell_{g^{-1}}$, which was done implicitly in Figure~\ref{fig:trivaction}.  In the more general case the starting point operator for a line $L$ should be isomorphic to the endpoint operator for its orientation reversal $\bar{L}$, with the isomorphism determined by (co)evaluation.}  We take this basis to be orthonormal in the sense that 
\be
\label{orthonormality}
\id = \sum_{i,j=1}^{|L|} \ell^j_L \circ \overline{\ell}^i_L, 
\hspace{1cm} \overline{\ell}^j_L \circ \ell^i_L = \delta_{i,j}\id,
\ee
where $\id$ is the local identity operator.  These two conditions are pictured in Figure~\ref{trax1} and~\ref{trax2}.

\begin{figure}[h]
	\begin{subfigure}{\textwidth}
		\centering
		\begin{tikzpicture}
			\draw[thick,->] (0,0) -- (1.5,0);
			\node at (1.5,-0.5) {$L$};
			\draw[thick] (1.5,0) -- (3,0);
			\node at (4,0) {$=\sum_{i,j=1}^{|L|}$};
			\draw[thick,->] (5,0) -- (5.5,0);
			\node at (5.5,-0.5) {$L$};
			\draw[thick] (5.5,0) -- (6,0);
			\filldraw[black] (6,0) circle (2pt);
			\node at (6,0.5) {$\bar{\ell}^i_L$};
			\draw[dashed] (6,0) -- (7,0);
			\filldraw[black] (7,0) circle (2pt);
			\node at (7,0.5) {$\ell^j_L$};
			\draw[thick,->] (7,0) -- (7.5,0);
			\node at (7.5,-0.5) {$L$};
			\draw[thick] (7.5,0) -- (8,0);
		\end{tikzpicture}
		\caption{}
		\label{trax1}
	\end{subfigure}
	\begin{subfigure}{\textwidth}
		\centering
		\begin{tikzpicture}
			\filldraw[black] (0,0) circle (2pt);
			\node at (0,0.5) {$\ell^i_L$};
			\draw[thick,->] (0,0) -- (1,0);
			\draw[thick] (1,0) -- (2,0);
			\filldraw[black] (2,0) circle (2pt);
			\node at (2,0.5) {$\bar{\ell}^j_L$};
			\node at (3,0) {$=$};
			\filldraw[black] (4,0) circle (2pt);
			\node at (4,0.5) {$\delta_{i,j}\id$};
		\end{tikzpicture}
		\caption{}
		\label{trax2}
	\end{subfigure}
	\caption{Relations for topological twist fields in a trivially-acting symmetry.}
	\label{trax}
\end{figure}

Using this setup to repeat the calculation in Figure~\ref{fig:trivaction}, we find that a trivially-acting (in general noninvertible) line $L$ acts on a local operator $\mcO$ to give
\be  
L\cdot\mcO = \sum_{i,j=1}^{|L|}\delta_{i,j}\mcO=|L|\mcO,
\ee
which is the behavior we would expect.  This matches our definition~(\ref{eq:defn-triv-action}) of a trivially-acting
noninvertible symmetry.

\begin{figure}[h]
	\begin{subfigure}{0.5\textwidth}
		\centering
		\begin{tikzpicture}
			\draw[thick,->] (-1,1) -- (-0.5,0.5);
			\draw[thick] (-0.5,0.5) -- (0,0);
			\node at (-0.25,0.75) {$L_1$};
			\filldraw[black] (-1,1) circle (2pt);
			\node at (-1.5,1) {$\ell_1$};
			\draw[thick,->] (-1,-1) -- (-0.5,-0.5);
			\draw[thick] (-0.5,-0.5) -- (0,0);
			\node at (-0.25,-0.75) {$L_2$};
			\filldraw[black] (-1,-1) circle (2pt);
			\node at (-1.5,-1) {$\ell_2$};
			\draw[thick,->] (0,0) -- (1,0);
			\draw[thick] (1,0) -- (2,0);
			\node at (1.5,0.5) {$L_3$};

			\draw[thick,color=blue,->,opacity=0.5] (0,0) [partial ellipse=135:360:0.2cm and 0.2cm];
		\end{tikzpicture}
		\caption{}
		\label{tpofuse1}
	\end{subfigure}
	\begin{subfigure}{0.5\textwidth}
		\centering
		\begin{tikzpicture}
			\draw[thick,->] (0,0) -- (1,0);
			\draw[thick] (1,0) -- (2,0);
			\filldraw[black] (0,0) circle (2pt);
			\node at (0,0.5) {$\ell_3$};
			\node at (1.5,0.5) {$L_3$};
		\end{tikzpicture}
		\vspace{1cm}
		\caption{}
		\label{tpofuse2}
	\end{subfigure}
	\caption{Fusion of twist fields on which TDLs begin/end.}
	\label{tpofuse}
\end{figure}
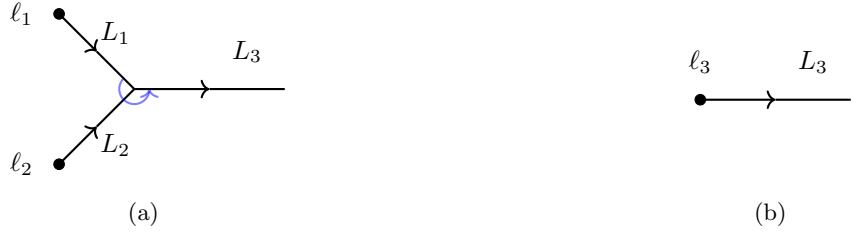

These local operators should also have a fusion operation inherited from the lines on which they live.  For instance, we ought to be able to deform Figure~\ref{tpofuse1} to Figure~\ref{tpofuse2}.  In the case of $\Rep(G)$, this map $\ell_1\otimes \ell_2\to \ell_3$ is given by the intertwiners.

The difference between $\ell$ and $\overline{\ell}$ is essentially the same as a choice of orientation for a TDL.  Note that we could choose to work only with the $\ell$, establishing our conventions such that all lines were outgoing from these operators (physically, there is indeed only one set of topological twist fields labeled by ${\cal C}$).  Doing so would require using evaluation to reverse the orientation of any ingoing line, at the cost of dualizing it.  We can instead choose (as we will) to work with both $\ell$ and $\overline{\ell}$, which are not independent of each other.  That is, $\ell_{\overline{L}}$ and $\overline{\ell}_{L}$ are isomorphic but not necessarily equal, in exactly the same way as $L$ and $\overline{L}$ with opposite orientation.  To translate back to the opposite convention, we can regard $\overline{\ell}$ as $\ell$ with a nearby insertion of the image of the evaluation map $F(\epsilon)\in\Hom_{\rm Vec}(F(L\otimes\overline{L}),F(1))$:
\be
	\begin{tikzpicture}
		\draw[thick] (0,0) -- (1,0);
		\draw[thick,->] (2,0) -- (1,0);
		\filldraw[black] (0,0) circle (2pt);
		\node at (0,-0.5) {$\overline{\ell}_L$};
		\node at (1.0,0.5) {$L$};
        \node at (2.5,0) {$\equiv$};
        \filldraw[black] (3,0) circle (2pt);
        \node at (3,-0.5) {$\ell_{\overline{L}}$};
        \draw[thick,->] (3,0) -- (3.5,0);
        \node at (3.5,0.5) {$\overline{L}$};
        \draw[thick] (3.5,0) -- (4,0);
        \filldraw[black] (4,0) circle (2pt);
        \draw[dashed] (4,0) -- (4,-0.5);
        \node at (4,0.5) {$\epsilon$};
        \draw[thick] (4,0) -- (5,0);
        \node at (5,0.5) {$L$};
        \draw[thick,->] (6,0) -- (5,0);

        \draw[thick,color=blue,->,opacity=0.5] (4,0) [partial ellipse=0:270:0.2cm and 0.2cm];
	\end{tikzpicture}.
\ee
Similarly, the coevaluation map $\gamma$ allows us to map $\overline{\ell}_L$ to $\ell_{\overline{L}}$.
(Given $\overline{\ell}_L: L \rightarrow 1$, note $\overline{\ell}_L \otimes \overline{L}: L \otimes \overline{L} \rightarrow \overline{L}$, and composing with coevaluation gives a map $1 \rightarrow \overline{L}$, giving an
$\ell_{\overline{L}}$.)

We can check that the above formulation of trivially-acting noninvertible symmetries is consistent with the action of noninvertible symmetries on boundaries.
The paper \cite{Choi:2023xjw} considers the related question of what it
means for a boundary to be invariant under the action of a noninvertible
symmetry.  For an ordinary, invertible, symmetry, defined by a collection
of lines $L_g$ associated to elements $g \in G$ in some finite group,
for a boundary state $| {\cal B} \rangle$ to be invariant\footnote{
The language of boundary states glosses over the distinction between
`invariant' and `equivariant.'  For example, because of the existence of
gauge transformations on worldvolume gauge fields, 
it is not quite right to speak of a D-brane ${\cal B}$
being `invariant' under
a group $G$, only `equivariant,' which means that there are isomorphisms
\begin{equation}
\psi_g: \: {\cal B} \: \longrightarrow \: {\cal B}
\end{equation}
associated to group elements $g \in G$, which respect the group
law:  $\psi_g \psi_h = \psi_{gh}$.  This structure is not visible
in the language of boundary states for two reasons:
(1) Boundary states are constructed from e.g.~Chern classes, for which
there are no gauge transformations, so `invariant' is well defined, and
(2) as states, they are elements of some vector space, which does not admit
the requisite higher categorical morphisms.
} means
\begin{equation}
L | {\cal B} \rangle \: = \: | {\cal B} \rangle.
\end{equation}
The reference
\cite[section 2.3]{Choi:2023xjw} provides two generalizations of this
notion to
to noninvertible symmetries.
Briefly,
\begin{itemize}
\item A boundary $| {\cal B} \rangle$ is defined to be weakly symmetric
under a fusion category symmetry ${\cal C}$ if for every simple line
$L \in {\cal C}$,
\begin{equation}
L | {\cal B} \rangle \: = \: | {\cal B} \rangle \: + \: \cdots,
\end{equation}
(see \cite[equ'n (2.17)]{Choi:2023xjw}).
\item A boundary $| {\cal B} \rangle$ is defined to be strongly symmetric
under a fusion category symmetry ${\cal C}$ if for every simple
line $L \in {\cal C}$,
\begin{equation} \label{eq:strongsymm-bc}
L | {\cal B} \rangle \: = \:
\langle L \rangle | {\cal B} \rangle
\end{equation}
(see \cite[equ'n (2.20)]{Choi:2023xjw})
where $\langle L \rangle$ is the quantum dimension of $L$.

They also note that a strongly symmetric boundary condition cannot exist
unless every topological line in ${\cal C}$ has an integer quantum dimension,
which is not true in general (see e.g.~the minimal models discussed
in \cite[section 4.2.1]{Choi:2023xjw}).

\end{itemize}
It is observed in \cite{Choi:2023xjw} that strongly symmetric implies
weakly symmetric; however, not all weakly symmetric boundaries are
strongly symmetric.
Further, in the special case of invertible lines, these
two notions are equivalent: 
$L \otimes | {\cal B} \rangle = | {\cal B} \rangle$,
reflecting both the indecomposability of the product, and the
fact that
the quantum dimension
of an invertible line is 1.

The second of these two boundary conditions, the strongly symmetric condition
in equation~(\ref{eq:strongsymm-bc}),
is clearly consistent with
our definition~(\ref{eq:defn-triv-action}) of trivial actions above.
This confirms that our definition~(\ref{eq:defn-triv-action}) is sensible.

\subsection{Decomposition conjecture}

In two-dimensional (ordinary) gauge theories in which a subgroup of the gauge group
acts trivially, it is by now well-known that the theory is equivalent to a disjoint union of effectively-acting gauge theories (see \cite{Hellerman:2006zs} for the original statement,
and \cite{Sharpe:2022ene} for a recent review).  In this section, we will propose a similar phenomenon when gauging a trivially-acting noninvertible symmetry.

Let us briefly review the ordinary case.  Suppose for simplicity $G$ is a finite group, with central subgroup $K \subset G$ acting trivially.  Then, as discussed in \cite{Hellerman:2006zs},
\begin{equation}
    {\rm QFT}\left( [X/G] \right) \: = \: {\rm QFT}\left( \coprod_{\hat{K}} [X/ (G/K)]_{\omega} \right),
\end{equation}
where the disjoint union is over irreducible representations of $K$, and $\omega$ denotes discrete torsion factors described in \cite{Hellerman:2006zs}.  A special case of this is
two-dimensional Dijkgraaf-Witten theory, which is the orbifold
$[{\rm point}/G]$, for $G$ a group (possibly twisted by discrete torsion).
It decomposes into a string on a disjoint union of points, as many copies
as irreducible (projective) representations of $G$.
(See e.g.~\cite{Robbins:2020msp} 
for the extension of decomposition to include discrete torsion in the
original orbifold.)  Formally, we could write this as\footnote{
Our notation glosses over extended objects, which can distinguish the `points' on the right-hand side as SPT phases.  The intent of the notation is to emphasize that a local QFT can be a disjoint union.
}
\begin{equation}
    {\rm QFT}\left( [{\rm point}/G] \right) \: = \:
    {\rm QFT}\left( \coprod_{\hat{G}} {\rm point} \right),
\end{equation}
where, since all of the orbifold group $G$ acts trivially, the disjoint union on the right is
indexed by irreducible (projective) representations of $G$, here denoted $\hat{G}$.

In this section we will consider the noninvertible analogue of Dijkgraaf-Witten theory,
meaning a gauge theory in which one gauges a trivially-acting noninvertible symmetry,
in the sense that all of the noninvertible symmetry acts trivially.
(More general actions of noninvertible symmetries should also exist, in which only a 
subcategory acts trivially; we leave such cases for future work.)
This might also be described as a two-dimensional
analogue of Turaev-Viro theory \cite{tv1} (which is a noninvertible 
generalization of three-dimensional Dijkgraaf-Witten theory \cite{Dijkgraaf:1989pz}).

Here we consider gauging a Frobenius algebra ${\cal A}$ in a trivially-acting ${\rm Rep}(G)$ symmetry,
meaning the orbifold\footnote{
To be clear, by `orbifold' we mean the result of gauging the noninvertible symmetry with the Frobenius
algebra ${\cal A}$.  The notation is not intended to indicate an ordinary quotient stack.
} $[{\rm point}/{\cal A}]$.
We assume the theory does not have any analogue of discrete torsion.

We conjecture that this theory
is equivalent to a disjoint union of trivial theories (or SPT phases),
as many as $|{\cal A}|$, and hence obeys an analogue of decomposition,
using the fact that
\begin{equation}
    |{\cal A}| \: = \: \sum_{\rho} |\rho|,
\end{equation}
where the sum is over simple objects $\rho$, counted with multiplicity.  (Recall in Rep$(G)$, the simple objects are the irreducible
representations $\rho$, and $|\rho|$ is the dimension of the representation $\rho$. The reader should also recall that we only define trivial actions in cases where fiber functors exist, and leave a more general definition for future work.)  
In terms of partition functions, we are predicting that
\begin{equation}  \label{eq:noninv-decomp-pred}
    Z_{ {\cal A}, {\cal A}} \: = \: | {\cal A}| \, Z_{1,1}^1.
\end{equation}
We will check this proposal in detail in examples later in this section.

Now, this proposal is incomplete:
\begin{itemize}
    \item For one, ideally one would like to understand decomposition not just in cases in which all of the simple objects act trivially, but also in cases in which only a subset of the simple objects act trivially.  We leave that for future work.
    \item Another issue is that the proposal above does not simply generalize to include ordinary orbifolds, for which the fusion category is Vec$(G) = {\rm Rep}( {\mathbb C}[G]^*)$.  The issue is that the number of simple objects is $|G|$, whereas in the decomposition of Dijkgraaf-Witten theory, the universes are counted by irreducible representations of $G$.
\end{itemize}
We leave these issues for future work.

In the remainder of this section, we will compute some examples.

\subsection{Computation of partial traces}

In this subsection, we will make a proposal for the computation of partial traces $Z_{L_1,L_2}^{L_3}$ in the special case that the noninvertible symmetry acts completely trivially.
This will be an analogue of two-dimensional Dijkgraaf-Witten theory, which describes the orbifold $[{\rm point}/G]$.

\begin{figure}[h]
	\begin{subfigure}{0.45\textwidth}
		\centering
		\begin{tikzpicture}
			\draw[thick] (-1,1) -- (-0.5,0.5);
			\draw[thick,->] (0,0) -- (-0.5,0.5);
			\node at (-0.25,0.75) {$L_1$};
			\draw[thick] (-1,-1) -- (-0.5,-0.5);
			\draw[thick,->] (0,0) -- (-0.5,-0.5);
			\node at (-0.25,-0.75) {$L_2$};
			\draw[thick] (0,0) -- (1,0);
			\draw[thick,->] (2,0) -- (1,0);
			\node at (1,0.5) {$L_3$};
			\draw[thick] (2,0) -- (2.5,0.5);
			\draw[thick,->] (3,1) -- (2.5,0.5);
			\node at (2.25,0.75) {$L_2$};
			\draw[thick] (2,0) -- (2.5,-0.5);
			\draw[thick,->] (3,-1) -- (2.5,-0.5);
			\node at (2.25,-0.75) {$L_1$};

            \draw[thick,color=blue,->,opacity=0.5] (2,0) [partial ellipse=-45:180:0.2cm and 0.2cm];
			\draw[thick,color=blue,->,opacity=0.5] (0,0) [partial ellipse=-0:225:0.2cm and 0.2cm];
		\end{tikzpicture}
		\caption{}
		\label{pt_lines1}
	\end{subfigure}
	\begin{subfigure}{0.45\textwidth}
		\centering
		\begin{tikzpicture}
			\draw[thick,->] (-0.5,0.5) -- (-0.75,0.75);
            \draw[thick] (-0.75,0.75) -- (-1,1);
            \filldraw[black] (-0.5,0.5) circle (2pt);
            \node at (-0.75,0.25) {$\gamma$};
            \draw[thick,->] (-0.5,0.5) -- (-0.25,0.25);
			\draw[thick] (-0.25,0.25) -- (0,0);
			\node at (-0.5,1) {$L_1$};
            \node at (0, 0.5) {$\bar{L}_1$};
			\draw[thick] (-1,-1) -- (-0.5,-0.5);
			\draw[thick,->] (0,0) -- (-0.5,-0.5);
			\node at (-0.25,-0.75) {$L_2$};
			\draw[thick] (0,0) -- (1,0);
			\draw[thick,->] (2,0) -- (1,0);
			\node at (1,0.5) {$L_3$};
			\draw[thick] (2,0) -- (2.5,0.5);
			\draw[thick,->] (3,1) -- (2.5,0.5);
			\node at (2.25,0.75) {$L_2$};
			\draw[thick] (2,0) -- (2.5,-0.5);
			\draw[thick,->] (3,-1) -- (2.5,-0.5);
			\node at (2.25,-0.75) {$L_1$};

            \draw[thick,color=blue,->,opacity=0.5] (2,0) [partial ellipse=-45:180:0.2cm and 0.2cm];
			\draw[thick,color=blue,->,opacity=0.5] (0,0) [partial ellipse=-0:225:0.2cm and 0.2cm];
		\end{tikzpicture}
		\caption{}
		\label{pt_lines2}
	\end{subfigure}
    \centering
    \begin{subfigure}{0.5\textwidth}
		\centering
		\begin{tikzpicture}
			\draw[thick,->] (-0.5,0.5) -- (-0.75,0.75);
            \draw[thick] (-0.75,0.75) -- (-1,1);
            \filldraw[black] (-0.5,0.5) circle (2pt);
            \node at (-0.75,0.25) {$\gamma$};
            \draw[thick,->] (-0.5,0.5) -- (-0.25,0.25);
			\draw[thick] (-0.25,0.25) -- (0,0);
			\filldraw[black] (-1,1) circle (2pt);
            \node at (-1.5,1) {$\overline{\ell}_1^i$};
			\draw[thick] (-1,-1) -- (-0.5,-0.5);
			\draw[thick,->] (0,0) -- (-0.5,-0.5);
			\filldraw[black] (-1,-1) circle (2pt);
			\node at (-1.5,-1) {$\bar{\ell}_2^j$};
			\draw[thick] (0,0) -- (0.4,0);
			\draw[thick,->] (0.8,0) -- (0.4,0);
			\filldraw[black] (0.8,0) circle (2pt);
			\node at (0.8,-0.5) {$\ell_3^k$};
			\draw[thick] (1.2,0) -- (1.6,0);
			\draw[thick,->] (2,0) -- (1.6,0);
			\filldraw[black] (1.2,0) circle (2pt);
			\node at (1.3,-0.5) {$\bar{\ell}_3^k$};
			\draw[thick] (2,0) -- (2.5,0.5);
			\draw[thick,->] (3,1) -- (2.5,0.5);
			\filldraw[black] (3,1) circle (2pt);
			\node at (3.5,1) {$\ell_2^j$};
			\draw[thick] (2,0) -- (2.5,-0.5);
			\draw[thick,->] (3,-1) -- (2.5,-0.5);
			\filldraw[black] (3,-1) circle (2pt);
			\node at (3.5,-1) {$\ell_1^i$};

            \draw[thick,color=blue,->,opacity=0.5] (2,0) [partial ellipse=-45:180:0.2cm and 0.2cm];
			\draw[thick,color=blue,->,opacity=0.5] (0,0) [partial ellipse=-0:225:0.2cm and 0.2cm];
		\end{tikzpicture}
		\caption{}
		\label{pt_lines3}
	\end{subfigure}
	\caption{Mapping $Z_{L_1,L_2}^{L_3}$ to a multiple of $Z_{1,1}^1$. Each of these diagrams should be understood as living on $T^2$, so that lines wrap around the edges.  Note that in diagram~\ref{pt_lines3}, all three of the lines $L_i$ have been broken in half.  The $L_1$ and $L_2$ lines wrap around the edges of the figure, and their breaking is indicated by the vertices $\bullet$ at the edges.  In each case, $\ell_i$ denote operators inserted at the endpoints at the breaks.}
	\label{pt_lines}
\end{figure}
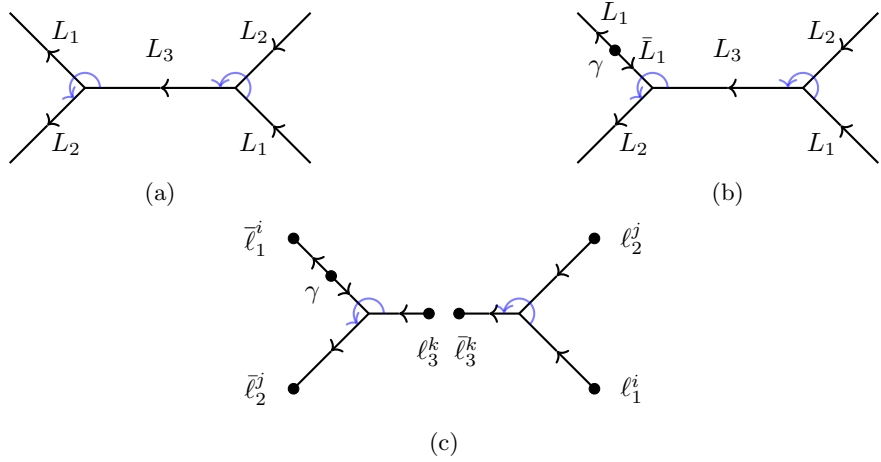

A generic torus partial trace will have the form shown in Figure~\ref{pt_lines1}.  We propose to evaluate it as follows.  Formally, we can rewrite this as a diagram involving two fusion operations by inserting a map $\gamma: \overline{\ell} \rightarrow \ell$, 
induced by the coevaluation map, along one of the lines, leading to Figure~\ref{pt_lines2}.  We can simplify this diagram using equation~(\ref{orthonormality}).  This allows us to deform Figure~\ref{pt_lines2} to Figure~\ref{pt_lines3}.

Compressing both halves of the resulting diagram, we are left only with multiples of the local identity operator.
This diagrammatic argument implies that we can write
\begin{equation}
    Z_{L_1,L_2}^{L_3} \: = \: C_{1,2}^3 \, Z_{1,1}^1
\end{equation}
where $C_{1,2}^3$ is a constant which we propose is formally given by
\be
\label{decomp_coefficient_formula}
C_{1,2}^3 \: = \: 
\sum_{i,j,k}\left[ \overline{\ell}_3^k \circ (\ell_1^i\otimes\ell_2^j)]\right]
\left[\overline{\ell}_2^j \circ (\ell_3^k\otimes \gamma(\overline{\ell}_1^i))\right].
\ee

As a quick consistency check, for an untwisted sector partial trace $Z_{1,L}^L$ the coefficient is
\be
C_{1,L}^L = 
\sum_{j,k=1}^{|L|}\left[\overline{\ell}_3^k \circ (\id\otimes\ell_2^j)\right]
\left[\overline{\ell}_2^j \circ (\ell_3^k\otimes \gamma(\id)) \right] 
= 
\sum_{j,k=1}^{|L|} \delta_{j,k}\delta_{k,j} = |L|,
\ee
which is what we expect from a single trivially-acting $L$ line wrapping a cycle of the torus.

The formula above represents our physically-motivated proposal for the relation between $Z_{L_1,L_2}^{L_3}$ and
$Z_{1,1}^1$ in the special case of a completely-trivially-acting noninvertible symmetry.  We leave a detailed mathematical understanding for future work.

In the case that the fusion category is group-like, the $C_{1,2}^3$ should reproduce known results for decomposition.  In this case all of the lines have dimension 1, which means that (\ref{decomp_coefficient_formula}) no longer contains any sums and becomes
\be
C_{g,h}^{gh} = \left[\overline{\ell}_{gh}\circ(\ell_g\otimes\ell_h)\right]\left[\overline{\ell}_h\circ(\ell_{gh}\otimes\gamma(\overline{\ell}_g))\right].
\ee
Here coevaluation is trivial, so $\gamma(\overline{\ell}_g)=\ell_{g^{-1}}$.  Fusion follows the group law, so we have
\be
C_{g,h}^{gh} = \left[\overline{\ell}_{gh}\circ\ell_{gh}\right]\left[\overline{\ell}_h\circ\ell_{ghg^{-1}}\right] = \delta_{h,ghg^{-1}}
\ee
where the last equality follows from the orthonormality (\ref{orthonormality}) of composition.  This reproduces the expected result for decomposition in a group-like orbifold, and in particular matches the topological operator-based formulation of decomposition given in \cite{Robbins:2022wlr}.

\subsection{Examples}

In this section we will compute partition functions for gauged trivially-acting Rep($G$) for the groups
$S_3$, $D_4$, and $Q_8$, for all Frobenius algebras discussed earlier, and compare to the decomposition conjecture.
In principle we expect that the same methods should also apply to Rep$({\cal H}_8)$ and other more general fusion
categories of the form Rep$({\cal H})$ for ${\cal H}$ any suitable Hopf algebra; however, as a practical matter, we do not have
sufficient information about e.g.~coevaluation maps to perform the computation in that case, and so it is left for
future work.

\subsubsection{Rep$(S_3)$}

In this section we will work through the prediction for the case of the Rep$(S_3)$ fusion category.
This will be an exercise in computing the constants $C_{1,2}^3$ described in the previous section, and then applying to simplify partition functions.

Let us work through a simple example of a computation of the constants $C_{1,2}^3$ from their definition~(\ref{decomp_coefficient_formula}).  Consider for example $C_{X,Y}^Y$.  From the definition,
\begin{equation}  \label{eq:cxyy}
    C_{X,Y}^Y \: = \: \sum_{i,j,k} \left[ \overline{\ell}_Y^k \circ \left( \ell_X^i \otimes \ell_Y^j \right) \right]
    \left[ \overline{\ell}_Y^j \circ \left( \ell_Y^k \otimes \gamma(\overline{\ell}_X^i) \right) \right],
\end{equation}
where $i$, $j$, $k$ run over the dimensions of the corresponding vector spaces.
(For example, since $X$ is a one-dimensional representation, $i=1$ only, but since $Y$ is a two-dimensional
representation, $j, k \in \{1, 2\}$.)

Using the Rep$(S_3)$ intertwiners (\ref{rs3int1})-(\ref{rs3int2}), the coevaluation maps induce
\begin{equation}
    \gamma : \begin{cases}
        \overline{\ell}_1 \mapsto \ell_1, \\
        \overline{\ell}_X \mapsto \beta_1^{-1} \ell_X, \\
        \overline{\ell}_{Y}^1 \mapsto \beta_4^{-1} \ell_Y^1, \\
        \overline{\ell}_Y^2 \mapsto \beta_4^{-1} \ell_Y^2.
    \end{cases}
\end{equation}
(In principle, the $\ell$'s on the right-hand side are computed with respect to $\overline{L}$, not $L$, but we are using the fact that in $S_3$, all representations obey $\overline{L} = L$.)

The intertwiners~(\ref{eq:s3:int:xy1}), (\ref{eq:s3:int:xy2}) induce
\begin{eqnarray}
     \ell_X^1 \otimes \ell_Y^j & = & \left\{ \begin{array}{cl}
     + \beta_2 \ell_Y^2 & j = 1, \\
     - \beta_2 \ell_Y^1 & j=2,
     \end{array} \right.
     \\
     {\ell}_Y^k \otimes \gamma\left( \overline{\ell}_X^1 \right)
     & = &
     \left\{ \begin{array}{cl}
     + \beta_3 \beta_1^{-1} \ell_Y^2 & k=1, \\
     - \beta_3 \beta_1^{-1} \ell_Y^1 & k=2.
     \end{array} \right.
\end{eqnarray}
Then, using the orthogonality relations~(\ref{orthonormality}), we see
\begin{eqnarray}
    \overline{\ell}_Y^k \circ \left( \ell_X^1 \otimes \ell_Y^j \right) 
    & = & \left\{ \begin{array}{cl}
    + \beta_2 & j=1, k=2, \\
    - \beta_2 & j=2, k=1, \\
    0 & {\rm else}.
    \end{array} \right.
\\
   \overline{\ell}_Y^j \circ \left( \ell_Y^k \otimes \gamma\left( \overline{\ell}_X^i \right) \right)
   & = &
   \left\{ \begin{array}{cl}
   - \beta_3 \beta_1^{-1} & j=1, k=2 \\
   + \beta_3 \beta_1^{-1} & j=2, k=1
   \end{array} \right.
\end{eqnarray}
so plugging into equation~(\ref{eq:cxyy}), we have
\begin{equation}
C_{X,Y}^Y \: = \:
   - 2 \beta_2 \beta_3 \beta_1^{-1}.
\end{equation}

Mapping the partial traces to the parent theory partition function is now a straightforward application of (\ref{decomp_coefficient_formula}).  We list here the results for the constants $C_{1,2}^3$:
\begin{eqnarray}
    C_{1,X}^X & = & 1 \: = \: C_{X,1}^X \: = \: C_{X,X}^1, \\
    C_{1,Y}^Y & = & 2 \: = \: C_{Y,1}^Y \: = \: C_{Y,Y}^1, \\
    C_{X,Y}^Y & = &  -2\frac{\beta_2 \beta_3}{\beta_1} , \\
    C_{Y,X}^Y & = &  -2\frac{\beta_3 \beta_6}{\beta_4} , \\
    C_{Y,Y}^X & = &  2\frac{\beta_2 \beta_6}{\beta_4} , \\
    C_{Y,Y}^Y & = & 4\frac{\beta^2_5}{\beta_4} .
\end{eqnarray}

We can now apply this result to the various Rep$(S_3)$ gaugings.
\begin{itemize}
    \item $1+X$:
    The $1+X$ partition function~(\ref{rs3_1+x_pf}) becomes 
    \begin{eqnarray}
        Z_{1+X} & = & \hlf\left[ Z_{1,1}^1+Z_{1,X}^X+Z_{X,1}^X+Z_{X,X}^1\right],
        \\
        & = & \frac{1}{2} \left( 1 + C_{1,X}^X + C_{X,1}^X + C_{X,X}^1 \right) Z_{1,1}^1, 
        \\
        & = & \frac{1}{2} \left( 1 + 1 + 1 + 1 \right) Z_{1,1}^1 \: = \: 2 \, Z_{1,1}^1,
    \end{eqnarray}
    exactly as we would expect for a trivially-acting $\Z_2$ orbifold.  Comparing to the conjecture~(\ref{eq:noninv-decomp-pred}), there are two simple objects ($1$, $X$),
    each of quantum dimension one, so indeed $|{\cal A}| = 2$, and so our our prediction~(\ref{eq:noninv-decomp-pred}) matches the computation.
    \item $1+Y$: The $1+Y$ partition function~(\ref{rs3_1+y_pf}) becomes  
    \begin{eqnarray}
        Z_{1+Y} & = & \frac{1}{3}\left[ Z_{1,1}^1 + Z_{1,Y}^Y + Z_{Y,1}^Y + Z_{Y,Y}^1 + \frac{\beta_4}{2\beta_5^2}Z_{Y,Y}^Y\right],
        \\
        & = & \frac{1}{3} \left[ 1 + C_{1,Y}^Y + C_{Y,1}^Y + C_{Y,Y}^1 + \frac{\beta_4}{2\beta_5^2}C_{Y,Y}^Y \right] Z_{1,1}^1, 
        \\
        & = & \frac{1}{3} \left[ 1 + 2 + 2 + 2 + \left( \frac{\beta_4}{2\beta_5^2} \right) \left( 4 \frac{\beta_5^2}{\beta_4} \right) \right] \: = \: 3 \, Z_{1,1}^1.
    \end{eqnarray}
    Comparing to the conjecture~(\ref{eq:noninv-decomp-pred}), there are two simple objects ($1$, $Y$).
    Of these, $|1| = 1$ but $|Y| = 2$, so our conjecture predicts a factor of $|{\cal A}| = 1+2 = 3$, and so our prediction~(\ref{eq:noninv-decomp-pred}) matches the result here.
    \item $1+X+2Y$:
    Finally the regular representation orbifold partition function~(\ref{rs3_rr_pf}) becomes 
    \begin{eqnarray}
        Z_{1+X+2Y} & = & \frac{1}{6} \biggl[  Z_{1,1}^1 \: + \: \left( Z_{1,X}^X + Z_{X,1}^X + Z^1_{X,X} \right) \: + \: 2 \left( Z_{1,Y}^Y + Z_{Y,1}^Y + Z_{Y,Y}^1 +  \frac{\beta_4}{2 \beta_5^2} Z_{Y,Y}^Y \right)
    \nonumber \\
    & & \hspace*{0.5in}
    \: - \:
    \frac{2 \beta_1}{\beta_2 \beta_3} \left(  Z^Y_{X,Y} +
    \frac{\beta_2 \beta_4}{\beta_1 \beta_6} Z^Y_{Y,X} - \frac{\beta_3 \beta_4}{\beta_1 \beta_6} Z_{Y,Y}^X - \frac{\beta_2 \beta_3 \beta_4}{2 \beta_1 \beta_5^2} Z_{Y,Y}^Y \right) \biggr],
    \\
    & = & \frac{1}{6} \biggl[ 1 + (1 + 1 + 1) + 2 \left( 2 + 2 + 2 + \left( \frac{\beta_4}{2 \beta_5^2} \right)  
    \left( 4 \frac{\beta_5^2}{\beta_4} \right) \right) 
    \nonumber \\
    & & \hspace*{0.5in}
    \: - \: \frac{2 \beta_1}{\beta_2 \beta_3} \biggl(  -2 \frac{\beta_2 \beta_3}{\beta_1} + \left(\frac{\beta_2 \beta_4}{\beta_1 \beta_6}  \right) \left( -2 \frac{\beta_3 \beta_6}{\beta_4} \right) - \left(\frac{\beta_3 \beta_4}{\beta_1 \beta_6}\right) \left( 2 \frac{\beta_2 \beta_6}{\beta_4} \right)
    \nonumber \\
    & & \hspace*{1.25in}
    \: - \: \left( \frac{\beta_2 \beta_3 \beta_4}{2 \beta_1 \beta_5^2} \right) \left( 4 \frac{\beta_5^2}{\beta_4} \right) \biggr) \biggr] Z_{1,1}^1,
    \\
    & = & 6 \, Z_{1,1}^1.
    \end{eqnarray}
    This is again in line with expectation, as we could have constructed the trivially-acting Rep$(S_3)$ symmetry from the free action of $S_3$ on six universes, and gauging the regular representation should undo the original $S_3$ orbifold.
    In terms of the conjecture, there are four simple objects ($1$, $X$, and two copies of $Y$), and taking
    into account their quantum dimensions, 
    \begin{equation}
        |{\cal A}| \: = \: |1| + | X | + | Y | + | Y | \: = \: 1 + 1 + 2 + 2 \: = \: 6,
    \end{equation}
    and so our prediction~(\ref{eq:noninv-decomp-pred}) matches the result here.

    We also note that the fact that the result is independent of the $\beta$'s is highly nontrivial in this case, 
    requiring intricate cancellations between different factors, 
    which is in itself a nontrivial consistency check on our methods.
\end{itemize}

\subsubsection{Rep$(D_4)$}

In Rep$(D_4)$, the constants $C_{1,2}^3$ relating the partial traces to $Z_{1,1}^1$ (for trivially-acting
noninvertible symmetry) are given by
\begin{equation}
    C_{1,a}^a \: = \: 1 \: = \: C_{a,1}^a \: = \: C_{a,a}^1 \: = \:
    C_{1,b}^b \: = \: C_{b,1}^b \: = \: C_{b,b}^1 \: = \: C_{1,c}^c \: = \: C_{c,1}^c \: = \: C_{c,c}^1 ,
\end{equation}
\begin{eqnarray}
    C_{a,b}^c & = & \frac{\beta_2 \beta_9}{\beta_1} , \\
    C_{b,a}^c & = &  \frac{\beta_5 \beta_{10}}{\beta_6} , \\
    C_{a,c}^b  & = & \frac{\beta_3 \beta_5}{\beta_1} , \\
    C_{c,a}^b & = &  \frac{\beta_7 \beta_{9}}{\beta_{11}} , \\
    C_{b,c}^a & = &  \frac{\beta_2 \beta_{7}}{\beta_6} , \\
    C_{c,b}^a & = & \frac{\beta_3 \beta_{10}}{\beta_{11}} ,
\end{eqnarray}
\begin{equation}
    C_{1,m}^m \: = \: 2 \: = \: C_{m,1}^m \: = \: C_{m,m}^1,
\end{equation}
\begin{eqnarray}
    C_{a,m}^m & = & 2\frac{\beta_4 \beta_{13}}{\beta_1} , \\
    C_{m,a}^m & = &  2\frac{\beta_{13} \beta_{19}}{\beta_{18}} , \\
    C_{m,m}^a & = &  -2\frac{\beta_{4} \beta_{19}}{\beta_{18}}, \\
    C_{b,m}^m & = & 2\frac{\beta_8 \beta_{14}}{\beta_6}, \\
    C_{m,b}^m & = &  2\frac{\beta_{14} \beta_{16}}{\beta_{18}}, \\
    C_{m,m}^b & = & 2\frac{\beta_{8} \beta_{16}}{\beta_{18}} , \\
    C_{c,m}^m & = &  -2\frac{\beta_{12} \beta_{15}}{\beta_{11}} , \\
    C_{m,c}^m & = & -2\frac{\beta_{15} \beta_{17}}{\beta_{18}} , \\
    C_{m,m}^c & = & -2\frac{\beta_{12} \beta_{17}}{\beta_{18}}, \\
\end{eqnarray}

We now calculate the partition functions of the various Rep$(D_4)$ orbifolds when the entire Rep$(D_4)$ symmetry acts trivially, and compare to the decomposition prediction~\ref{eq:noninv-decomp-pred}).
\begin{itemize}
    \item $1+a$:  The $1+a$ partition function~(\ref{d4_orb_1a}) becomes
    \begin{eqnarray}
        Z_{1+a} & = & \frac{1}{2}\left(Z_{1,1}^1 + Z_{1,a}^a + Z_{a,1}^a + Z^1_{a,a}\right),
        \\
        & = & \frac{1}{2} \left( 1 + C_{1,a}^a + C_{a,1}^a + C_{a,a}^1 \right) Z_{1,1}^1,
        \\
        & = & \frac{1}{2} \left( 1 + 1 + 1 + 1 \right) Z_{1,1}^1 \: = \: 2 \, Z_{1,1}^1,
    \end{eqnarray}
     exactly as we would expect for a trivially-acting ${\mathbb Z}_2$ orbifold.  Comparing to the conjecture~(\ref{eq:noninv-decomp-pred}), there are two simple objects ($1$, $a$), each of quantum dimension one, so indeed $| {\cal A}| = 2$,
     and so our prediction~(\ref{eq:noninv-decomp-pred}) matches the computation.
    \item $1+b$:  The $1+b$ partition function~(\ref{d4_orb_1b}) becomes
    \begin{eqnarray}
        Z_{1+b} & = & \frac{1}{2}\left( Z_{1,1}^1 + Z_{1,b}^b + Z_{b,1}^b + Z_{b,b}^1 \right),
        \\
        & = & \frac{1}{2} \left( 1 + C_{1,b}^b + C_{b,1}^b + C_{b,b}^1 \right) Z_{1,1}^1,
        \\
        & = & \frac{1}{2} \left( 1 + 1 + 1 + 1 \right) Z_{1,1}^1 \: = \: 2 \, Z_{1,1}^1,
    \end{eqnarray}
    again exactly as we would expect for a trivially-acting ${\mathbb Z}_2$ orbifold.  Comparing to the conjecture~(\ref{eq:noninv-decomp-pred}), there are two simple objects ($1$, $b$), each of quantum dimension one, so indeed $| {\cal A}| = 2$, and so our prediction~(\ref{eq:noninv-decomp-pred}) matches the computation.
    \item $1+a+b+c$:  The $1+b+b+c$ partition function~(\ref{d4_orb_1abc}) becomes
    \begin{eqnarray}
        Z_{1+a+b+c} & = & \frac{1}{4} \biggl[  Z_{1,1}^1+Z_{1,a}^a+Z_{1,b}^b + Z_{1,c}^c + Z_{a,1}^a + Z_{a,a}^1+\frac{\beta_1}{\beta_2\beta_9}Z_{a,b}^c + \frac{\beta_1}{\beta_3\beta_5}Z_{a,c}^b + Z_{b,1}^b
        \\
        & & \qquad + \frac{\beta_6}{\beta_5\beta_{10}}Z_{b,a}^c+Z_{b,b}^1+\frac{\beta_6}{\beta_2\beta_7}Z_{b,c}^a+Z_{c,1}^c+\frac{\beta_{11}}{\beta_7\beta_9}Z_{c,a}^b+\frac{\beta_{11}}{\beta_3\beta_{10}}Z_{c,b}^a+Z_{c,c}^1\biggr]  ,
        \nonumber \\
        & = &  \frac{1}{4} \biggl[  1 + C_{1,a}^a + C_{1,b}^b + C_{1,c}^c + C_{a,1}^a + C_{a,a}^1 + \frac{\beta_1}{\beta_2\beta_9} C_{a,b}^c + \frac{\beta_1}{\beta_3\beta_5} C_{a,c}^b + C_{b,1}^b
        \\
        & & \qquad  + \frac{\beta_6}{\beta_5\beta_{10}} C_{b,a}^c + C_{b,b}^1 + \frac{\beta_6}{\beta_2\beta_7} C_{b,c}^a + C_{c,1}^c + \frac{\beta_{11}}{\beta_7\beta_9} C_{c,a}^b + \frac{\beta_{11}}{\beta_3\beta_{10}} C_{c,b}^a + C_{c,c}^1\biggr] Z_{1,1}^1 ,
        \nonumber \\
        & = & \frac{1}{4} \biggl[ 1 + 1 + 1 + 1 + 1 + 1 +  \frac{\beta_1}{\beta_2\beta_9} \frac{ \beta_2 \beta_9}{\beta_1} + \frac{\beta_1}{\beta_3\beta_5} \frac{\beta_3 \beta_5}{\beta_1} + 1
        \\
        & & \qquad + \frac{\beta_6}{\beta_5\beta_{10}} \frac{ \beta_5 \beta_{10}}{\beta_6} + 1 + \frac{\beta_6}{\beta_2\beta_7} \frac{\beta_2 \beta_7}{\beta_6} + 1 +  \frac{\beta_{11}}{\beta_7\beta_9}
        \frac{ \beta_7 \beta_9 }{\beta_{11}} +  \frac{\beta_{11}}{\beta_3\beta_{10}} \frac{ \beta_3 \beta_{10}}{\beta_{11}} + 1 \biggr] Z_{1,1}^1,
        \nonumber \\
        & = & 4 \, Z_{1,1},
    \end{eqnarray}
    again exactly as we would expect for a trivially-acting ${\mathbb Z}_2 \times {\mathbb Z}_2$ orbifold.  Comparing to the conjecture~(\ref{eq:noninv-decomp-pred}), there are four simple objects ($1$, $a$, $b$, $c$), each of quantum dimension one, so indeed $| {\cal A} | = 4$, and so our prediction~(\ref{eq:noninv-decomp-pred}) matches the computation.
    \item $1+b+m$:  The $1+b+m$ partition function~(\ref{d4_orb_1bm}) becomes
    \begin{eqnarray}
        Z_{1+b+m} & = &
         \frac{1}{4}\biggl[ Z_{1,1}^1 + Z_{1,b}^b + Z_{1,m}^m + Z_{b,1}^b + Z_{b,b}^1 + \frac{\beta_6}{\beta_8\beta_{14}} Z_{b,m}^m
         \\
    & & \qquad + Z_{m,1}^m + \frac{\beta_{18}}{\beta_{14}\beta_{16}} Z_{m,b}^m + Z_{m,m}^1 + \frac{\beta_{18}}{\beta_8\beta_{16}} Z_{m,m}^b\biggr],
    \nonumber \\
    & = &  \frac{1}{4}\biggl[ 1 + C_{1,b}^b + C_{1,m}^m + C_{b,1}^b + C_{b,b}^1 + \frac{\beta_6}{\beta_8\beta_{14}} C_{b,m}^m
         \\
    & & \qquad + C_{m,1}^m + \frac{\beta_{18}}{\beta_{14}\beta_{16}} C_{m,b}^m + C_{m,m}^1 + \frac{\beta_{18}}{\beta_8\beta_{16}} C_{m,m}^b\biggr] Z_{1,1}^1,
    \nonumber \\
    & = & \frac{1}{4} \biggl[ 1 + 1 + 2 + 1 + 1 + \frac{\beta_6}{\beta_8\beta_{14}} \left( 2\frac{\beta_8 \beta_{14}}{\beta_6} \right)
    \\
    & & \qquad + 2 + \frac{\beta_{18}}{\beta_{14}\beta_{16}} \left(  2\frac{\beta_{14} \beta_{16}}{\beta_{18}} \right)
    + 2 + \frac{\beta_{18}}{\beta_8\beta_{16}} \left(  2\frac{\beta_{8} \beta_{16}}{\beta_{18}} \right)
    \biggr] Z_{1,1}^1 ,
    \\
    & = & 4 \, Z_{1,1}^1.
    \end{eqnarray}
    Comparing to the conjecture~(\ref{eq:noninv-decomp-pred}), there are three simple objects ($1$, $b$, $m$),
    of quantum dimensions $1$, $1$, $2$, respectively, so $| {\cal A}| = 1 + 1 + 2 = 4$, and so our 
    prediction~(\ref{eq:noninv-decomp-pred}) matches the computation.  We also observe that the factor of $4$ is consistent with the fact that this is dual to a ${\mathbb Z}_2 \times {\mathbb Z}_2$ subgroup of the original $D_4$.
    \item $1+a+b+c+2m$: The partition function~(\ref{d4_orb_1abc2m}) for the regular representation becomes
    \begin{eqnarray}
        Z_{1+a+b+c+2m} & = &
        \frac{1}{8} \biggl[  Z_{1,1}^1 + Z_{1,a}^a + Z_{1,b}^b + Z_{1,c}^c + 2Z_{1,m}^m + Z_{a,1}^a + Z_{a,a}^1 + \frac{\beta_1}{\beta_2\beta_9} Z_{a,b}^c + \frac{\beta_1}{\beta_3\beta_5} Z_{a,c}^b
        \nonumber \\
    & & \qquad + \frac{2\beta_1}{\beta_4\beta_{13}} Z_{a,m}^m + Z_{b,1}^b + \frac{\beta_6}{\beta_5\beta_{10}} Z_{b,a}^c + Z_{b,b}^1 + \frac{\beta_6}{\beta_2\beta_7} Z_{b,c}^a + \frac{2\beta_6}{\beta_8\beta_{14}} Z_{b,m}^m 
    \nonumber \\
    & & \qquad + Z_{c,1}^c + \frac{\beta_{11}}{\beta_7\beta_9} Z_{c,a}^b + \frac{\beta_{11}}{\beta_3\beta_{10}} Z_{c,b}^a + Z_{c,c}^1 - \frac{2\beta_{11}}{\beta_{12}\beta_{15}} Z_{c,m}^m 
    \nonumber \\
    & & \qquad + 2Z_{m,1}^m + \frac{2\beta_{18}}{\beta_{13}\beta_{19}} Z_{m,a}^m + \frac{2\beta_{18}}{\beta_{14}\beta_{16}} Z_{m,b}^m - \frac{2\beta_{18}}{\beta_{15}\beta_{17}} Z_{m,c}^m
    \nonumber \\
    & & \qquad + 2Z_{m,m}^1 - \frac{2\beta_{18}}{\beta_4\beta_{19}} Z_{m,m}^a + \frac{2\beta_{18}}{\beta_8\beta_{16}} Z_{m,m}^b - \frac{2\beta_{18}}{\beta_{12}\beta_{17}} Z_{m,m}^c \biggr] ,
    \\
    & = &
     \frac{1}{8} \biggl[  1 + 1 + 1 + 1 + 2 (2) + 1 + 1 + \frac{\beta_1}{\beta_2\beta_9} \left( \frac{\beta_2 \beta_9}{\beta_1}  \right)  + \frac{\beta_1}{\beta_3\beta_5} \left( \frac{\beta_3 \beta_5}{\beta_1} \right) 
        \nonumber \\
    & & \qquad + \frac{2\beta_1}{\beta_4\beta_{13}}\left( 2\frac{\beta_4 \beta_{13}}{\beta_1}\right)  + 1 + \frac{\beta_6}{\beta_5\beta_{10}} \left(\frac{\beta_5 \beta_{10}}{\beta_6} \right)  + 1 + \frac{\beta_6}{\beta_2\beta_7} \left(  \frac{\beta_2 \beta_{7}}{\beta_6} \right)
    \nonumber \\
    & & \qquad 
    + \frac{2\beta_6}{\beta_8\beta_{14}}\left( 2\frac{\beta_8 \beta_{14}}{\beta_6} \right) 
    + 1 + \frac{\beta_{11}}{\beta_7\beta_9}\left(  \frac{\beta_7 \beta_{9}}{\beta_{11}} \right)  + \frac{\beta_{11}}{\beta_3\beta_{10}} \left(  \frac{\beta_3 \beta_{10}}{\beta_{11}} \right)  + 1 
    \nonumber \\
    & & \qquad 
    - \frac{2\beta_{11}}{\beta_{12}\beta_{15}} \left(  -2\frac{\beta_{12} \beta_{15}}{\beta_{11}}  \right)  
    + 2 \left( 2 \right)  + \frac{2\beta_{18}}{\beta_{13}\beta_{19}} \left(  2\frac{\beta_{13} \beta_{19}}{\beta_{18}}  \right)  + \frac{2\beta_{18}}{\beta_{14}\beta_{16}} \left(  2\frac{\beta_{14} \beta_{16}}{\beta_{18}} \right) 
    \nonumber \\
    & & \qquad 
     - \frac{2\beta_{18}}{\beta_{15}\beta_{17}} \left(  -2\frac{\beta_{15} \beta_{17}}{\beta_{18}}  \right) 
    + 2 (2)  - \frac{2\beta_{18}}{\beta_4\beta_{19}} \left( -2\frac{\beta_{4} \beta_{19}}{\beta_{18}}  \right)  + \frac{2\beta_{18}}{\beta_8\beta_{16}} \left(  2\frac{\beta_{8} \beta_{16}}{\beta_{18}} \right)
    \nonumber \\
    & & \qquad \qquad \qquad 
    - \frac{2\beta_{18}}{\beta_{12}\beta_{17}} \left(  -2\frac{\beta_{12} \beta_{17}}{\beta_{18}} \right)  \biggr] Z_{1,1}^1,
    \\
    & = & 8 \, Z_{1,1}^1.
    \end{eqnarray}
    Comparing to the conjecture~(\ref{eq:noninv-decomp-pred}), there are six simple objects ($1$, $a$, $b$, $c$,
    and two copies of $m$), four of which ($1$, $a$, $b$, $c$) have quantum dimension one,
    and two of which (the copies of $m$) have quantum dimension two, so
    \begin{equation}
        | {\cal A}| \: = \: 1 + 1 + 1 + 1 + 2 + 2 \: = \: 8,
    \end{equation}
    and so our prediction~(\ref{eq:noninv-decomp-pred}) matches the computation.  This also reflects the fact that we could have constructed the theory with trivially-acting Rep$(D_4)$ symmetry from the free action of $D_4$ on eight trivial theories -- gauging the regular representation returns us to this disjoint union of eight objects.
\end{itemize}

We also observe that the fact that the intertwiner $\beta$'s all cancel out, in a rather intricate fashion, is a strong consistency check on our methods.

\subsubsection{Rep$(Q_8)$}

The results for Rep$(Q_8)$ will of course closely mirror those of Rep$(D_4)$.  We find
\begin{equation}
    C_{1,a}^a \: = \: 1 \: = \: C_{a,1}^a \: = \: C_{a,a}^1 \: = \: C_{1,b}^b \: = \: C_{b,1}^b \: = \: C_{b,b}^1 
    \: = \: C_{1,c}^c \: = \: C_{c,1}^c \: = \: C_{c,c}^1,
\end{equation}
\begin{eqnarray}
    C_{a,b}^c  & = & \frac{\beta'_2 \beta'_9}{\beta'_1}, \\
    C_{b,a}^c & = & \frac{\beta'_5 \beta'_{10}}{\beta'_6}, \\
    C_{a,c}^b & = & \frac{\beta'_3 \beta'_5}{\beta'_1} , \\
    C_{c,a}^b & = & \frac{\beta'_7 \beta'_{9}}{\beta'_{11}} , \\
    C_{b,c}^a & = &  \frac{\beta'_2 \beta'_{7}}{\beta'_6} , \\
    C_{c,b}^a & = & \frac{\beta'_3 \beta'_{10}}{\beta'_{11}} ,
\end{eqnarray}
\begin{equation}
    C_{1,m}^m \: = \: 2 \: = \: C_{m,1}^m \: = \: C_{m,m}^1,
\end{equation}
\begin{eqnarray}
    C_{a,m}^m & = &  2\frac{\beta'_4 \beta'_{13}}{\beta'_1}  , \\
    C_{m,a}^m & = & 2\frac{\beta'_{13} \beta'_{19}}{\beta'_{18}} , \\
    C_{m,m}^a & = &  -2\frac{\beta'_{4} \beta'_{19}}{\beta'_{18}} , \\
    C_{b,m}^m & = &  -2\frac{\beta'_8 \beta'_{14}}{\beta'_6} , \\
    C_{m,b}^m & = &  2\frac{\beta'_{14} \beta'_{16}}{\beta'_{18}} , \\
    C_{m,m}^b & = & -2\frac{\beta'_{8} \beta'_{16}}{\beta'_{18}}  , \\
    C_{c,m}^m & = & 2\frac{\beta'_{12} \beta'_{15}}{\beta'_{11}} , \\
    C_{m,c}^m & = & -2\frac{\beta'_{15} \beta'_{17}}{\beta'_{18}} , \\
    C_{m,m}^c & = &  2\frac{\beta'_{12} \beta'_{17}}{\beta'_{18}} ,
\end{eqnarray}
where the only change relative to Rep$(D_4)$ is that the coefficients $C_{b,m}^m$, $C_{m,m}^b$, $C_{c,m}^m$ and $C_{m,m}^c$ have flipped sign.

We now calculate the partition functions of the various Rep$(Q_8)$ orbifolds when the entire Rep$(Q_8)$ symmetry acts trivially, and in each case compare to the decomposition prediction~(\ref{eq:noninv-decomp-pred}).
\begin{itemize}
    \item $1+a$: The $1+a$ partition function~(\ref{eq:q8-orb-1a}) becomes
    \begin{eqnarray}
        Z_{1+a} & = & \frac{1}{2} \left( Z_{1,1}^1 + Z_{1,a}^a + Z_{a,1}^a + Z_{a,a}^1 \right], 
        \\
        & = & \frac{1}{2} \left(  1 + C_{1,a}^a + C_{a,1}^a + C_{a,a}^1 \right) Z_{1,1}^1 ,
        \\
        & = & 2 \, Z_{1,1}^1,
    \end{eqnarray}
    exactly as we would expect for a trivially-acting ${\mathbb Z}_2$ orbifold.  Comparing to the conjecture~(\ref{eq:noninv-decomp-pred}), there are two simple objects ($1$, $a$), each of quantum dimension one, so indeed $| {\cal A}| = 2$, and so our prediction~(\ref{eq:noninv-decomp-pred}) matches the computation.
    \item $1+a+b+c$:  The $1+a+b+c$ partition function~\ref{eq:q8-orb-1abc}) becomes
    \begin{eqnarray}
        Z_{1+a+b+c} & = & \frac{1}{4}\biggl[  Z_{1,1}^1 + Z_{1,a}^a + Z_{1,b}^b + Z_{1,c}^c + Z_{a,1}^a + Z_{a,a}^1 + \frac{\beta_1'}{\beta_2'\beta_9'} Z_{a,b}^c + \frac{\beta_1'}{\beta_3'\beta_5'} Z_{a,c}^b 
         \\
        & & \qquad + Z_{b,1}^b + \frac{\beta_6'}{\beta_5'\beta_{10}'} Z_{b,a}^c + Z_{b,b}^1 + \frac{\beta_6'}{\beta_2'\beta_7'} Z_{b,c}^a + Z_{c,1}^c + \frac{\beta_{11}'}{\beta_7'\beta_9'} Z_{c,a}^b + \frac{\beta_{11}'}{\beta_3'\beta_{10}'} Z_{c,b}^a + Z_{c,c}^1 \biggr] ,
        \nonumber \\
        & = & \frac{1}{4}\biggl[  1 + C_{1,a}^a + C_{1,b}^b + C_{1,c}^c + C_{a,1}^a + C_{a,a}^1 + \frac{\beta_1'}{\beta_2'\beta_9'} C_{a,b}^c + \frac{\beta_1'}{\beta_3'\beta_5'} C_{a,c}^b 
         \\
        & & \qquad + C_{b,1}^b + \frac{\beta_6'}{\beta_5'\beta_{10}'} C_{b,a}^c + C_{b,b}^1 + \frac{\beta_6'}{\beta_2'\beta_7'} C_{b,c}^a + C_{c,1}^c + \frac{\beta_{11}'}{\beta_7'\beta_9'} C_{c,a}^b + \frac{\beta_{11}'}{\beta_3'\beta_{10}'} C_{c,b}^a + Z_{c,c}^1 \biggr] Z_{1,1}^1,
        \nonumber \\
        & = & \frac{1}{4}\biggl[  1 + 1 + 1 + 1 + 1 + 1 + \frac{\beta_1'}{\beta_2'\beta_9'} \left( \frac{\beta'_2 \beta'_9}{\beta'_1} \right)  + \frac{\beta_1'}{\beta_3'\beta_5'} \left(  \frac{\beta'_3 \beta'_5}{\beta'_1} \right)  
         \\
        & & \qquad + 1 + \frac{\beta_6'}{\beta_5'\beta_{10}'} \left( \frac{\beta'_5 \beta'_{10}}{\beta'_6} \right)   + 1 + \frac{\beta_6'}{\beta_2'\beta_7'} \left(  \frac{\beta'_2 \beta'_{7}}{\beta'_6}  \right) + 1 + \frac{\beta_{11}'}{\beta_7'\beta_9'} \left( \frac{\beta'_7 \beta'_{9}}{\beta'_{11}}  \right)  
        \nonumber \\
        & & \qquad \qquad \qquad + \frac{\beta_{11}'}{\beta_3'\beta_{10}'} \left( \frac{\beta'_3 \beta'_{10}}{\beta'_{11}} \right)  + 1 \biggr] Z_{1,1}^1,
        \nonumber \\
        & = & 4 \, Z_{1,1}^1,
    \end{eqnarray}
    exactly as we would expect for a trivially-acting ${\mathbb Z}_2 \times {\mathbb Z}_2$ orbifold.  Comparing to the conjecture~(\ref{eq:noninv-decomp-pred}), there are four simple objects ($1$, $a$, $b$, $c$), each of quantum dimension one, so indeed $|{\cal A}| = 4$, and so our prediction~(\ref{eq:noninv-decomp-pred}) matches the computation.
    \item $1+a+b+c+2m$: The $1+a+b+c+2m$ partition function~(\ref{eq:q8-orb-1abc2m}) becomes
    \begin{eqnarray}
        Z_{1+a+b+c+2m} & = &
        \frac{1}{8} \biggl[  Z_{1,1}^1 + Z_{1,a}^a + Z_{1,b}^b + Z_{1,c}^c + 2Z_{1,m}^m + Z_{a,1}^a + Z_{a,a}^1 + \frac{\beta_1'}{\beta_2'\beta_9'} Z_{a,b}^c + \frac{\beta_1'}{\beta_3'\beta_5'} Z_{a,c}^b
        \nonumber \\
        & & \qquad + \frac{2\beta_1'}{\beta_4'\beta_{13}'} Z_{a,m}^m + Z_{b,1}^b + \frac{\beta_6'}{\beta_5'\beta_{10}'} Z_{b,a}^c + Z_{b,b}^1 + \frac{\beta_6'}{\beta_2'\beta_7'} Z_{b,c}^a - \frac{2\beta_6'}{\beta_8'\beta_{14}'} Z_{b,m}^m
        \nonumber \\
        & & \qquad + Z_{c,1}^c + \frac{\beta_{11}'}{\beta_7'\beta_9'} Z_{c,a}^b + \frac{\beta_{11}'}{\beta_3'\beta_{10}'} Z_{c,b}^a + Z_{c,c}^1 + \frac{2\beta_{11}'}{\beta_{12}'\beta_{15}'} Z_{c,m}^m
        \nonumber \\
        & & \qquad + 2Z_{m,1}^m + \frac{2\beta_{18}'}{\beta_{13}'\beta_{19}'} Z_{m,a}^m + \frac{2\beta_{18}'}{\beta_{14}'\beta_{16}'} Z_{m,b}^m - \frac{2\beta_{18}'}{\beta_{15}'\beta_{17}'} Z_{m,c}^m
        \nonumber \\
        & & \qquad + 2Z_{m,m}^1 - \frac{2\beta_{18}'}{\beta_4'\beta_{19}'} Z_{m,m}^a - \frac{2\beta_{18}'}{\beta_8'\beta_{16}'} Z_{m,m}^b + \frac{2\beta_{18}'}{\beta_{12}'\beta_{17}'} Z_{m,m}^c \biggr] ,
        \\
        & = & 
         \frac{1}{8} \biggl[  1 + 1 + 1 + 1 + 2 (2)  + 1 + 1 + \frac{\beta_1'}{\beta_2'\beta_9'} \left( \frac{\beta'_2 \beta'_9}{\beta'_1} \right)  + \frac{\beta_1'}{\beta_3'\beta_5'}\left(  \frac{\beta'_3 \beta'_5}{\beta'_1} \right) 
        \nonumber \\
        & & \qquad + \frac{2\beta_1'}{\beta_4'\beta_{13}'} \left(  2\frac{\beta'_4 \beta'_{13}}{\beta'_1}  \right)  + 1 + \frac{\beta_6'}{\beta_5'\beta_{10}'} \left( \frac{\beta'_5 \beta'_{10}}{\beta'_6} \right)   + 1 + \frac{\beta_6'}{\beta_2'\beta_7'} \left(  \frac{\beta'_2 \beta'_{7}}{\beta'_6} \right)
        \nonumber \\
        & & \qquad
        - \frac{2\beta_6'}{\beta_8'\beta_{14}'} \left( -2\frac{\beta'_8 \beta'_{14}}{\beta'_6}  \right) 
         + 1 + \frac{\beta_{11}'}{\beta_7'\beta_9'} \left( \frac{\beta'_7 \beta'_{9}}{\beta'_{11}} \right)  + \frac{\beta_{11}'}{\beta_3'\beta_{10}'} \left( \frac{\beta'_3 \beta'_{10}}{\beta'_{11}} \right)  + 1 
         \nonumber \\
         & & \qquad 
         + \frac{2\beta_{11}'}{\beta_{12}'\beta_{15}'} \left( 2\frac{\beta'_{12} \beta'_{15}}{\beta'_{11}}  \right) 
         + 2 (2)  + \frac{2\beta_{18}'}{\beta_{13}'\beta_{19}'} \left( 2\frac{\beta'_{13} \beta'_{19}}{\beta'_{18}}  \right)  + \frac{2\beta_{18}'}{\beta_{14}'\beta_{16}'} \left(  2\frac{\beta'_{14} \beta'_{16}}{\beta'_{18}} \right) 
         \nonumber \\
         & & \qquad 
         - \frac{2\beta_{18}'}{\beta_{15}'\beta_{17}'} \left( -2\frac{\beta'_{15} \beta'_{17}}{\beta'_{18}} \right) 
         + 2 (2)  - \frac{2\beta_{18}'}{\beta_4'\beta_{19}'} \left(  -2\frac{\beta'_{4} \beta'_{19}}{\beta'_{18}}  \right)  - \frac{2\beta_{18}'}{\beta_8'\beta_{16}'} \left( -2\frac{\beta'_{8} \beta'_{16}}{\beta'_{18}}  \right) 
         \nonumber \\
         & & \qquad \qquad \qquad 
         + \frac{2\beta_{18}'}{\beta_{12}'\beta_{17}'} \left(  2\frac{\beta'_{12} \beta'_{17}}{\beta'_{18}} \right)  \biggr] Z_{1,1}^1 ,
        \\
        & = & 8 \, Z_{1,1}^1.
    \end{eqnarray}
    Comparing to the conjecture~(\ref{eq:noninv-decomp-pred}), there are six simple objects ($1$, $a$, $b$, $c$, and two copies of $m$), four of which ($1$, $a$, $b$, $c$) have quantum dimension one, and two of which (the copies of $m$) have quantum dimension two, so indeed
    \begin{equation}
        | {\cal A}| \: = \: 1 + 1 + 1 + 1 + 2 + 2 \: = \: 8,
    \end{equation}
    and so our prediction~(\ref{eq:noninv-decomp-pred}) matches the computation.
\end{itemize}

As in previous examples, the fact that all of the $\beta'$ cancel out, often in a rather intricate fashion, is a solid self-consistency test of our methods.

\subsection{TQFT interpretation}

The field theory of a trivially-acting symmetry is a symmetry-protected topological (SPT) phase, so we can regard the above results as gauging SPTs for Rep$(G)$ symmetries.  More specifically, the coefficients we obtain when applying decomposition to the genus one partition function tell us the ground state degeneracy of the resulting theory.  Below we give slightly more detail on some of the theories appearing in the above examples:
\begin{itemize}
    \item SPT$(\Rep(S_3))/(1+X)$, 2 ground states.  We expect that gauging the $\Z_2$ subgroup generated by $X$ of a theory with Rep$(S_3)$ symmetry will produce a theory with $S_3$ symmetry, so we can identify this as the $S_3$-symmetic $\Z_2$ SSB phase discussed in \cite[section 4.6.3]{Bhardwaj:2023idu}.
    \item SPT$(\Rep(S_3))/(1+Y)$, 3 ground states.  If we present our Rep$(S_3)$-symmetric theory as an $S_3$ orbifold, gauging the $1+Y$ subsymmetry should return the $\Z_2$ orbifold of the $S_3$-symmetric theory.  In particular this means that the resulting theory still carries Rep$(S_3)$ symmetry.  We can identify this theory as the Rep$(S_3)/\Z_2$ SSB phase of \cite[section 5.3.3]{Bhardwaj:2023idu}.
    \item SPT$(\Rep(S_3))/(1+X+2Y)$, 6 ground states.  This is Rep$(S_3)$ gauge theory, also known as Rep$(S_3)$ Dijkgraaf-Witten theory.  The six ground states carry a freely-acting $S_3$ symmetry that is the quantum dual to the original Rep$(S_3)$.  This is the $S_3$ SSB phase of \cite[section 4.6.1]{Bhardwaj:2023idu}.
    \item SPT$(\Rep(D_4))/(1+a+b+c+2m)$ and SPT$(\Rep(Q_8))/(1+a+b+c+2m)$, 8 ground states each.  These are the gauge theories for Rep($D_4$) and Rep($Q_8$), and the story is similar to Rep$(S_3)$ gauge theory.  In both cases the resulting theories consist of eight copies of the trivial theory (i.e.~an SPT for the trivial group) with a free action of $D_4$ or $Q_8$, respectively.
\end{itemize}

\section{Conclusions}

In this paper we have explicitly gauged some examples of multiplicity-free
noninvertible symmetries in two dimensions.  We began with a general overview of the procedure.  For a noninvertible symmetry defined by a fusion category of the
form Rep$({\cal H})$, for ${\cal H}$ a Hopf algebra, we put the structure of a special symmetric Frobenius algebra on ${\cal H}^*$ in order to construct modular-invariant partition functions.  We checked that the general procedure correctly
reproduced results for ordinary group orbifolds, described in present language by the fusion category Vec$(G) = {\rm Rep}( {\mathbb C}[G]^*)$.
We then did explicit computations in
Rep$(S_3)$, Rep$(D_4)$, Rep$(Q_8)$, and Rep$({\cal H}_8)$, and discussed applications in $c=1$ CFTs.  We also discussed decomposition arising in cases in which the gauged noninvertible symmetry acts trivially.

We intend to return to these matters in upcoming work \cite{toappear},
for example generalizing to non-multiplicity-free cases.

\section*{Acknowledgements}

We would like to thank N.~Carqueville, C.~M.~Chang, Y.~Choi, H.~T.~Lam, J.~McNamara, T.~Pantev, I.~Runkel, S.~H.~Shao, Y.~Tachikawa, Y.~Wang, and Y.~Zheng for useful discussions. We also thank the authors of \cite{Wang:2023} for communicating about their upcoming work and coordinating arXiv submission.
E.S. was partially supported by NSF grant PHY-2310588.  X.Y. was partially supported by NSF grant PHY-2014086. 

\appendix

\section{Algebras}  \label{app:algebras}

This paper will frequently make use of both Hopf algebras and Frobenius algebras -- representations of Hopf algebras will arise when describing pertinent noninvertible symmetries, and Frobenius algebras will be used to define their gauging.  To make this paper self-contained, we outline their definitions in this appendix.

\subsection{Definition of Hopf algebra}
\label{app:Hopf}

A Hopf algebra $({\cal H}, \mu_H, u_H, \Delta_H, u_H^o, S)$ is
defined by
\begin{itemize}
    \item a multiplication $\mu_H: {\cal H} \otimes {\cal H} \rightarrow {\cal H}$,
    \item a unit $u_{H}: {\mathbb C} \rightarrow {\cal H}$,
    \item a comultiplication $\Delta_H: {\cal H} \rightarrow {\cal H} \otimes {\cal H}$,
    \item a counit $u_H^o: {\cal H} \rightarrow {\mathbb C}$,
    \item an antipode $S: {\cal H} \rightarrow {\cal H}$
\end{itemize}
satisfying several identities, of which we list the key ones below:
\begin{itemize}
    \item associativity:
    \be
\label{assoc-hopf}
\mu_H\circ \left(\mu_H\otimes\text{Id}_{\cal H}\right) = \mu_H\circ \left(\text{Id}_{\cal H}\otimes\mu_H\right),
\ee
 \item unit axiom:
    \begin{equation}  \label{eq:id:unit-hopf}
        \mu_H\circ\left( \text{Id}_{\cal H} \otimes u_H(1) \right) \: = \: \text{Id}_{\cal H} \: = \:
        \mu_H\circ\left( u_H(1) \otimes \text{Id}_{\cal H}\right),
    \end{equation}
     \item coassociativity:
\begin{equation}
        (\text{Id}_{\mathcal{H}}\otimes \Delta_H)\circ \Delta_H = (\Delta_H\otimes \text{Id}_{\mathcal{H}})\circ\Delta_H,
    \end{equation}
        \item counit axiom:
    \begin{equation}
        (\text{Id}_{\cal H} \otimes u_H^o) \circ \Delta_H \: = \: \text{Id}_{\cal H} \: =  \:
        (u_H^o \otimes \text{Id}_{\cal H}) \circ \Delta_H,
    \end{equation}
    \item antipode axiom:
    \begin{equation}
    \xymatrix{
    & {\cal H} \otimes {\cal H} \ar[rr]^{S \otimes 1} & & {\cal H} \otimes {\cal H} \ar[dr]^{\mu_H} & \\
    {\cal H} \ar[ur]^{\Delta_H} \ar[rr]^{u_H^o} \ar[dr]_{\Delta_H} & & {\mathbb C}
    \ar[rr]^{u_H} & & {\cal H} \\
    & {\cal H} \otimes {\cal H} \ar[rr]^{1 \otimes S} & & {\cal H} \otimes {\cal H} \ar[ur]_{\mu_H} &
    }
\end{equation}
\end{itemize}
More formally, this means that a Hopf algebra is an associative and coassociative bialgebra,
together with a compatible antipode map.
In this paper we also take ${\cal H}$ to be finite-dimensional and semisimple.

A common example is the special case ${\cal H} = {\mathbb C}[G]$, the group
algebra of a finite group $G$, for which the algebra structure is a linear extension of the product on $G$, and where the co-algebra structure, as well as the antipode map, is given by
\begin{equation}
    \Delta_H(g) \: = \: g \otimes g, \: \: \: u_H^o(g) \: = \: 1,
    \: \: \:
    S(g) \: = \: g^{-1},
\end{equation}
for $g \in G$, and extended linearly over ${\mathbb C}[G]$, where $\Delta_H$ is the comultiplication and $u_H^o$ the counit.

Because group algebras are a special case, Hopf algebras generalize group algebras.

\subsection{Definition of Frobenius algebra} \label{app:Frobenius}

Briefly, a Frobenius algebra $({\cal A}, \mu_F, u_F, \Delta_F, u_F^o)$ is defined by
\begin{itemize}
    \item a multiplication $\mu_F: {\cal A} \otimes {\cal A} \rightarrow {\cal A}$,
    \item a unit $u_F: {\mathbb C} \rightarrow {\cal A}$,
    \item a comultiplication $\Delta_F: {\cal A} \rightarrow {\cal A} \otimes {\cal A}$,
    \item a counit $u_F^o: {\cal A} \rightarrow {\mathbb C}$,
\end{itemize}
satisfying several identities, of which we list the key ones below:
\begin{itemize}
    \item associativity:
    \be
\label{assoc}
\mu_F\circ \left(\mu_F\otimes\text{Id}_{\cal A}\right) = \mu_F\circ \left(\text{Id}_{\cal A}\otimes\mu_F\right),
\ee
 \item unit axiom:
    \begin{equation}  \label{eq:id:unit}
        \mu_F\circ\left( \text{Id}_{\cal A} \otimes u_F(1) \right) \: = \: \text{Id}_{\cal A} \: = \:
        \mu_F\circ\left( u_F(1) \otimes \text{Id}_{\cal A}\right),
    \end{equation}
     \item coassociativity:
\begin{equation}
        (\text{Id}_{\mathcal{A}}\otimes \Delta_F)\circ \Delta_F = (\Delta_F\otimes \text{Id}_{\mathcal{A}})\circ\Delta_F,
    \end{equation}
        \item counit axiom:
    \begin{equation}
        (\text{Id}_{\cal A} \otimes u_F^o) \circ \Delta_F \: = \: \text{Id}_{\cal A} \: =  \:
        (u_F^o \otimes \text{Id}_{\cal A}) \circ \Delta_F,
    \end{equation}
        \item Frobenius identities~(\ref{eq:Frobenius-identities}):
    \begin{equation}  
        \xymatrix{
        {\cal A} \otimes {\cal A} \ar[rr]^{\Delta_F \otimes \text{Id}_{\cal A}} \ar[d]_{\mu_F} 
        & & {\cal A} \otimes {\cal A} \otimes {\cal A} \ar[d]^{\text{Id}_{\cal A} \otimes \mu_F}
        \\
        {\cal A} \ar[rr]^{\Delta_F} & & {\cal A} \otimes {\cal A}
        },
        \: \: \: \: \:
        \xymatrix{
        {\cal A} \otimes {\cal A} \ar[rr]^{\text{Id}_{\cal A} \otimes \Delta_F} \ar[d]_{\mu_F} 
        & & {\cal A} \otimes {\cal A} \otimes {\cal A} \ar[d]^{\mu_F \otimes \text{Id}_{\cal A}}
        \\
        {\cal A} \ar[rr]^{\Delta_F} & & {\cal A} \otimes {\cal A}
        }
    \end{equation}
\end{itemize}

In this paper we usually work with special symmetric Frobenius algebras.
These satisfy the following additional key axioms:
\begin{itemize}
        \item special~(\ref{eq:conds}):
    \begin{equation}
    u^o_F\circ u_F \propto \text{Id}_1, \: \: \: \mu_F \circ \Delta_F = \text{Id}_{\cal A},
\end{equation}
    \item symmetric~(\ref{eq:Fr:symmetric}):
    \begin{equation}
        \begin{tikzcd}
            {\cal A} \arrow[rr, maps to, "\tilde{\gamma}_{\cal A} \otimes \text{Id}_{\cal A}"]
            \arrow[dd,  "\text{Id}_{\cal A} \otimes \gamma_{\cal A}"]
            & &
            {\cal A}^* \otimes {\cal A} \otimes {\cal A} 
            \arrow[dd, "\text{Id}_{{\cal A}^*} \otimes ( u_F^o \circ \mu_*)"]
            \\
            & & 
            \\
            {\cal A} \otimes {\cal A} \otimes {\cal A}^* 
            \arrow[rr, "(u_F^o \circ \mu_*) \otimes \text{Id}_{{\cal A}^*}"]
            & & 
            {\cal A}^*
        \end{tikzcd}
    \end{equation}
    where $\tilde{\gamma}_{\cal A}:\C\to {\cal A^*}\otimes {\cal A}$ and $\gamma_{\cal A}:\C\to {\cal A}\otimes {\cal A^*} $ are coevaluation maps that exist by definition of $\mathcal{A}^*$ being dual to $\mathcal{A}$ (and vice-versa) as vector spaces.
\end{itemize}

\section{$\Z_2\times\Z_2$ Tambara-Yamagami modular transformations}   \label{app:ty}

We compute the relevant transformations with $p,q\in\{1,a,b,c\}$, and $x\in\{a,b,c\}$. We will drop the bars since all objects are self-dual.

The associators are listed at \cite[def'n (3.1)]{ty} as F-symbols.
They are related to the $\tilde{K}$'s we primarily use in this paper by
\begin{equation}
    F^{2,1,5}_{4,3,6} \: = \: \tilde{K}^{1,\overline{4}}_{2,3}(5,6).
\end{equation}
The reader may also consult \cite[fig. 7]{Wang:2016rzy}, which lists slightly different $F$ symbols for $D_4$
and $Q_8$, corresponding to the matrix elements of the $a$ of \cite{ty}.

In this paper we give first-principles derivations of associators for Rep$(D_4)$ and Rep$(Q_8)$, listing results for general intertwiners.  A standard result for associators for a fixed choice of intertwiners, for all of Rep$(D_4)$,
Rep$(Q_8)$, and Rep$({\cal H}_8)$,
can be found in \cite[fig. 7]{Wang:2016rzy},
which lists
\begin{equation}
    F^{q,p,pq}_{pqr,r,qr} \: = \: F^{q,p,pq}_{m,m,m} \: = \: F^{p,m,m}_{m,q,pq} \: = \: F^{m,p,m}_{pq,m,q} \: = \: F^{m,m,q}_{pq,p,m} \: = \: 1,
\end{equation}
\begin{equation}
    F^{m,p,m}_{m,q,m} \: = \: F^{p,m,m}_{q,m,m} \: = \: \chi(p,q),
\end{equation}
\begin{equation}
    F^{m,m,p}_{m,m,q} \: = \: n \chi(p,q)/2,
\end{equation}
where $p, q, r$ index one-dimensional representations, and $m$ is the two-dimensional
representation.
\begin{enumerate}
    \item For $D_4$, $n=+1$, $\chi(1,p) = \chi(p,1) = +1$, $\chi(x,x) = +1$, $\chi(x,y) = -1$, for $x \neq y$ corresponding to nontrivial one-dimensional irreps.
    \item For $Q_8$, $n=-1$, $\chi(1,p) = \chi(p,1) = +1$, $\chi(x,x) = +1$, $\chi(x,y) = -1$, for $x \neq y$ corresponding to nontrivial one-dimensional irreps.
    \item For ${\cal H}_8$, $n=+1$, $\chi(1,p) = \chi(p,1) = \chi(c,c)=+1$, $\chi(a,a)=\chi(b,b)=\chi(a,c)=\chi(b,c)=-1$
\end{enumerate}
where $a,b$ not arbitrary irreps but are the particular one-dimensional irreps that generate the Klein group.

In the sums below, $p, q \in \{1, a, b, c\}$, whereas the index $L \in \{1, a, b, c, m\}$.

\begin{eqnarray}
    Z_{p,q}^{pq}(\tau+1,\bar{\tau}+1) &=& \tilde{K}^{p,q}_{q,p}(pq,pq) Z^q_{p,pq}(\tau,\bar{\tau}) = F^{q,p,pq}_{q,p,pq}Z^q_{p,pq}(\tau,\bar{\tau})= Z^q_{p,pq}(\tau,\bar{\tau})
    \\
    Z^{m}_{p,m}(\tau+1,\bar{\tau}+1) &=& \tilde{K}^{p,m}_{m,p}(m,m) Z^m_{p,m}(\tau,\bar{\tau}) = F^{m,p,m}_{m,p,m}Z^m_{p,m}(\tau,\bar{\tau})= \chi(p,p)Z^m_{p,m}(\tau,\bar{\tau})
    \\
    Z^p_{m,m}(\tau+1,\bar{\tau}+1) &=& \sum_q \tilde{K}^{m,m}_{m,m}(p,q) Z^m_{m,q}(\tau,\bar{\tau}) = \sum_{q}F^{m,m,p}_{m,m,q}Z^m_{m,q}(\tau,\bar{\tau})\\
    &=&\frac{1}{2}\sum_{q}n\chi(p,q)Z^m_{m,q}(\tau,\bar{\tau})
    \\
    Z^m_{m,p}(\tau+1,\bar{\tau}+1) &=& \tilde{K}^{m,p}_{p,m}(m,m) Z^p_{m,m}(\tau,\bar{\tau}) =  F^{p,m,m}_{p,m,m}Z^p_{m,m}(\tau,\bar{\tau})=\chi(p,p)Z^p_{m,m}(\tau,\bar{\tau})
\end{eqnarray}
\begin{eqnarray}
    Z^{pq}_{p,q}(-1/\tau,-1/\bar{\tau}) &=& \tilde{K}^{p,1}_{pq,q}(q,p) \tilde{K}^{p,q}_{q,p}(pq,pq) Z^{pq}_{q,p}(\tau,\bar{\tau})= F^{pq,p,q}_{1,q,p} F^{q,p,pq}_{q,p,pq} Z^{pq}_{q,p}(\tau,\bar{\tau})
    \\
    &=& Z^{pq}_{q,p}(\tau,\bar{\tau})
    \\
    Z^{m}_{p,m}(-1/\tau,-1/\bar{\tau}) &=& \tilde{K}^{p,1}_{m,m}(m,p)\tilde{K}^{p,m}_{m,p}(m,m) Z^m_{m,p}(\tau,\bar{\tau})=F^{m,p,m}_{1,m,p}F^{m,p,m}_{m,p,m}Z^m_{m,p}(\tau,\bar{\tau})\\ 
    &=& \chi(p,p)Z^m_{m,p}(\tau,\bar{\tau})
    \\
    Z^m_{m,p}(-1/\tau,-1/\bar{\tau}) &=& \tilde{K}^{m,1}_{m,p}(p,m) \tilde{K}^{m,p}_{p,m}(m,m) Z^m_{p,m}(\tau,\bar{\tau})= F^{m,m,p}_{1,p,m}F^{p,m,m}_{p,m,m} Z^m_{p,m}(\tau,\bar{\tau}) \\
    &=& \chi(p,p) Z^m_{p,m}(\tau,\bar{\tau})
    \\
     Z^p_{m,m}(-1/\tau,-1/\bar{\tau}) &=& \sum_{q}\tilde{K}^{m,1}_{q,m}(m,m)\tilde{K}^{m,m}_{m,m}(p,q)Z^q_{m,m}=\sum_{q}F_{q,m,m}^{1,m,m}F^{m,m,p}_{m,m,1}Z^1_{m,m}\\
&=&\frac{1}{2}\sum_{q}n\chi(p,q)Z^{q}_{m,m}
\end{eqnarray}
We now show that the partition functions previously obtained are indeed modular invariant. This serves as a useful consistency check. In all cases, the sum
\begin{equation*}
    Z_{\Z_2\times\Z_2} = \sum_{g,h\in\{1,a,b,c\}} Z_{g,h}^{gh}
\end{equation*}
always appears, which corresponds to gauging a non-anomalous $\Z_2\times\Z_2$ group. This is known to be modular invariant, so we will only focus on the part of the partition functions that involve the noninvertible object $m$.

For $D_4$, we have the combination 
\begin{eqnarray*}
    Z_{\text{Rep}(D_4)}'&:=& Z_{1,m}^m-Z_{a,m}^m+Z_{b,m}^m+Z_{c,m}^m+Z_{m,1}^m-Z_{m,a}^m
     +Z_{m,b}^m+Z_{m,c}^m \\ & &+Z_{m,m}^1
    -Z^a_{m,m}
    +Z^b_{m,m}+Z^c_{m,m} 
\end{eqnarray*}
Under $T$-transformation this becomes
\begin{eqnarray*}
    Z_{\text{Rep}(D_4)}'(\tau+1) &=& Z_{1,m}^m-Z_{a,m}^m+Z_{b,m}^m+Z_{c,m}^m+Z_{m,m}^1-Z_{m,m}^a+Z_{m,m}^b+Z_{m,m}^c\\
   & &+\frac{1}{2}\left(Z_{m,1}^m+Z^m_{m,a}+Z^m_{m,b}+Z^m_{m,c}\right) 
-\frac{1}{2}\left(Z^m_{m,1}+Z^m_{m,a}-Z^m_{m,b}-Z^m_{m,c}\right)  \\
   & & +\frac{1}{2}\left(Z^m_{m,1}-Z^m_{m,a}+Z^m_{m,b}-Z^m_{m,c}\right) +\frac{1}{2}\left(Z^m_{m,1}-Z^m_{m,a}-Z^m_{m,b}+Z^m_{m,c}\right) 
    \\
    &=& Z_{\text{Rep}(D_4)}'(\tau),
\end{eqnarray*}
while under $S$-transformation it becomes
\begin{eqnarray*}
    Z_{\text{Rep}(D_4)}'(-1/\tau) &=& Z_{m,1}^m-Z_{m,a}^m+Z^m_{m,b}+Z^m_{m,c} +Z^m_{1,m}-Z^m_{a,m}+Z^m_{b,m}+Z_{c,m}^m 
    \\ & &
    +\frac{1}{2}\left(Z_{m,m}^1+Z^a_{m,m}+Z^b_{m,m}+Z^c_{m,m}\right)
    -\frac{1}{2}\left(Z^1_{m,m}+Z^a_{m,m}-Z^b_{m,m}-Z^c_{m,m}\right)
    \\ & &+\frac{1}{2}\left(Z^1_{m,m}-Z^a_{m,m}+Z^b_{m,m}-Z^c_{m,m}\right)+\frac{1}{2}\left(Z^1_{m,m}-Z^a_{m,m}-Z^b_{m,m}+Z^c_{m,m}\right)
    \\
    &=& Z_{\text{Rep}(D_4)}'(\tau),
\end{eqnarray*}
both of which match the original term $Z_{\text{Rep}(D_4)}'$.

For $Q_8$ the relevant term is
\begin{eqnarray*}
     Z_{\text{Rep}(Q_8)}'&:=& Z_{1,m}^m-Z_{a,m}^m-Z_{b,m}^m-Z_{c,m}^m+Z_{m,1}^m-Z_{m,a}^m
     -Z_{m,b}^m-Z_{m,c}^m \\ & &+Z_{m,m}^1
    -Z^a_{m,m}
    -Z^b_{m,m}-Z^c_{m,m} 
\end{eqnarray*}
Under $T$-transformation this becomes
\begin{eqnarray*}
    Z_{\text{Rep}(Q_8)}'(\tau+1) &=& Z_{1,m}^m-Z_{a,m}^m-Z_{b,m}^m-Z_{c,m}^m+Z_{m,m}^1-Z_{m,m}^a
     -Z_{m,m}^b-Z_{m,m}^c
    \\
    & & -\frac{1}{2}\left(Z^m_{m,1}+Z^m_{m,a}+Z^m_{m,b}+Z^m_{m,c}\right)  
    +\frac{1}{2}\left(Z^m_{m,1}+Z^m_{m,a}-Z^m_{m,b}-Z^m_{m,c}\right)\\ 
    & & +\frac{1}{2}\left(Z^m_{m,1}-Z^m_{m,a}+Z^m_{m,b}-Z^m_{m,c}\right)
   +\frac{1}{2}\left(Z^m_{m,1}-Z^m_{m,a}-Z^m_{m,b}+Z^m_{m,c}\right)
    \\
    &=& Z_{\text{Rep}(Q_8)}'(\tau),
\end{eqnarray*}
while under $S$-transformation it becomes
\begin{eqnarray*}
    Z_{\text{Rep}(Q_8)}'(-1/\tau) &=& Z_{m,1}^m-Z_{m,a}^m-Z^m_{m,b}-Z^m_{m,c} +Z^m_{1,m}-Z^m_{a,m}-Z^m_{b,m}-Z_{c,m}^m 
    \\ & &
    -\frac{1}{2}\left(Z_{m,m}^1+Z^a_{m,m}+Z^b_{m,m}+Z^c_{m,m}\right)
    +\frac{1}{2}\left(Z^1_{m,m}+Z^a_{m,m}-Z^b_{m,m}-Z^c_{m,m}\right)
    \\ & &+\frac{1}{2}\left(Z^1_{m,m}-Z^a_{m,m}+Z^b_{m,m}-Z^c_{m,m}\right) + \frac{1}{2}\left(Z^1_{m,m}-Z^a_{m,m}-Z^b_{m,m}+Z^c_{m,m}\right)
    \\
    &=& Z_{\text{Rep}(Q_8)}'(\tau),
\end{eqnarray*}
both of which match the original term $Z_{\text{Rep}(D_4)}'$.

Finally, for ${\cal H}_8$, the combination is
\begin{eqnarray*}
    Z_{\text{Rep}(H_8)}'&:=& Z_{1,m}^m+Z_{c,m}^m+Z_{m,1}^m+Z_{m,c}^m+Z_{m,m}^1
    +Z^c_{m,m} .
\end{eqnarray*}
Under $T$-transformation this becomes
\begin{eqnarray*}
    Z_{\text{Rep}(H_8)}'(\tau+1) &=& Z_{1,m}^m+Z_{c,m}^m+Z_{m,m}^1+Z_{m,m}^c
    \\ 
    & & +\frac{1}{2}\left(Z_{m,1}^m+Z_{m,a}^m+Z_{m,b}^m+Z_{m,c}^m \right)+\frac{1}{2}\left(Z_{m,1}^m-Z_{m,a}^m-Z_{m,b}^m+Z_{m,c}^m \right),
    \\
    &=& Z_{\text{Rep}(H_8)}'(\tau),
\end{eqnarray*}
and under $S$-transformation one gets
\begin{eqnarray*}
    Z_{\text{Rep}(H_8)}'(-1/\tau)&=& Z_{m,1}^m+Z_{m,c}^m+Z_{1,m}^m+Z_{c,m}^m + \frac{1}{2}\left(Z^1_{m,m}+Z^a_{m,m}+Z^b_{m,m}+Z^c_{m,m}\right)
    \\
    & &+\frac{1}{2}\left(Z_{m,m}^1-Z_{m,m}^a-Z_{m,m}^b+Z^c_{m,m}\right),
    \\ &=& Z_{\text{Rep}(H_8)}'(\tau),
\end{eqnarray*}
matching the original expression $Z_{\text{Rep}(H_8)}'$. Note that modular invariance provides a further consistency check for the absence of partial traces involving the noninvertible object along with either $a$ or $b$. For example, under $T$-transformation one has that
\begin{eqnarray*}
    Z_{a,m}^m\longmapsto -Z_{a,m}^m,\\
    Z_{b,m}^m\longmapsto -Z_{b,m}^m,
\end{eqnarray*}
and since no other partial traces map to $Z_{a,m}^m$ or $Z_{a,m}^m$, they cannot be present in the gauged theory partition function. Furthermore, under $S$-transformation the partial traces
\begin{eqnarray*}
    Z_{a,m}^m\longleftrightarrow -Z_{m,a}^m,\\
    Z_{b,m}^m\longleftrightarrow -Z_{m,b}^m,
\end{eqnarray*}
are interchanged, meaning the partial traces $Z_{m,a}^m$ and $Z_{m,a}^m$ should also be absent. But the latter implies that the partial traces $Z_{m,m}^a$ and $Z_{m,m}^b$ vanish by $T$-transformation. Thus, modular invariance confirms that partial traces involving $m$ with $a$ or $b$ must vanish, precisely as what was obtained.

\section{Gauging actions on disjoint unions}   \label{app:disjoint-union}

Disjoint unions of spaces can be simple playgrounds for actions of both abelian and nonabelian symmetry groups, as well as simple examples in which to see quantum symmetries of the form Rep$(G)$ at work.
In this section we will give an overview of orbifolds $[X/G]$ in which $X$ is a disjoint union of copies of some other space $Y$,
\begin{equation}
    X \: = \: \coprod_i Y,
\end{equation}
and the group $G$ acts by interchanging copies.

Let us begin with a few examples of this form.
Suppose first that $X$ is a disjoint union of three copies of $Y$,
and $G = {\mathbb Z}_3$, acting by interchanging those copies.
In this case, $G$ acts transitively on $X$, and $[X/G] = Y$.

Ordinarily, in an orbifold, there are twisted sectors.  If the orbifold group acts freely, those twisted sectors may only encode massive states,
but they are still present.  One way to see this is the existence of the quantum symmetry:  in order for an orbifold by the quantum symmetry to return the original theory, all of the information of the original theory must be present in the orbifold.  

Here, by contrast, since the group ${\mathbb Z}_3$ interchanges elements of a disjoint union, no twisted sectors can exist -- there are no (connected) string worldsheets that start on one element of a disjoint union, and end on a different element of the disjoint union.

Nevertheless, orbifolding by the quantum symmetry does restore the original theory in this case.  This is a result of decomposition
(see e.g.~\cite{Hellerman:2006zs,Sharpe:2022ene}).  Because there are no twisted sectors at all, the quantum symmetry (here, ${\mathbb Z}_3$) acts trivially, hence when we orbifold by the trivially-acting quantum symmetry,
we get three copies.  In other words, 
\begin{equation}
    [X/{\mathbb Z}_3] \: = \: Y,
\end{equation}
and since the quantum symmetry $\hat{\mathbb Z}_3$ acts trivially,
\begin{eqnarray}
    [ [X/{\mathbb Z}_3] / \hat{\mathbb Z}_3] & = & \coprod_3 [X/{\mathbb Z}_3],
    \\
    & = & \coprod_3 Y, \\
    & = & X.
\end{eqnarray}

Now, let us compare the orbifold $[X/S_3]$, where $S_3$ is the symmetric group on three objects, interchanging the three copies of $Y$.  In this case, ${\mathbb Z}_3$ is a normal subgroup of $S_3$, with quotient
${\mathbb Z}_2 = S_3/{\mathbb Z}_3$.  In particular,
${\mathbb Z}_2 = S_3/{\mathbb Z}_3$ acts (trivially) on $Y = [X/{\mathbb Z}_3]$,
with quotient
\begin{equation}
    [ [X/{\mathbb Z}_3] / {\mathbb Z}_2] \: = \: [X/S_3].
\end{equation}
Furthermore, since ${\mathbb Z}_2$ acts trivially, we see that
$[X/S_3]$ is a trivial ${\mathbb Z}_2$ gerbe on $Y = [X/{\mathbb Z}_3]$,
and hence from decomposition,
\begin{equation}
    [X/S_3] \: = \: \coprod_2 Y.
\end{equation}

Let us generalize from three objects to $n$ objects.
Define
\begin{equation}
    X \: = \: \coprod_n Y.
\end{equation}
The group ${\mathbb Z}_n$ acts transitively on $X$, hence $[X/{\mathbb Z}_n] = Y$.  For $n > 3$, ${\mathbb Z}_n$ is not a normal subgroup of $S_n$.  Technically, $[X/S_n]$ is a non-banded $S_{n-1}$ gerbe on $Y$.  The space $Y$ is an atlas for that gerbe, but the gerbe is not a quotient $[Y/S_{n-1}]$.


\begin{thebibliography}{299}

\addcontentsline{toc}{section}{References}

\bibitem{Cordova:2022ruw}
C.~Cordova, T.~T.~Dumitrescu, K.~Intriligator and S.~H.~Shao,
``Snowmass white paper: Generalized symmetries in quantum field theory and Beyond,''
{\tt arXiv:2205.09545 [hep-th]}.

\bibitem{Schafer-Nameki:2023jdn}
S.~Schafer-Nameki,
``ICTP Lectures on (non-)invertible generalized symmetries,''
{\tt arXiv:2305.18296 [hep-th]}.

\bibitem{Shao:2023gho}
S.~H.~Shao,
``What's done cannot be undone: TASI lectures on non-invertible symmetry,''
{\tt arXiv:2308.00747 [hep-th]}.

\bibitem{Carqueville:2012dk}
N.~Carqueville and I.~Runkel,
``Orbifold completion of defect bicategories,''
Quantum Topol. \textbf{7} (2016) 203-279,
{\tt arXiv:1210.6363 [math.QA]}.

\bibitem{Carqueville:2017aoe}
N.~Carqueville, I.~Runkel and G.~Schaumann,
``Orbifolds of n-dimensional defect TQFTs,''
Geom. Topol. \textbf{23} (2019) 781-864,
{\tt arXiv:1705.06085 [math.QA]}.

\bibitem{Carqueville:2018sld}
N.~Carqueville, I.~Runkel and G.~Schaumann,
``Orbifolds of Reshetikhin-Turaev TQFTs,''
Theor. Appl. Categor. \textbf{35} (2020) 513-561,
{\tt arXiv:1809.01483 [math.QA]}.


\bibitem{Carqueville:2023aak}
N.~Carqueville and L.~M\"uller,
``Orbifold completion of 3-categories,''
{\tt arXiv:2307.06485 [math.QA]}.

\bibitem{Carqueville:2023qrk}
N.~Carqueville,
``Orbifolds of topological quantum field theories,''
{\tt arXiv:2307.16674 [math-ph]}.



\bibitem{Fuchs:2004dz}
J.~Fuchs, I.~Runkel and C.~Schweigert,
``TFT construction of RCFT correlators. 3. Simple currents,''
Nucl. Phys. B \textbf{694} (2004) 277-353,
{\tt arXiv:hep-th/0403157 [hep-th]}.

\bibitem{Frohlich:2004ef}
J.~Frohlich, J.~Fuchs, I.~Runkel and C.~Schweigert,
``Kramers-Wannier duality from conformal defects,''
Phys. Rev. Lett. \textbf{93} (2004) 070601,
{\tt arXiv:cond-mat/0404051 [cond-mat]}.

\bibitem{Frohlich:2009gb}
J.~Frohlich, J.~Fuchs, I.~Runkel and C.~Schweigert,
``Defect lines, dualities, and generalised orbifolds,''
{\tt arXiv:0909.5013 [math-ph]}.

\bibitem{Fuchs:2007tx}
J.~Fuchs, M.~R.~Gaberdiel, I.~Runkel and C.~Schweigert,
``Topological defects for the free boson CFT,''
J. Phys. A \textbf{40} (2007) 11403,
{\tt arXiv:0705.3129 [hep-th]}.



















\bibitem{Fuchs:2002cm}
J.~Fuchs, I.~Runkel and C.~Schweigert,
``TFT construction of RCFT correlators 1. Partition functions,''
Nucl. Phys. B \textbf{646} (2002) 353-497,
{\tt arXiv:hep-th/0204148 [hep-th]}.

\bibitem{Bhardwaj:2017xup}
L.~Bhardwaj and Y.~Tachikawa,
``On finite symmetries and their gauging in two dimensions,''
JHEP \textbf{03} (2018) 189,
{\tt arXiv:1704.02330 [hep-th]}.




\bibitem{Kaidi:2023maf}
J.~Kaidi, E.~Nardoni, G.~Zafrir and Y.~Zheng,
``Symmetry TFTs and anomalies of non-invertible symmetries,''
JHEP \textbf{10}, 053 (2023)
doi:10.1007/JHEP10(2023)053
{\tt arXiv:2301.07112 [hep-th]}.


\bibitem{Zhang:2023wlu}
C.~Zhang and C.~C\'ordova,
``Anomalies of $(1+1)D$ categorical symmetries,''
{\tt arXiv:2304.01262 [cond-mat.str-el]}.


\bibitem{Cordova:2023bja}
C.~Cordova, P.~S.~Hsin and C.~Zhang,
``Anomalies of non-invertible symmetries in (3+1)d,''
{\tt arXiv:2308.11706 [hep-th]}.

\bibitem{Antinucci:2023ezl}
A.~Antinucci, F.~Benini, C.~Copetti, G.~Galati and G.~Rizi,
``Anomalies of non-invertible self-duality symmetries: fractionalization and gauging,''
{\tt arXiv:2308.11707 [hep-th]}.

\bibitem{Mulevicius:2023jhg}
V.~Mulevicius, I.~Runkel and T.~Vo\ss{},
``Internal Levin-Wen models,''
{\tt arXiv:2309.05755 [cond-mat.str-el]}.




\bibitem{toappear} A. Perez-Lona, D. Robbins, E. Sharpe, T. Vandermeulen, X. Yu, to appear.

\bibitem{CLS23}
Y. Choi, D.C. Lu, and Z. Sun, ``Self-duality under gauging a non-invertible symmetry'', {\tt arXiv:2310.19867 [hep-th]}.


\bibitem{Wang:2023}
O.~Diatlyk, C.~Luo, Y.~Wang, and Q.~Weller, 
``Gauging non-invertible symmetries: Topological interfaces and generalized orbifold groupoid in 2d QFT,''
{\tt arXiv:2311.17044 [hep-th]}.


\bibitem{eno} P. Etingof, D. Nikshych, V. Ostrik, ``On fusion categories,''
Ann. Math. {\bf 162} (2005) 581-642.

\bibitem{Kel82}
M. Kelly, \textit{Basic concepts of enriched category theory}, Lecture Notes in Mathematics \textbf{64},
Cambridge University Press, 1982.

\bibitem{JF22}
T. Johnson-Freyd, ``On the classification of topological orders,'' Commun. Math. Phys. \textbf{393} (2022) 989–1033.

\bibitem{Usher19}
R. Usher, ``On some notions of cohomology for fusion categories,'' 2019. (Order No. 13898585). Available From ProQuest Dissertations \& Theses Global; SciTech Premium Collection, (2293989931). \url{https://www.proquest.com/docview/2293989931}.


\bibitem{Chang:2018iay}
C.~M.~Chang, Y.~H.~Lin, S.~H.~Shao, Y.~Wang and X.~Yin,
``Topological defect lines and renormalization group flows in two dimensions,''
JHEP \textbf{01} (2019) 026,
{\tt arXiv:1802.04445 [hep-th]}.


\bibitem{Robbins:2019zdb}
D.~Robbins and T.~Vandermeulen,
``Orbifolds from modular orbits,''
Phys. Rev. D \textbf{101} (2020)  106021,
{\tt arXiv:1911.05172 [hep-th]}.

\bibitem{ty} D. Tambara, S. Yamagami,
``Tensor categories with fusion rules of self-duality for finite
abelian groups,'' J. Algebra {\bf 209} (1998) 692-707.

\bibitem{Moore:1988qv}
G.~W.~Moore and N.~Seiberg,
``Classical and quantum conformal field theory,''
Commun. Math. Phys. \textbf{123} (1989)
177-254.

\bibitem{Thorngren:2019iar}
R.~Thorngren and Y.~Wang,
``Fusion category symmetry I: Anomaly in-flow and gapped phases,''
{\tt arXiv:1912.02817 [hep-th]}.


\bibitem{Huang:2021zvu}
T.~C.~Huang, Y.~H.~Lin and S.~Seifnashri,
``Construction of two-dimensional topological field theories with non-invertible symmetries,''
JHEP \textbf{12} (2021) 028,
{\tt arXiv:2110.02958 [hep-th]}.


\bibitem{ostrik1} V. Ostrik, ``Module categories, weak Hopf algebras and modular invariants,''
Transform. Groups {\bf 8} (2003) 177-206,
{\tt arXiv:math/0111139 [math.QA]}.



\bibitem{FMT22}
D. S. Freed, G. W. Moore, and C. Teleman, ``Topological symmetry in quantum field theory''. {\tt arXiv:2209.07471 [hep-th]}.


\bibitem{tannaka}
P.~Etingof, S.~Gelaki, ``Isocategorical groups,''
Internat. Math. Res. Notices (2001) 59-76,
{\tt arXiv:math/0007196 [math.QA]}.


\bibitem{ENO02}
P. Etingof, D. Nikshych, V. Ostrik, ``On fusion categories,''  
Ann. of Math. (2) {\bf 162} (2005) 581-642,
{\tt arxiv:math/0203060 [math.QA]}.



\bibitem{IK08}
M. C. Iovanov, L. Kadison, ``When weak Hopf algebras are Frobenius,'' 
Proc. Amer. Math. Soc. {\bf 138} (2010) 837-845,
{\tt arxiv:0810.4777 [math.QA]}.


\bibitem{Nik04}
D. Nikshych, ``Semisimple weak Hopf algebras,'' J. Algebra \textbf{275} (2004) 639-667.


\bibitem{FSS11}
J. Fuchs, C. Schweigert, C. Stigner, ``Modular invariant Frobenius algebras from ribbon Hopf algebra automorphisms,'' J. Algebra \textbf{363} (2012) 29-72,
{\tt arXiv:1106.0210 [math.QA]}.



\bibitem{LS69}
R. Larson, M. Sweedler, ``An associative orthogonal bilinear form for Hopf algebras,''
Amer. J. Math. \textbf{91} (1969) 75-94.


\bibitem{Sweedler69}
M. E. Sweedler, \textit{Hopf algebras}, Mathematics Lecture Note Series, Benjamin, 1969.


\bibitem{Vafa:1986wx}
C.~Vafa,
``Modular invariance and discrete torsion on orbifolds,''
Nucl. Phys. B \textbf{273} (1986) 592-606.


\bibitem{Sharpe:2000ki}
E.~R.~Sharpe,
``Discrete torsion,''
Phys. Rev. D \textbf{68} (2003) 126003,
{\tt arXiv:hep-th/0008154 [hep-th]}.




\bibitem{Aspinwall:2000xv}
P.~S.~Aspinwall,
``A note on the equivalence of Vafa's and Douglas's picture of discrete torsion,''
JHEP \textbf{12} (2000) 029,
{\tt arXiv:hep-th/0009045 [hep-th]}.


\bibitem{Alex30}
J.W. Alexander, ``The combinatorial theory of complexes,'' Ann. Math. (2)
\textbf{31} (1930) 292-320. 

\bibitem{Fukuma:1993hy}
M.~Fukuma, S.~Hosono and H.~Kawai,
``Lattice topological field theory in two-dimensions,''
Commun. Math. Phys. \textbf{161} (1994) 157-176,
{\tt arXiv:hep-th/9212154 [hep-th]}.


  \bibitem{Bhardwaj:2023idu}
L.~Bhardwaj, L.~E.~Bottini, D.~Pajer and S.~Schafer-Nameki,
``Gapped phases with non-invertible symmetries: (1+1)d,''
{\tt arXiv:2310.03784 [hep-th]}.


\bibitem{Brunner:2013ota}
I.~Brunner, N.~Carqueville and D.~Plencner,
``Orbifolds and topological defects,''
Commun. Math. Phys. \textbf{332} (2014) 669-712,
{\tt arXiv:1307.3141 [hep-th]}.

\bibitem{se}  {\tt https://math.stackexchange.com/questions/3258286/grading-of-module-in-hmod}


\bibitem{CRV14}
T. Crespo, A. Rio, M. Vela, ``On the Galois correspondence theorem in separable Hopf Galois theory,'' 
Publ. Mat. {\bf 60} (2016) 221-234,
{\tt arXiv:1405.0881 [math.GR]}.


\bibitem{mo-subalg} {\tt https://mathoverflow.net/questions/328396/subalgebra-of-a-group-algebra}



\bibitem{Bantay:2000eq}
P.~Bantay,
``Symmetric products, permutation orbifolds and discrete torsion,''
Lett. Math. Phys. \textbf{63} (2003) 209-218,
{\tt arXiv:hep-th/0004025 [hep-th]}.

\bibitem{Barter_2022}
D.~Barter, J.~Bridgeman, and R.~Wolf, ``Computing associators of endomorphism
  fusion categories,'' {\em {SciPost} Phys.} {\bf 13} (2022) 029, 
  {\tt arXiv:2110.03644 [math.QA]}.





\bibitem{Hellerman:2006zs}
S.~Hellerman, A.~Henriques, T.~Pantev, E.~Sharpe and M.~Ando,
``Cluster decomposition, T-duality, and gerby CFT's,''
Adv. Theor. Math. Phys. \textbf{11} (2007) 751-818,
{\tt arXiv:hep-th/0606034 [hep-th]}.

\bibitem{Sharpe:2022ene}
E.~Sharpe,
``An introduction to decomposition,''
{\tt arXiv:2204.09117 [hep-th]}.


\bibitem{Yu:2023nyn}
X.~Yu,
``Non-invertible symmetries in 2d from type IIB string theory,''
{\tt  arXiv:2310.15339 [hep-th]}.


\bibitem{kp66}
G. I. Kac and V. G. Paljutkin, ``Finite ring groups,'' Trans. Moscow Math. Soc. \textbf{15} (1966)
251–294.

\bibitem{Alaoui03}
A.~El Alaoui, ``The character table for a Hopf algebra arising from the Drinfel’d double,'' J.~Algebra \textbf{265} (2003) 478–495.



\bibitem{Burciu17}
S.~Burciu, ``Representations and conjugacy classes of semisimple quasitriangular Hopf algebras,'' SIGMA {\bf 16} (2020) 039, {\tt arxiv:1709:02176 [math.QA]}.

\bibitem{buerschaper-thesis} G.~O.~Buerschaper, {\it The structure of nonchiral topological order},
Ph.D.~thesis, Technische Universit\"at M\"unchen, 2011, available from
{\tt https://mediatum.ub.tum.de/doc/1072458/1072458.pdf}


\bibitem{GKKL11}
R.~M.~Guralnick, W.~M.~Kantor, M.~Kassabov, A.~Lubotzky, ``Presentations of finite simple groups: a computational approach,'' J. Eur. Math. Soc. (JEMS) \textbf{13} 2 (2011) 391-458. 


\bibitem{Dixon:1985jw}
L.~J.~Dixon, J.~A.~Harvey, C.~Vafa and E.~Witten,
`Strings on orbifolds,''
Nucl. Phys. B \textbf{261} (1985) 678-686.


\bibitem{Hamidi:1986vh}
S.~Hamidi and C.~Vafa,
``Interactions on orbifolds,''
Nucl. Phys. B \textbf{279} (1987) 465-513.

\bibitem{Dixon:1986qv}
L.~J.~Dixon, D.~Friedan, E.~J.~Martinec and S.~H.~Shenker,
``The conformal field theory of orbifolds,''
Nucl. Phys. B \textbf{282} (1987) 13-73.


\bibitem{Becker:2017zai}
M.~Becker, Y.~Cabrera and D.~Robbins,
``Conformal interfaces between free boson orbifold theories,''
JHEP \textbf{09} (2017) 148,
{\tt arXiv:1706.03802 [hep-th]}.


\bibitem{lpr} Z.~Liu, S.~Palcoux, Y.~Ren, ``Classification of Grothendieck rings of complex fusion categories of multiplicity one up to rank six,''
Lett. Math. Phys. {\bf 112} (2022) 54,
{\tt arXiv:2010.10264 [math.CT]}.


\bibitem{Thorngren:2021yso}
R.~Thorngren and Y.~Wang,
``Fusion category symmetry II: Categoriosities at $c$ = 1 and beyond,''
{\tt arXiv:2106.12577 [hep-th]}.


\bibitem{Brunner:2014lua}
I.~Brunner, N.~Carqueville and D.~Plencner,
``Discrete torsion defects,''
Commun. Math. Phys. \textbf{337}, no.1, 429-453 (2015)
doi:10.1007/s00220-015-2297-9
{\tt arXiv:1404.7497 [hep-th]}.


\bibitem{Nagoya:2023zky}
Y.~Nagoya and S.~Shimamori,
``Non-invertible duality defect and non-commutative fusion algebra,''
{\tt arXiv:2309.05294 [hep-th]}.


\bibitem{Chang:202x}
C.~M.~Chang, unpublished note.

\bibitem{Gepner:1986hr}
D.~Gepner and Z.~A.~Qiu,
``Modular invariant partition functions for parafermionic field theories,''
Nucl. Phys. B \textbf{285} (1987) 423-453.


\bibitem{Choi:2021kmx}
Y.~Choi, C.~Cordova, P.~S.~Hsin, H.~T.~Lam and S.~H.~Shao,
``Noninvertible duality defects in 3+1 dimensions,''
Phys. Rev. D \textbf{105} (2022) 125016,
{\tt arXiv:2111.01139 [hep-th]}.


\bibitem{Kadanoff:1978ve}
L.~P.~Kadanoff,
``Lattice Coulomb gas representations of two-dimensional problems,''
J. Phys. A \textbf{11} (1978) 1399-1417.

\bibitem{Nienhuis:1984wm}
B.~Nienhuis,
``Critical behavior of two-dimensional spin models and charge asymmetry in the Coulomb gas,''
J. Statist. Phys. \textbf{34} (1984) 731-761.



\bibitem{DiFrancesco:1987gwq}
P.~Di Francesco, H.~Saleur and J.~B.~Zuber,
``Modular invariance in nonminimal two-dimensional conformal theories,''
Nucl. Phys. B \textbf{285} (1987) 454-480.


\bibitem{Choi:2023xjw}
Y.~Choi, B.~C.~Rayhaun, Y.~Sanghavi and S.~H.~Shao,
``Comments on boundaries, anomalies, and non-invertible symmetries,''
{\tt arXiv:2305.09713 [hep-th]}.

\bibitem{Lin:2019hks}
Y.~H.~Lin and S.~H.~Shao,
``Duality defect of the Monster CFT,''
J. Phys. A \textbf{54} (2021)  065201,
{\tt arXiv:1911.00042 [hep-th]}.


\bibitem{Lin:2019kpn}
Y.~H.~Lin and S.~H.~Shao,
``Anomalies and bounds on charged operators,''
Phys. Rev. D \textbf{100} (2019)  025013,
{\tt arXiv:1904.04833 [hep-th]}.



\bibitem{Dijkgraaf:1987vp}
R.~Dijkgraaf, E.~P.~Verlinde and H.~L.~Verlinde,
``c=1 conformal field theories on Riemann surfaces,''
Commun. Math. Phys. \textbf{115} (1988) 649-690.

\bibitem{Dijkgraaf:1989hb}
R.~Dijkgraaf, C.~Vafa, E.~P.~Verlinde and H.~L.~Verlinde,
``The operator algebra of orbifold models,''
Commun. Math. Phys. \textbf{123} (1989) 485-526.


\bibitem{Hsin:2020nts}
P.~S.~Hsin and H.~T.~Lam,
``Discrete theta angles, symmetries and anomalies,''
SciPost Phys. \textbf{10} (2021)  no.2, 032,
{\tt arXiv:2007.05915 [hep-th]}.




\bibitem{kadanoff1979multicritical}
L.~P.~Kadanoff, 
``Multicritical behavior at the Kosterlitz-Thouless critical point,''
Annals Phys. \textbf{120} (1979) 39-71.



\bibitem{Lin:2022dhv}
Y.~H.~Lin, M.~Okada, S.~Seifnashri and Y.~Tachikawa,
``Asymptotic density of states in 2d CFTs with non-invertible symmetries,''
JHEP \textbf{03} (2023) 094,
{\tt arXiv:2208.05495 [hep-th]}.


\bibitem{Dijkgraaf:1989pz}
R.~Dijkgraaf and E.~Witten,
``Topological gauge theories and group cohomology,''
Commun. Math. Phys. \textbf{129} (1990) 393-429.

\bibitem{Robbins:2022wlr}
D.~G.~Robbins, E.~Sharpe and T.~Vandermeulen,
``Decomposition, trivially-acting symmetries, and topological operators,''
Phys. Rev. D \textbf{107} (2023)  085017,
{\tt arXiv:2211.14332 [hep-th]}.


\bibitem{Robbins:2020msp}
D.~Robbins, E.~Sharpe and T.~Vandermeulen,
``A generalization of decomposition in orbifolds,''
JHEP \textbf{21} (2020) 134,
{\tt arXiv:2101.11619 [hep-th]}.


\bibitem{tv1} V.~Turaev, O.~Viro, ``State sum invariants of 3-manifolds
and quantum $6j$-symbols,'' Topology {\bf 31} (1992) 865-902.


\bibitem{Wang:2016rzy}
Z.~Wang and X.~Chen,
``Twisted gauge theories in three-dimensional Walker-Wang models,''
Phys. Rev. B \textbf{95} (2017)  115142,
{\tt arXiv:1611.09334 [cond-mat.str-el]}.



\end{thebibliography}
\end{document}